\documentclass{emulateapj}

\newcommand{\noprint}[1]{}

\submitted{}

\shorttitle{Spitzer IRS Spectral Mapping of Major Mergers}
\shortauthors{Haan et al.}

\begin{document}

\title{Spitzer IRS Spectral Mapping of the Toomre Sequence: Spatial Variations of PAH, Gas, and Dust Properties in Nearby Major Mergers}

\author{S. Haan\altaffilmark{1}, L. Armus\altaffilmark{1}, S. Laine\altaffilmark{1},  V. Charmandaris\altaffilmark{2,3},  J.D. Smith\altaffilmark{4}, F. Schweizer\altaffilmark{5}, B. Brandl\altaffilmark{6},  A.S. Evans\altaffilmark{7,8},  J.A. Surace\altaffilmark{1}, T. Diaz-Santos\altaffilmark{2,1},  P. Beir\~ao\altaffilmark{1}, E.J. Murphy\altaffilmark{5}, S. Stierwalt\altaffilmark{1}, J.E. Hibbard\altaffilmark{7,8}, M. Yun\altaffilmark{9}, T.H. Jarrett\altaffilmark{10}}

\altaffiltext{1}{Spitzer Science Center, California Institute of Technology, Pasadena, CA 91125, USA}
\altaffiltext{2}{Department of Physics and Institute of Theoretical and Computational Physics, University of Crete, GR-71003, Heraklion, Greece}
\altaffiltext{3}{IESL/Foundation for Research and Technology - Hellas, GR-71110, Heraklion, Greece and Chercheur Associ\'{e}, Observatoire de Paris, F-75014, Paris, France}
\altaffiltext{4}{Ritter Astrophysical Observatory, University of Toledo, Toledo, OH 43606, USA}
\altaffiltext{5}{Observatories of the Carnegie Institution, 813 Santa Barbara Street, Pasadena, CA 91101, USA}
\altaffiltext{6}{Leiden Observatory, Leiden University, P.O. Box 9513, 2300 RA Leiden, The Netherlands}
\altaffiltext{7}{National Radio Astronomy Observatory, Charlottesville, VA 22903, USA}
\altaffiltext{8}{Department of Astronomy, University of Virginia, Charlottesville, VA 22904, USA}
\altaffiltext{9}{Astronomy Department, University of Massachusetts, 710 North Pleasant Street, Amherst, MA 01003, USA }
\altaffiltext{10}{Infrared Processing and Analysis Center, California Institute of Technology, Pasadena, CA 91125, USA}

\begin{abstract}
We have mapped the key mid-IR diagnostics in eight major merger systems of the Toomre Sequence (NGC~4676, NGC~7592, NGC~6621, NGC~2623, NGC~6240, NGC~520, NGC~3921, and NGC~7252) using the Spitzer Infrared Spectrograph (IRS). With these maps, we explore the variation of the ionized-gas, PAH, and warm-gas (H$_2$) properties across the sequence and within the galaxies. While the global PAH interband strength and ionized gas flux ratios ([Ne III]/[Ne II]) are similar to those of normal star forming galaxies, the distribution of the spatially resolved PAH and fine structure line flux ratios is significant different from one system to the other. Rather than a constant H$_2$/PAH flux ratio, we find that the relation between the H$_2$ and PAH fluxes is characterized by a power law with a roughly constant exponent ($0.61\pm0.05$) over all merger components and spatial scales. While following the same power law on local scales, three galaxies have a factor of ten larger integrated (i.e. global) H$_2$/PAH flux ratio than the rest of the sample, even larger than what it is in most nearby AGNs. These findings suggest a common dominant excitation mechanism for H$_2$ emission over a large range of global H$_2$/PAH flux ratios in major mergers. Early merger systems show a different distribution between the cold (CO J=1--0) and warm (H$_2$) molecular gas component, which is likely due to the merger interaction. Strong evidence for buried star formation in the overlap region of the merging galaxies is found in two merger systems (NGC~6621 and NGC~7592) as seen in the PAH, [Ne II], [Ne III], and warm gas line emission, but with no apparent corresponding CO (J=1--0) emission. The minimum of the 11.3/7.7~$\mu$m PAH interband strength ratio is typically located in the nuclei of galaxies, while the [Ne III/[Ne II] ratio increases with distance from the nucleus. Our findings also demonstrate that the variations of the physical conditions within a merger are much larger than any systematic trends along the Toomre Sequence.
\end{abstract}


\section{Introduction}

Interactions and mergers are important drivers of galaxy evolution and are responsible for the formation of the most luminous infrared galaxies observed locally and many at high redshift. In particular, major mergers can completely disrupt the stellar and gaseous morphologies of the merging galaxies, transform spirals into massive ellipticals, and fuel both powerful starbursts and massive nuclear black holes \citep[e.g.,][]{Bar92}. In the most extreme cases, mergers can become Ultra-Luminous Infrared Galaxies (ULIRGs) which are in many cases powered by massive starbursts with star formation rates sufficient to form the entire stellar population of an L$^*$ galaxy in only a few dynamical times \citep{Arm87,San88, Hec90, Mur96}. While most mergers do not produce ULIRGs, they invariably trigger large-scale starbursts and lead to a large-scale redistribution of the progenitors' gaseous and stellar disks. At high redshift, mergers are responsible for the rapid evolution of the IR luminosity function seen in deep imaging surveys, and the formation of some of the most luminous galaxies detected at both UV and far-infrared wavelengths \citep[e.g.,][]{Ste96, Bla02,Cha05}.  At low redshifts, interactions and mergers provide us a detailed picture of the most rapid and intense star formation known, and the generation and fueling of active nuclei.\par 

The last decades have brought immense progress in the observations and modeling of both the stellar and the gaseous components of interacting galaxies. From the pioneering studies of \cite{Too72}, and \cite{Sch82} through the work of \cite{Bar92} to the more recent work of \cite{Fer06}, \cite{Cot07}, \cite{Kor09}, and \cite{Hop09}, it is now clear that interactions and mergers produce large-scale (tens of kpc) stellar tidal tails, distorted and /or overlapping disks, central starbursts, and in some cases, stellar remnants that have the properties of elliptical galaxies. The gas, being dissipative, can rapidly form dense molecular disks, fueling both starbursts and an active nucleus \citep[e.g.,][]{Sco91,Mih94, Cha02, Arm04}. Hubble Space Telescope observations of nuclei in merging galaxies have shown an increasing nuclear luminosity density in the optical/near-infrared light with advancing merger stage \citep{Lai03, Ros07, Vei09, Haa11}, suggesting an increase of nuclear starburst activity. However, the fate of the dust in galactic interactions and mergers, and its direct relation to the gas and the young stars is much less well understood. For instance, a change in dust grain size, ionization stage, neutral to ionized gas fraction, or gas temperatures as function of merger stage could be expected. The dust responsible for re-radiating much of the UV light into the far-infrared can optically obscure the most energetic regions of star formation in interacting galaxies, which makes it challenging to explore the physical connection between the stars, gas and dust in the nuclei, disks, and overlap regions where the brightest, most intense starbursts are triggered. To address the physical conditions in major mergers and excitation mechanisms such as shocks and starbursts, it is necessary to resolve the spatial variations of the warm dust, molecular gas, and ionized gas on subgalactic scales.\par

Here we present Spitzer IRS spectral mapping observations of a sample of eight nearby, IR-bright galaxies that span the range from early through mid- to late-stage major mergers. Our maps of the mid-IR features reveal the spatial variations of gas temperatures, masses, dust grain sizes, and the ionization state of the interstellar medium. In \S~\ref{sec:obs} we describe our sample, the observational setup, and the data reduction. The extracted spectra, mid-IR feature maps, dust and warm gas properties (e.g., H$_2$ temperature maps and masses) are presented in \S~\ref{sec:results}, compared to normal star-forming galaxies, and discussed in the context of the physical properties of the ISM. Finally, we provide a brief overview of our conclusions in \S~\ref{sec:sum}.

\section{Observations and Data Reduction}
\label{sec:obs}

\subsection{Sample}
By employing the Spitzer IRS in Spectral Mapping mode \citep{Hou04}, we measure the gas and dust properties within a set of eight interacting/merging galaxies representing the early (NGC~4676, NGC~7592, NGC~6621 -- distinguishable parent disks and large tidal tails), middle (NGC~2623, NGC~6240, NGC~520 -- highly overlapping disk and/or double nuclei) and late (NGC~3921, NGC~7252 -- single nucleus, elliptical-like light distribution) stages of the major merger process. In total, we obtained 44.9~hrs of Spitzer observing time in cycle 2 (P.I. Lee Armus) to map these eight galaxies with the SL and LL modules of the IRS. All eight galaxies have Spitzer Infrared Array Camera (IRAC) and Multiband Imaging Photometer (MIPS) imaging, and are, except for NGC 6240, part of the classical ''Toomre sequence'' of merging galaxies \citep{Too72,Too77}. Note that the classical ''Toomre Sequence'' lists galaxies only in rough order of completeness of the merger, and  is not meant to represent a precise order in time sequence of the merging process.
An overview of the sample is given in Table~\ref{tab_sample}. The galaxies are all nearby (mean distance of 79~Mpc), IR bright (log $L_{IR}$[L$_{\sun}$]=10.3--11.9\footnote{This refers to estimates of the 8--1000~$\mu$m infrared luminosity constructed from IRAS flux densities as defined by \cite{San03}}.), and have a wealth of available ancillary data, from radio through x-rays (HI, CO, HST WFPC2 and ACS, NICMOS, VLA, Chandra). 
To illustrate the stellar distribution of these major mergers, we show in Fig.~\ref{maps_IRAC3} the IRAC 3.6~$\mu$m images of the sample.

\subsection{IRS Observations and Data Reduction} 
\label{subsec:reduction}
\begin{figure*}
\begin{center}
\includegraphics[scale=0.54]{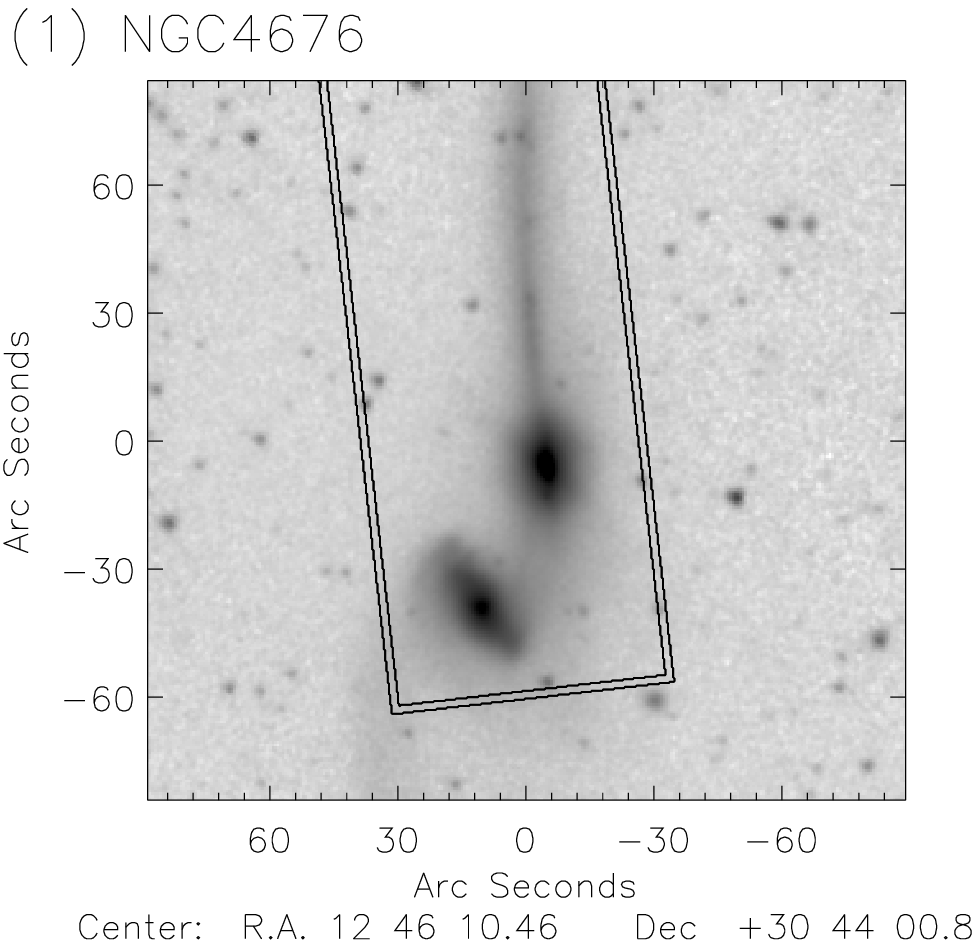}
\includegraphics[scale=0.54]{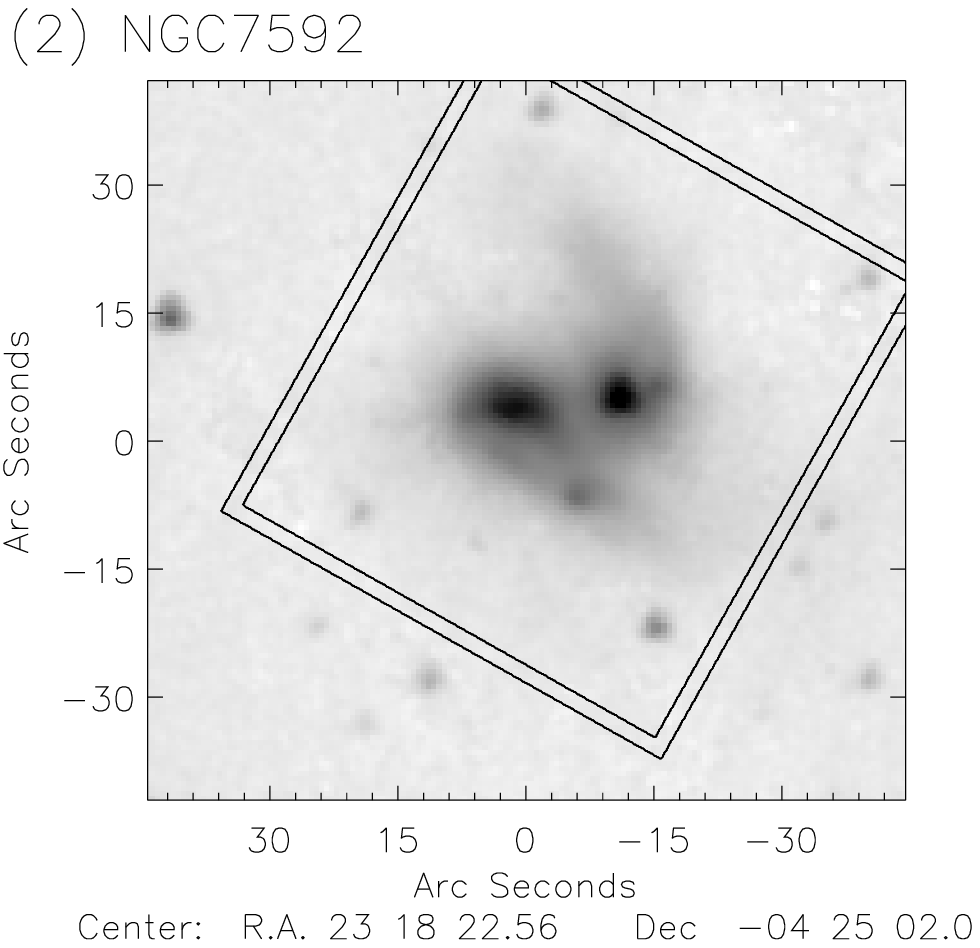}
\includegraphics[scale=0.54]{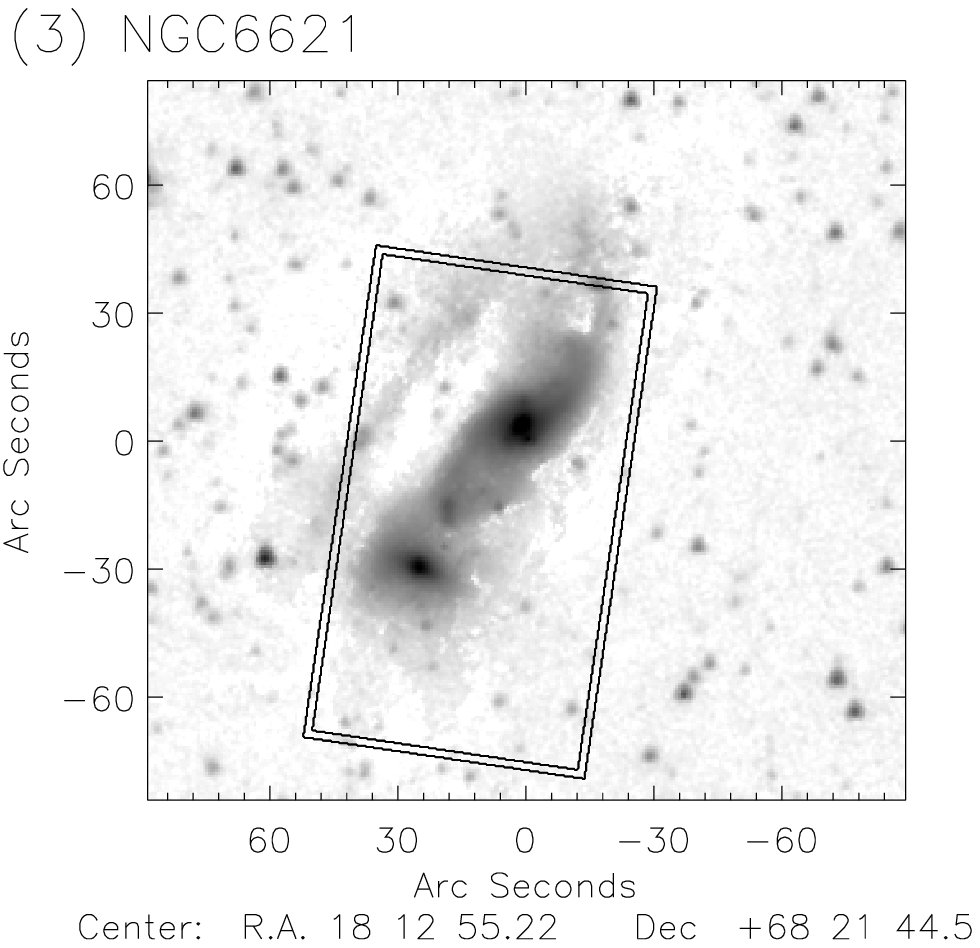}
\includegraphics[scale=0.54]{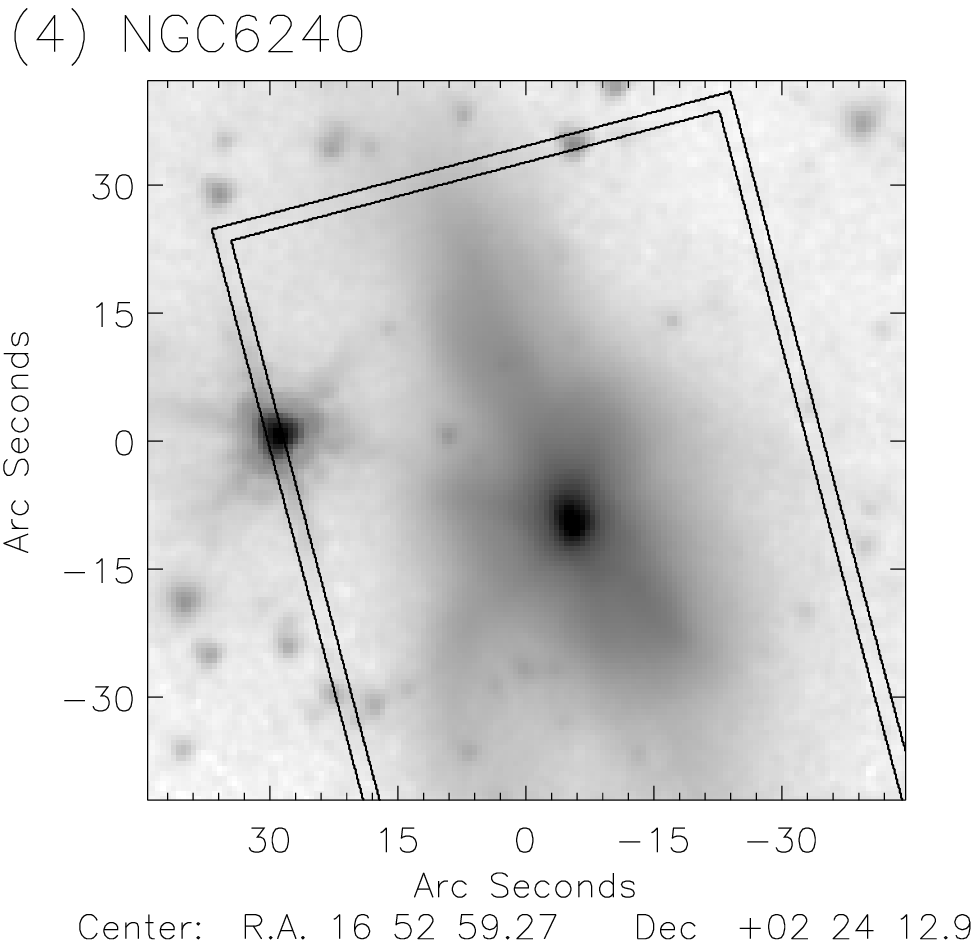}
\includegraphics[scale=0.54]{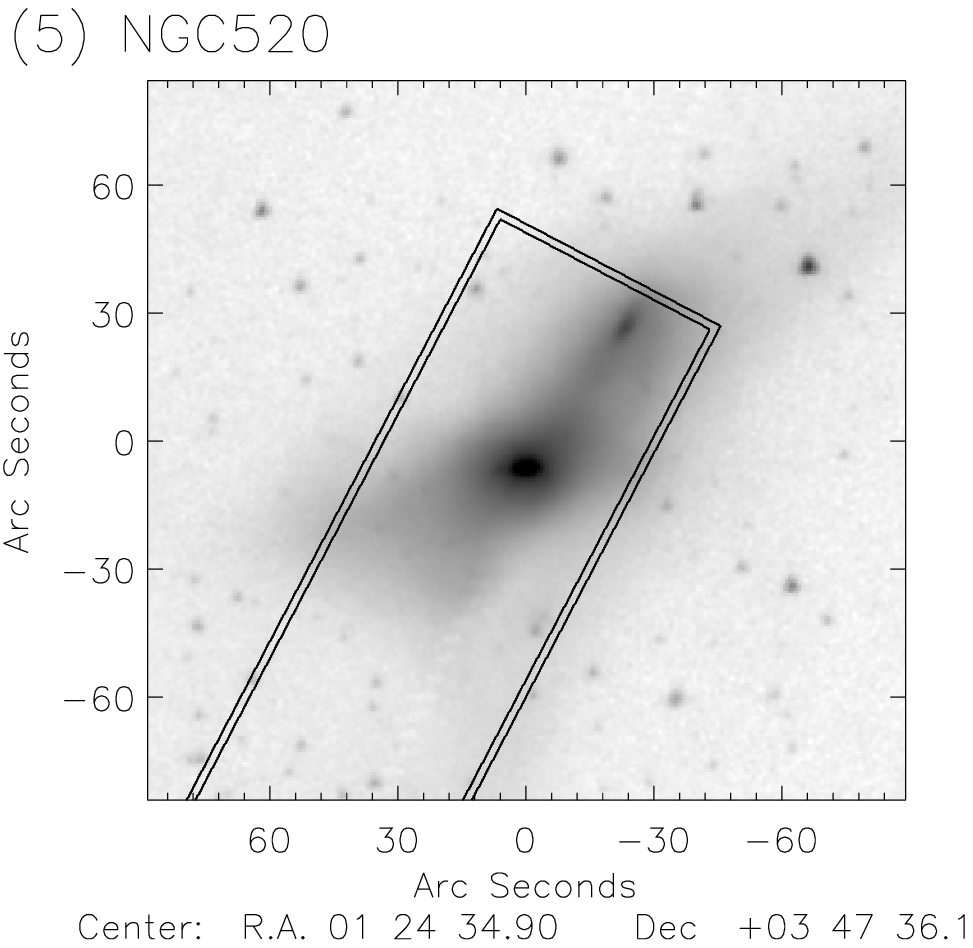}
\includegraphics[scale=0.54]{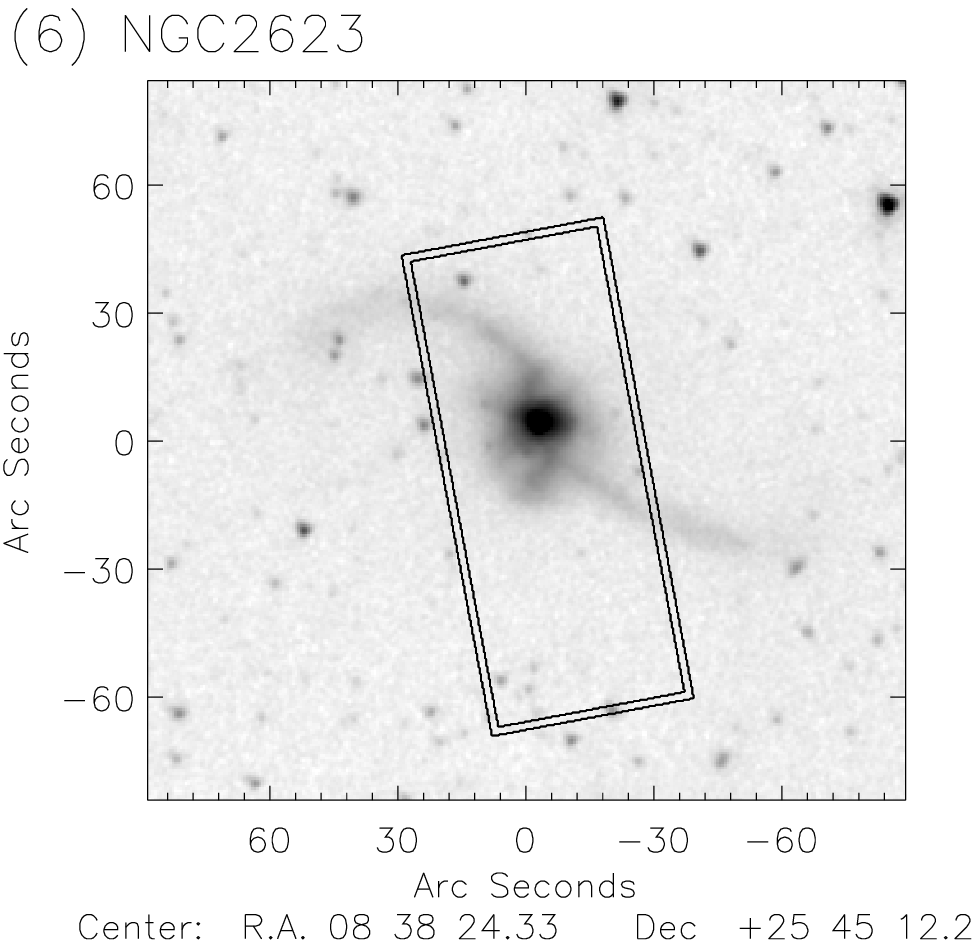}
\includegraphics[scale=0.54]{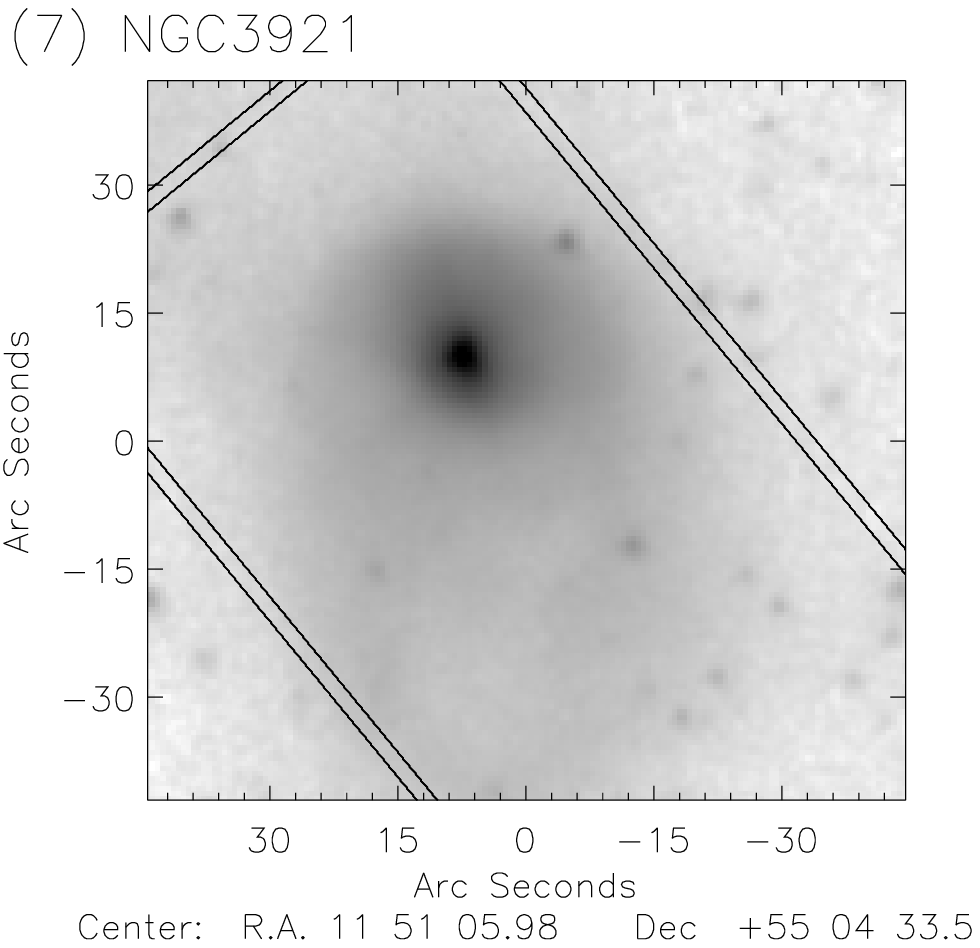}
\includegraphics[scale=0.54]{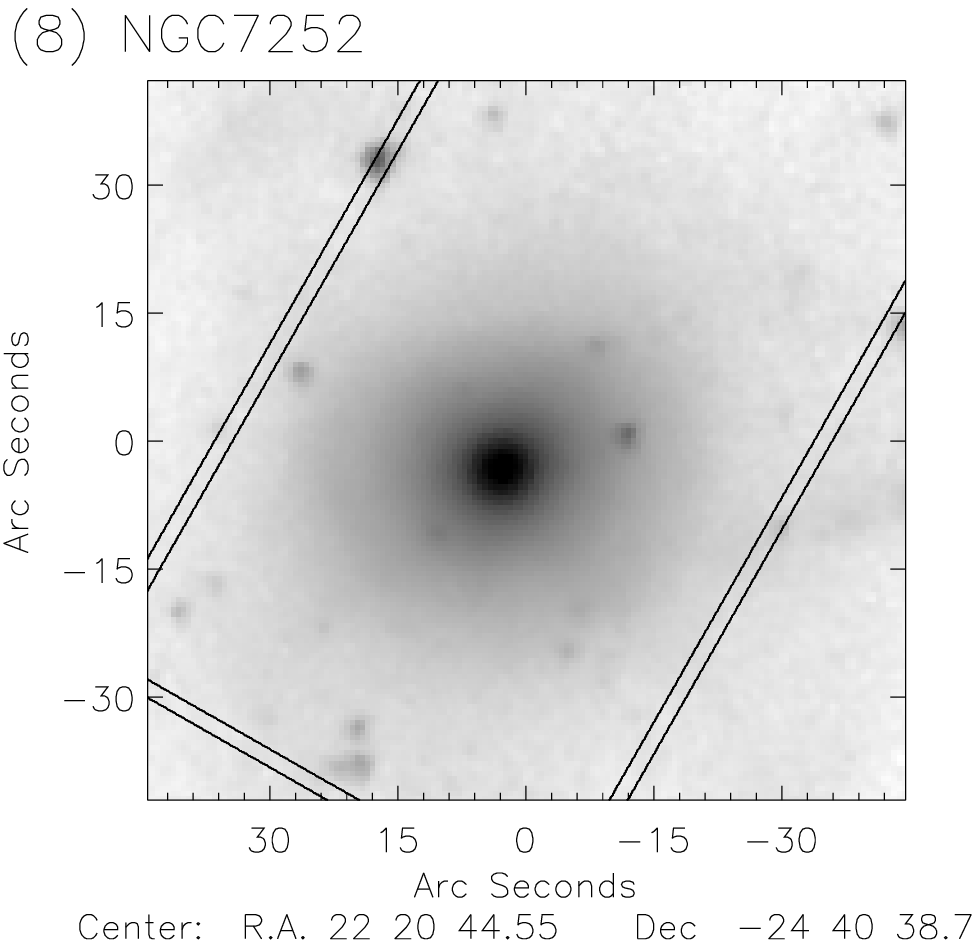}
\end{center}
\caption{\footnotesize{Spitzer IRAC 3.6~$\mu$m images of our sample along the Toomre Sequence (from 1--8, top-left to bottom-right). The Field-of-View of the Spitzer IRS SL images is overlaid in rectangles and the figures are oriented such that north is up and east left.}} 
\label{maps_IRAC3}
\end{figure*}

The Short-Low (SL) and Long-Low (LL) modules of the IRS are used to create 3D spectral data cubes over the central 1--2 square arcminutes in each system (corresponding to physical sizes of $\sim$8--28 $\times$ 22--110~kpc) and over the entire wavelength interval (5--38~$\mu$m). In all cases the maps are large enough to cover the entire main bodies (as defined by the optical and NIR images) and overlap regions. All SL and LL maps have half slit-width steps in the dispersion direction to avoid gaps and produce filled maps. The SL integrations are all 60~sec with two cycles per position. The LL integrations are 120~sec (one cycle), or 30~sec (3--4 cycles) per position to maximize Signal/Noise (S/N) and avoid saturation near the nuclei. The SL maps use sub-slit steps in the spatial direction to cover the intended area.
We have designed the maps to cover the central regions of each galaxy, and at the same time provide multiple sky positions (from the non-primary slit) for background subtraction (see Fig.~\ref{maps_IRAC3}). 
The IRS data reduction was performed using the interactive IDL software package CUBISM \citep{Smi07b} that takes IRS Basic Calibrated Data (BCD) produced via spectral mapping mode, and constructs spectral data cubes suitable for detailed spatial and spectral analysis. CUBISM allows the user to construct 2D line and/or continuum images, and also extract 1D spectra at any spatial position in the map (see Fig.~\ref{maps_IRAC3}).  Background spectra are constructed from our sky observations off the galaxies (when the primary slit is mapping the target, the secondary slit is looking at the sky next to the galaxy). The background is subtracted from the BCDs before the cubes are constructed.  At a first step,  line ([NeII], [NeIII], H$_2$ (S(0)--S(7)), [Ar II], [S iII]) and continuum (dust and PAH feature) maps and 1D spectra are extracted and fit within CUBISM. These maps are later used as a benchmark for our maps created with PAHFIT (see \S~\ref{sec:PAHFITmaps}).

\clearpage

\section{Results and Discussion}
\label{sec:results}
To investigate the properties of the warm gas and dust in our galaxies, we have studied the spatial distribution and global properties of the main mid-IR features, which are: the primary PAH complex features at 6.2~$\mu$m, 7.7~$\mu$m,  11.3~$\mu$m, 12.7~$\mu$m, 16.6~$\mu$m, and 17~$\mu$m, the primary diagnostics for the ionization state of the low-density gas ([Ne II] 12.81~$\mu$m, [Ne III] 15.55~$\mu$m, [Ar II] 6.99~$\mu$m,  [Si II] 34.82~$\mu$m, [O IV] 25.9~$\mu$m,  and [Fe II] 25.99~$\mu$m emission lines\footnote{Note that the [O IV] 25.9~$\mu$m  and [Fe II] 25.99~$\mu$m emission lines are  blended in the LL data}), as well as the pure rotational lines of H$_2$ S(0) to S(7) which probe the warm molecular gas (a detailed study of the H$_2$ temperature and masses is elaborated in \S~\ref{subsubsec:H2_temp}). Two different approaches are conducted. First, we study the global properties of the merger components, which has the advantage of high signal-to-noise ratios and allows us to compare our results with spatially integrated mid-IR properties of normal star-forming galaxies. Second, to investigate in detail the distribution of the warm gas and dust, we map all mid-IR line features using a customized PAHFIT routine and also compare these maps with the cold gas distribution (using radio interferometer CO and HI emission lines) of these major mergers.

\subsection{1D Spectra and Spectral Features}
\label{sec:spectra}
\begin{figure*}
\begin{center}
\includegraphics[scale=0.40]{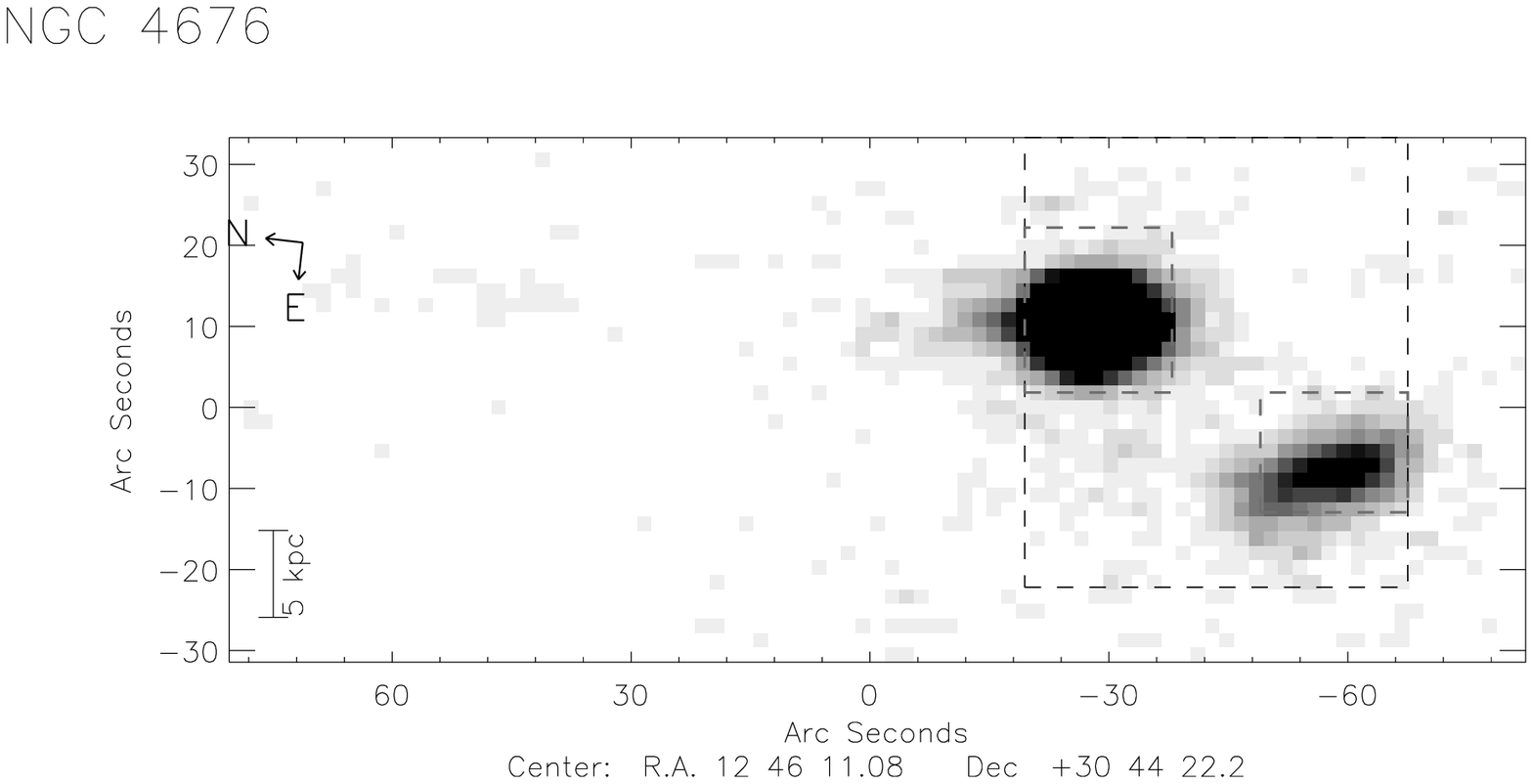}
\includegraphics[scale=0.40]{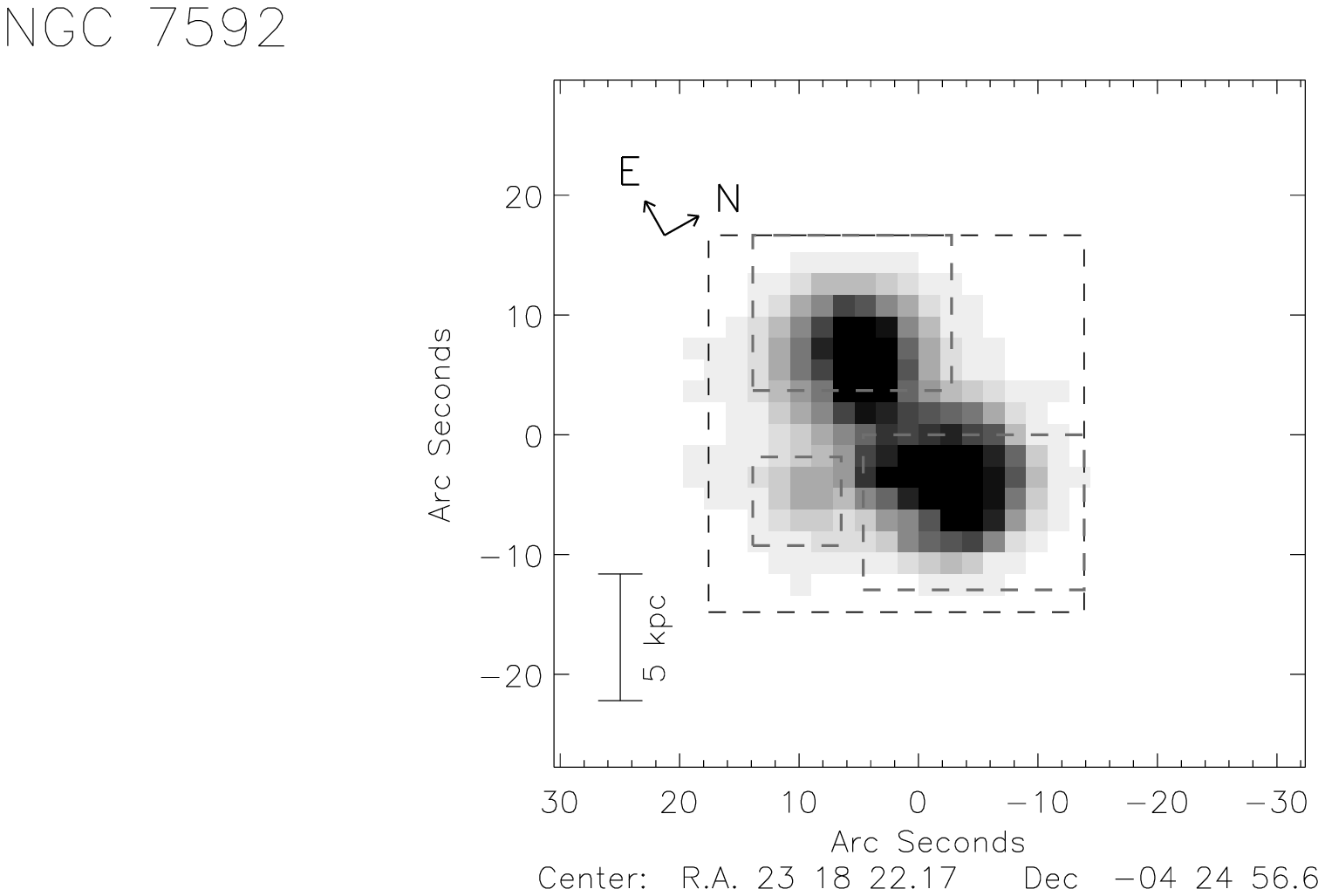}
\includegraphics[scale=0.40]{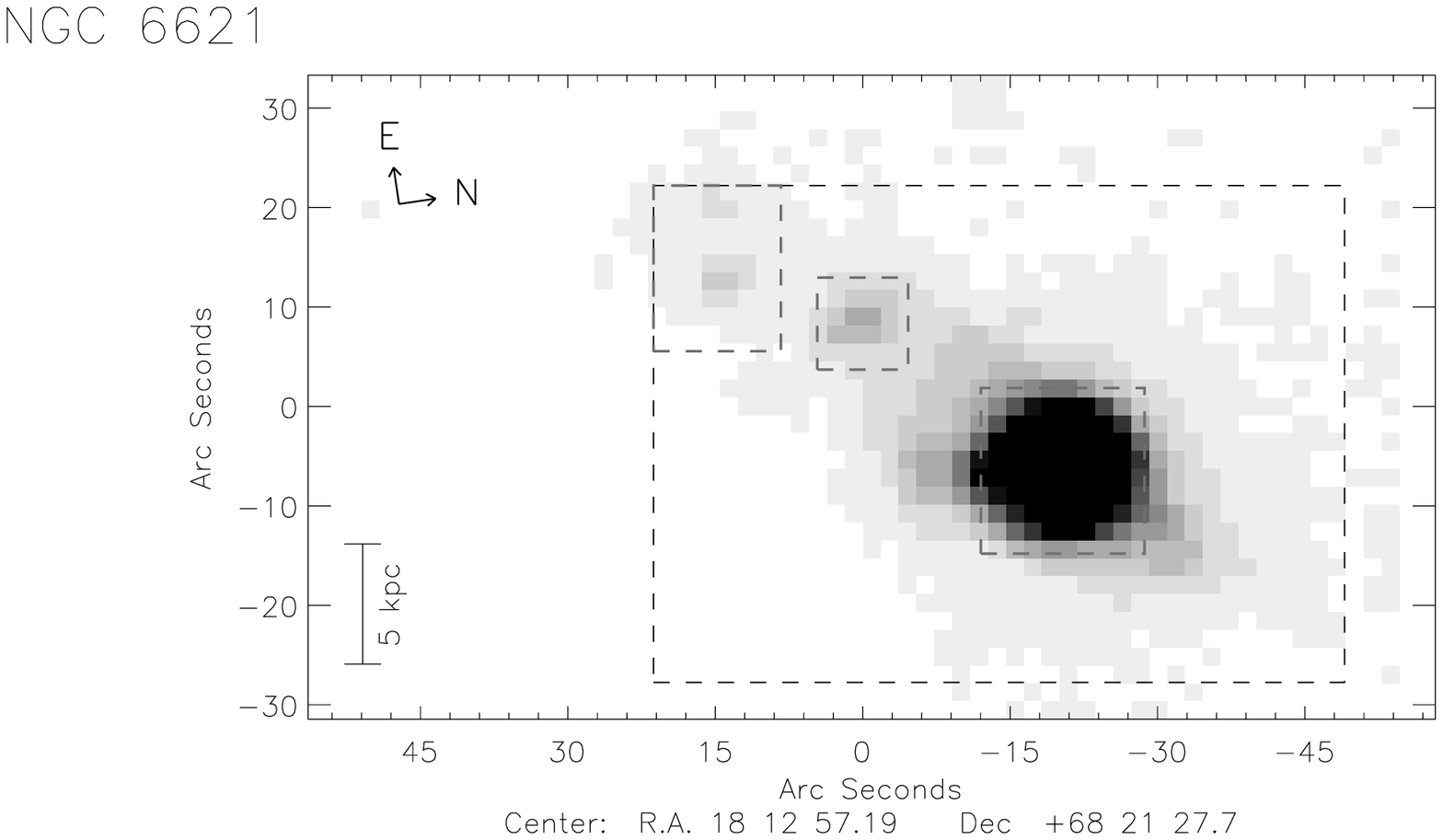}
\includegraphics[scale=0.40]{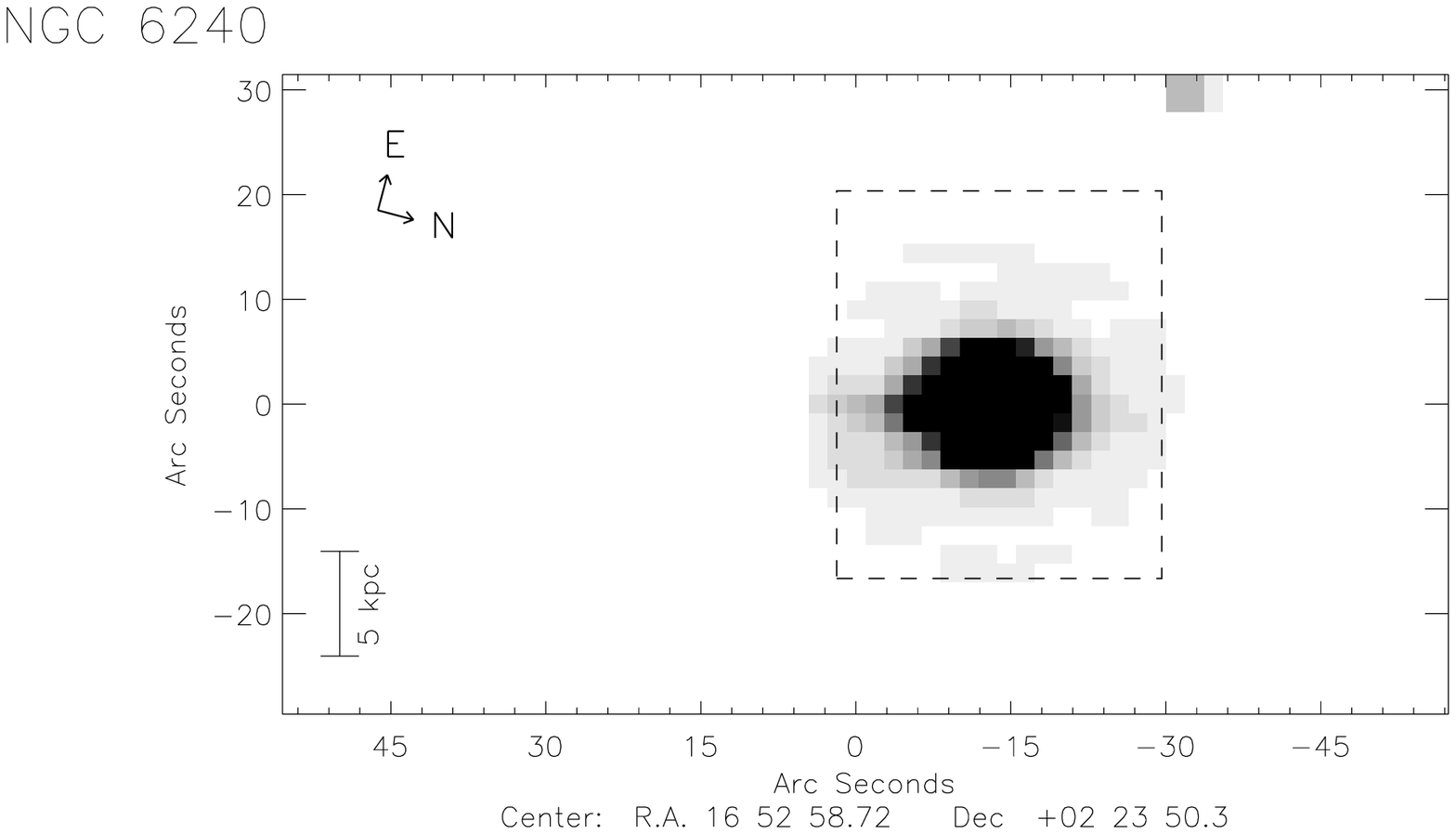}
\includegraphics[scale=0.40]{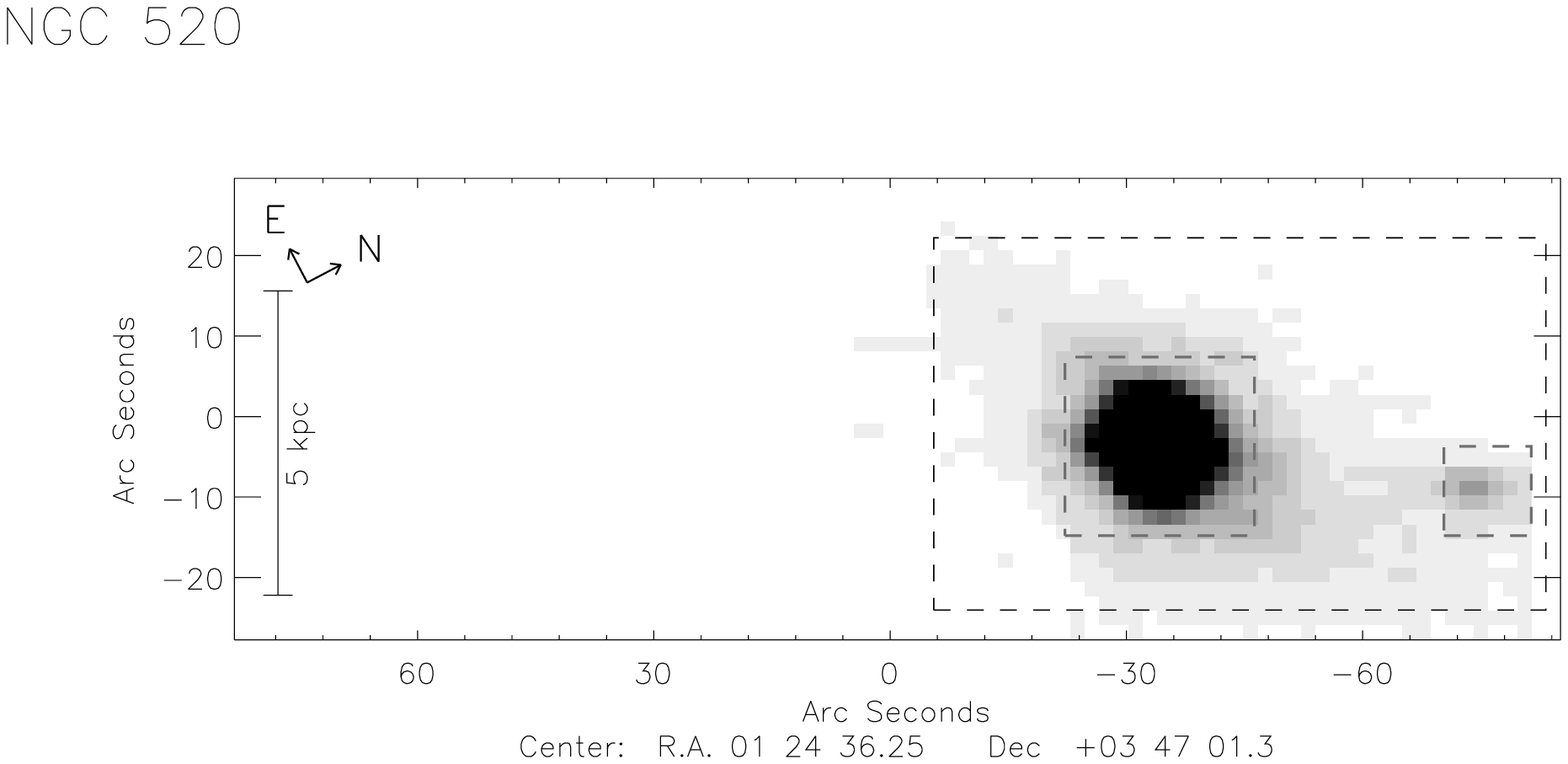}
\includegraphics[scale=0.40]{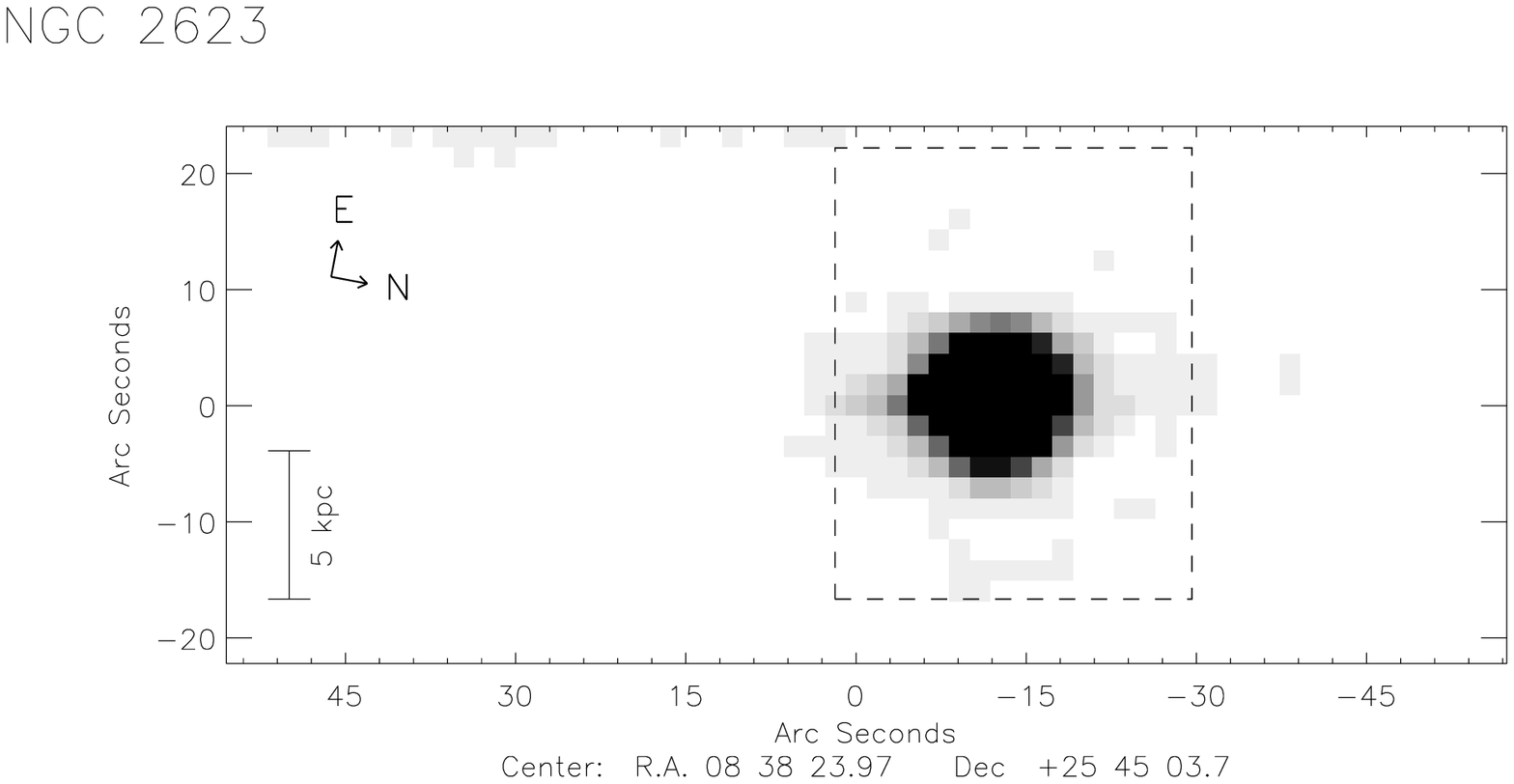}
\includegraphics[scale=0.40]{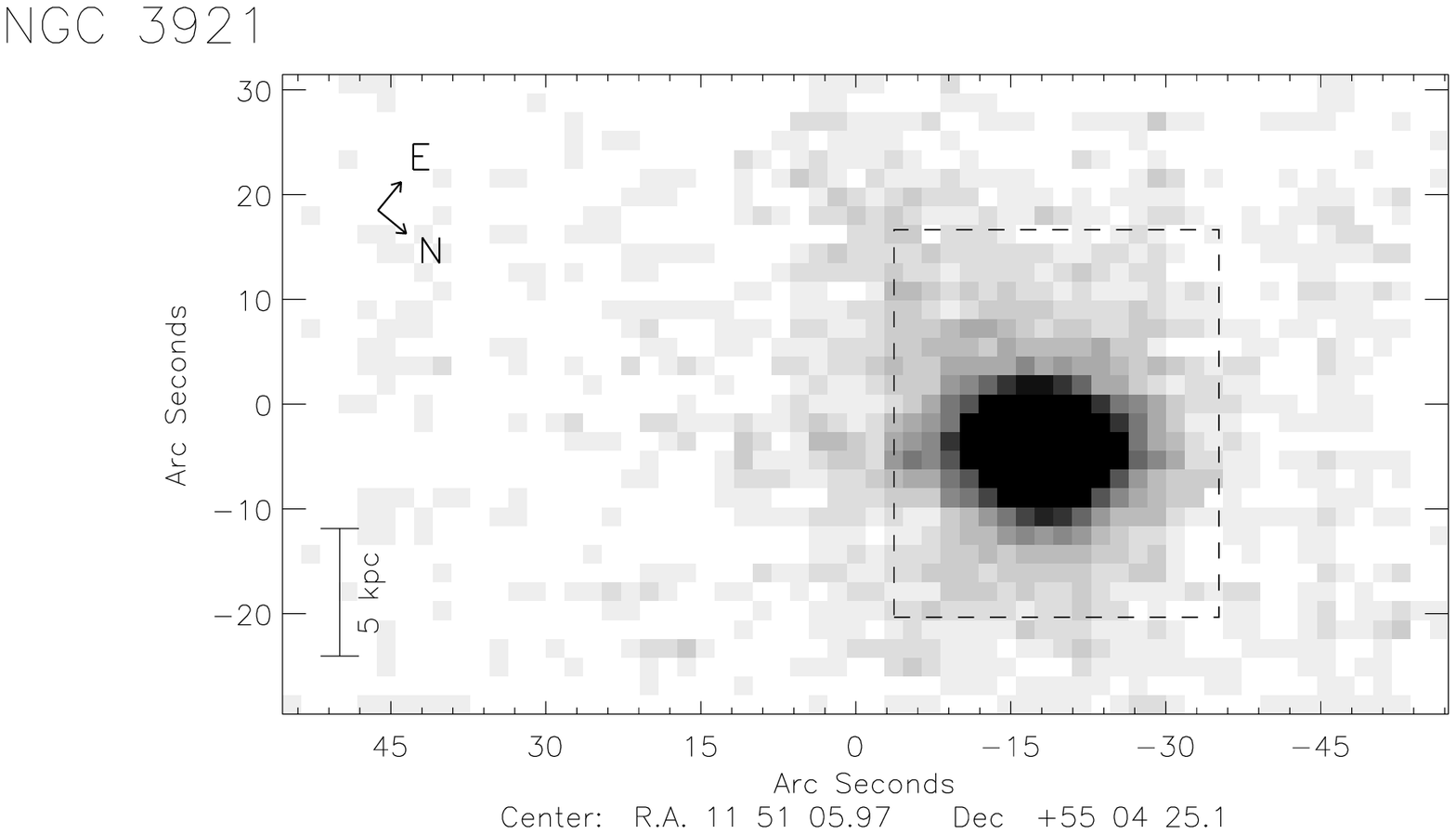}
\includegraphics[scale=0.40]{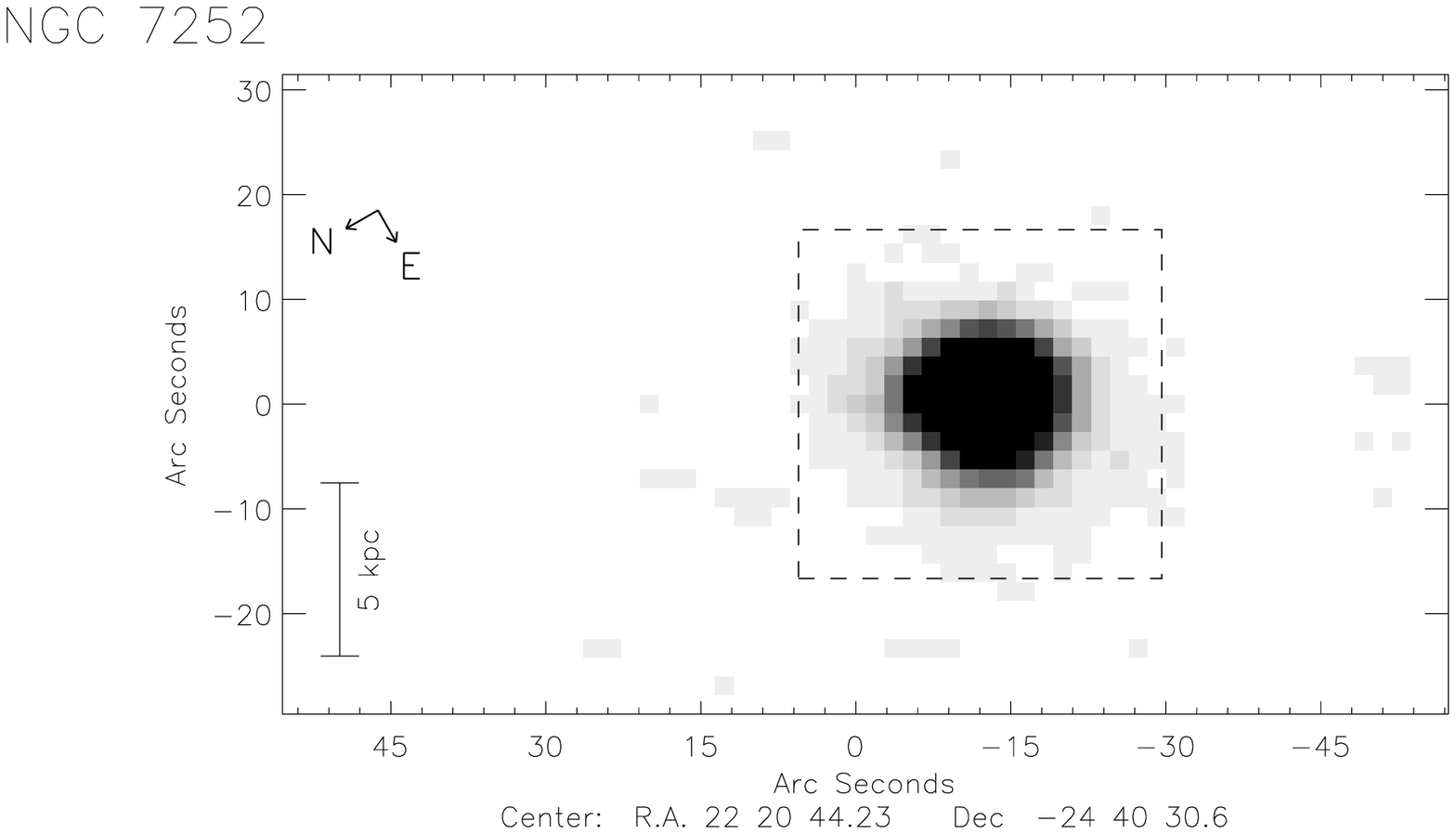}
\caption{\footnotesize{Mid-IR continuum maps of our sample (see \S~\ref{sec:spectra} for more details). The dashed-line rectangles show the individual regions for which spectra have been extracted. North and east are marked by arrows in each frame and the bar corresponds to 5~kpc.}}
\label{maps_mid-IR}
\end{center}
\end{figure*}

To extract high signal-to-noise spectra and line features, we have integrated the flux over those regions of each merger system that show significant emission in the spectral maps. These regions for each system (ranging from 2--20~kpc) are outlined on the IRS continuum maps in Fig.~\ref{maps_mid-IR} . The continuum maps are constructed using the mean flux at 5.3 --- 5.7~$\mu$m and 13.6 ---13.9~$\mu$m, which are wavelength ranges with no significant emission line features. Spectra were extracted from the IRS SL (5--15~$\mu$m) and LL (15--37~$\mu$m) modules in spatially matched extraction regions using CUBISM . Each spectrum is composed of data from the four low-resolution orders of IRS: SL2 (5.25--7.6~$\mu$m), SL1 (7.5--14.5~$\mu$m), LL2 (14.5--20.75~$\mu$m), and LL1 (20.5--38.5~$\mu$m). Slight mismatches between order segments, resulting from small residual photometric and astrometric uncertainties, were addressed by fitting the continuum near the overlap region, then scaling SL2 to match SL1, LL1 to match LL2, and finally SL to match LL. The scaled spectra were then concatenated and averaged in the overlap region. Typical scaling adjustments were $\sim$10\%. To measure the line flux of the mid-IR features, we fit the combined SL/LL spectra  using PAHFIT \citep{Smi07}. This spectral fitting routine decomposes IRS spectra into broad PAH features, unresolved line emission, and grain continuum, with the main advantage to recover the full line flux of any blended features. 
Reliable redshifts of our objects are taken from NED. The best-fit solution of the observed spectrum is computed using a $\chi^2$ minimization method. The code returns the best-fit parameters for each of the components, including the line or feature fluxes and the optical depth of the silicate absorption feature at 9.7~$\mu$m.

PAHFIT provides several options to take dust extinction into account. We employed PAHFIT assuming a uniform foreground screen extinction (\textit{SCREEN} model, i.e. a screen of dust between the galaxy and observer), dereddening the emitted line intensities with the extinction corrections estimated from the depth of the 9.7~$\mu$m silicate absorption feature \citep{Smi07}.  The choice of a uniform screen model (over, e.g., a fully mixed dust geometry model or a galactic center extinction) is based on observations which have shown that a simple screen attenuation is able to account reasonably well for the diffuse ISM attenuation in normal and starburst galaxies \citep[e.g.][]{Cha00, Smi07, Wil11}.  The silicate extinction as given by the fitted $\tau$ at 9.7~$\mu$m ranges from 0.08 to 1.4 (mean $\tau_{9.7}=0.7\pm0.5$, median: 0.6) for our sample, which is slightly larger than in normal galaxies \citep[typically $\tau_{9.7}<0.2$][]{Smi07} and starburst galaxies \citep[median $\tau_{9.7}=0.23$][]{Bra06}, but smaller than in LIRGs \citep[median $\tau_{9.7}=1.4$][]{Sti11} and ULIRGs \citep[median $\tau_{9.7}=1.6$][]{Arm07}. To estimate the effect of dust extinction, we fixed in test-runs the extinction optical depth to zero and compared both results (extinction and extinction-free) using the H$_2$ line flux at 9.66~$\mu$m. Since the H$_2$ line at 9.66~$\mu$m sits on top of the silicate absorption feature, we expect that this line is most affected by the extinction correction. We find that  the models with no extinction have $\sim$20\% lower H$_2$ fluxes than the \textit{SCREEN} models, indicating fairly small silicate absorption. Only for NGC~6240 and NGC~520 we find a three times smaller H$_2$ flux  with no-extinction than for the screen model, which is due to the large silicate absorption in these galaxies ($\tau_{9.7}\simeq 1.3$). For a more thorough discussion of how PAHFIT determines quantities such as the uncertainty in the integrated line flux, line FWHM, line equivalent width, and the used mid-IR extinction law see \cite{Smi07}.\par

The combined SL/LL spectra and PAHFIT model are shown in Fig.~\ref{spectra} for each region. The line flux  and errors of the main mid-IR features are listed in Table~\ref{tab_PAH}.  
The 5--38~$\mu$m wavelength range contains many important diagnostic lines, such as unresolved atomic fine-structure lines of Ar, Ne, O, Si, Fe, covering a large range in ionization potential.
In addition to the fine-structure lines, numerous PAH emission features are easily detectable in our spectra; the fluxes of the strongest features at 6.2, 7.7, 11.3,  and the 17~$\mu$m complex are listed in Table~\ref{tab_PAH}. We see a large range in global line intensities and PAH interband strengths for our galaxies as described in more detail in the following sections.
 Furthermore, the pure rotational lines of H$_2$ [S(0)--S(7)] are often very strong in starburst galaxies (Table~\ref{tab_H2}) and are used in \S~\ref{subsubsec:H2_temp} to determine the temperature and mass of the  (warm) molecular gas as listed in Table~\ref{tab_temp}.

\subsection{Mapping Spectral Lines and PAH Emission}
\label{sec:PAHFITmaps}
In Fig.~\ref{maps_feature_4627} we present maps of the main mid-IR features for the eight major mergers in our sample. 
These continuum-subtracted maps were created using our own customized code with PAHFIT as core-algorithm. At first, we scale SL2 to match SL1, LL1 to match LL2, and finally SL to match LL by fitting the continuum near the overlap region (similar to the 1D spectra fitting procedure). The combined LL cube is rotated and interpolated in order to match the SL cube, conserving the flux. The scale factors are obtained using the entire range of pixels that match the LL and SL aperture, assuming a uniform scale-factor rather than independent scale-factors for each pixel. The aligned SL and LL cubes are concatenated on a common wavelength grid and averaged in the overlap wavelength regions. The maps in the combined SL/LL cubes (5.3--38.5~$\mu$m) are  smoothed by a 3 pixel $\times$ 3 pixel box ($5.55\arcsec \times 5.55\arcsec$), conserving the flux, to increase the signal-to-noise ratio of the spectra. Then, for each pixel, the spectrum is extracted and PAHFIT is used to decompose it. Finally, we obtain maps of the properties of the fitted mid-IR features (i.e., integrated line flux, line FWHM, line equivalent-width, the uncertainty in the line flux, the fit to the entire spectrum and the fit to the continuum).\par

As a consistency check, we compared our line flux maps of the strongest line features with continuum-subtracted line flux maps created with CUBISM (see \S~\ref{subsec:reduction}). The results show a very similar distribution of the strong line features, but CUBISM is less sensitive to reveal faint extended emission in the outer disk. Overall, the main advantage of fitting each pixel is that faint local variations can be taken into account that would not be visible by subtracting only the continuum averaged over all the pixels.  Another important advantage of this method over CUBISM is to fit the full features before creating the maps, which would otherwise tend to lose wings of the broad PAH features and having difficulties to deblend the PAH features from each other, and other contiunuum features (e.g., the silicate absorption). In some mid-IR feature maps (Fig.~\ref{maps_feature_4627}) we find faint rings which are artifacts of the continuum subtraction (also present in CUBISM derived maps, although in a smaller degree).

\subsubsection{Continuum Fluxes}
\label{subsubsec:cont}
To characterize the basic properties of the spectral continuum we have mapped the flux densities for three rest-frame wavelength ranges: 5.3--5.7, 14.75--15.25, and 29.5--30.5~$\mu$m.  These wavelengths were originally chosen to cover a large baseline in wavelength while being least affected by PAH emission features, silicate absorption, or strong emission lines. The flux densities derived for 5.5, 15, and 30~$\mu$m are the median values in the above wavelength ranges, respectively. Although the continuum fit derived with PAHFIT is not affected by features, we have chosen these wavelength ranges to be consistent with CUBISM-derived maps and previous IRS studies for comparison.
All of our galaxies show a decrease of the 30 to 15~$\mu$m continuum flux ratio towards their center. In general, also a smaller 15/5~$\mu$m ratio in the galaxies' center is found, but in some cases we find strong spatial deviations from the azimuthal averaged profile (NGC~2623, NGC~4676, NGC~520, and NGC~3921). \par

The spatially integrated 30/15~$\mu$m flux ratio of our sample ranges from 5 (NGC~6621, NGC~3921, NGC~7252) to 16 (NGC~2623) with a mean value of 7.3, which is similar to measurements of starburst galaxies \citep[mean 30/15~$\mu$m flux ratio of 8.56; ][]{Bra06}. The same behaviour is seen for the spatially integrated 15/5.5~$\mu$m flux ratio which ranges from 1.5 (NGC~6621, NGC~3921) to 6.3 (NGC~2623) with a mean value of 3.4 \citep[3.8 for starburst galaxies; ][]{Bra06}. Two merger components in our sample show significantly smaller ratios than the ratios averaged over the entire merger system: NGC~4676  South (30/15~$\mu$m=2.2, 15/5.5~$\mu$m=0.8)  and NGC~6621 South (30/15~$\mu$m=2.0, 15/5.5~$\mu$m=0.5).

\subsubsection{Polycyclic Aromatic Hydrocarbons}
PAHs are thought to be responsible for a series of broad emission features that dominate the mid-infrared spectra of starbursts \citep[see, e.g.,][]{Pet04}. They are observed in a diverse range of sources, with their strongest emission originating in photo-dissociation regions (PDRs), the interfaces between H II regions and molecular clouds. The relative strength of the different PAH bands is expected to vary with the size and the ionization state of the PAH molecule \citep{Dra01, Smi07}. The 6.2~$\mu$m and 7.7~$\mu$m PAH bands dominate the emission of ionized PAH, whereas the 11.3~$\mu$m PAH emission is more intense for neutral PAHs. Variations of PAH features in nearby galaxies have been studied recently by \cite{Gal08}, \cite{Alo10}, \cite{Per10}, and \cite{Dia11}, including the effects of the radiation field hardness and intensity, the size distribution of the PAH carriers, extinction effects, and the dependency on the age of the stellar population.\par

Our spectra in Fig.~\ref{spectra} show that the spectral shape is usually dominated by strong emission features from PAHs. Here, we concentrate our analysis on several of the strongest PAH features at 6.2~$\mu$m, 7.7~$\mu$m (which is three blended sub-features centered at 7.42, 7.60, and 7.85~$\mu$m), 11.3~$\mu$m (two blended sub-features at 11.23 and 11.33~$\mu$m),  and around 17~$\mu$m (two blended PAH sub-features centered at 16.4 and 17.1~$\mu$m). Other PAH features can be seen in the spectra (e.g., at 8.6 or 12.7~$\mu$m), but they have in general lower S/N or are blended with other features. The fluxes of the PAH bands at 6.2, 7.7, and 11.3~$\mu$m are all affected by dust extinction, but the 11.3~$\mu$m is likely more affected by the 9.7~$\mu$m silicate feature than the 6.2~$\mu$m feature. The 7.7~$\mu$m PAH feature is affected by extinction to a similar level as the 6.2~$\mu$m feature \citep{Dra01, Bei08}. \par 

\begin{figure*}
\begin{center}
\includegraphics[scale=0.43]{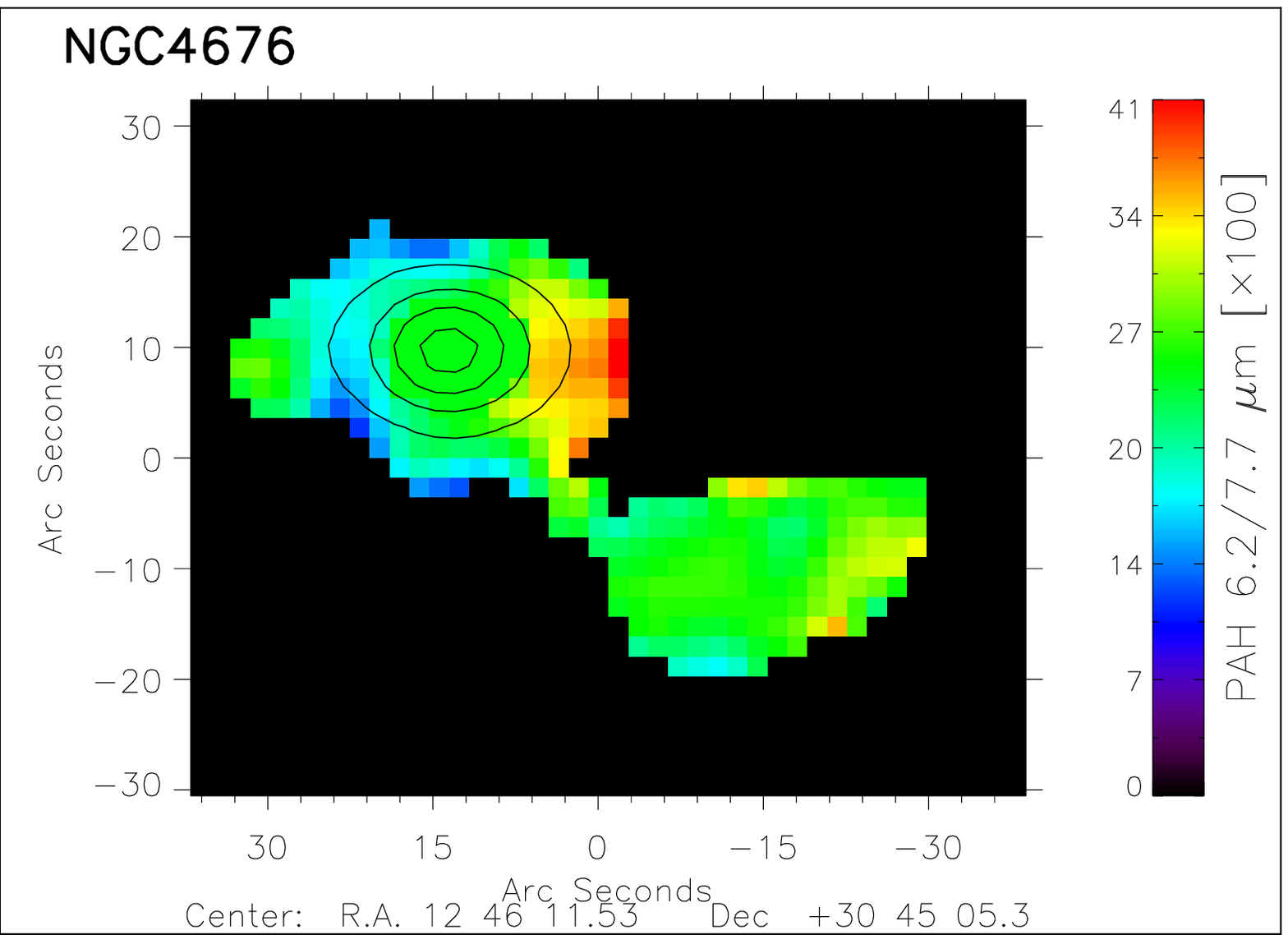}
\includegraphics[scale=0.43]{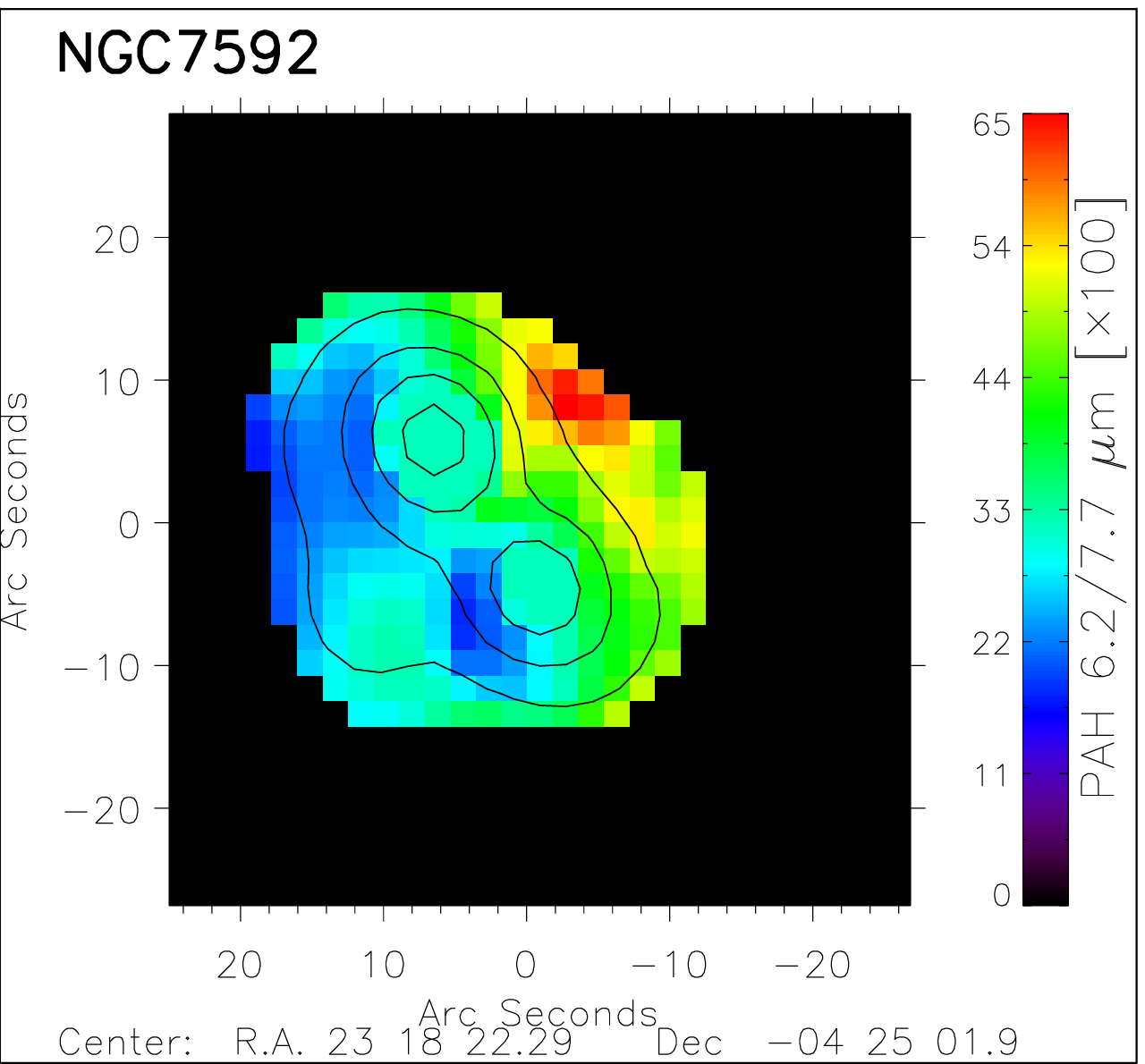}
\includegraphics[scale=0.43]{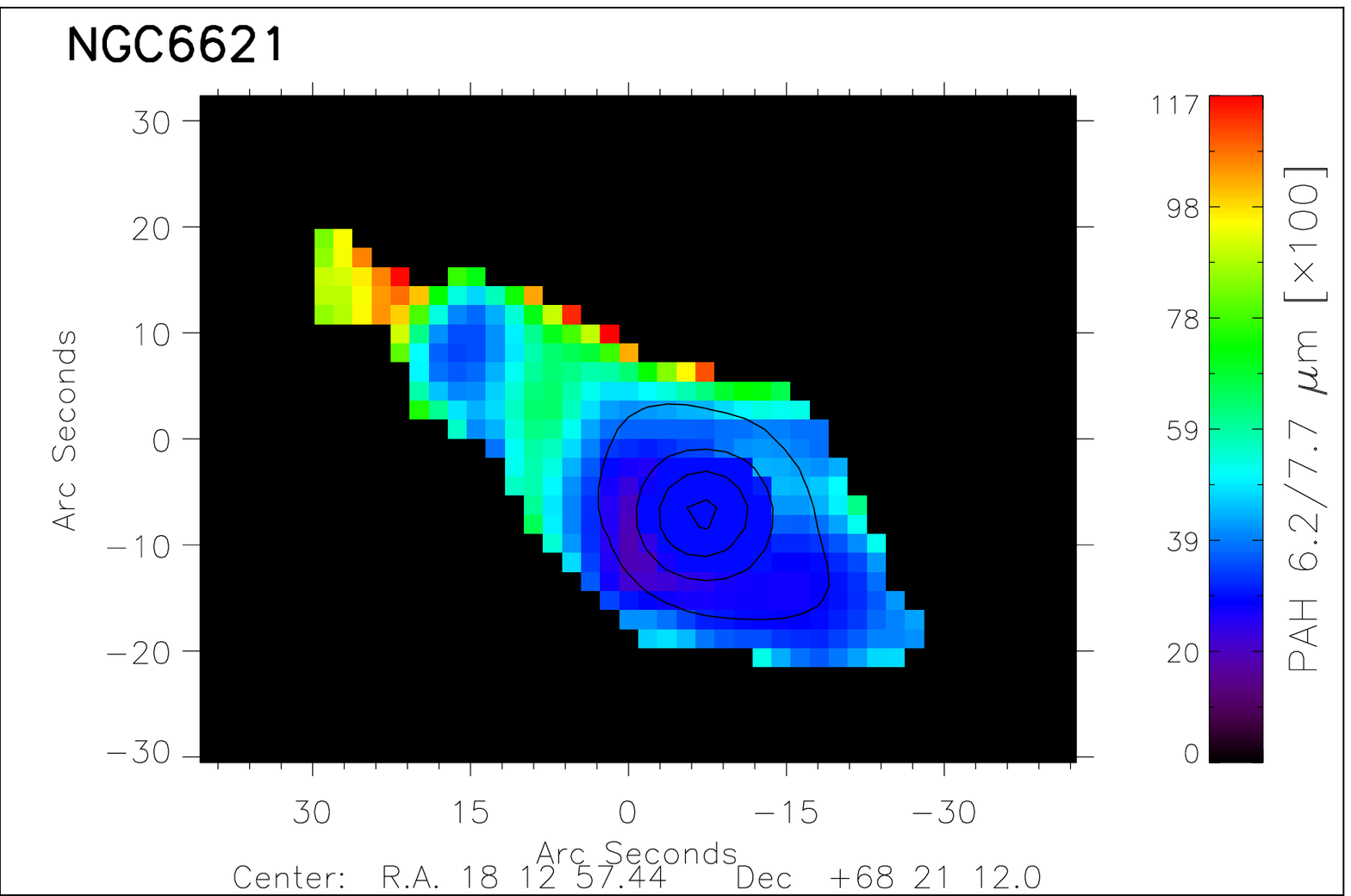}
\includegraphics[scale=0.43]{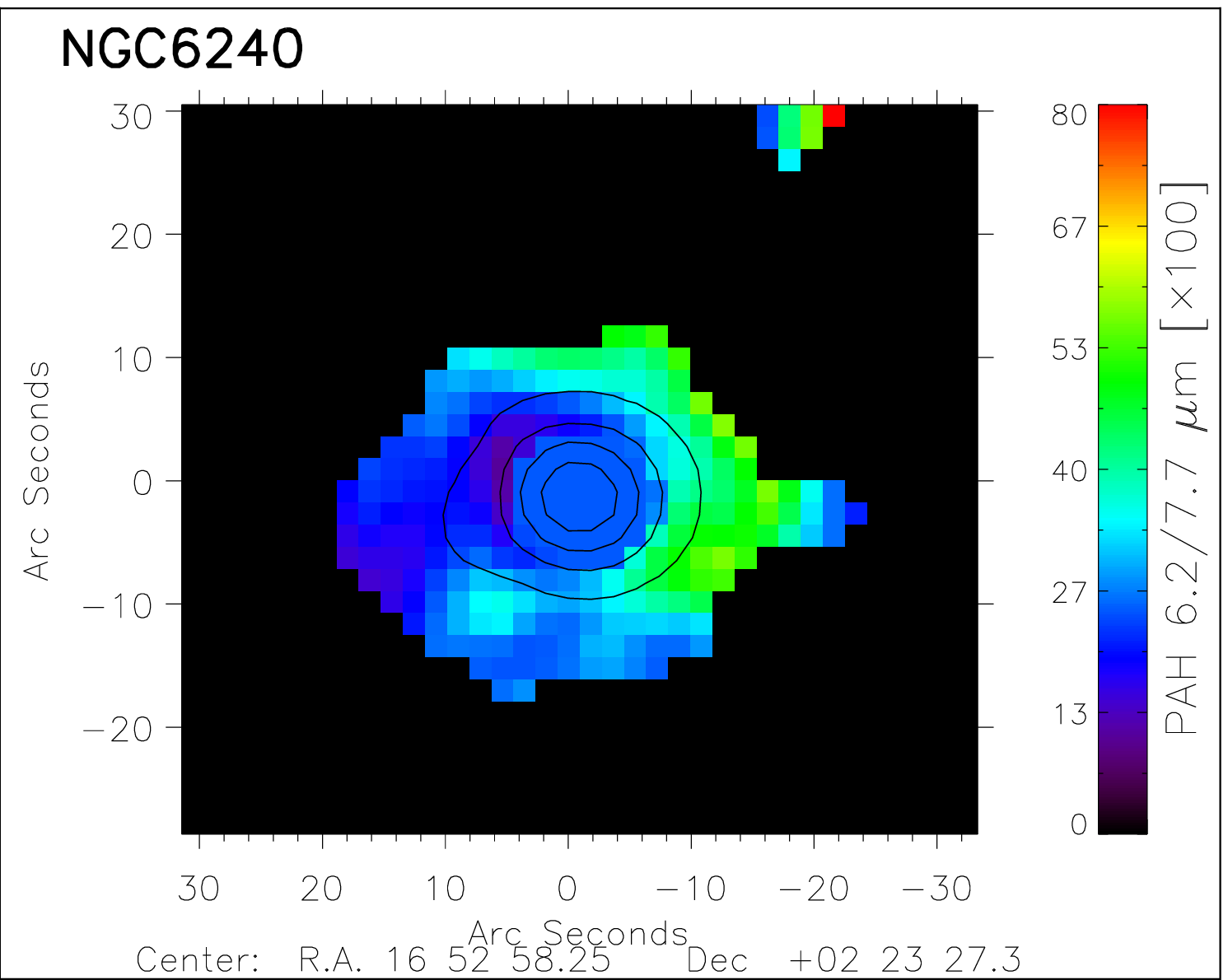}
\includegraphics[scale=0.43]{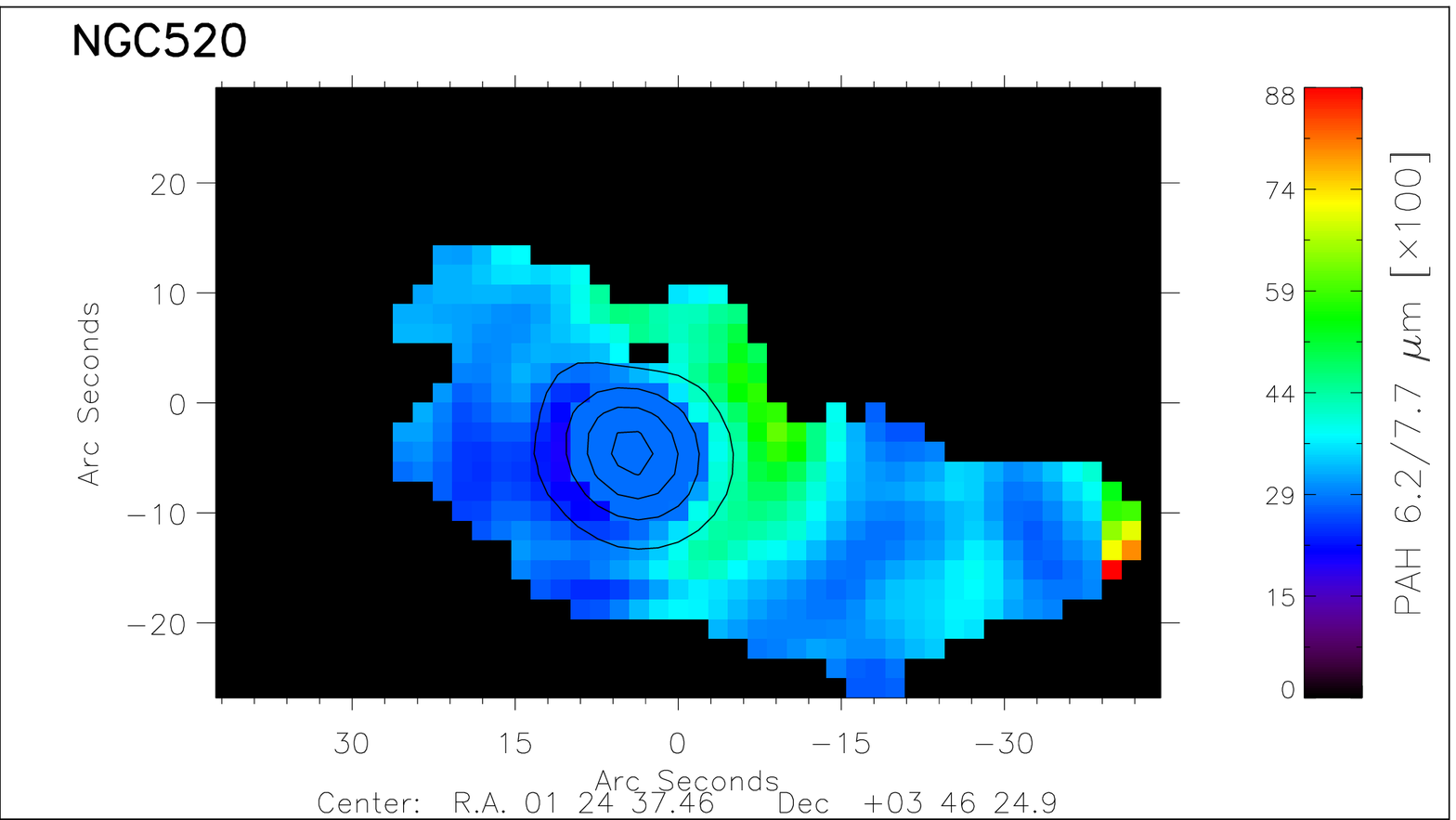}
\includegraphics[scale=0.43]{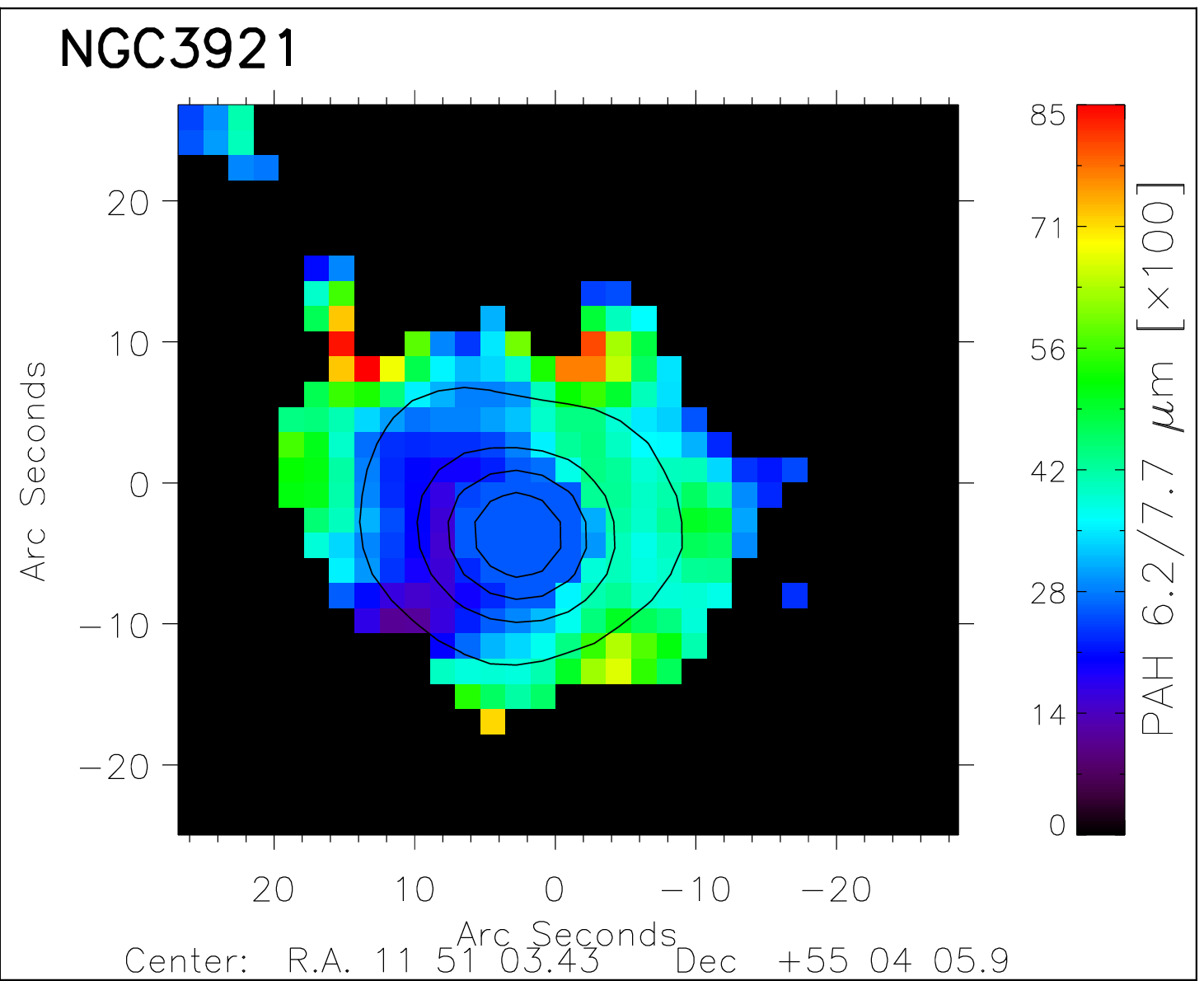}
\includegraphics[scale=0.43]{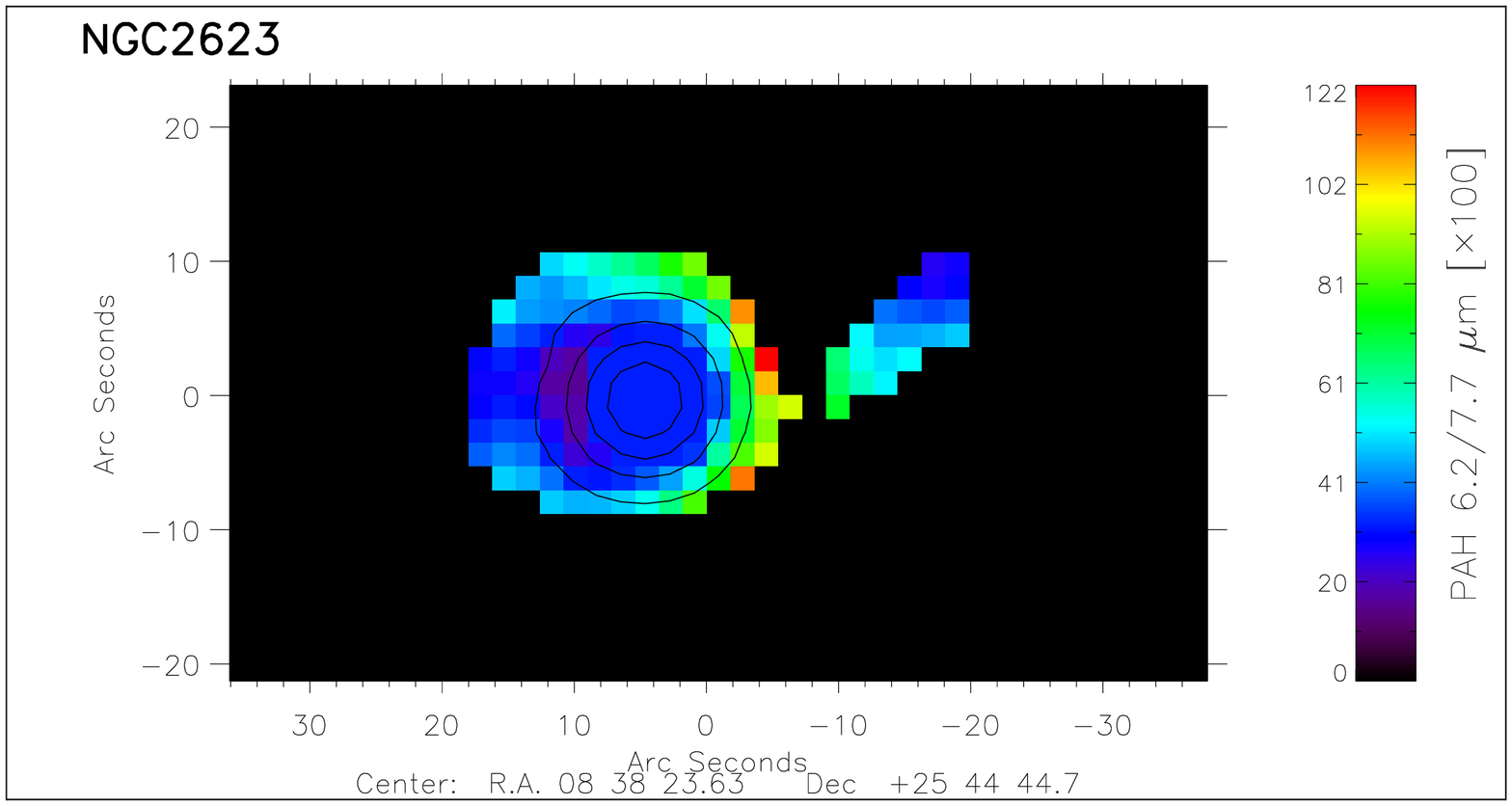}
\includegraphics[scale=0.43]{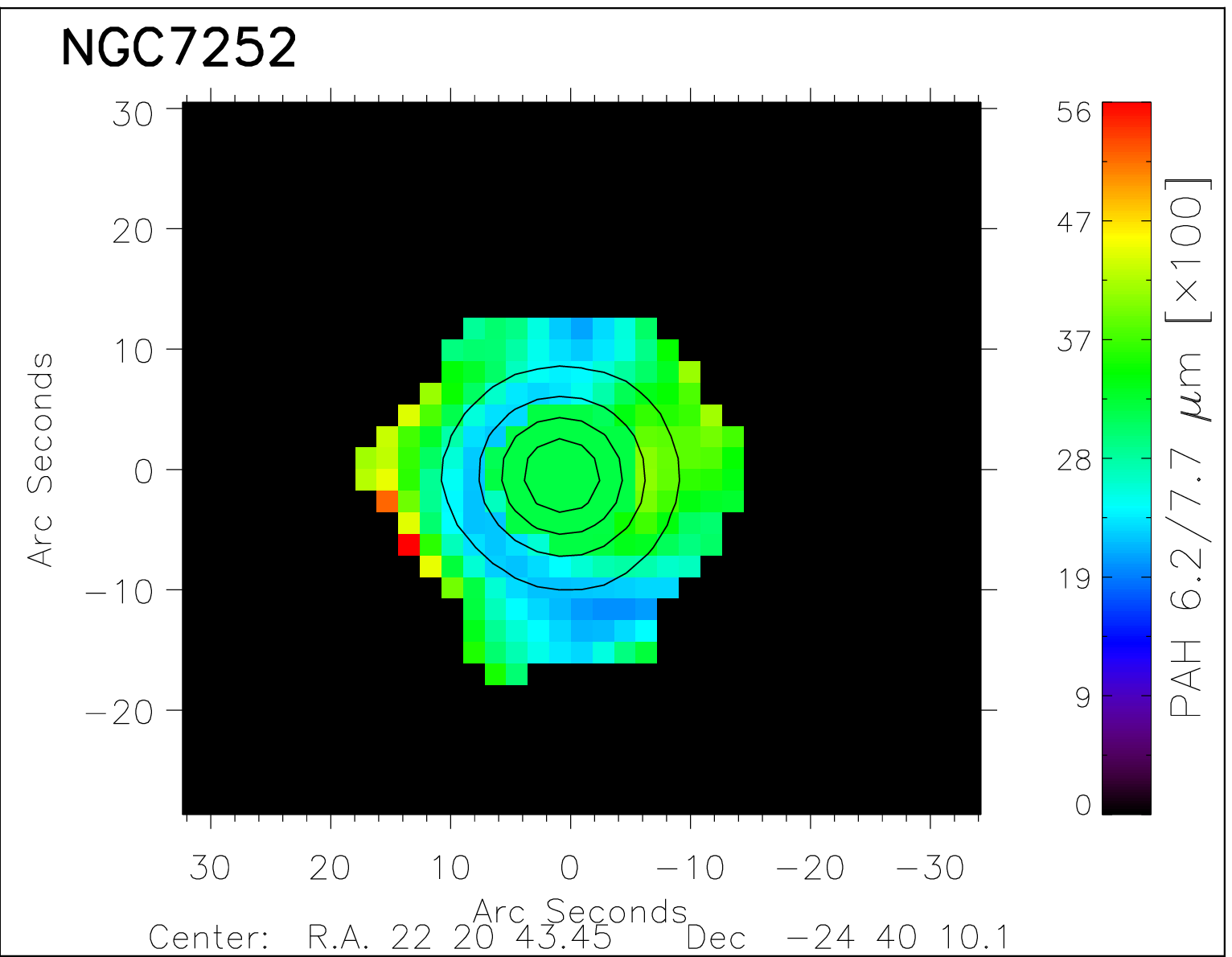}
\end{center}
\caption{\footnotesize{Ratio map of PAH 6.2/7.7~$\mu$m overlaid with the combined PAH emission (6.2,7.7 and 11.3~$\mu$m) in contours. The North and East directions in each frame of this figure are the same as in the corresponding frame of Fig.~\ref{maps_mid-IR}.}}
\label{maps_PAH6/7}
\end{figure*}

\begin{figure*}
\begin{center}
\includegraphics[scale=0.43]{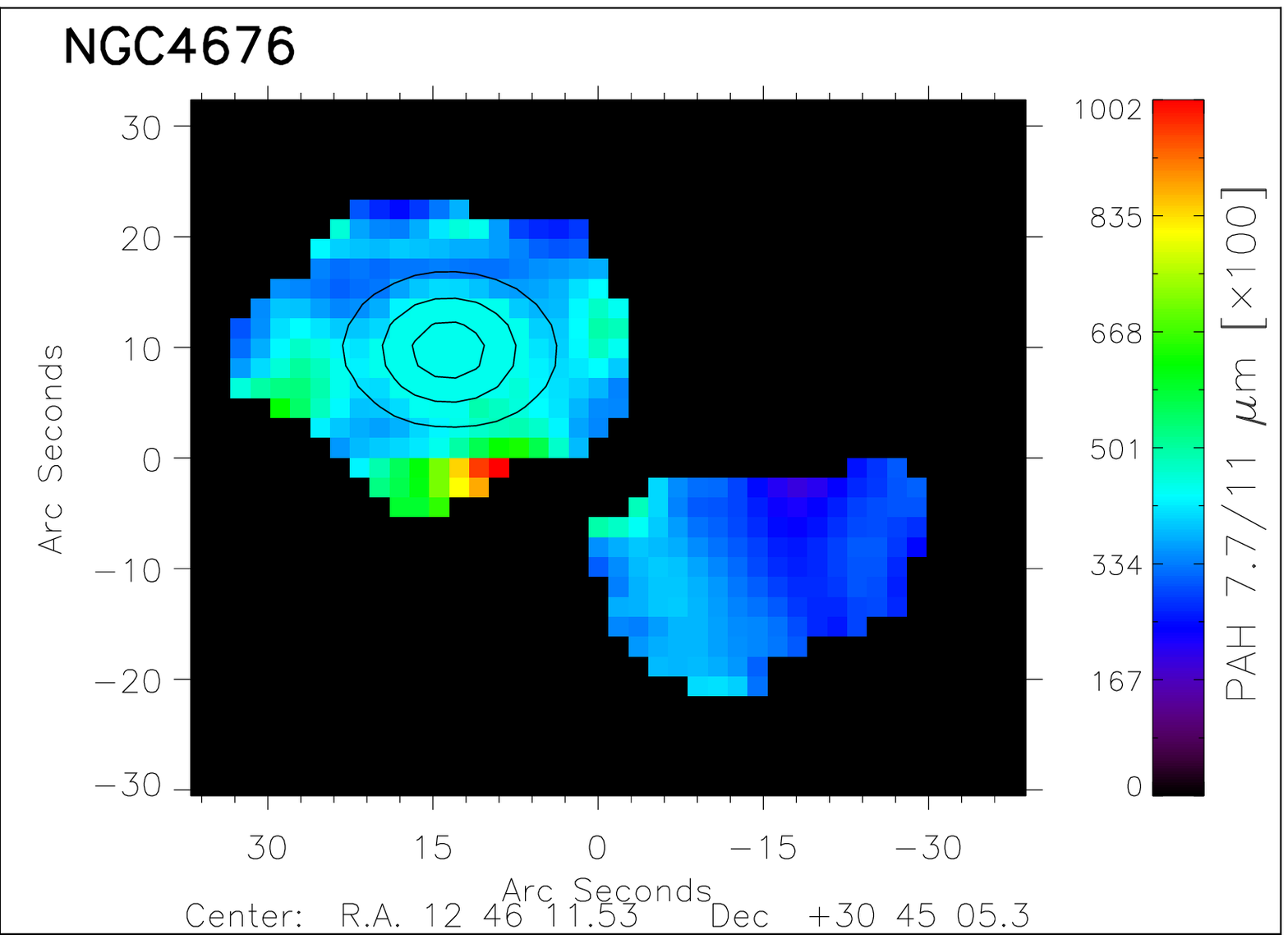}
\includegraphics[scale=0.43]{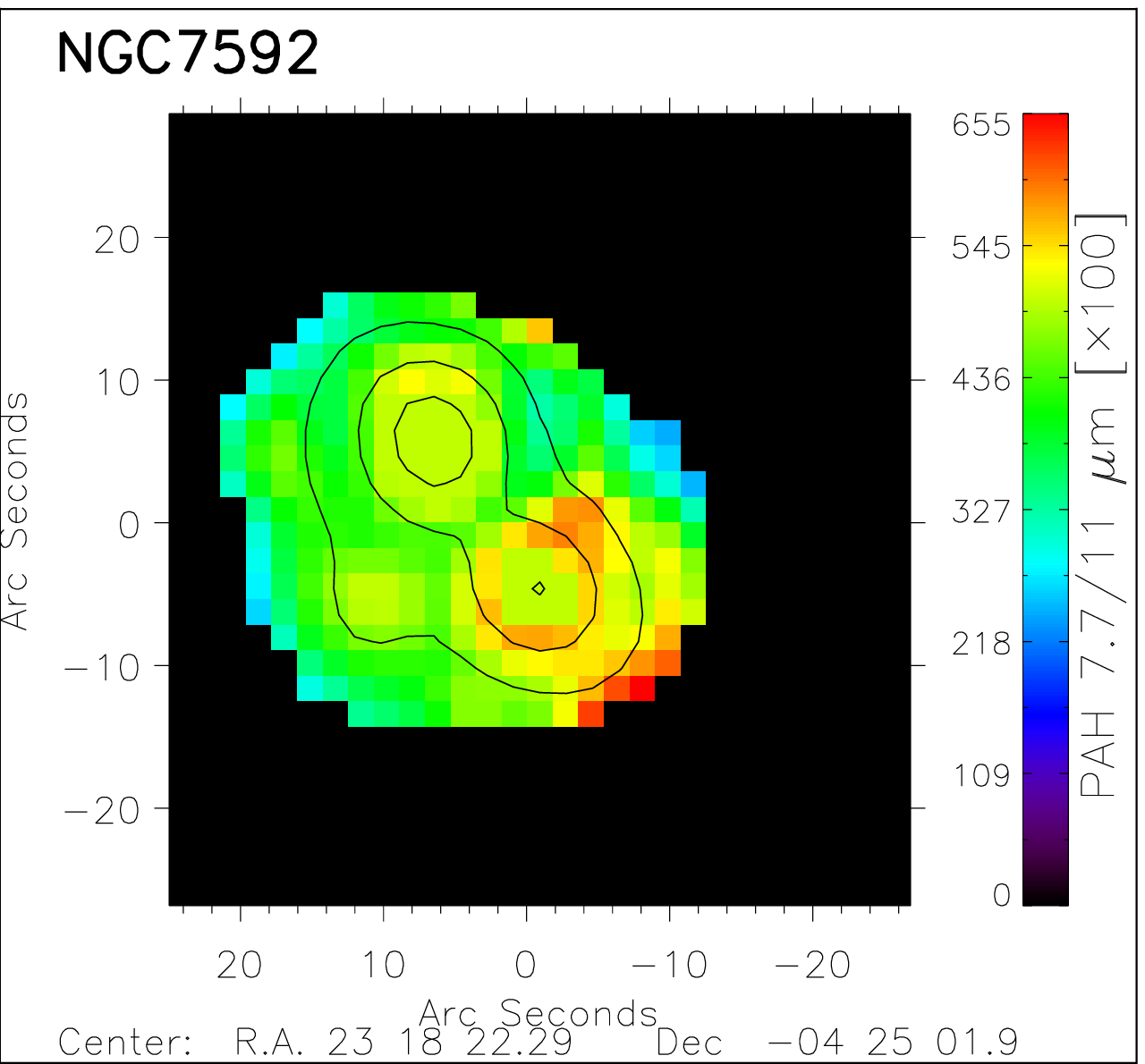}
\includegraphics[scale=0.43]{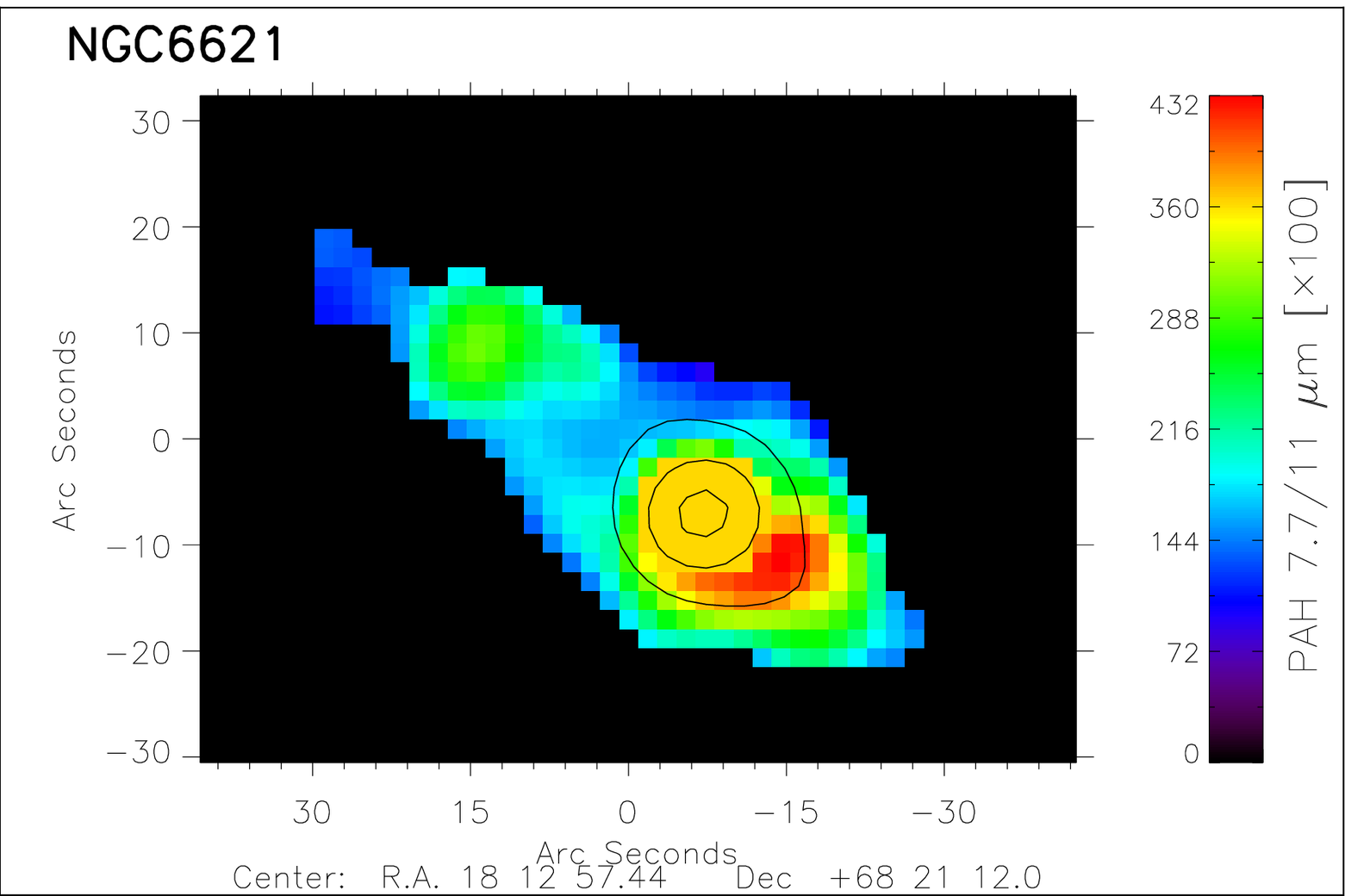}
\includegraphics[scale=0.43]{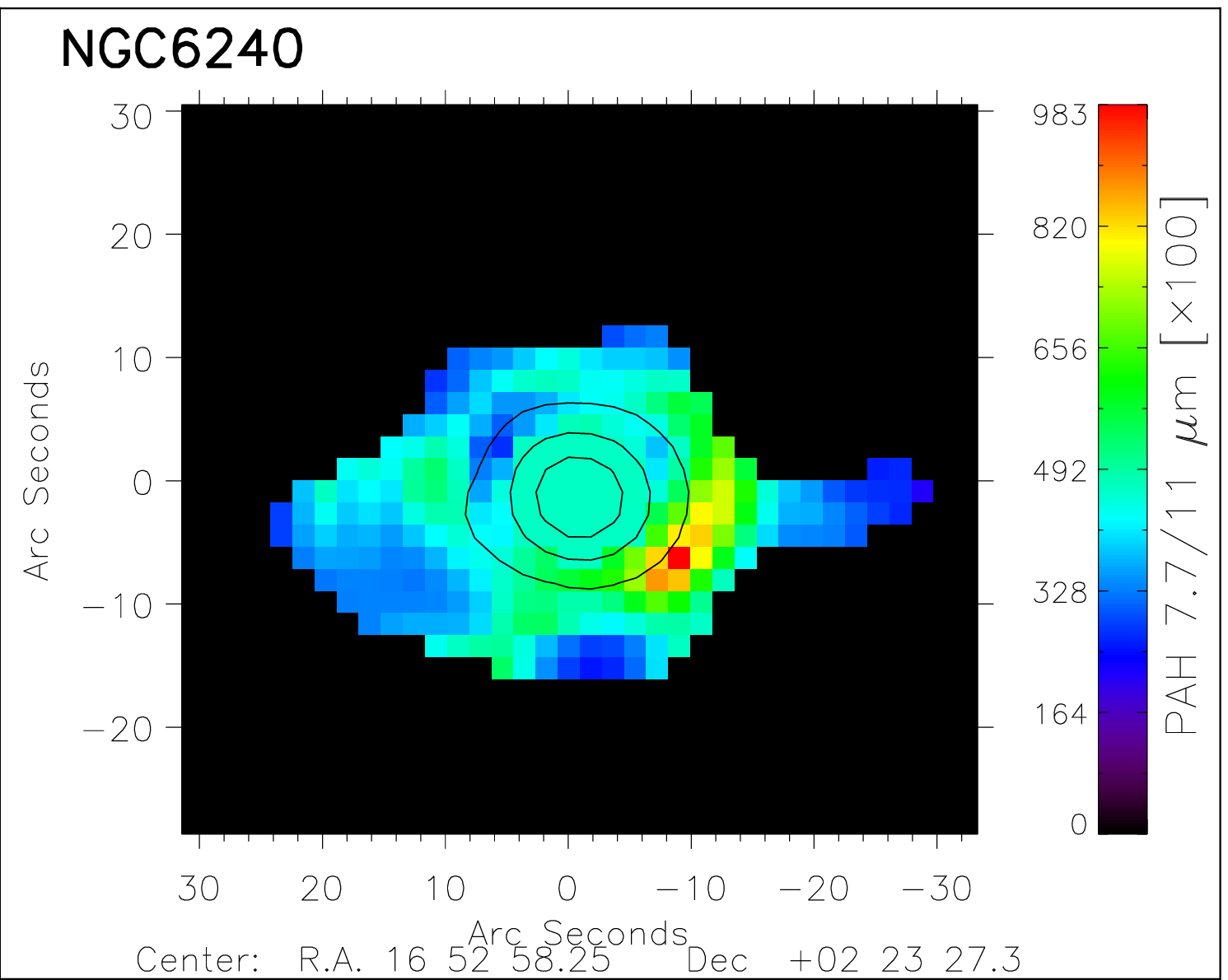}
\includegraphics[scale=0.43]{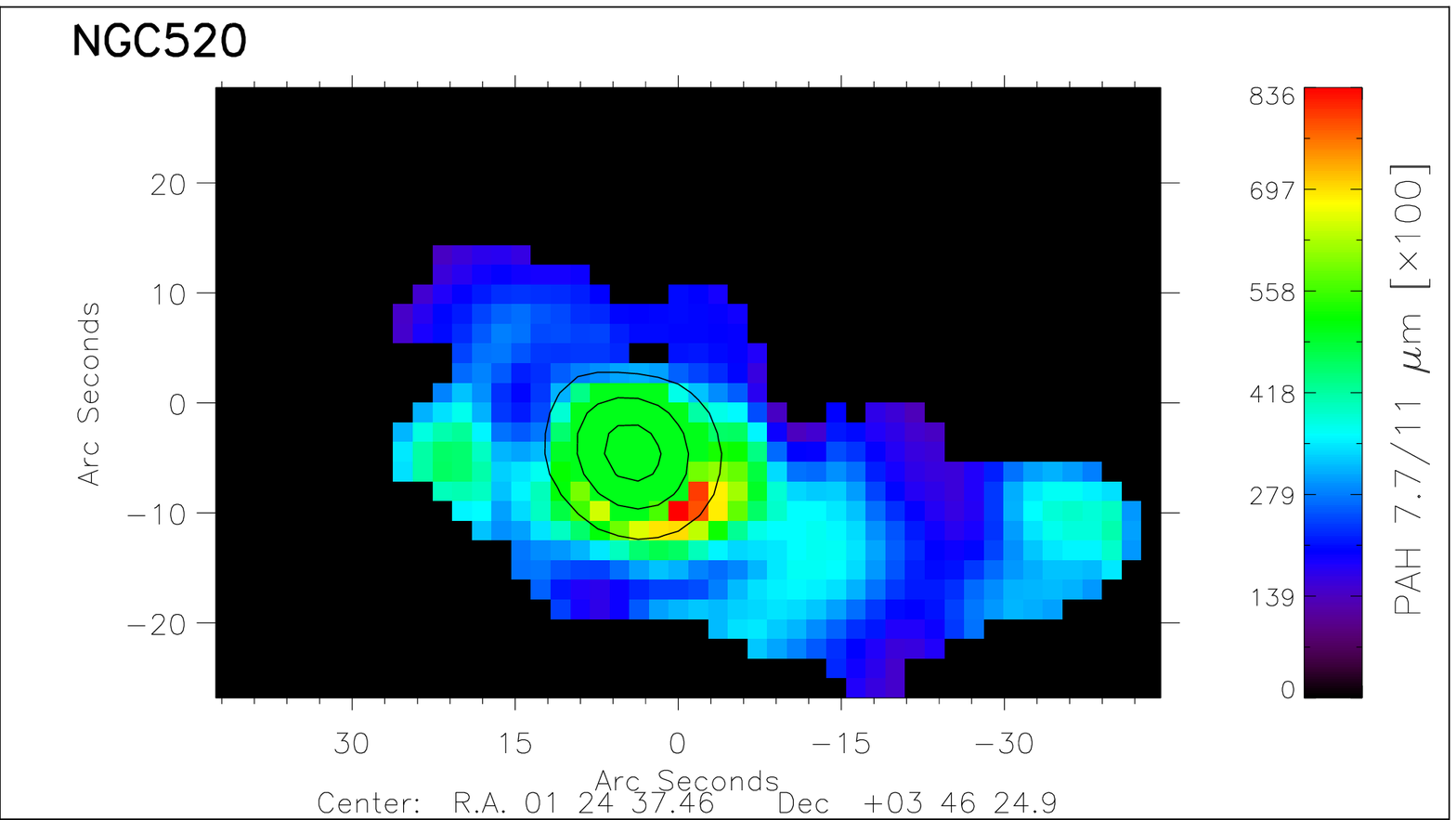}
\includegraphics[scale=0.43]{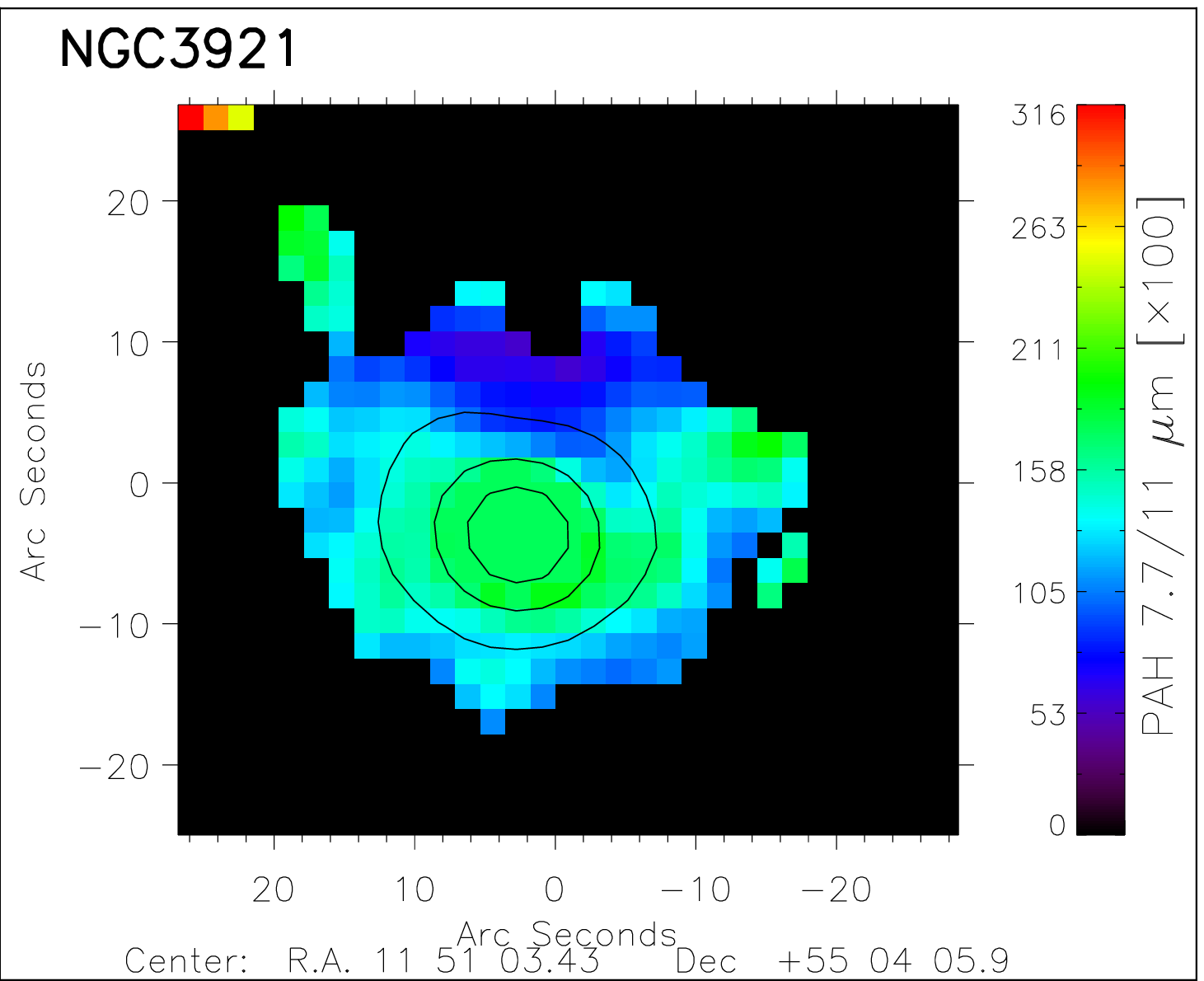}
\includegraphics[scale=0.43]{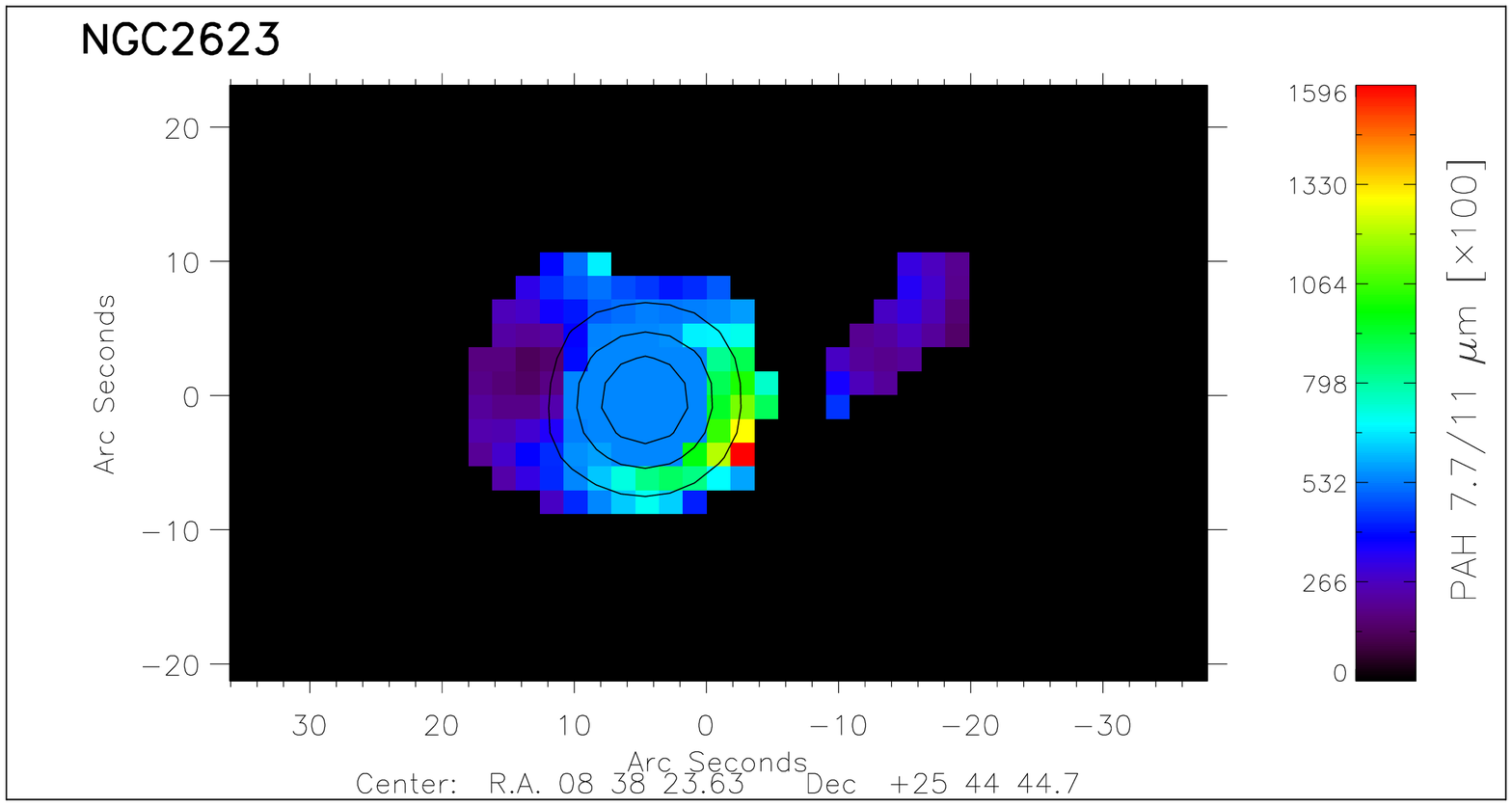}
\includegraphics[scale=0.43]{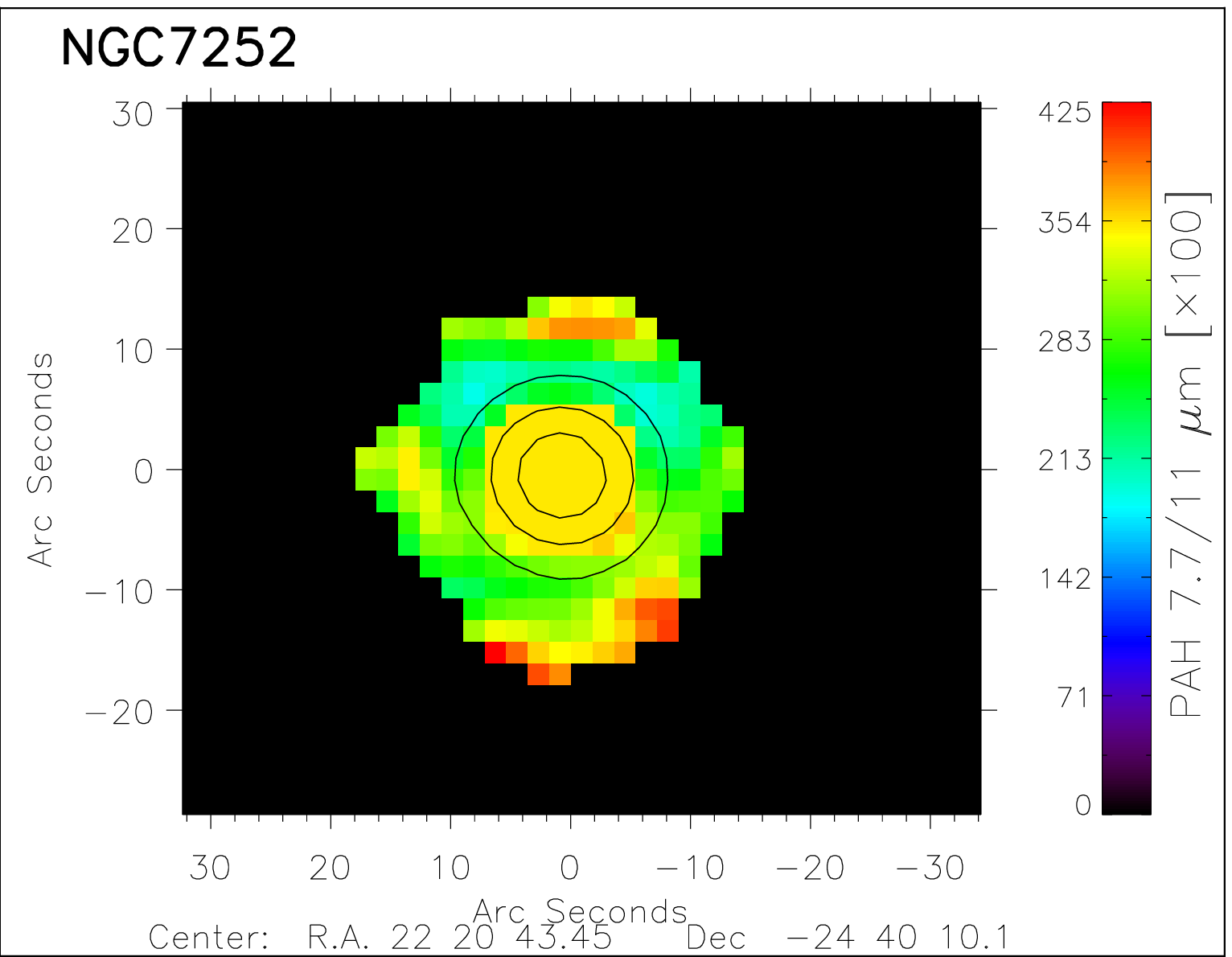}
\end{center}
\caption{\footnotesize{Ratio map of PAH 7.7/11.3~$\mu$m overlaid with the combined PAH emission (6.2,7.7 and 11.3~$\mu$m) in contours. The North and East directions in each frame of this figure are the same as in the corresponding frame of Fig.~\ref{maps_mid-IR}.}}
\label{maps_PAH7/11}
\end{figure*}

In Fig.~\ref{maps_PAH6/7} and Fig.~\ref{maps_PAH7/11} we present the PAH 6.2/7.7~$\mu$m and 7.7/11.3~$\mu$m ratio maps for the Toomre Sequence. For most of our galaxies the PAH feature at 11.3~$\mu$m is more extended than the PAH feature at 7.7~$\mu$m (factor of $\sim$2, ranging from 1.3--3.3), equivalent to an increase of the PAH 7.7/11.3~$\mu$m ratio towards the center, which is in agreement with previous studies \citep{Gal08, Per10, Dia11}. This is expected, since the 7.7~$\mu$m PAH feature is related to the ionized PAHs (which are more likely located in the central starburst region) rather than neutral PAHs, which are traced by PAH 11.3~$\mu$m and are located more likely in diffuse regions throughout the galaxies. However, for three galaxies (NGC~4676 North, NGC~6240, and NGC~2623) we find the opposite behavior, namely, a decreasing PAH  7.7/11.3~$\mu$m ratio towards the center, which has been also found in the center of weak AGN host galaxies \citep[e.g.,]{Smi07, Odo09, Dia10, Sal10} and suggested to be attributed to the selective destruction of the 7.7~$\mu$m PAH molecule with the increasing hardness of the radiation field. This picture is supported by our observations since two of the galaxies mentioned above (NGC~6240 and NGC~2623) are characterized as low luminosity AGN hosts (see Table~\ref{tab_sample}), wheras none of the rest of the sample seems to host an AGN.
We find for most of our galaxies a decreasing PAH  6.2/7.7~$\mu$m ratio towards the center (typically a factor of ~2). This is rather surprising since both bands, 6.2~$\mu$m and 7.7~$\mu$m, are related to ionized PAHs, and have shown only very small variations in previous studies \citep{Gal08, Per10}. Since the PAH 6.2 and 7.7~$\mu$m ratio decreases towads larger grain sizes (larger number of PAH carbon atoms), an intuitive explanation is that the PAHs with larger grain sizes are more likely located in the outskirts rather than in the center, which is in agreement with models of PAHs in PDRs \citep{Dra01}. A possible correlation between PAH grain size  and ionization state is discussed in the next section.

\subsubsection{PAH interband strengths and ionization}

To characterize the distribution and population of the PAH grain sizes and ionization states for each merger system, we present in Fig.~\ref{PAH_scatter} the variation of interband strength ratios of the PAH features and [NeIII]/[NeII] ratios (integrated over apertures of $3\times3$ pixels, or $5.6\arcsec \times 5.6\arcsec$) where significant emission is present, and compared to PAH models by \cite{Dra01}. In general, neutral PAHs have large ratios of PAH 11.3/7.7~$\mu$m, while PAH ions have much smaller values. For both neutrals and ions, there is a regular progression of PAH 6.2/7.7 to smaller values with increasing number of PAH carbon atoms \citep{Dra01}. \par

We find a large variety of PAH distributions, ranging from ionized to neutral PAHs over a large range in grain sizes (from a few ten to thousands of carbon atoms). For instance, NGC~4676 (primarily ionized PAHs) and NGC~520 (primarily neutral PAHs) show a very concentrated distribution in the PAH interband strength ratios, while, e.g., NGC~6621 (ionized and neutral PAHs) and NGC~3921 (primarily neutral PAHs, including the nucleus) exhibit more widespread PAH populations. We find no significant correlation  between PAH grain size (PAH 6.2/7.7~$\mu$m) and ionization state (PAH 11.3/7.7~$\mu$m). Instead, we find a large variety of PAH grain sizes for a given charge state, e.g., for NGC~7592, NGC~6240, and  NGC~7252. NGC~3921 is mainly populated by neutral PAHs.  For most of our galaxies, the central apertures (center of galaxies) have low PAH 11.3/7.7~$\mu$m ratios, representing typically the most ionized PAH population for each galaxy, as one would expect from a central starburst or a low-luminosity active galactic nucleus (AGN). This finding is in agreement with other recent studies of PAH populations in ULIRGs \citep{Dia11} which show that the PAH 11.3~$\mu$m is spatially more extended than the PAH 7.7~$\mu$m feature.

\subsubsection{The [Ne III]/[Ne II] line ratios: Probing the hardness of the radiation field}

In general the ratio of the 12.81~$\mu$m [Ne II] and 15.55~$\mu$m [Ne III] lines provides a diagnostic of ionization that is relatively insensitive to extinction, gas-phase abundances, and gas temperature (being low excitation lines in warm gas). Since the two ionization potentials of Ne are 21.6~eV [Ne II] and 41.0~eV [Ne III], the measured [Ne III]/[Ne II] line flux ratio is sensitive to the average effective temperature of the ionizing stars, and hence might provide a measure of the hardness of the radiation field.\par

\begin{figure*}
\begin{center}
\includegraphics[scale=0.43]{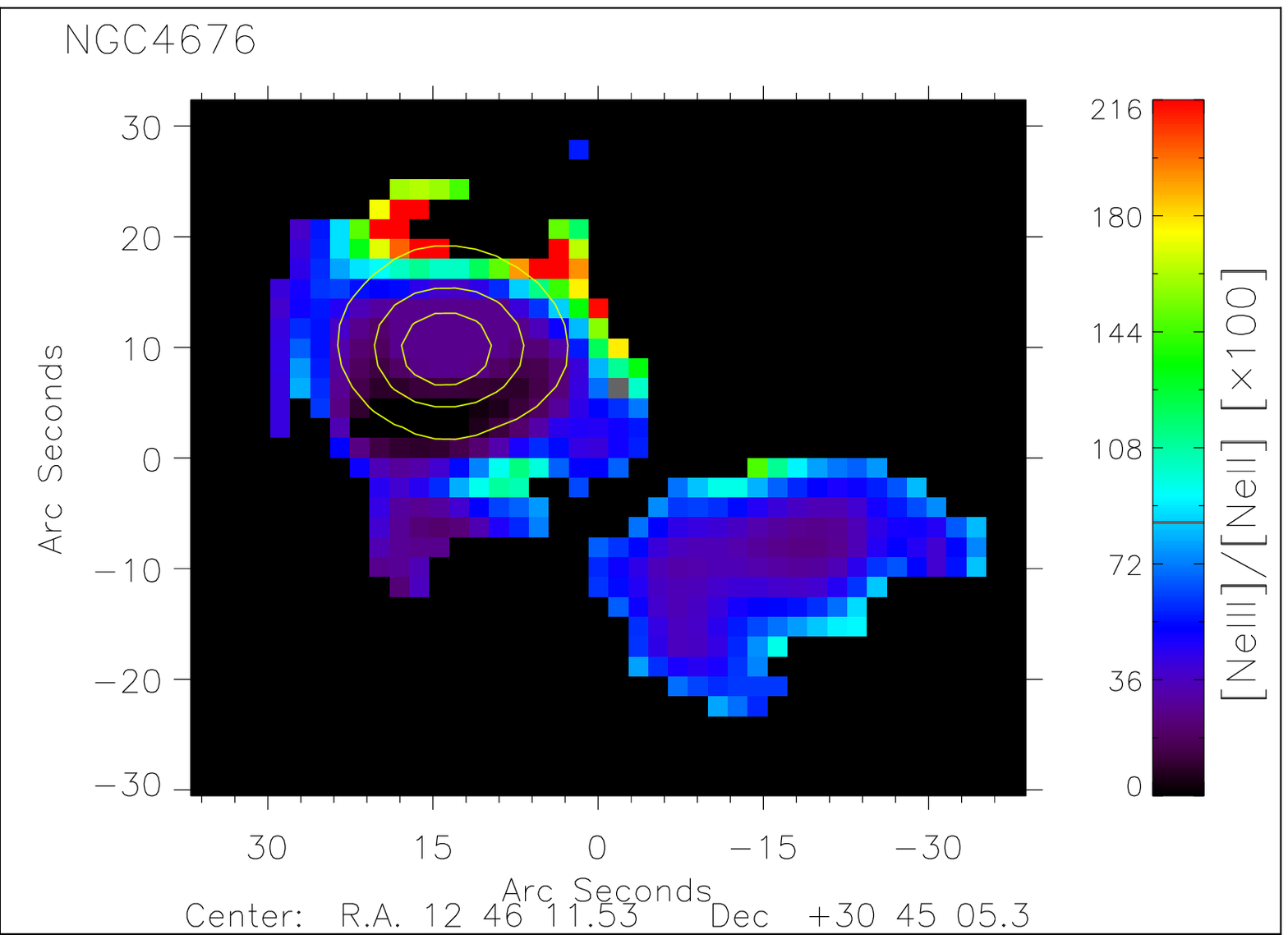}
\includegraphics[scale=0.43]{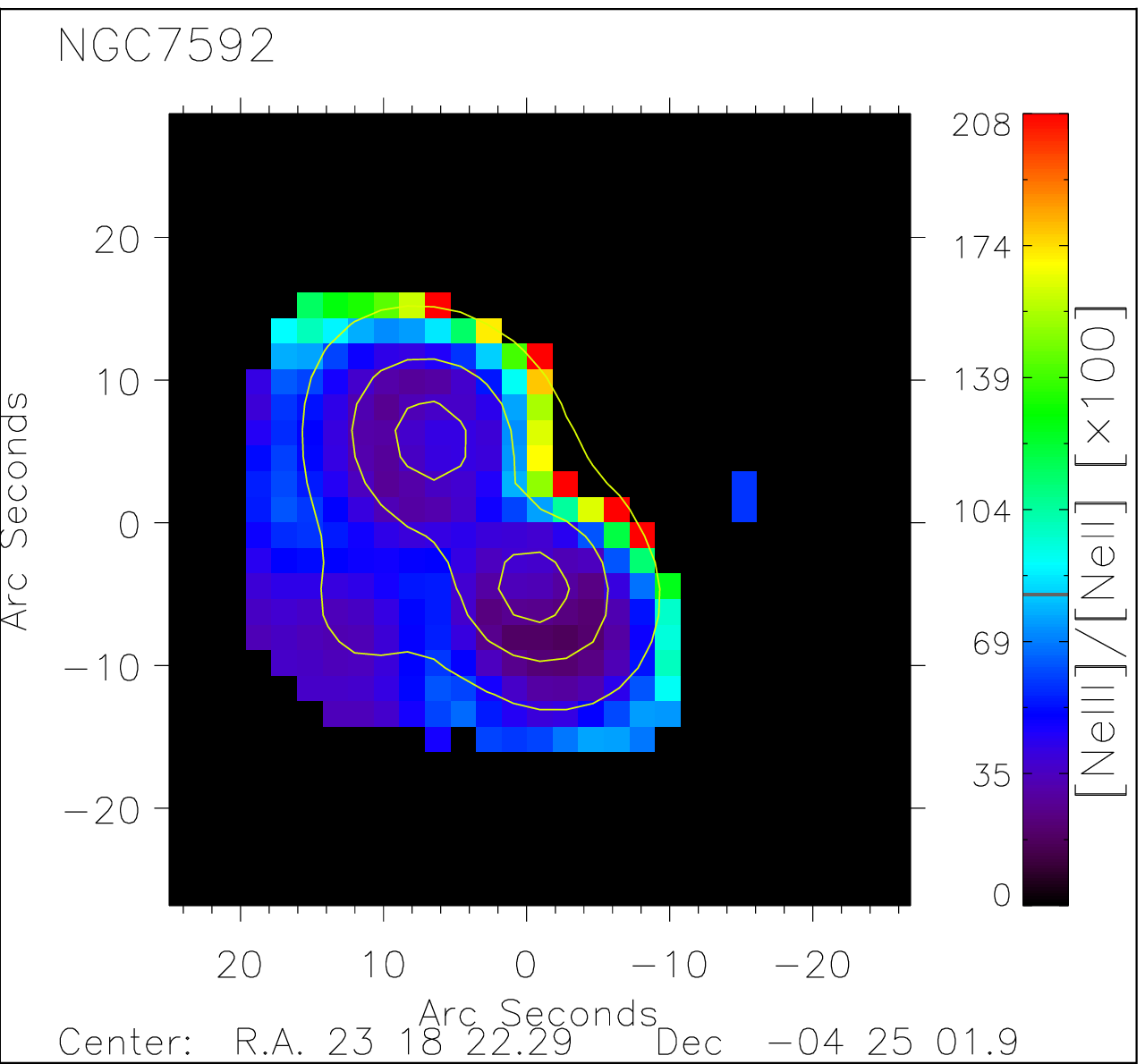}
\includegraphics[scale=0.43]{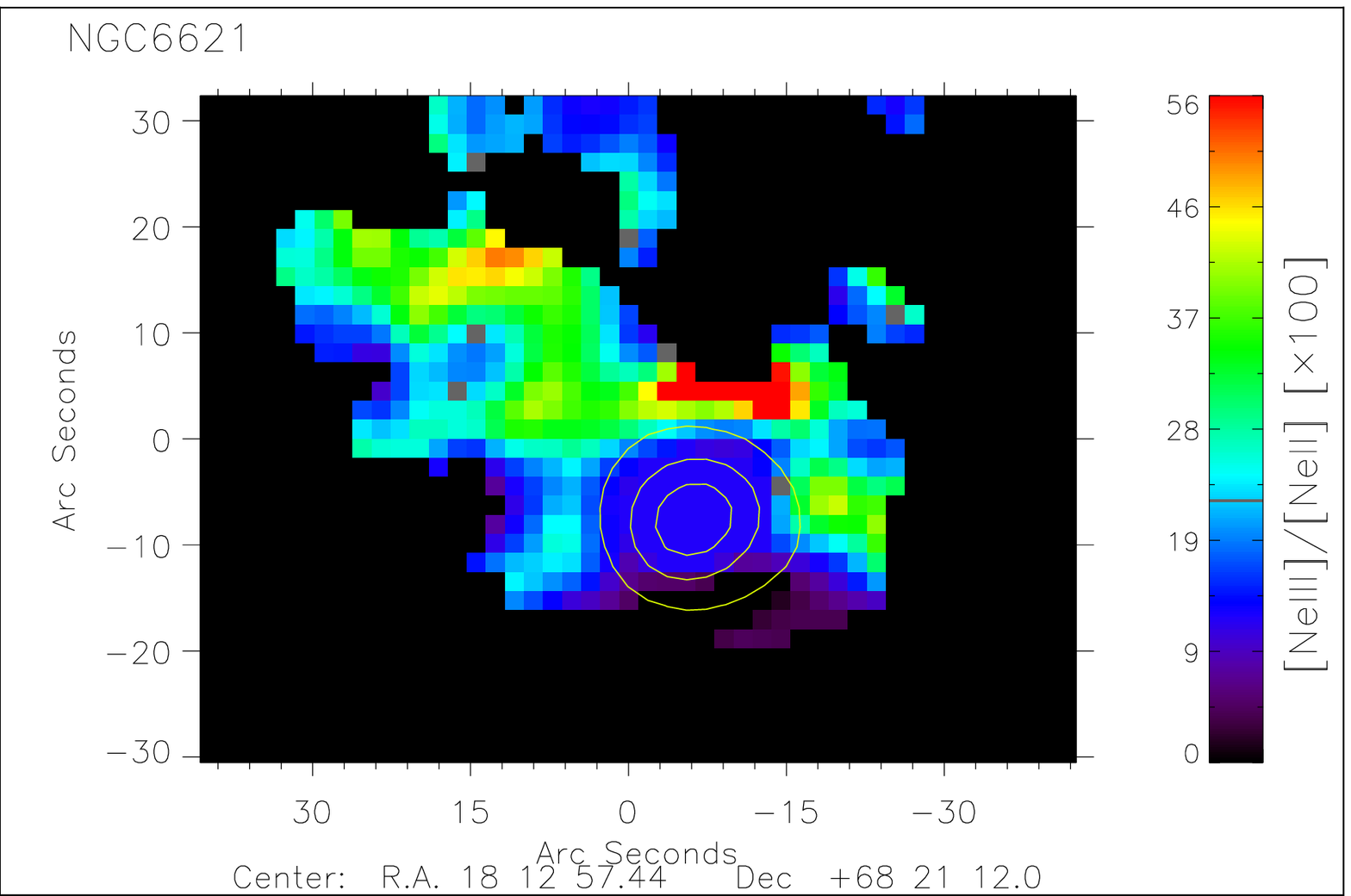}
\includegraphics[scale=0.43]{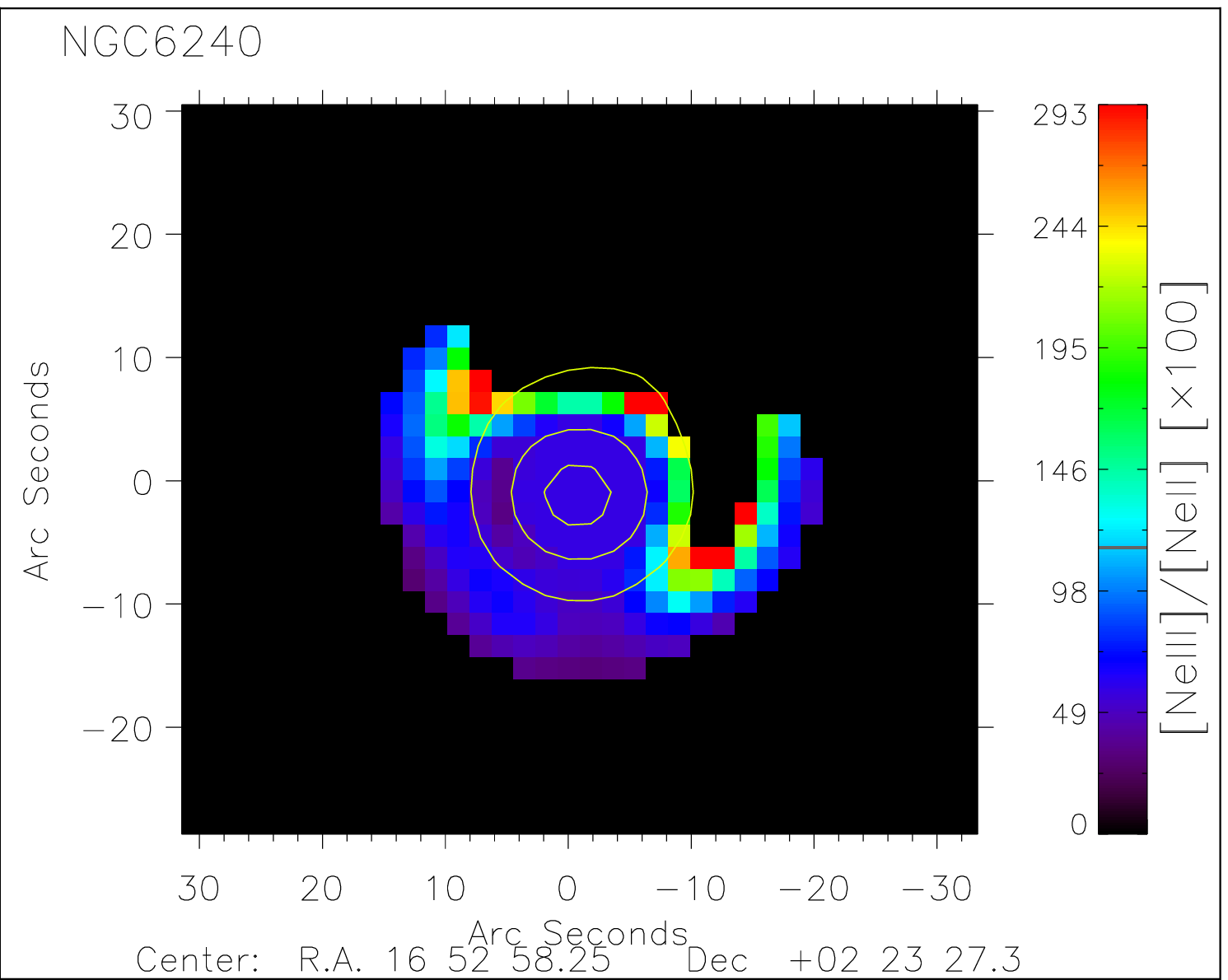}
\includegraphics[scale=0.43]{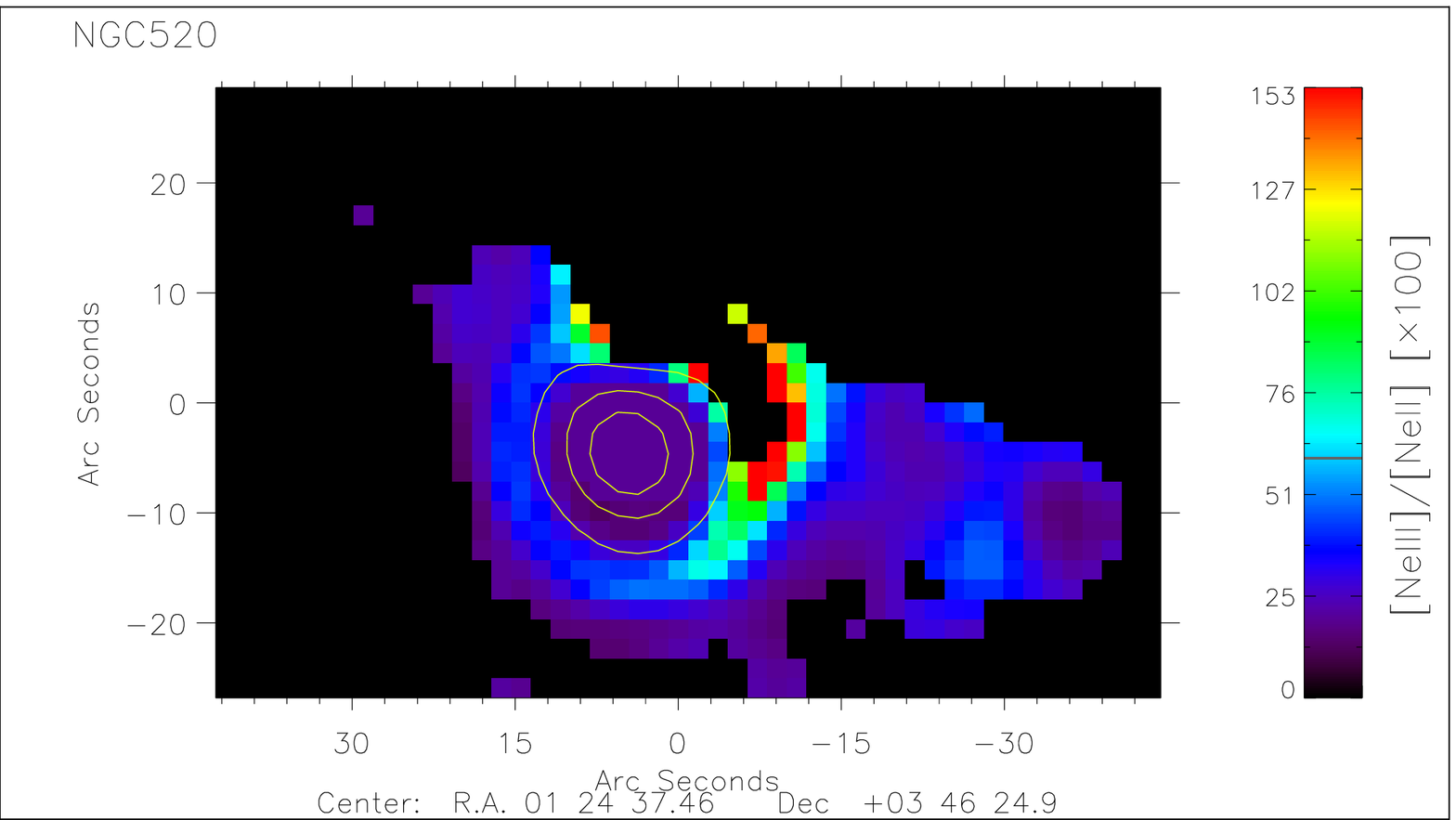}
\includegraphics[scale=0.43]{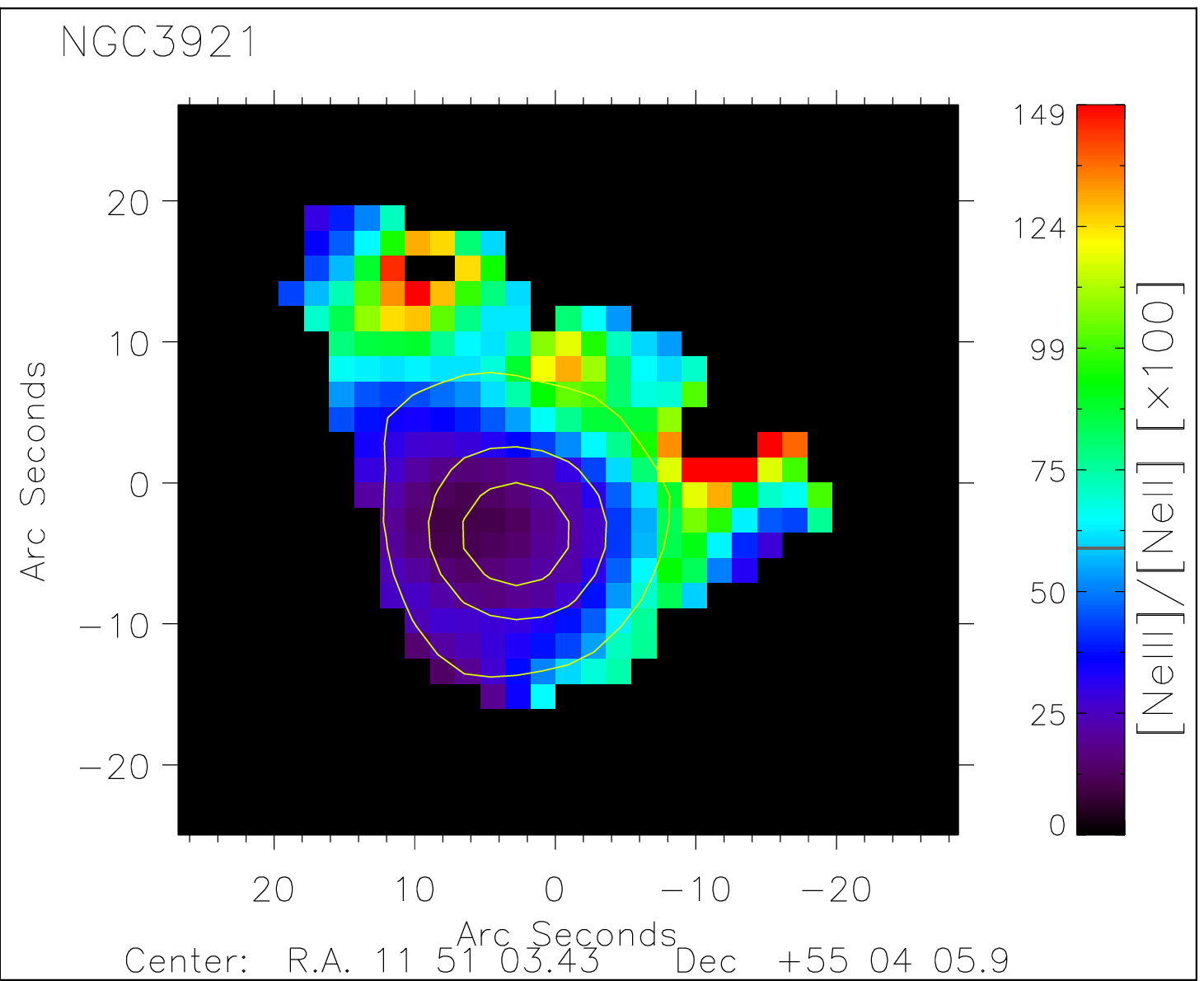}
\includegraphics[scale=0.43]{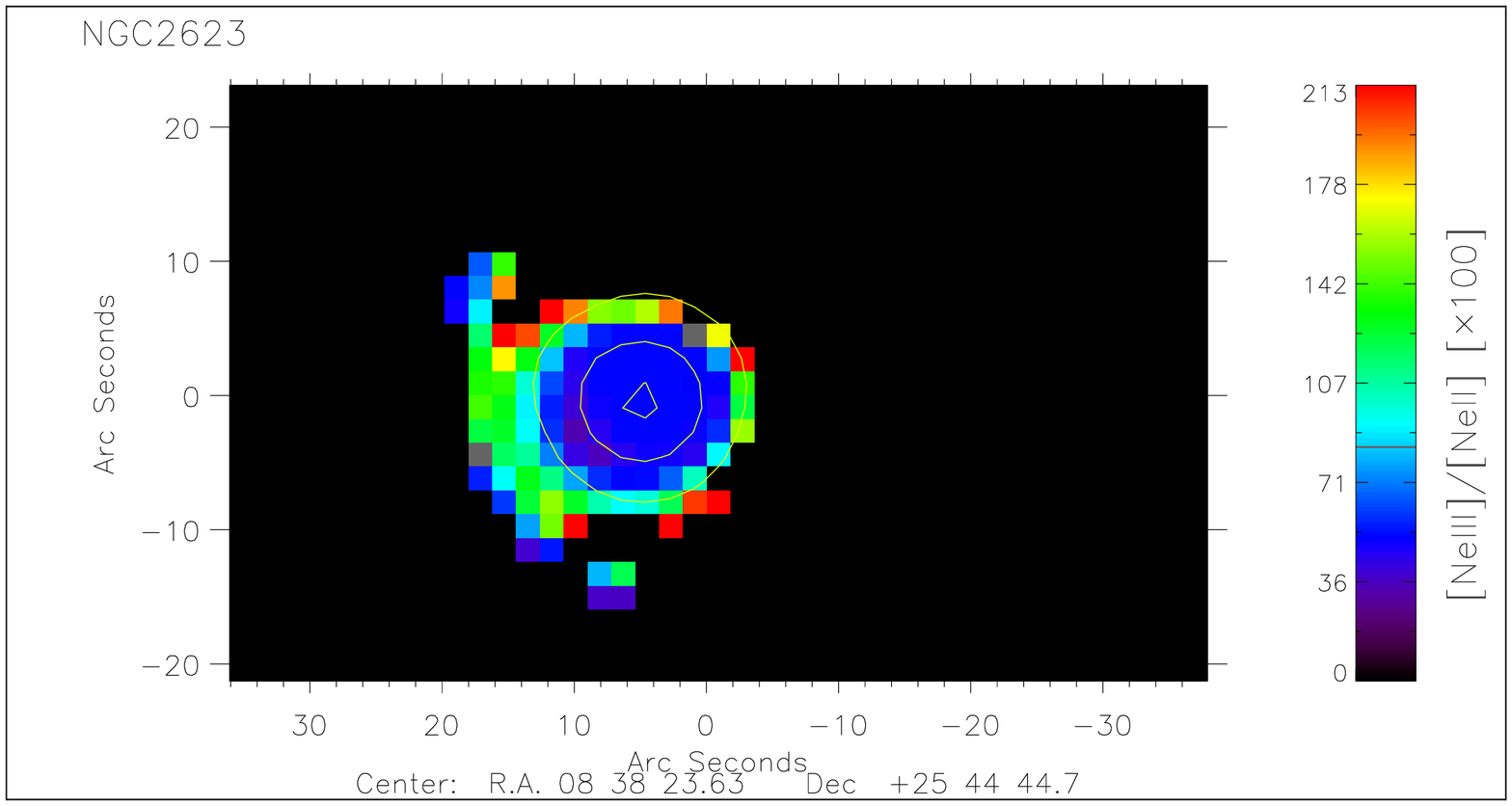}
\includegraphics[scale=0.43]{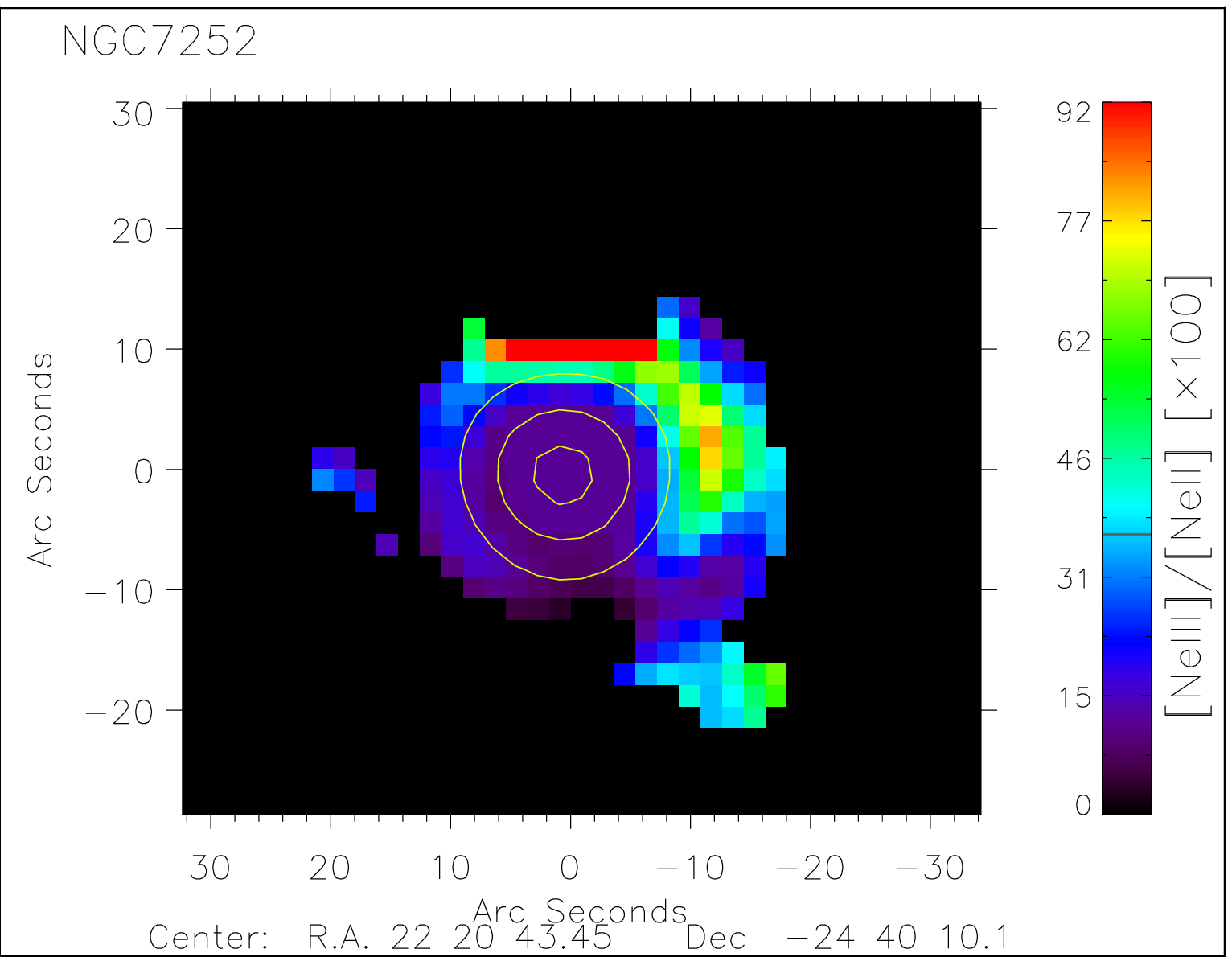}
\end{center}
\caption{\footnotesize{[Ne III]/[Ne II] line flux ratio maps for the Toomre sequence galaxies. The contours indicate the sum of the [Ne III] and [Ne II] flux. The North and East directions in each frame of this figure are the same as in the corresponding frame of Fig.~\ref{maps_mid-IR}.}}
\label{Ne_ratio}
\end{figure*}

The spectral maps of the [Ne III]/[Ne II] line ratios for our sample are shown in Fig.~\ref{Ne_ratio}. For all our galaxies the nuclei have the smallest [Ne III]/[Ne II] ratios (see also right panel of Fig.~\ref{PAH_scatter}), while the circumnuclear regions tend to show (a factor of 2--4) larger ratios. Intuitively, this is in contrast to the expectation of a decrease of the hardening of the radiation field with distance to the center due to the small number of high energy photons available at large distances to ionize Ne$^+$ to Ne$^{2+}$. However, our result is in agreement with other recent observations: \cite{Per10} have found that the minimum of the [Ne III]/[Ne II] ratio tends to be located at the nuclei for a sample of 15 LIRGs observed with Spitzer IRS; \cite{Tho00} have shown that H~II galaxies have a nuclear [Ne III]/[Ne II] ratio of a factor $\sim$2--3 smaller than those measured in the extra-nuclear regions. A similar behavior of icreasing [Ne III]/[Ne II] ratios with increasing galactocentric distances has been observed in the nearby galaxies NGC~253 \citep{Dev04}, M82 \citep{Bei08}, and NGC~891 \citep{Ran08}. Galactic HII regions follow the same trend, with HII regions at increasing distances from the galactic nucleus showing larger [Ne III]/[Ne II] ratios \citep{Giv02}. Our results confirm this trend, even if dust extinction corrections are taken into account as a function of position in the galaxies (using the \textit{SCREEN} model in PAHFIT). Metallicity-related effects are unlikely to be the dominant cause of the relative increase of [Ne III] strength outside the central regions of these galaxies \citep{Sni07, Per10}. A more likely explanation may be that a fraction of the most massive stars are hidden in ultra-compact HII regions \citep{Rig04}) due to high dust densities in the nuclear regions, leading, e.g., to an decrease of the [Ne III]/[Ne II] ratio by a factor of $\sim$2 if the density increases from 10$^3$ to 10$^5$ cm$^{-3}$ \citep{Sni07}. Another explanation for this behaviour might be attributed to the variation in the age of the stellar populations, with a larger contribution of the old stellar population in the nuclei (e.g. due to stellar bulges), subsequently decreasing the average stellar age (and ionization) towards the center. \par

\subsubsection{Comparison of Global Mid-IR Properties to Normal Star Formation Galaxies}

The SINGS galaxies, being mostly isolated, star-forming systems with a large range in IR power, are a perfect sample against which to compare the mid-IR properties of our major mergers. In particular one might expect to find different PAH populations between these two samples due to, e.g., a different interplay between radiation field and the ISM. Since the PAH 11.3~$\mu$m feature is thought to be produced by neutral PAHs, whereas the 7.7~$\mu$m feature arises primarily from PAH ions \citep{All99, Dra01}, a change in the neutral fraction should change the observed flux ratio. Harder radiation fields are also more efficient at destroying small PAH grains, an effect that may dominate the center of our galaxies. 
For all of our objects the range of the integrated (i.e., global) PAH strength ratios  6.2/7.7~$\mu$m and 11.3/17~$\mu$m are in the range of [0.2--0.4] and [1--2.5], respectively. 
We find a range in  the ratio 7.7/11.3 PAH bands of [1--6] and a range in [Ne III]/[Ne II] of [0.1--1]. The range and distribution of the ratios PAH 6.2/7.7~$\mu$m, 11.3/17~$\mu$m, and [Ne III]/[Ne II] is very similar to interband strength ratios found in the nuclei of 22 starburst galaxies \citep{Bra06} and for a large sample of SINGS galaxies \citep{Smi07}, including normal star formation galaxies and low-luminosity AGN (Seyfert and LINER). This suggests that the physical properties in major mergers are, at least on global scales, similar to those in normal starburst galaxies, but can be very different on local scales within the individual merger systems as shown in the following sections. However, we do not see any global ratios of [Ne III]/[Ne II]$>2$ in our sample, which is the case for five lower metallicity galaxies with H~II-dominated nuclei in the SINGS sample \citep{Smi07} as well as in most blue compact dwarfs and in galaxies with a strong AGN component \cite{Hao09}.

\subsection{Properties and Distribution of the Warm Molecular Gas Component}

\subsubsection{The Distribution of the Warm Molecular Gas}
\label{subsubsec:H2distro}
Excitation of the warm molecular gas observed in the rotational emission lines of H$_2$ (S(0)--S(7)) can commonly arise from at least three different mechanisms: UV-excitation in PDRs surrounding or adjacent to the H II regions; shocks that collisionally excite the H$_2$ molecules; or hard X-ray photons capable of penetrating the molecular clouds and heating large columns of gas \citep{Lep83,Mal96}. Observationally separating these primary excitation mechanisms is not trivial in other galaxies: high UV radiation as well as high-velocity gas motions are both present in the nuclei of active galaxies and, to a lesser extent, in starbursts. UV radiation can originate either from the existing stellar population, from current star-forming processes, or the AGN accretion disk. On the other hand, shock excitation can originate from the central engine, an outflow due to the large number of supernova remnants, or, finally from gas motions and cloud collisions due to galaxy-galaxy interactions  \citep{Shu78, Dra83}. In addition, the low rotational transitions arise from collisionally excited warm gas which complicates the distinction even further.
However, if H$_2$ emission is excited mainly by UV-radiation in PDRs, the fluxes of H$_2$ lines and PAH features should correlate closely, since PAH emission features arise from the same mechanism in PDRs.\par 

\begin{figure*}
\begin{center}
\includegraphics[scale=0.4]{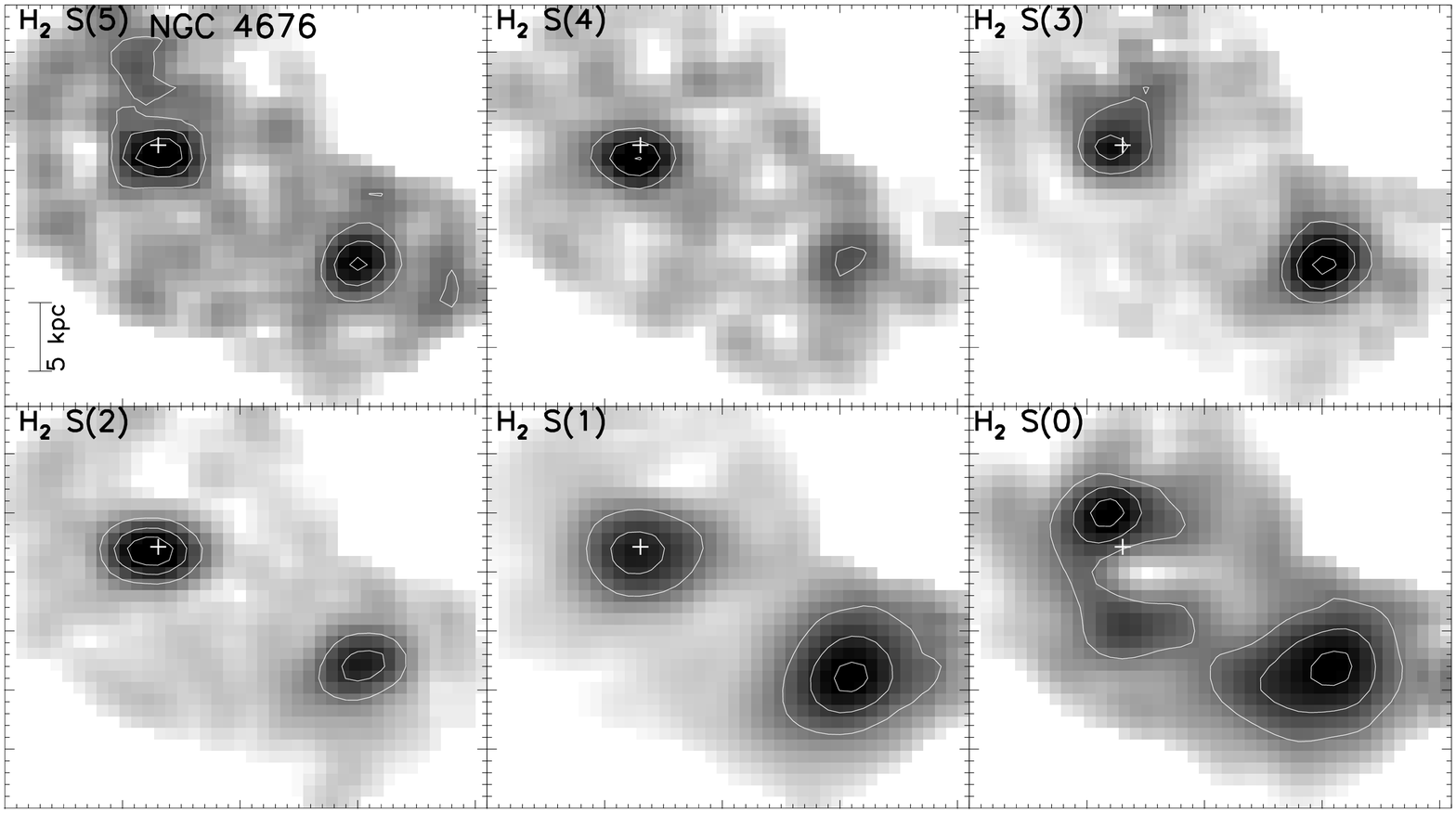}
\end{center}
\caption{\footnotesize{H$_2$ feature maps with contours (from S(5) at top left to S(0) at bottom right). The plus marker indicates the central peak in the mid-IR continuum emission. Figures for the rest of the sample are available in the online version of the Journal. The North and East directions in each frame of this figure are the same as in the corresponding frame of Fig.~\ref{maps_mid-IR}.}}
\label{H2_feature_4627}
\end{figure*}

In Fig.~\ref{H2_feature_4627} we have mapped the emission from several H$_2$ rotational lines for each merger system. Our comparison between the flux distribution of the strongest H$_2$ emission line  (the extinction corrected 9.7~$\mu$m H$_2$ emission line) and PAH (using the combined main PAH features in the SL wavelength range at 6.1, 7.7, and 11.3~$\mu$m) shows that (1)  H$_2$ has typically a larger extent than PAH with an average ratio of H$_2$/PAH radius of 1.5 (ranging from 1.1 to 1.9, including PSF size corrections), (2) some galaxies show a significantly different H$_2$/PAH flux ratio between different merger components, most significantly in the early merger systems NGC~4676 and NGC~7592 (see \S~\ref{sec:indgalaxies}), and (3) at least one of our galaxies (NGC~3921) shows signs of a tail of H$_2$ emission not associated with a strong PAH feature (see Fig.~\ref{maps_feature_4627}).\par

Five galaxies (NGC~7592, NGC~6621, NGC~6240, NGC~520, NGC~2623) show in Fig.~\ref{H2_feature_4627} an offset of the peak H$_2$ emission at 9.66~$\mu$m from the stellar nucleus, which is not seen in any PAH or other H$_2$ features (e.g., S(2) and S(4)). One possible explanation for this offset is an underestimation of dust extinction (silicate absorption at 9.7~$\mu$m) which is strongest at the center of the galaxy, and hence suppresses the apparent H$_2$ flux at 9.66~$\mu$m. However, even without the 9.7~$\mu$m H$_2$ emission line, we find spatial deviations from a constant relation between the warm molecular gas (H$_2$) and PAH (as described above), indicating that, at least in some merger galaxies, different physical conditions are present in the ISM \citep[e.g., electron density, gas temperature, or UV radiation field intensity; for a review see][]{Gal08}. The nature of the variation in the H$_2$/PAH ratio will be discussed in more detail in \S~\ref{subsec:PDR}.

\subsubsection{Spatially Integrated Temperatures and Masses}
\label{subsubsec:H2_temp}
\begin{figure*}
\begin{center}
\includegraphics[scale=0.6]{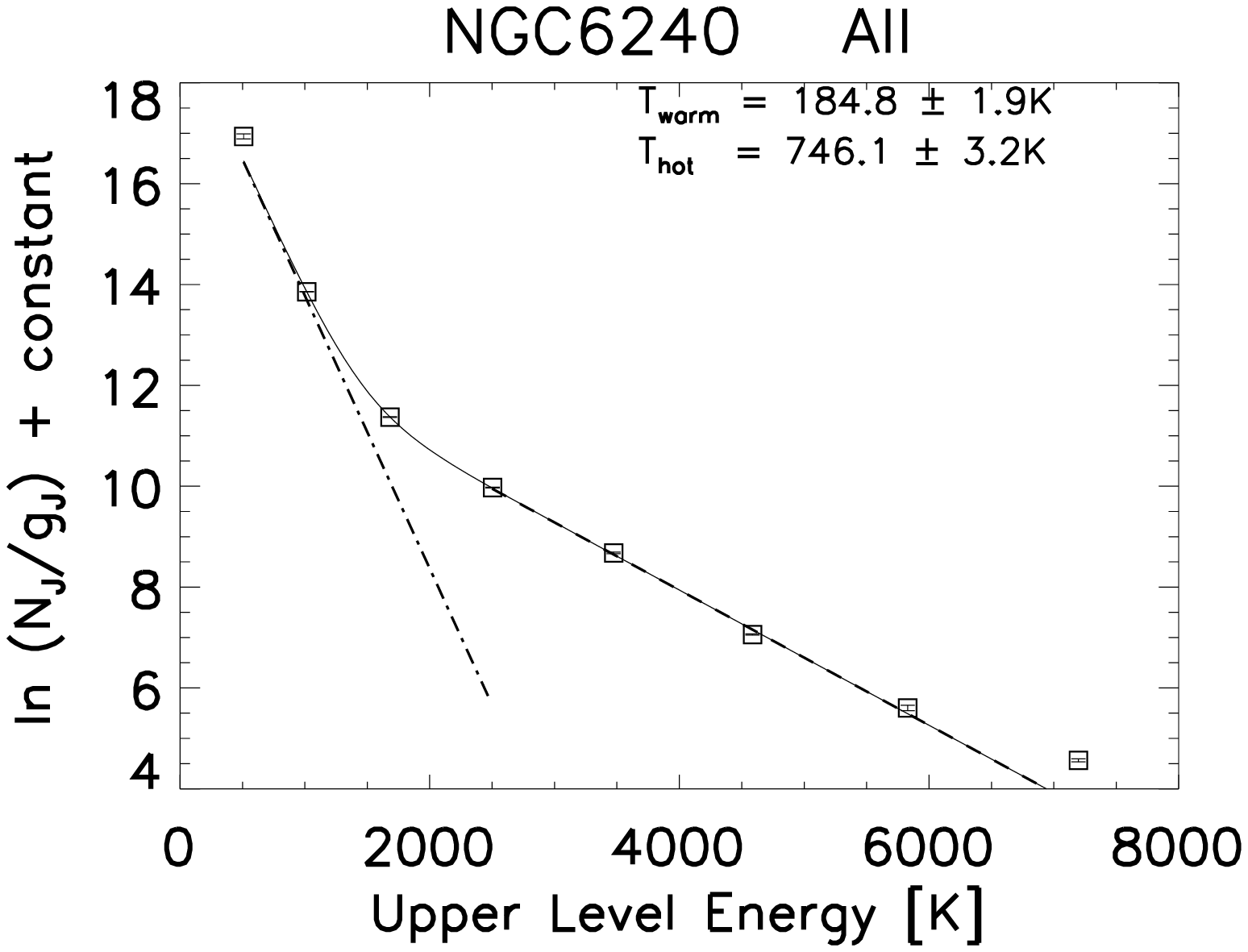}
\end{center}
\caption{\footnotesize{H$_2$ excitation diagram for NGC~6240, showing the upper level population divided by the level degeneracy as a function of upper level energy (for the lines observed with the IRS). The lines are measured with PAHFIT, including a correction for extinction. Uncertainties are indicated as vertical bars for each line. A double exponential profile fit (solid line) takes into account two temperature components (see text), separately indicated as dashed and dashed dotted lines. Figures for the rest of the sample are available in the online version of the Journal.}
\label{temp_diag_6240}}
\end{figure*}

Temperatures and masses of the warm molecular hydrogen (T$<$2000~K) are derived from the pure rotational H$_2$ lines. Fig.~\ref{temp_diag_6240} shows the excitation diagrams using the extinction corrected line fluxes (see Table~\ref{tab_H2}) for different regions of the merger systems. In all cases an ortho-to-para ratio of 3 is assumed \citep{Rig02,Rou07}. The excitation diagram is basically a plot of the natural logarithm of the column density $N_i$ divided by the statistical weight $g(i)$ in the upper level of each transition against the energy level $E_i$. A detailed description can be found in \cite{Rou07}, including a discussion about the implications of using different ortho-to-para ratios.\par 

The excitation temperature of the line-emitting gas is the reciprocal of the slope of the excitation diagram. If the H$_2$ gas is characterized by a single temperature (and assuming a single ortho-to-para ratio, the data will lie on a straight line, corresponding to a single component fit. We find that a simple single temperature fit does not work for most of our galaxies, since our data show a change of the slope with energy level. This is due to the fact that the gas phase is in reality more complex and consists of various components at different temperatures. The S(4) to S(7) lines (upper energy level from $\sim$3500 to 7200 K) show in general a much smaller slope than the slope inferred from the S(0) and S(1) lines (upper energy level of $\sim$500 to 1000 K), representing a "hot" and "warm" temperature component of the gas, respectively. To take into account these two temperature components, we fitted a double exponential profile to our data, given as
\begin{equation}
\frac{N_i}{g_i} = c_0 + c_1\mathrm{exp}(E_i /T_{warm}) + c_2\mathrm{exp}(E_i /T_{hot})
\end{equation}
where $N_i$ is the number of molecules, $g_i$ the statistical weight of each state, $E_i$ the upper Energy level, c$_{0-2}$ constants, and $T_{warm}$ and $T_{hot}$  the two excitation temperatures. This fit has the advantage that it simultaneously fits two components, rather than starting with two independent temperature estimates using different lines and subsequently subtracting the contribution of the hot component from the warm component.\par 

Fig.~\ref{temp_diag_6240} shows the excitation diagrams and the fits to the warm and hot H$_2$ phases for the different regions of our merger systems. For all our objects, except for NGC~6621, the  excitation diagram of the entire merger system could be well approximated by a two temperature component fit. Only the warm component could be measured for NGC~6621 because the H$_2$ lines S(4)-S(6) are not detected.
We find for our sample a mean temperature of the warm H$_2$ phase of 255K, but with a large variance between different regions, ranging from 87~K (NGC~520 South) to 348~K (NGC~7592 South). The mean temperature of the hot phase for our sample is 830~K, ranging from 550K (NGC~520 South) to 1079~K (NGC~7592 South). In comparison to other studies, our average excitation temperature of the warm component seems to be slightly larger than in the nuclear region of a sample of SINGS galaxies \citep[$\sim$200~K;][]{Rou07}, but smaller than in the nuclear region of starburst galaxies \citep[$\sim$360~K;][]{Ber09} and ULIRGs \citep[$\sim$336~K;][]{Hig06}. Note that this comparison does neither take into account the different sizes over which the excitation temperature is averaged nor the large differences in warm molecular masses involved.\par 

We used the two excitation temperatures $T_{warm}$ and $T_{hot}$ to estimate the total warm and hot gas masses \citep{Rig02}, $M_{warm}$ and $M_{hot}$, in our sources, which are presented in Table~\ref{tab_temp}. The total mass of the warm component (median $M_{warm}=3.3 \times 10^7$~M$_\sun$) is in general at least an order of magnitude larger than the total mass of the hot component (median $M_{hot}=0.1 \times 10^7$~M$_\sun$).

\subsubsection{Temperature maps}
\begin{figure*}
\begin{center}
\includegraphics[scale=0.6]{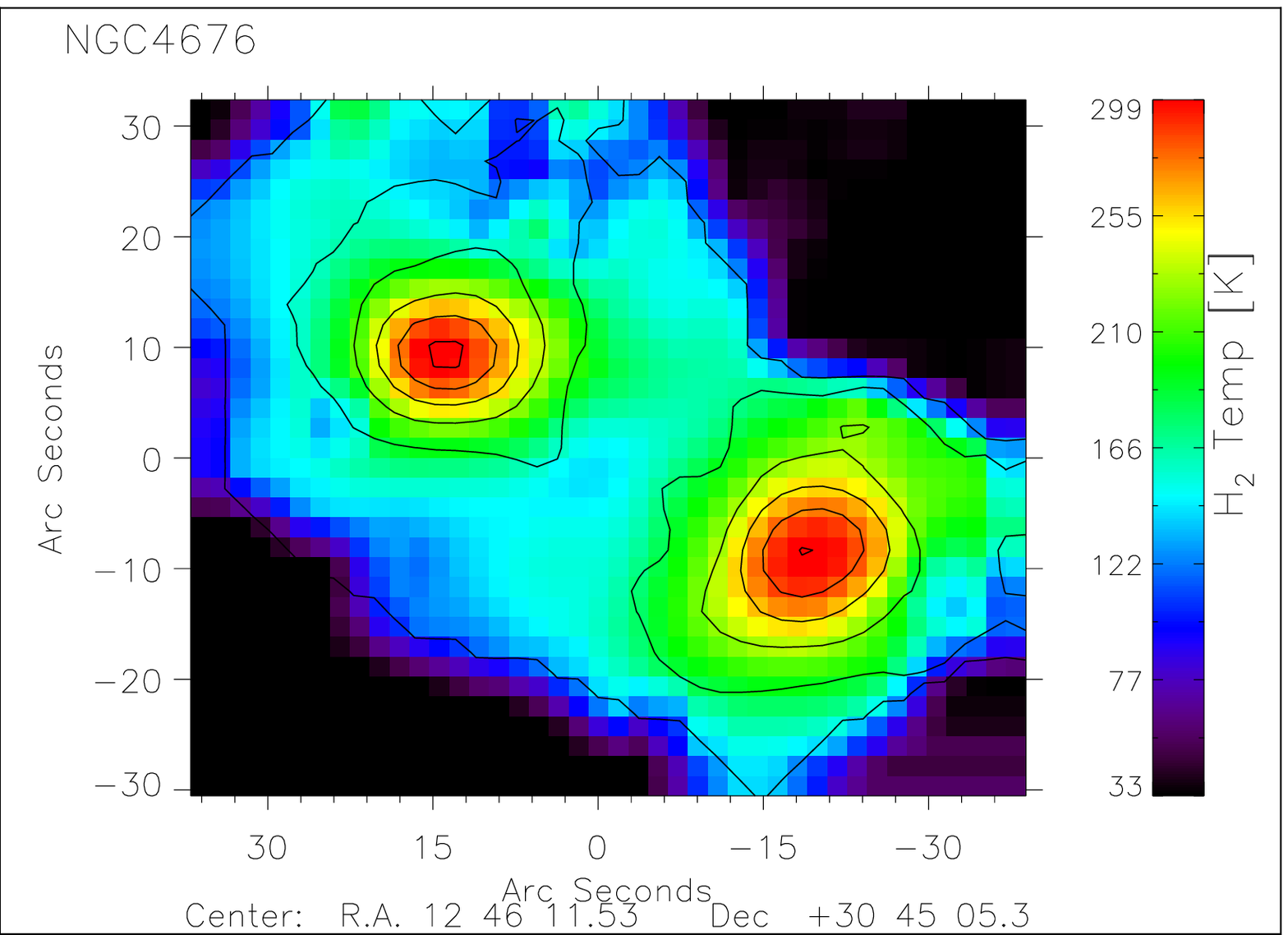}
\end{center}
\caption{\footnotesize{The single-component H$_2$ temperature map (in colors) with the combined H$_2$ emission (in contours) for NGC~4676. Figures for the rest of the sample are available in the online version of the Journal.}}
\label{temp_maps_4676}
\end{figure*}

\begin{figure*}
\begin{center}
\includegraphics[scale=0.5]{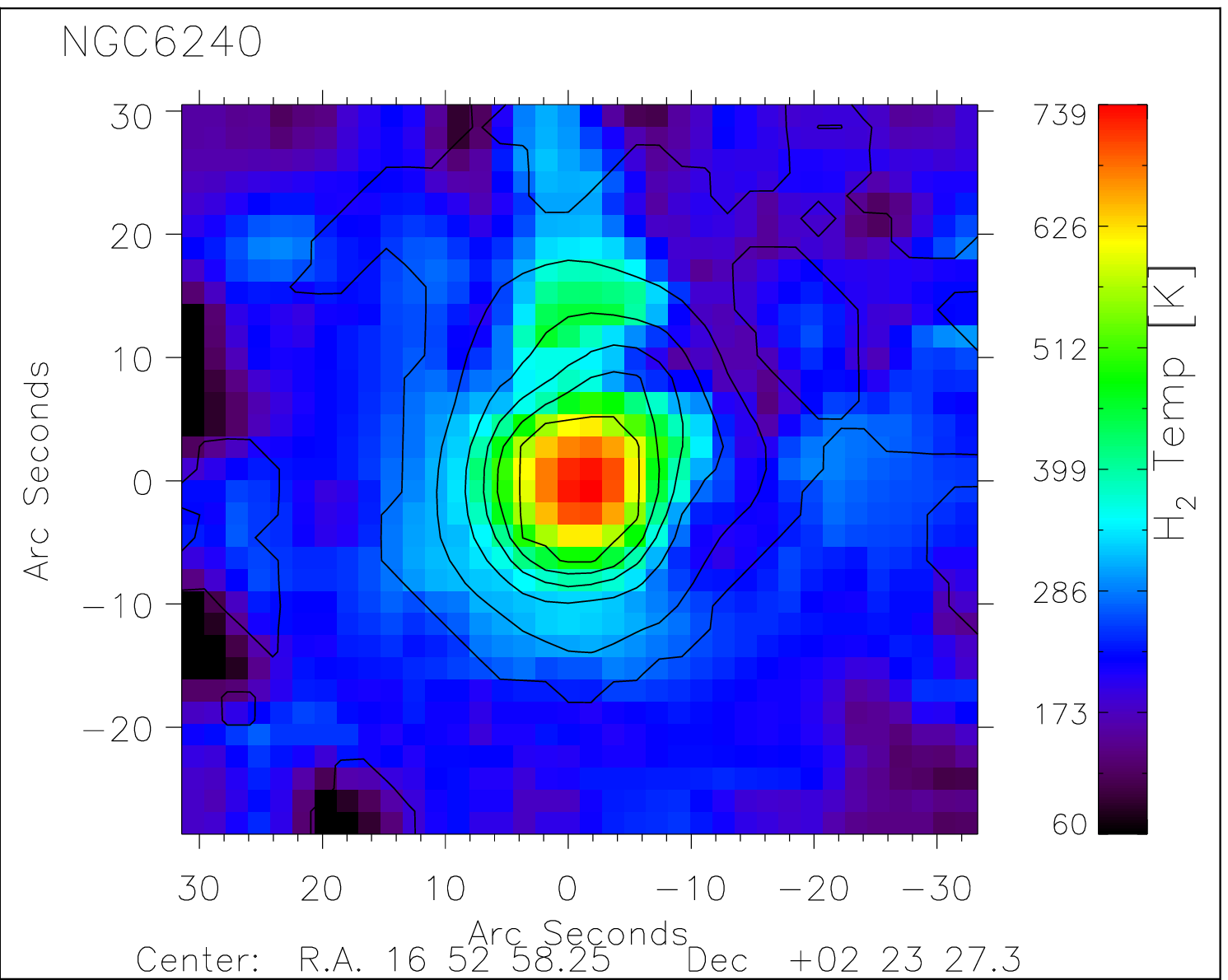}
\includegraphics[scale=0.5]{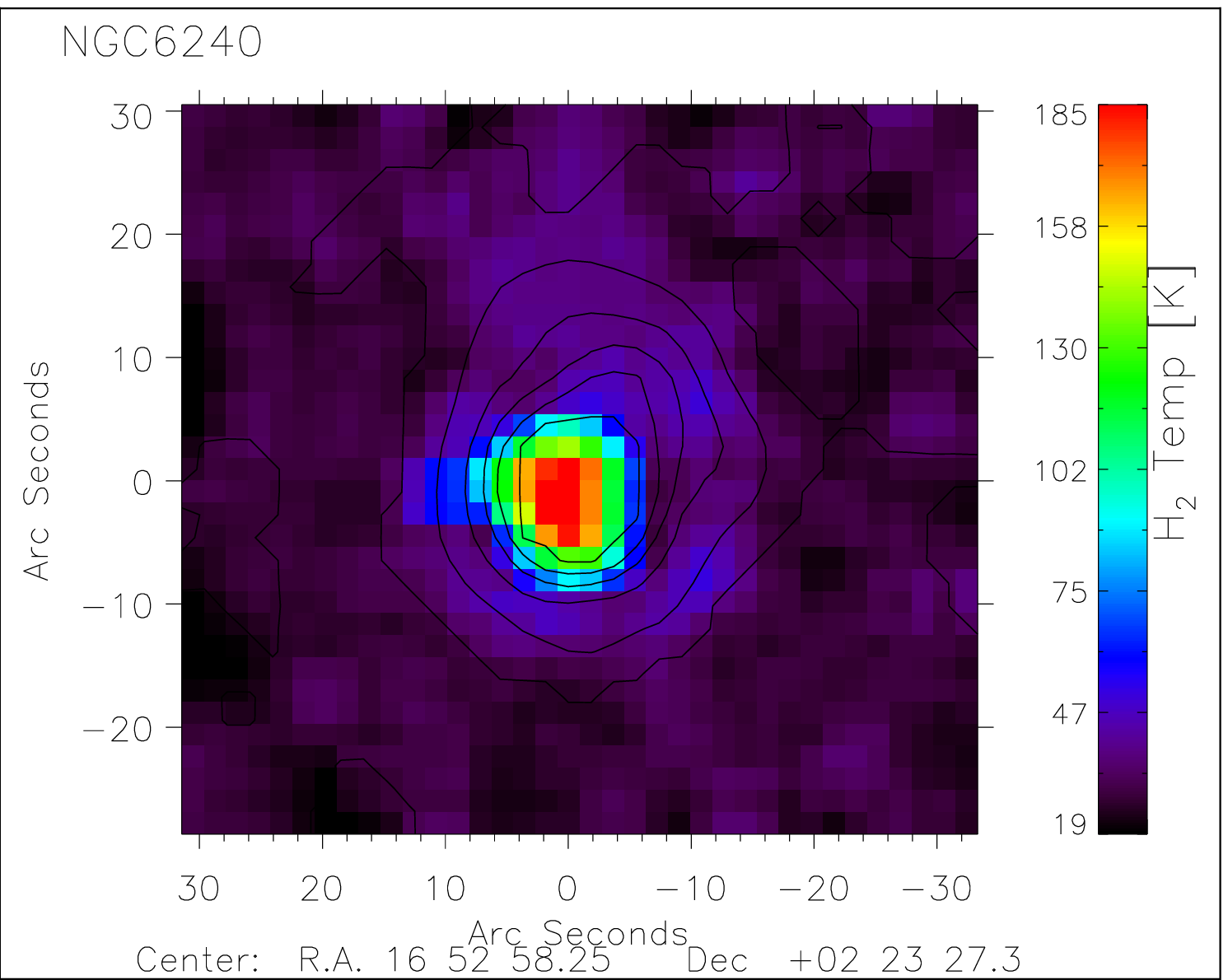}
\end{center}
\caption{\footnotesize{The two H$_2$ temperature maps (left: hot, right: warm component) for NGC~6240, with the combined H$_2$ emission in contours.}}
\label{temp_maps_6240}
\end{figure*}

To reveal spatial variations of the H$_2$ temperature in the merger components, we compute H$_2$ temperature maps as shown in Fig.~\ref{temp_maps_4676}. In only one galaxy, NGC~6240, does the H$_2$ line emission from the the hot component (S(4)-S(7)) have a sufficient signal-to-noise ratio per pixel to employ a two component temperature fit per pixel. (Fig.~\ref{temp_maps_6240}). For all other galaxies a one-component fit (per pixel) reveals similar results on the warm component as the two-component fit (per pixel) due to a limited number of significant H$_2$ lines detected per pixel. Note that this does not imply that these galaxies have only one temperature component per pixel, since our spatially integrated flux measurements (with higher signal-to-noise) show that almost all galaxies have at least two temperature components (\S~\ref{subsubsec:H2_temp}). Hence, the absolute temperature scale is not neccessarily equal to the spatially integrated temperature, which is due to smaller signal-to-noise ratios on pixel-scale. The absolute temperatures are given in the previous section (see also Table~\ref{tab_temp}) and we focus instead on the temperature distribution in this section. In general, we find an increase in temperature towards the peak of the combined H$_2$ flux, which is in most cases aligned with the center of the main galaxy component. We find no $H_2$ hotspots between merger galaxies. An exception is NGC~6240 where we find a faint second temperature peak (500~K), $\sim$7~kpc eastwards, not in alignment with the $H_2$ flux intensity nor any other mid-IR feature or NIR concentration. Although both gas temperature components are mapped, only the ``hot'' component exhibits this secondary temperature peak.

\subsubsection{Excitation Mechanisms: Evidence for an Intrinsic Relation between H$_2$ and PAH Emission}
\label{subsec:PDR}

One of the major excitation sources in starburst galaxies is the far-ultraviolet (FUV) radiation of massive stars in photo-dissociation regions (PDRs), with photon energies between 6 and 13.6 eV \citep{Hol97}. PDRs are attenuated by the absorption and scattering of UV photons from a hot star, generating several layers with different temperatures and material components.
In general, PAH and H$_2$ can be expected to be co-spatial since both species can be excited by FUV photons (as discussed in \S~\ref{subsubsec:H2distro}). Galactic studies of the molecular hydrogen emission in PDRs suggest that the formation of H$_2$ occurs on the surfaces of PAHs \citep[see, e.g.,][]{Hab03, Vel08}. A similar correlation between H$_2$ and PAH emission is seen in external galaxies \citep{Rig02, Rou07}. In particular, \cite{Rou07} found for normal star-forming galaxies, on average, a constant flux ratio of H$_2$(S0-S2)/PAH$\sim$0.008 over a large range of  total H$_2$ fluxes, suggesting relatively constant physical conditions in PDRs. However, this study was limited a) to the average emission of only the H II nuclei and complexes within the kpc-scale circumnuclear region for each galaxy, and b) by the fact that the power emitted in the aromatic bands was estimated from the IRAC4 filter rather than via direct spectroscopic measurement of PAH emission lines. On the other hand, a recent study of ULIRGs suggest that the H$_2$ emission does not come from PAH regions (PDR's) and originates outside the dust obscured regions due to shocks in the surrounding material \citep{Zak10}.\par

Since H$_2$ emission can arise from at least three different mechanisms (UV-excitation in PDRs, shocks, or hard X-ray radiation from an AGN), one important question is whether there is one single excitation mechanism that dominates the radiation power emitted in the mid-IR or a mix of various excitation mechanisms in mergers. In principle, studying the relation between PAH and rotational H$_2$ emission over various spatial scales allows us to constrain the role of PDRs, as both, PAH and H$_2$, are tightly correlated with the physical conditions in PDRs (in particular with the radiation field intensity $G_0$ and the hydrogen density $n_H$). Since most of the warm H$_2$, at the lowest rotational temperatures, emits primarily in the S(0) to S(2) lines, whereas the S(3) line emission has a noticeably higher contribution from hotter H$_2$, probably indicating more mixed excitation, a useful measure of the warm H$_2$ in star-forming galaxies is the sum of the S(0) to S(2) lines \citep{Rou07}.\par

\begin{figure*}
\begin{center}
\includegraphics[scale=0.6]{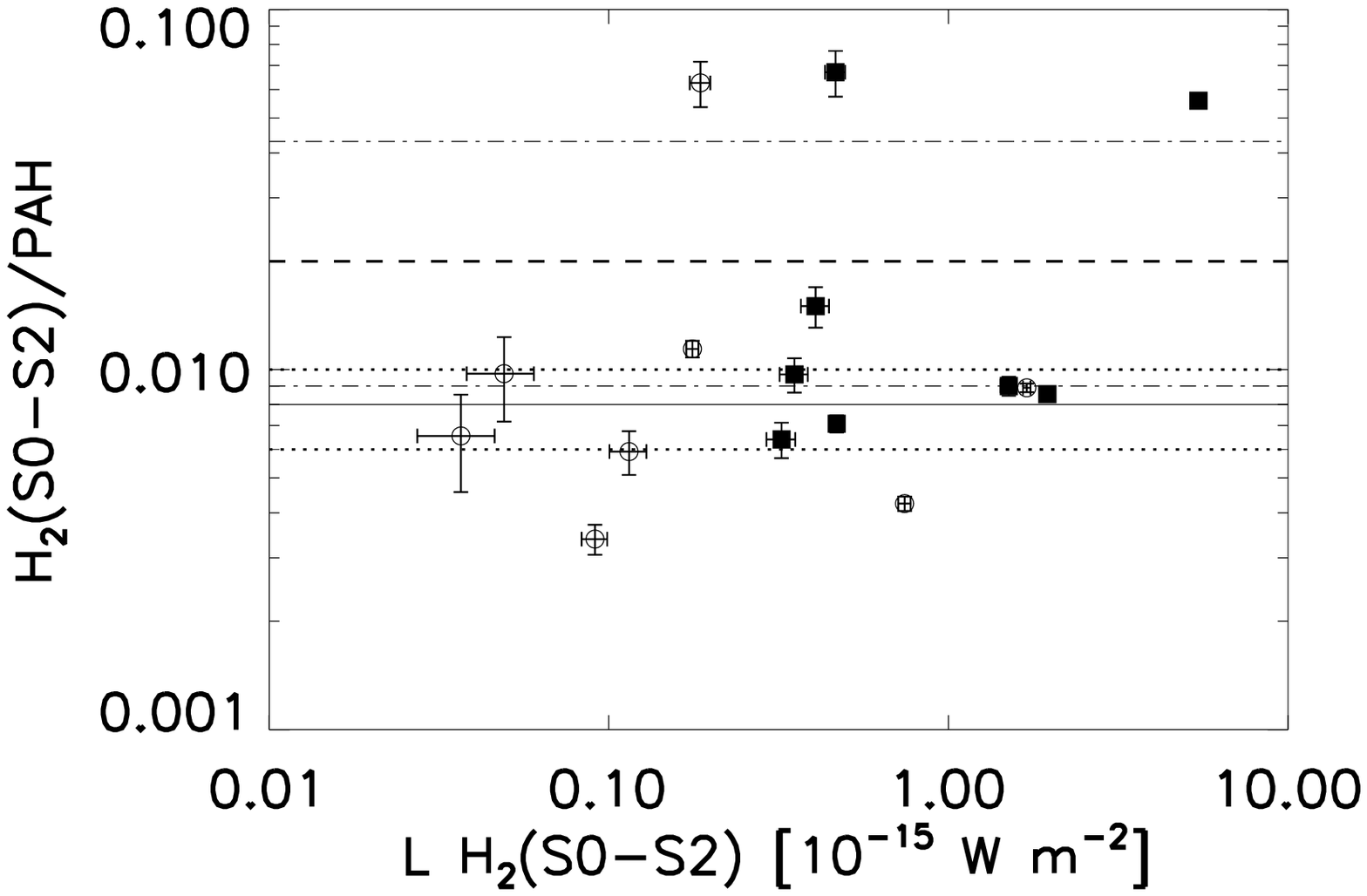}
\end{center}
\caption{\footnotesize{Ratio of the power emitted in the sum of the H$_2$ S(0) to S(2) transitions to the power emitted in the main PAH bands (at 6.2, 7.7, 11.3, and 17~$\mu$m) over a large range of H$_2$ luminosities. Measurements of the entire merger system are marked with filled squares while the individual merger components are marked with open circles. The solid (dotted) and dashed (dashed-dotted) line represents the mean (uncertainty) H$_2$(S0--S2)/PAH value measured for the center of normal star-forming galaxies and LINER/Seyfert nuclei in the SINGS sample, respectively. \citep{Rou07}.}}
\label{H2-PAH_all}
\end{figure*}

\begin{figure*}
\begin{center}
\includegraphics[scale=0.36]{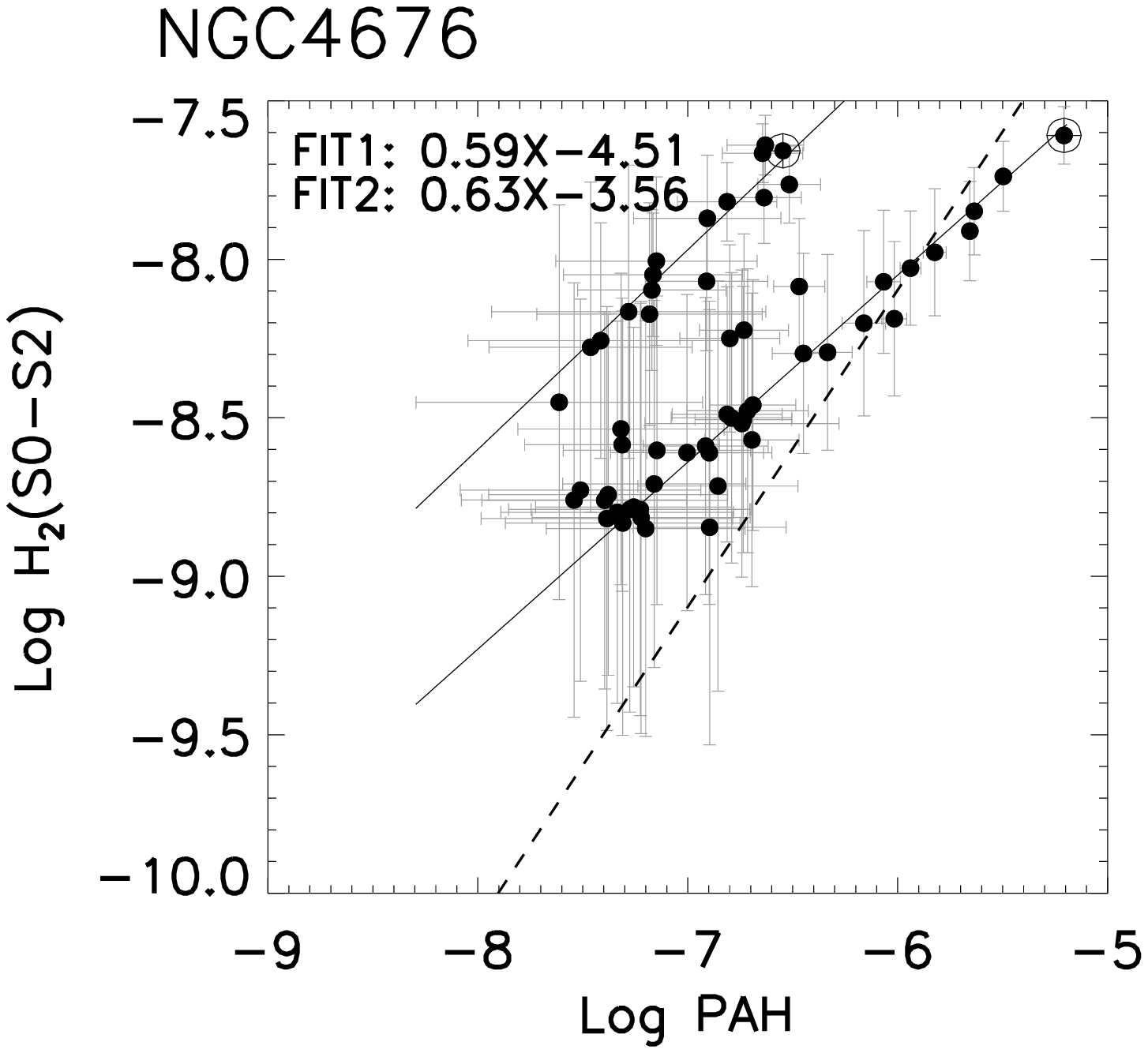}
\includegraphics[scale=0.36]{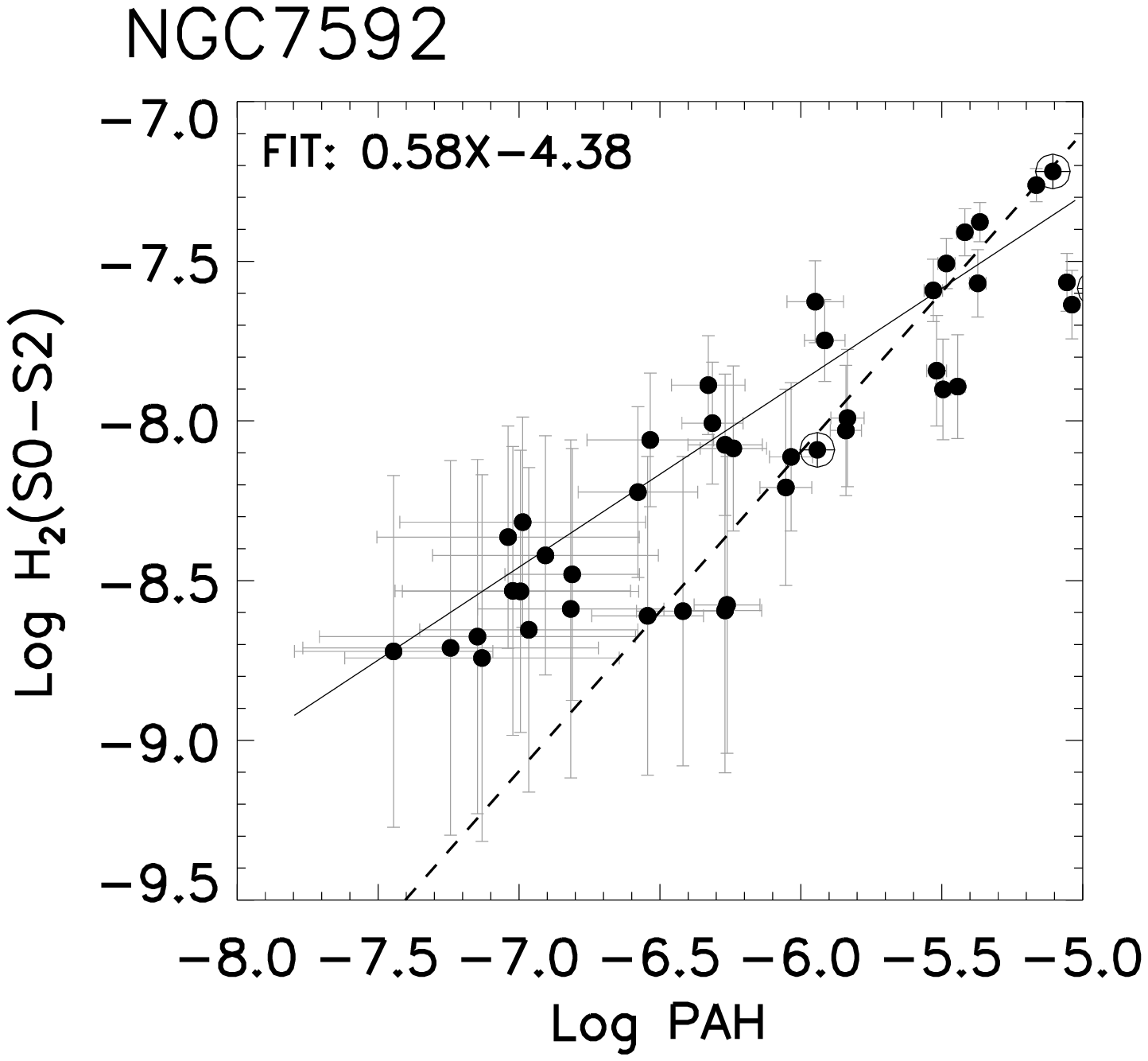}
\includegraphics[scale=0.36]{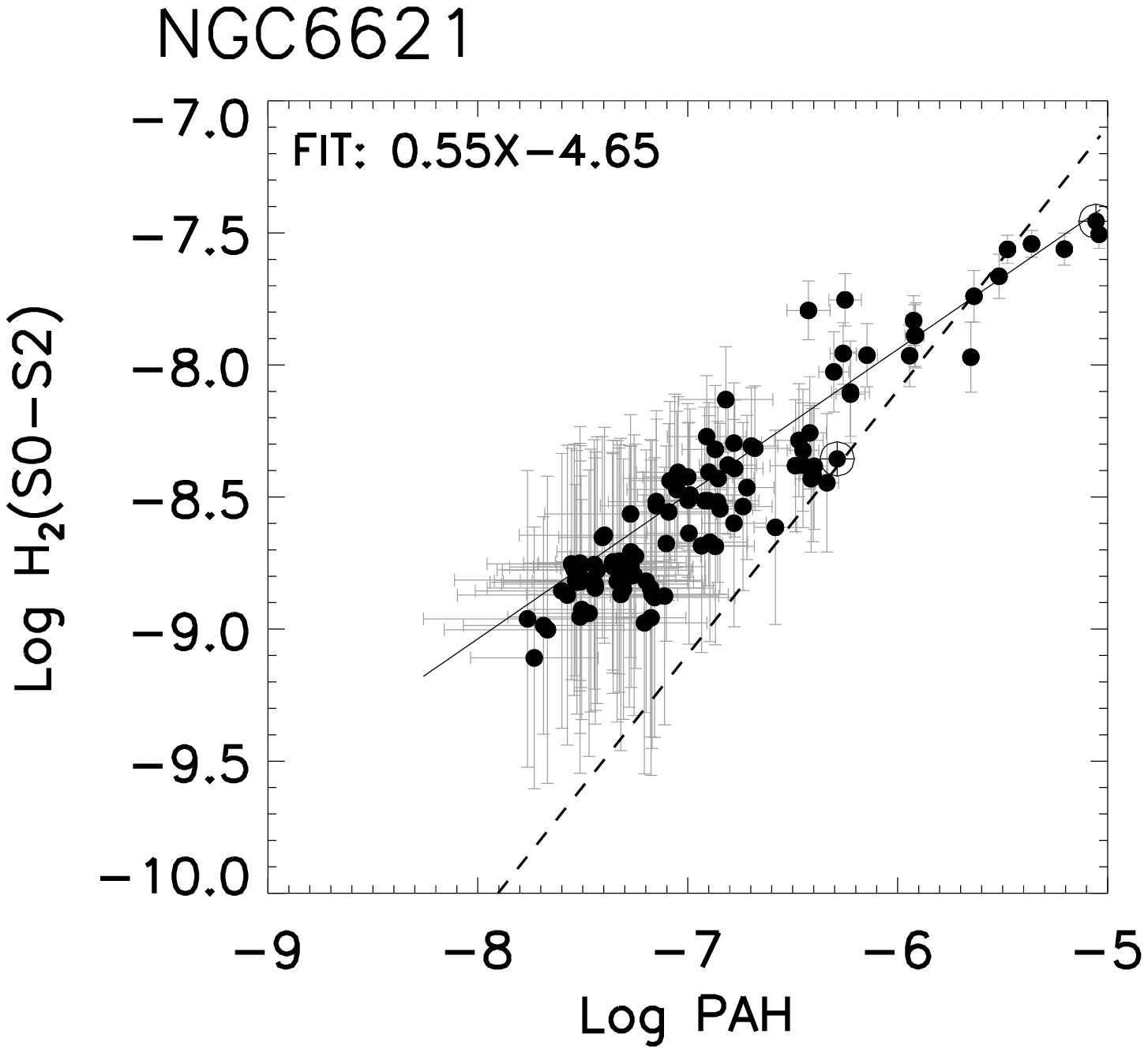}
\includegraphics[scale=0.36]{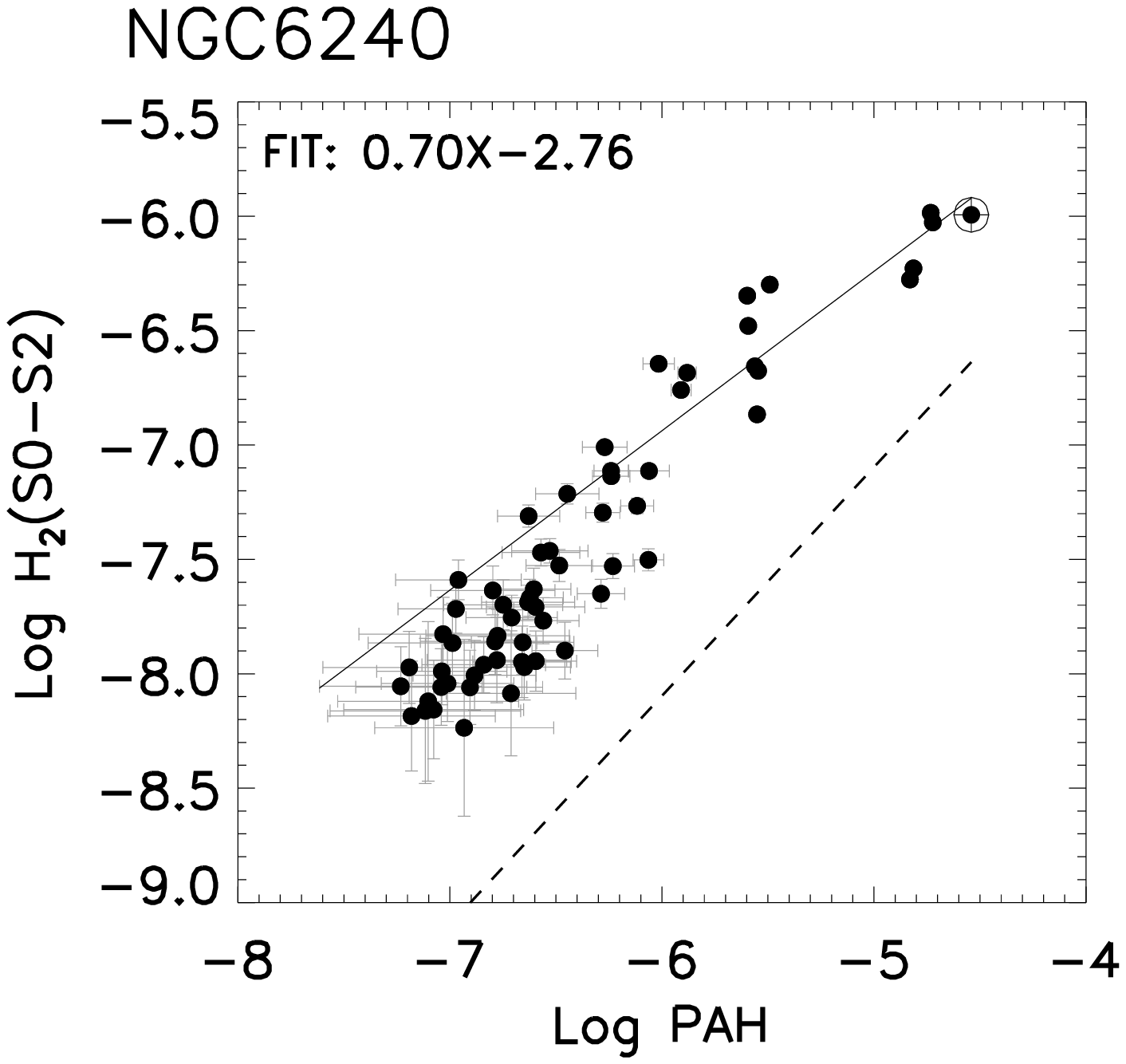}
\includegraphics[scale=0.36]{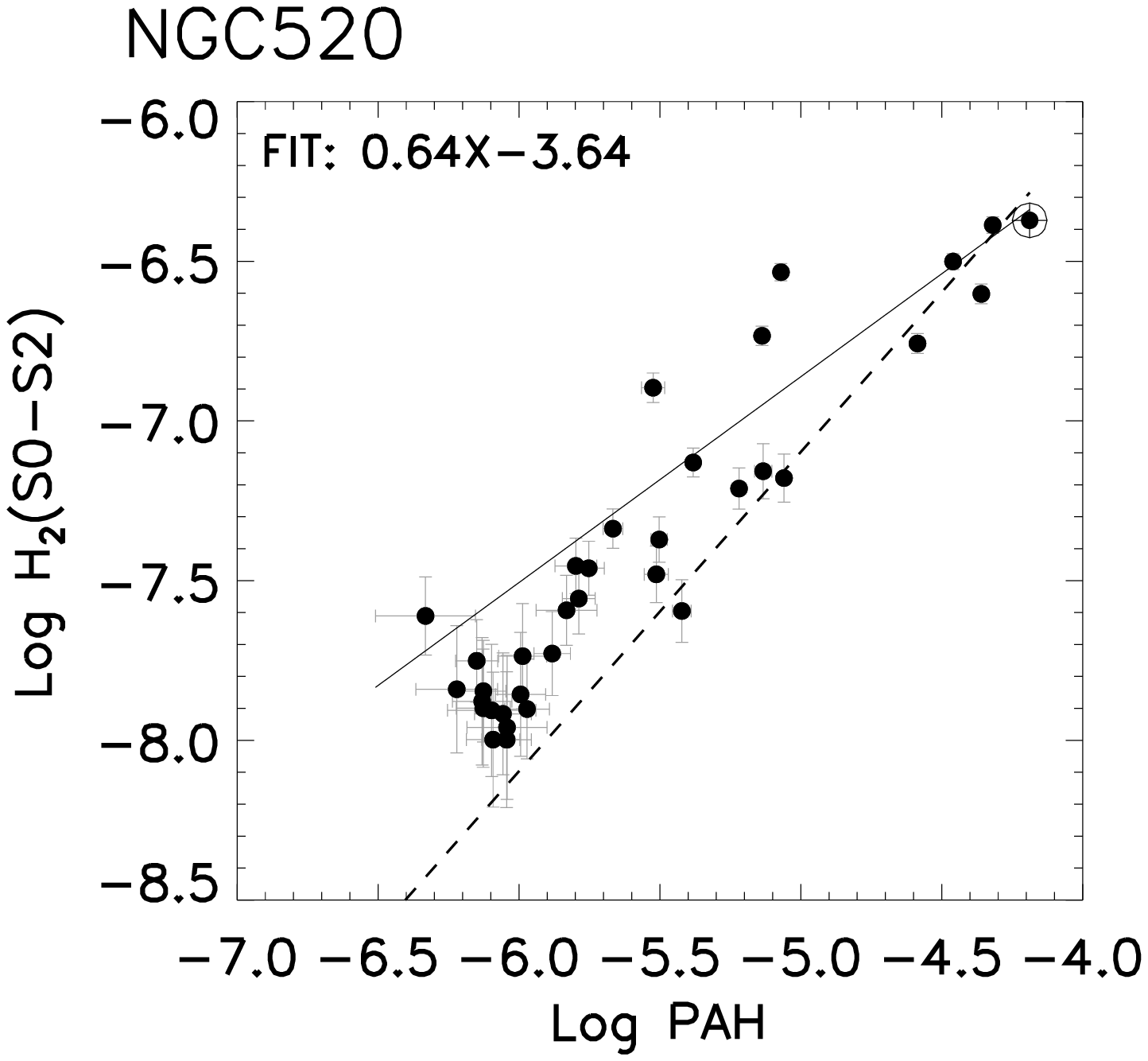}
\includegraphics[scale=0.36]{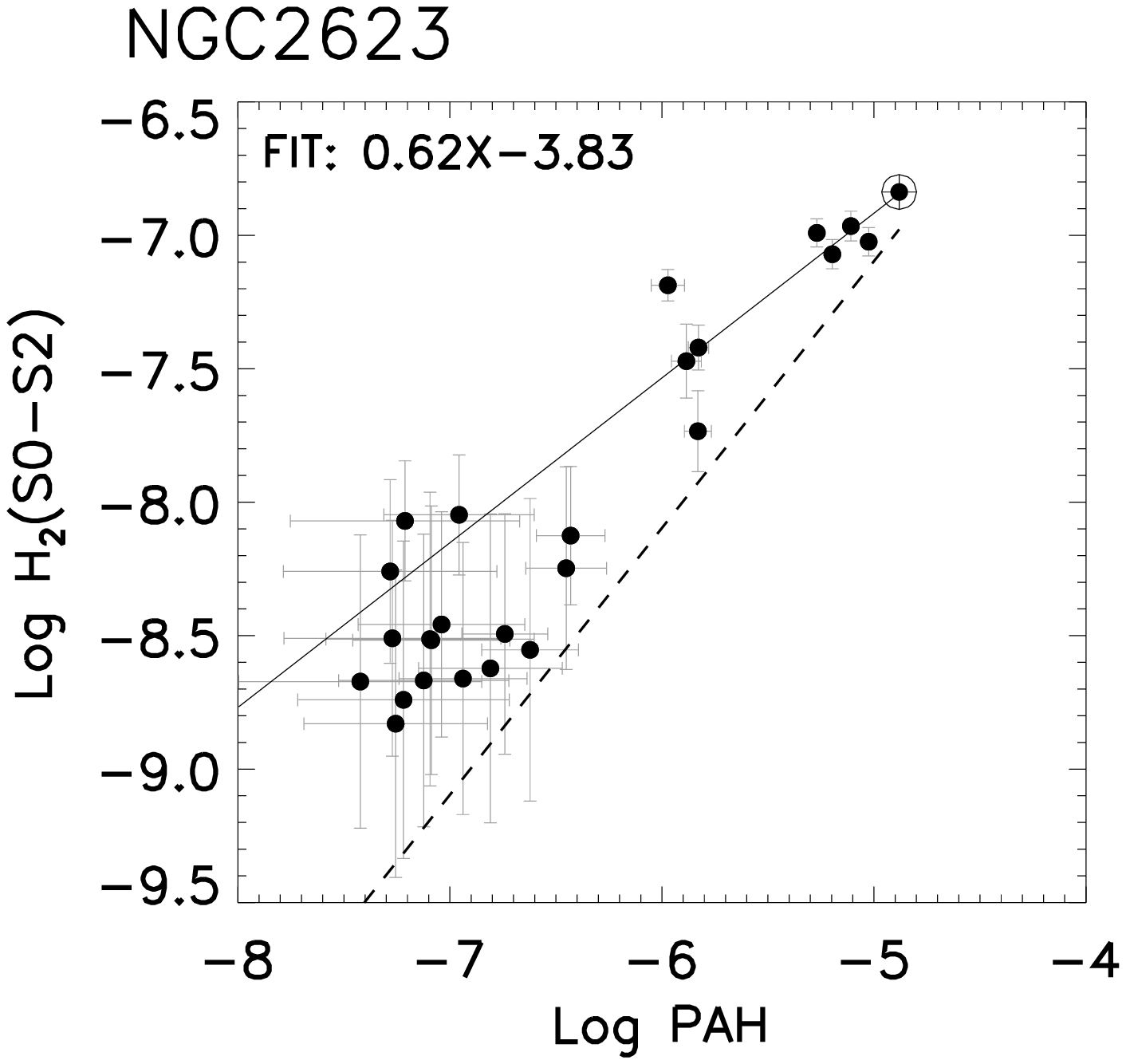}
\includegraphics[scale=0.36]{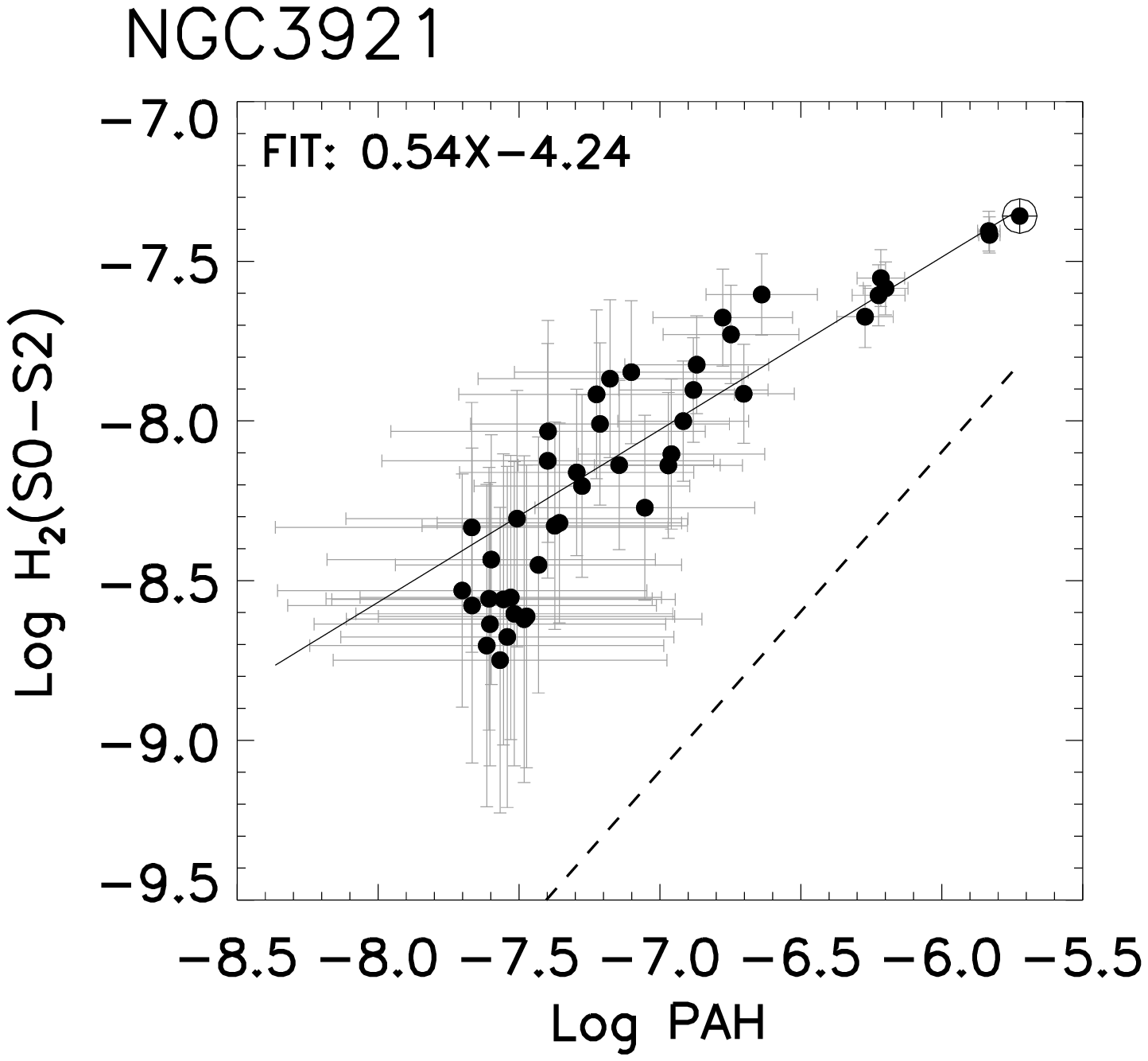}
\includegraphics[scale=0.36]{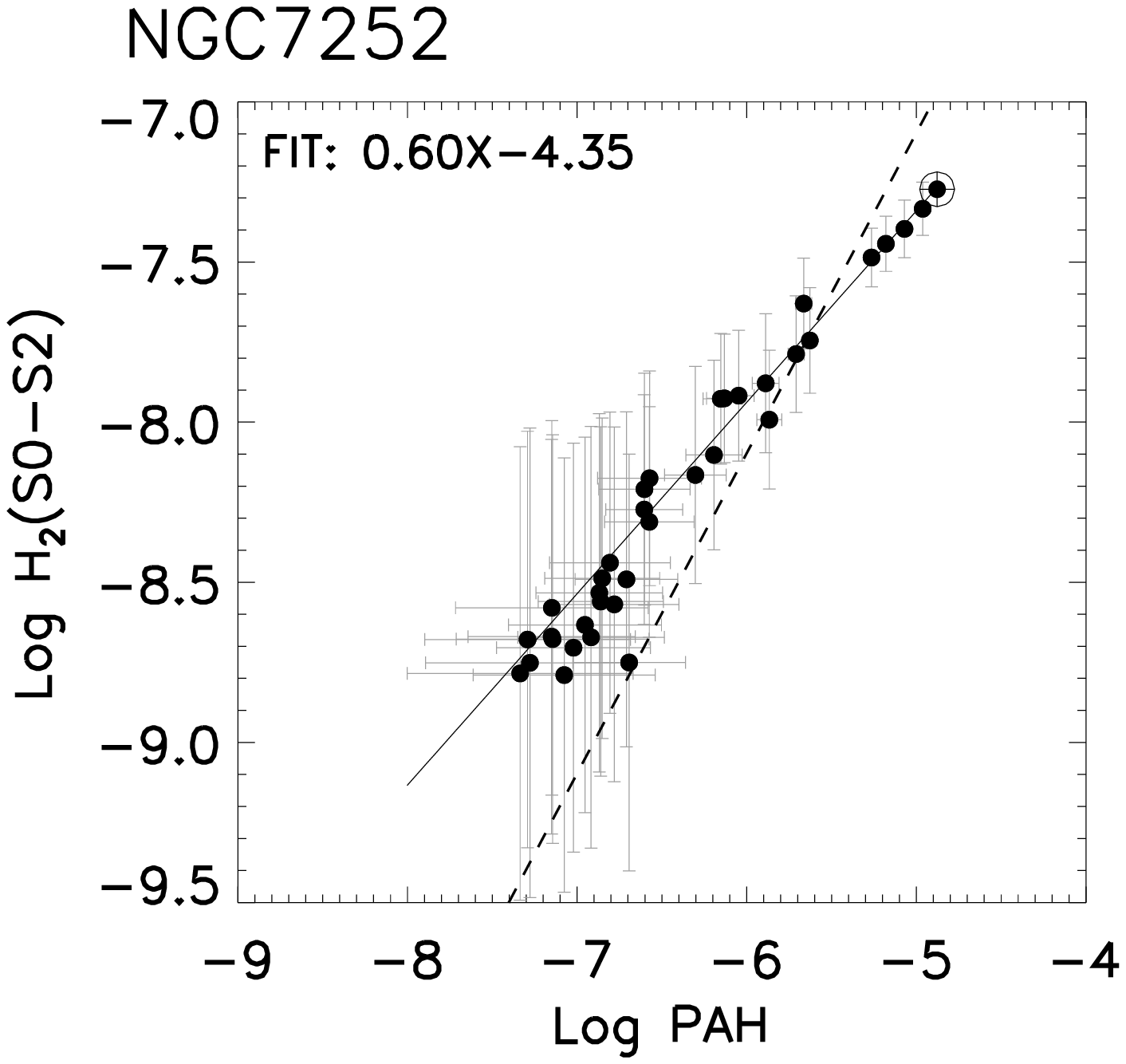}
\end{center}
\caption{\footnotesize{The power emitted in the sum of the H$_2$ S(0) to S(2) transitions as a function of the power emitted in the main PAH bands (at 6.2, 7.7, 11.3, and 17~$\mu$m) in logarithmic scales and in units of W~m$^{-2}$sr$^{-1}$. The crossed circles represent the central aperture of a merger component. The distribution is fitted using a power law (solid line), given as log[H$_2$(S0--S2]$=k\times$log[PAH]$+ c$ with the slope $k$ and the constant $c$. The dashed line represents the constant ratio model for the center of normal star-forming galaxies, given as H$_2$(S0--S2)/PAH=0.008 \citep{Rou07}.}}
\label{H2-PAH_ind}
\end{figure*}

Here we study the relation between the combined H$_2$ flux (from S0 to S2) and the sum of the four main PAH bands (at 6.2, 7.7., 11.3, and 17~$\mu$m) for various spatial locations in major merger systems (integrated over apertures of $3\times3$ pixels, or $5.6\times5.6\arcsec$, corresponding to $\sim$2~kpc at a mean distance of 79~Mpc). In Fig.~\ref{H2-PAH_all} we present the total flux ratios H$_2$(S0--S2)]/PAH as a function of the total H$_2$(S0--S2) luminosity, while Fig.~\ref{H2-PAH_ind} shows the logarithmic relation between H$_2$(S0-S2) and PAH for each galaxy. The following results are derived, and compared to the center of normal star-forming galaxies \citep{Rou07}:
\begin{enumerate}
\item On global scales (integrated over the entire merger component), the majority of galaxies in our merger sample have similar total H$_2$(S0-S2)/PAH ratios (mean: $0.008\pm0.004$) as normal star-forming galaxies (H$_2$(S0--S2)/PAH$\sim$0.008). However, three galaxies in our sample (NGC~4676 South, NGC~6240, and NGC~3921) have a significant (a factor of ten) larger global ratio (H$_2$(S0-S2)/PAH=0.08), clearly separated from the rest of our sample and normal star-forming galaxies, and even larger than most AGN galaxies in the SINGS sample \citep[mean H$_2$(S0--S2)/PAH=0.016 for Seyfert and LINER host galaxies][]{Rou07}.
\item On local scales (measuring different apertures within each merger component), we find an intrinsic relation between H$_2$ and PAH  flux as given by a power law, 
\begin{equation}
\mathrm{log[H}_2\mathrm{(S0-S2)]}= k\;\mathrm{log[PAH] + log[const.]}
\end{equation}
All galaxies in our sample follow this power law with a similar slope component $k=0.61\pm0.05$ but a different constant (up to a factor of 10 difference). 
\item Although the center of galaxies has the strongest radiation field and hosts an AGN for at least two galaxies in our sample (see Table~\ref{tab_sample}), it follows the same power law as regions in the outer disk. We find no clear correlation between the presence of an AGN and the central H$_2$/PAH ratio.
\end{enumerate}  
\par
The presence of a power law suggests that H$_2$(S0--S2) and PAH emission are tightly correlated, given as H$_2$(S0--S2)$\simeq$ PAH$^k$, where $k$ is the scaling exponent and $a$ a constants. The scale invariance of the power law becomes clearer in the logarithmic form, log[H$_2$(S0--S2)]=$k$ log[PAH]$+$log[$a$], with a mean slope (exponent) $k=0.61(\pm0.05)$ for our sample and a mean constant of log$[a]=-4.0\pm0.6$. Note that the scatter around this slope typically increases towards the outer disk (smaller signal-to-noise) and that we can not rule out a possible dependence of the power law on the resolved spatial scales. In a number of galaxies (NGC~6240, NGC~520, NGC~2623, and NGC~3921) the faint regions typically fall below our best fit power law for the entire galaxy.  These regions may have a steeper slope, closer to a constant value, implying the bright regions may have systematically less H$_2$ emission for a given PAH flux than would be extrapolated from a fit to the faint regions in the outer disk. A higher extinction for the H$_2$ lines might explain this, but higher signal to noise data is required to confirm this change in slope. \par 

To ensure that the slopes in Fig.~\ref{H2-PAH_ind} and the fact that the nuclei display lower H$_2$/PAH ratios than the  galactic disks was not the result of having higher spatial resolution in the dominant PAH feature (7.7~$\mu$m) compared to the dominant H$_2$ feature (17~$\mu$m) because of the diffraction-limited nature of Spitzer, we have smoothed the SL PAH maps by a factor of three to match SL to LL, and re-calculated the H$_2$/PAH ratios for all galaxies. After this process, the slopes in Fig.~\ref{H2-PAH_ind} and the ratios in the nuclei show a change of less than 3\%, suggesting the effects we are seeing, namely slopes significantly less than unity and lower H$_2$/PAH ratios in the nuclei than in the disk, are real. Moreover, we find a similar trend for most of our galaxies if we calculate the H$_2$/PAH ratios using the H$_2$ S(3) line (9.7~$\mu$m) instead of H$_2$(S0--S2), but with a typically much larger scatter and uncertainty (likely due to the fact that the S(3) emission line is more affected by uncertainties in the dust exctinction correction and that it has a noticeably higher contribution from hotter H$_2$ with a more mixed excitation as mentioned above).\par 

\begin{figure*}
\begin{center}
\includegraphics[scale=0.7]{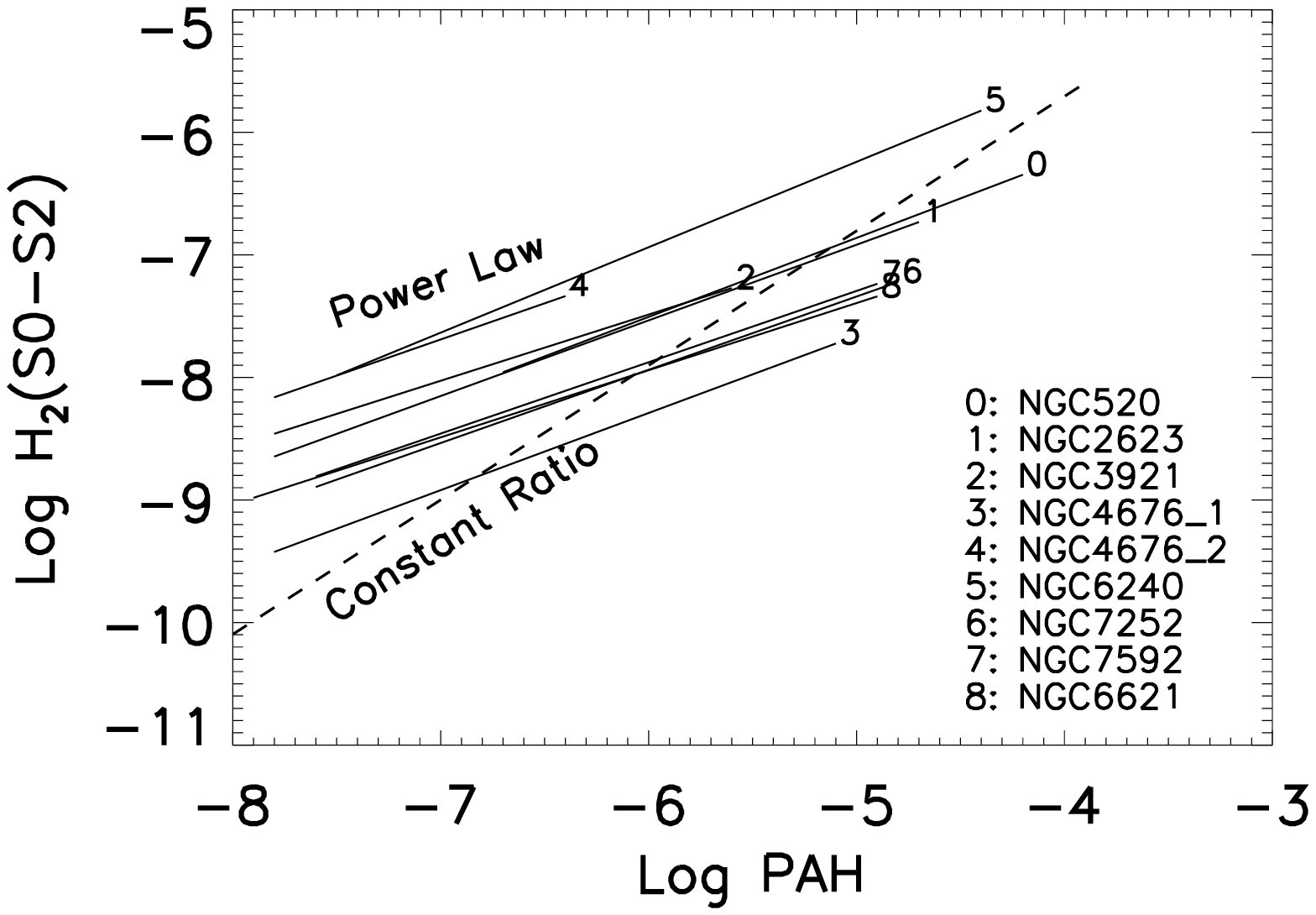}
\end{center}
\caption{\footnotesize{Overview of the power laws (log[H$_2$(S0--S2]$=k\times$log[PAH]$+c$) as derived in Fig~\ref{H2-PAH_ind} for each galaxy (solid lines). The dashed line represents the constant ratio model for the center of normal star-forming galaxies, given as H$_2$(S0--S2)/PAH=0.008 \citep{Rou07}, clearly demonstrating that a constant ratio (i.e. $k=1$) can not describe the measured relation between PAH and H$_2$ (mean $k=0.61\pm0.05$).} }
\label{H2-PAH_comp}
\end{figure*}

However, compared to the study of \cite{Rou07}, which suggests a linear relation with a constant ratio H$_2$(S0--S2)/PAH=0.008 for the center of normal star-forming galaxies, we find that a constant ratio (i.e. $k=1$) can not describe the relation between PAH and H$_2$ emission over the entire galaxies' disk as demonstrated in Fig.~\ref{H2-PAH_comp}. The nuclei of the galaxies tend to be closest to the \cite{Rou07} value while the regions in the disk tend to have higher H$_2$/PAH ratios \citep[increased H$_2$/PAH ratios at radii of 5 to 15~kpc have been also found in the edge-on disk galaxies NGC~4565 and NGC~5907;][]{Lai10}. Instead, the intrinsic relation between PAH and H$_2$ emission follows a power law from the center to the outskirts of major mergers. This tight correlation and the fact that we find a similar slope (exponent) for diverse merger systems suggest a common underlying excitation mechanism. The most likely mechanism is FUV excitation of PAH and H$_2$ gas in PDRs, since other mechanisms, such as shocks due to supernovae or AGN, would a) not show the same behavior over all spatial scales and b) are not expected to follow a tight relation such as a power law. The coupling between PAH and H$_2$ implies that both are excited predominantly in PDRs, although they may not come from the exact same layers (at the same optical depths within the clouds).\par 

A possible explanation for the large range of the constant $a$ may be caused by the different physical properties of the merger systems, e.g. a different abundance of warm molecular gas to PAHs. This would imply that any correlation between H$_2$ and PAH emission based only on global measurements for a sample of different galaxies would be washed out, simply because the global measurement reflects only the static physical properties rather than the behaviour of the H$_2$ and PAH excitation within the galaxies.
In princible, a possible contribution from mechanical heating through shocks may play a role as well on some local scales (e.g., for the center of NGC~6240, see \S~\ref{subsec:toomre}). However, this is likely not the dominant contribution in our galaxies since observations of extreme H$_2$-luminous galaxies \citep{Ogl10, Nes10, Gui11} have shown that shock-excited galaxies have typically a larger H$_2$/PAH ratio\footnote{Note that in some studies the H$_2$/PAH ratio is defined as H$_2$(S(0)--S(3))/PAH 7.7~$\mu$m which is for our sample typically a factor of $\sim$3 larger than H$_2$(S(0)--S(2))/PAH(6.2, 7.7, 11.3, 17~$\mu$m), but the former ratio shows typically a larger variation on local scales and a less significant power law than the latter one.} than in our sample \citep[e.g., the Stephan’s Quintet with H$_2$(S(0)--S(3))/PAH 7.7~$\mu$m$>$1][]{Gui10}.

\subsection{Local Variations versus Evolutionary Trends along the Toomre Sequence}
\label{subsec:toomre}

\begin{figure*}
\begin{center}
\includegraphics[scale=0.4]{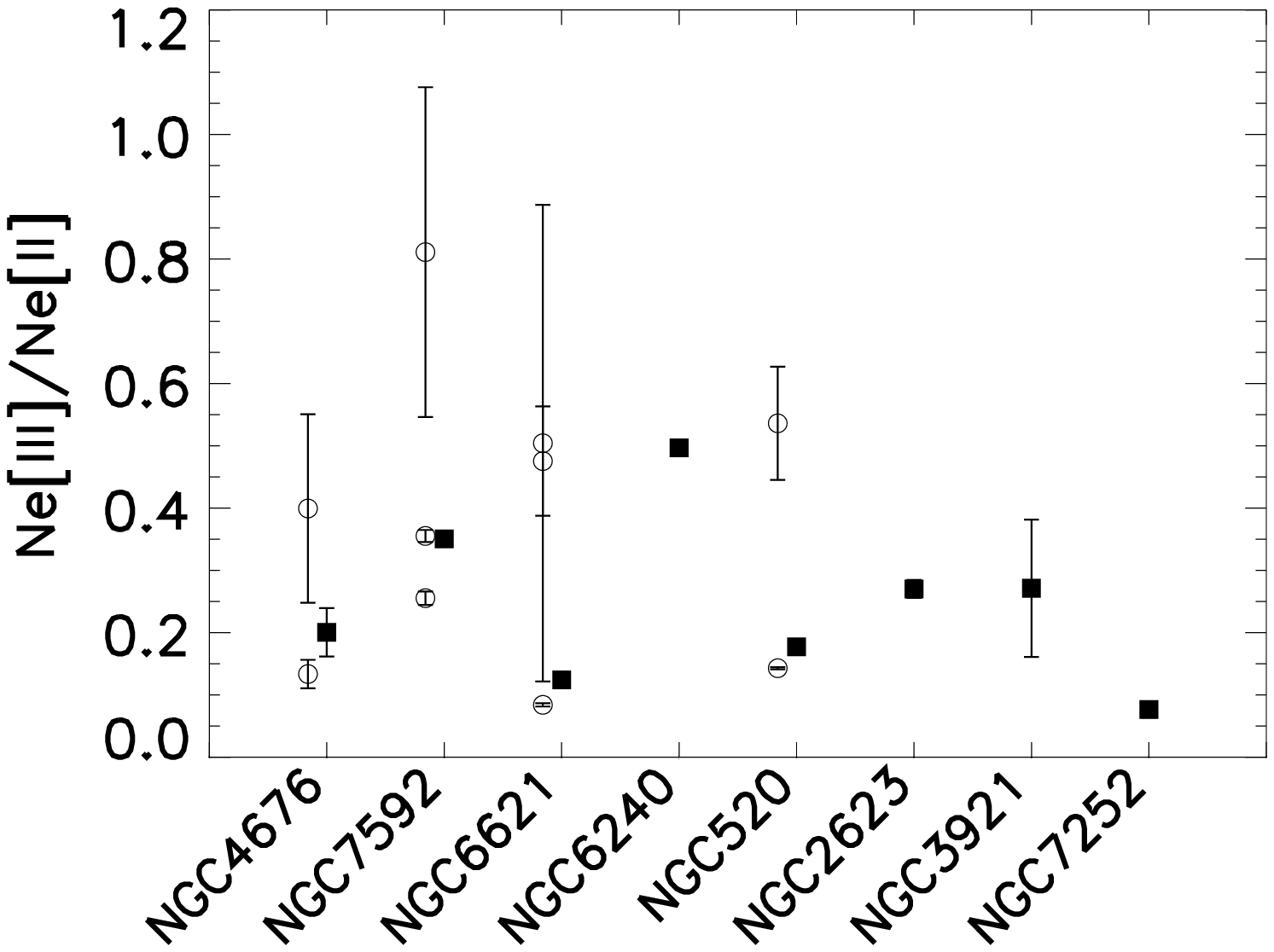}
\includegraphics[scale=0.4]{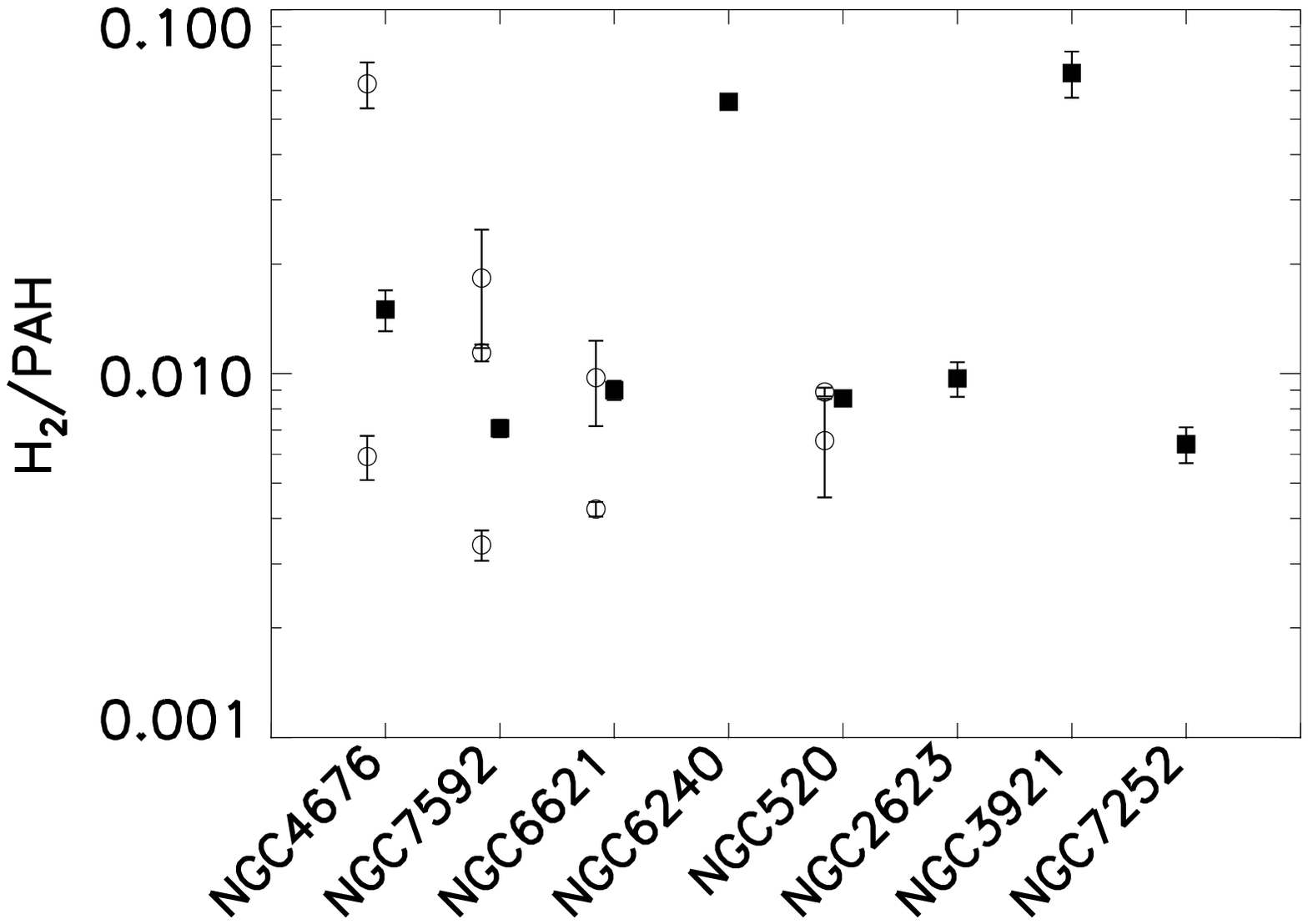}
\includegraphics[scale=0.4]{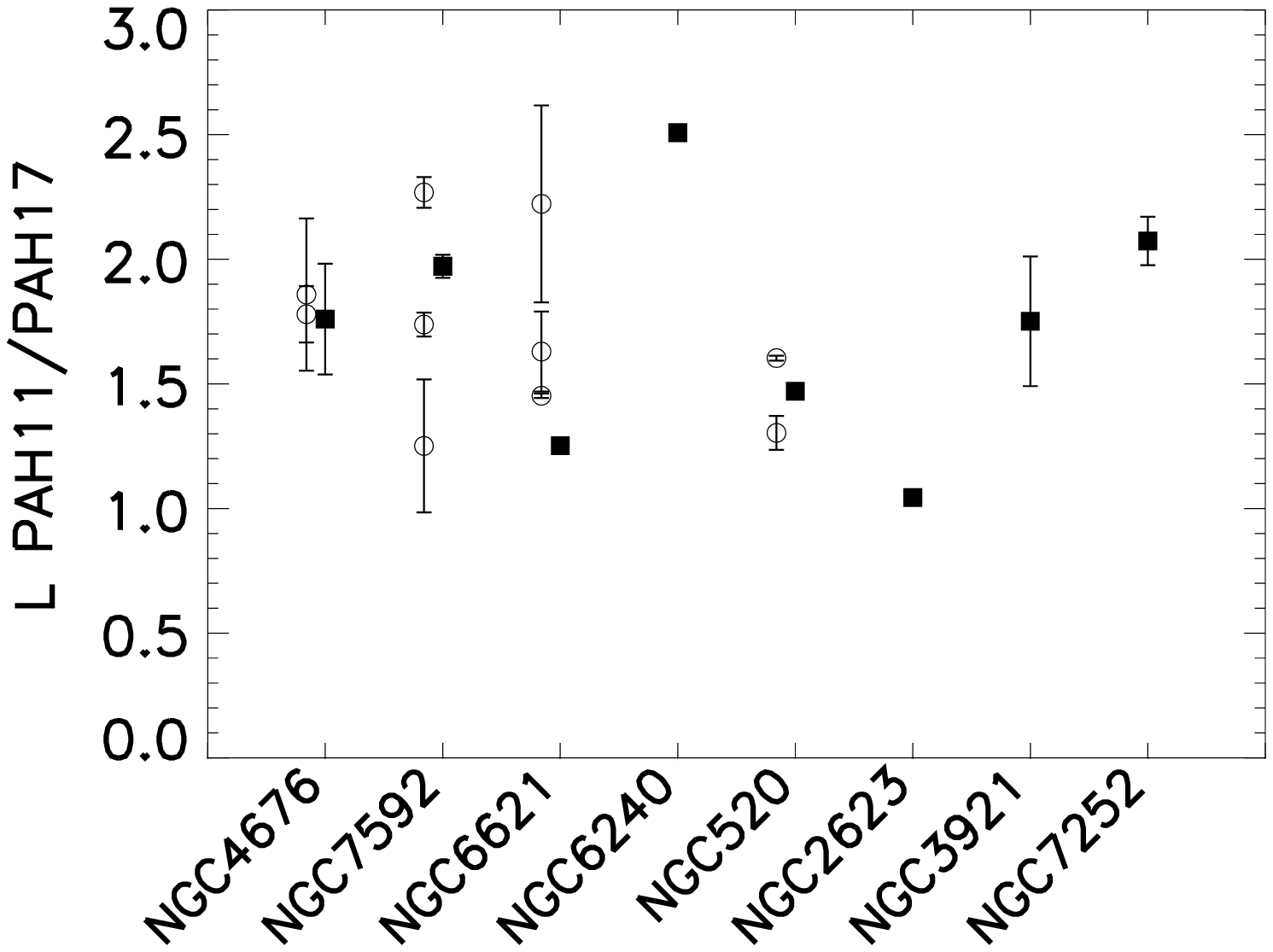}
\includegraphics[scale=0.4]{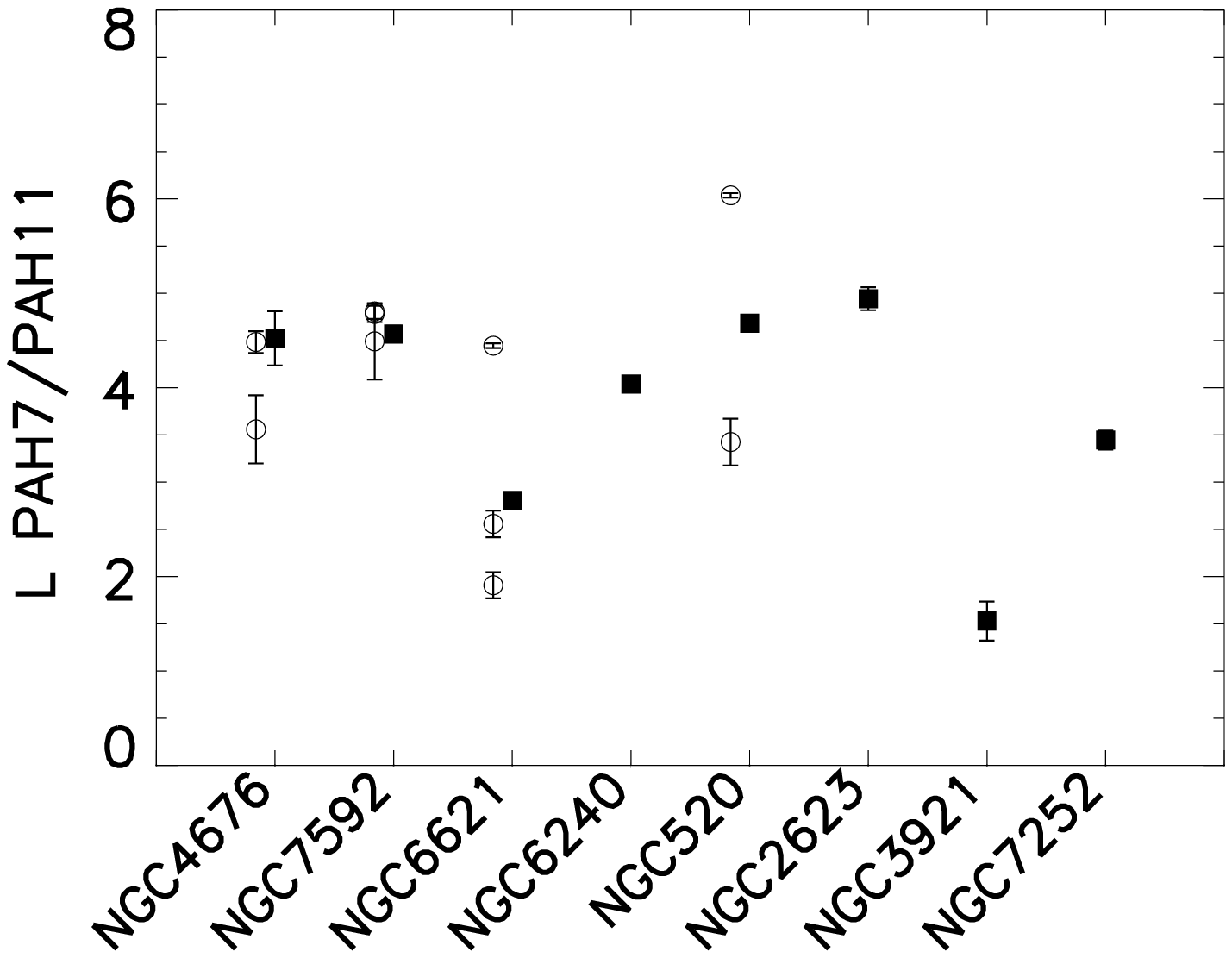}
\includegraphics[scale=0.4]{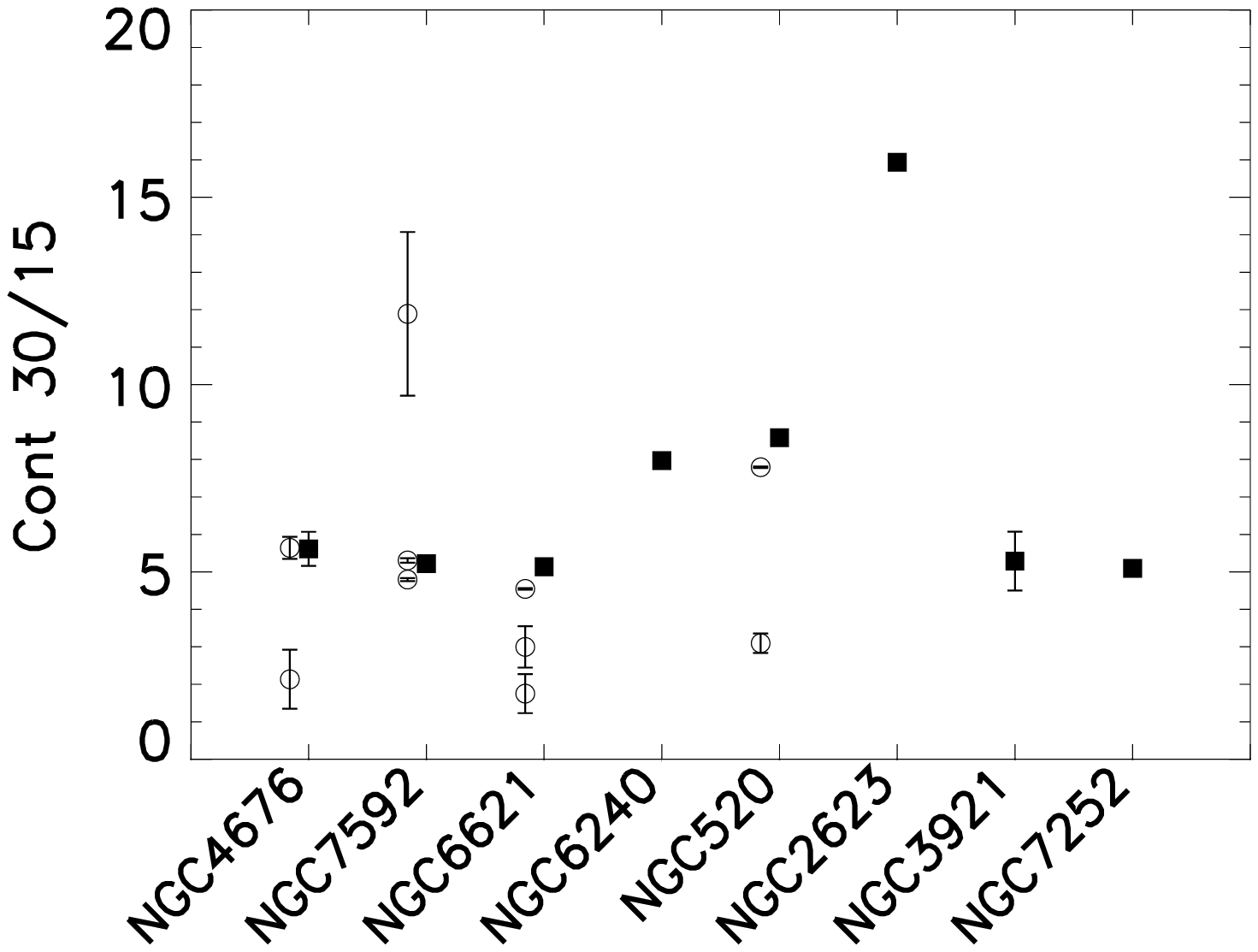}
\includegraphics[scale=0.4]{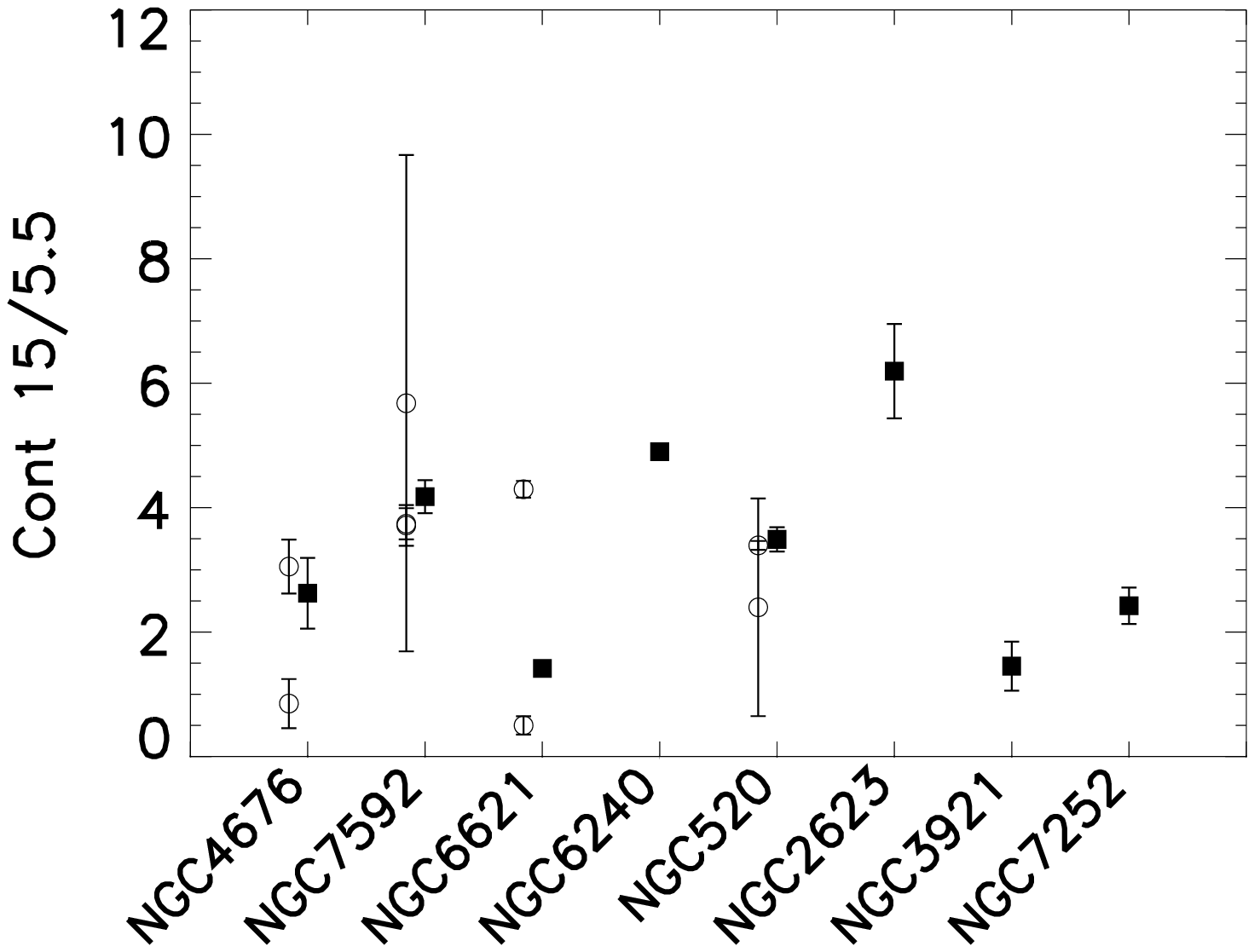}
\includegraphics[scale=0.4]{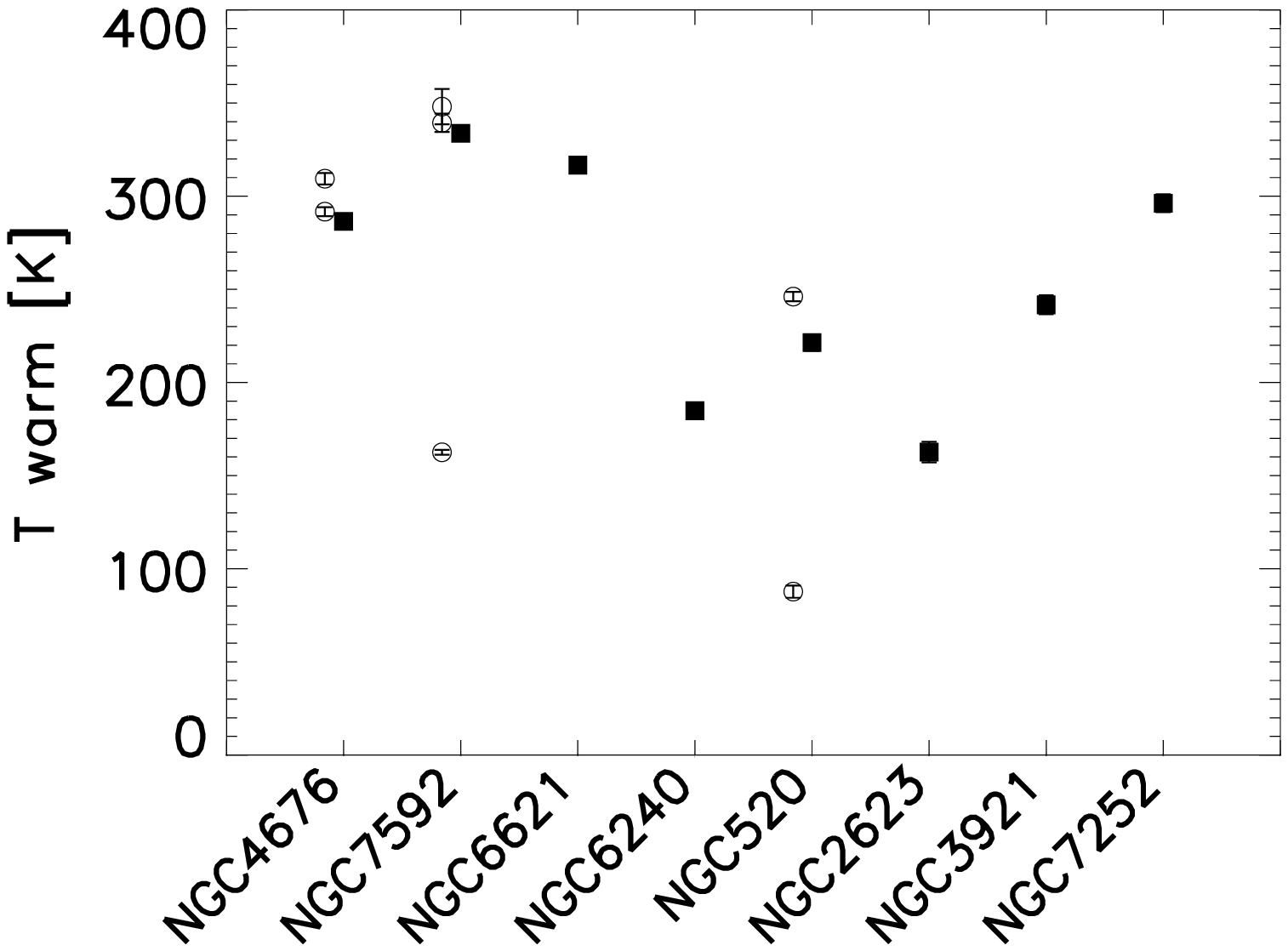}
\includegraphics[scale=0.4]{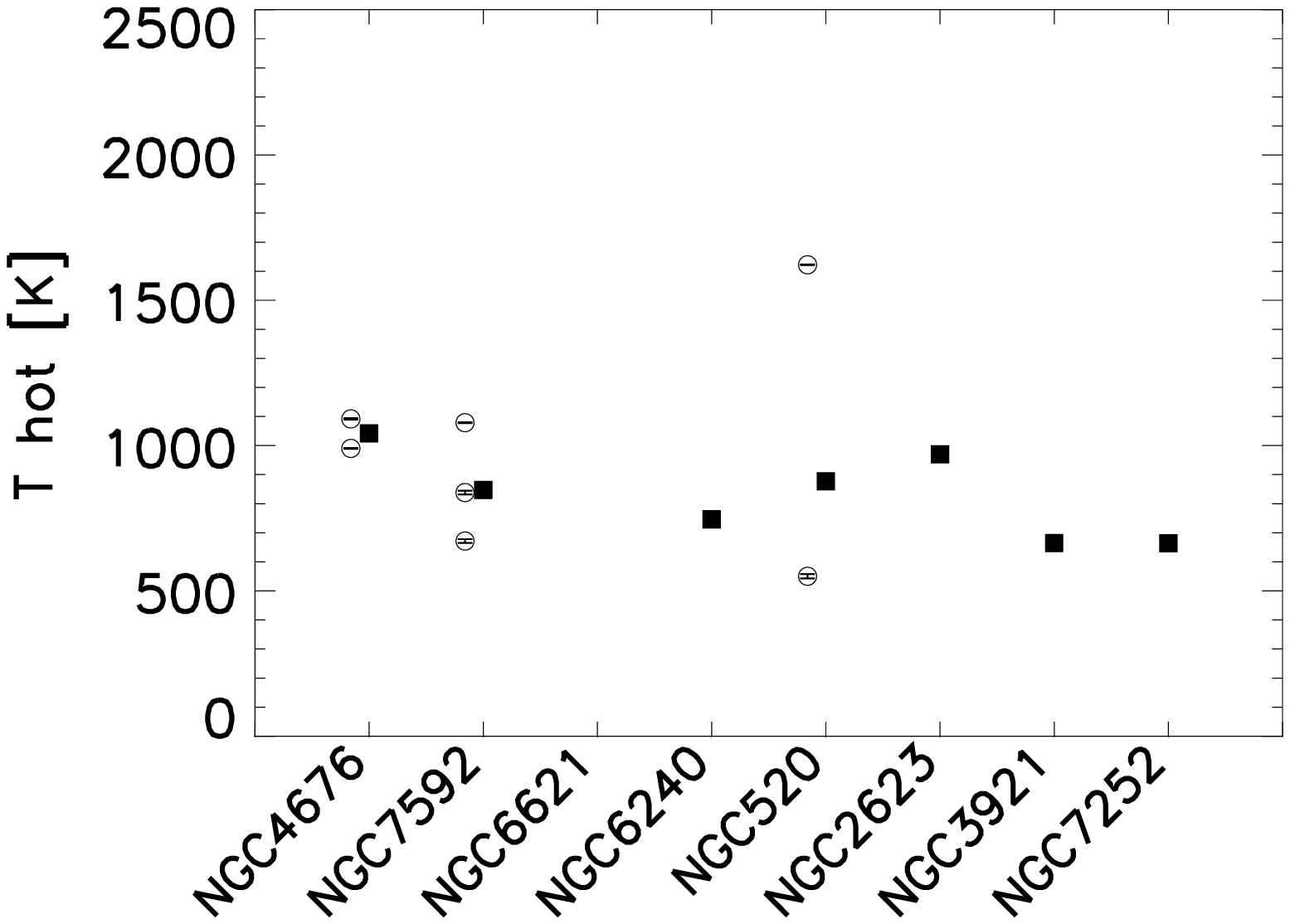}
\end{center}
\caption{\footnotesize{Dust and gas properties as a function of the Toomre Sequence (from the early merger stage NGC~4676 to the late stage NGC~7252). The entire merger systems and individual merger components are marked as filled squares and open circles, respectively. From left to right and top to bottom: ratio [Ne III]/[Ne II], ratio of the sum of the H$_2$ S(0) to S(2) transitions to the main PAH bands, PAH ratio 11.3/7.7~$\mu$m, PAH ratio 7.7/11.3~$\mu$m, the mid-IR continuum ratio 30/15~$\mu$m, 15/5.5~$\mu$m, the ''warm'' H$_2$ temperature, and ''hot'' H$_2$ temperature component (see text for details).}}
\label{prop_toomre}
\end{figure*}

\begin{figure*}
\begin{center}
\includegraphics[scale=0.4]{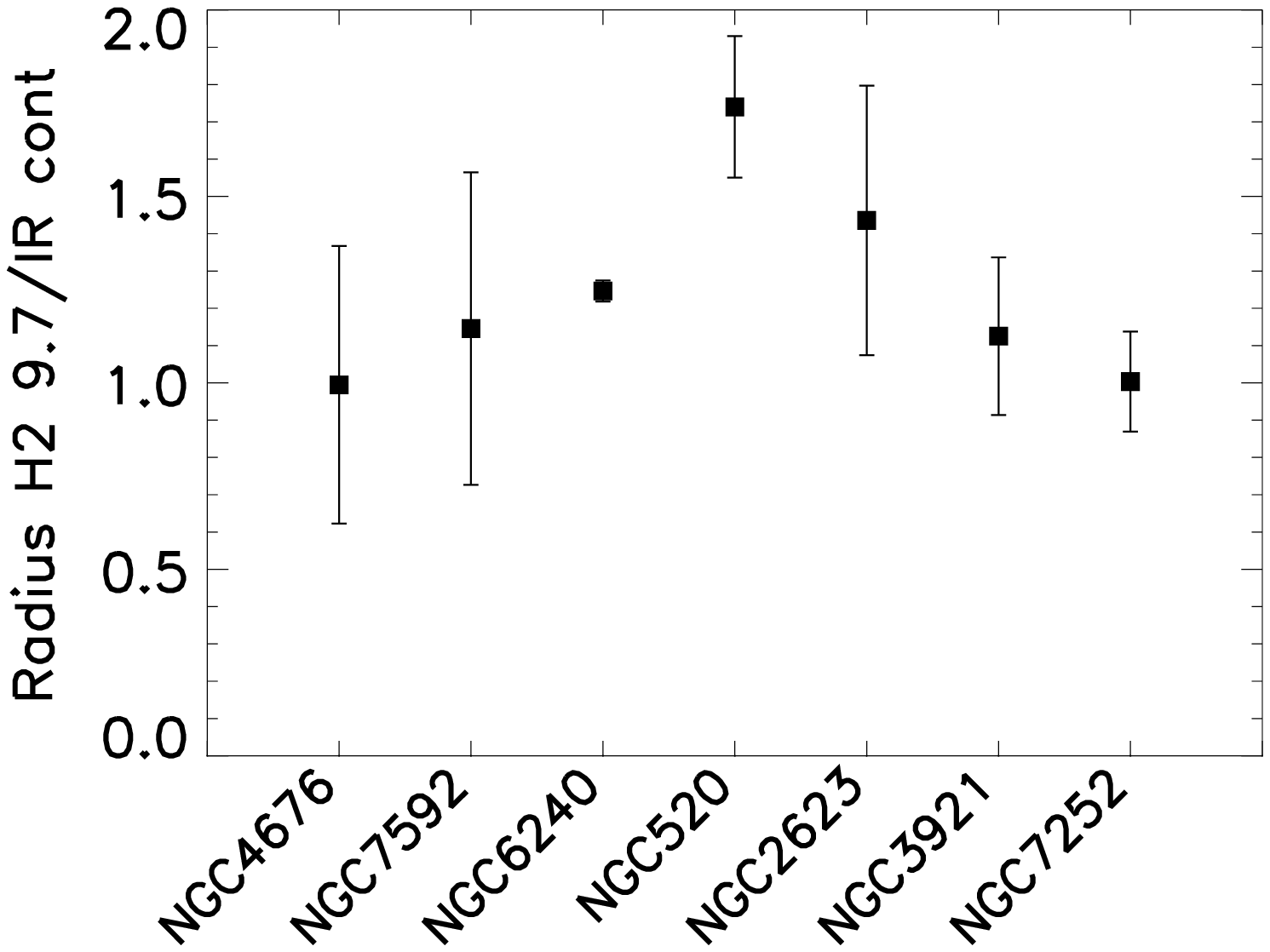}
\includegraphics[scale=0.4]{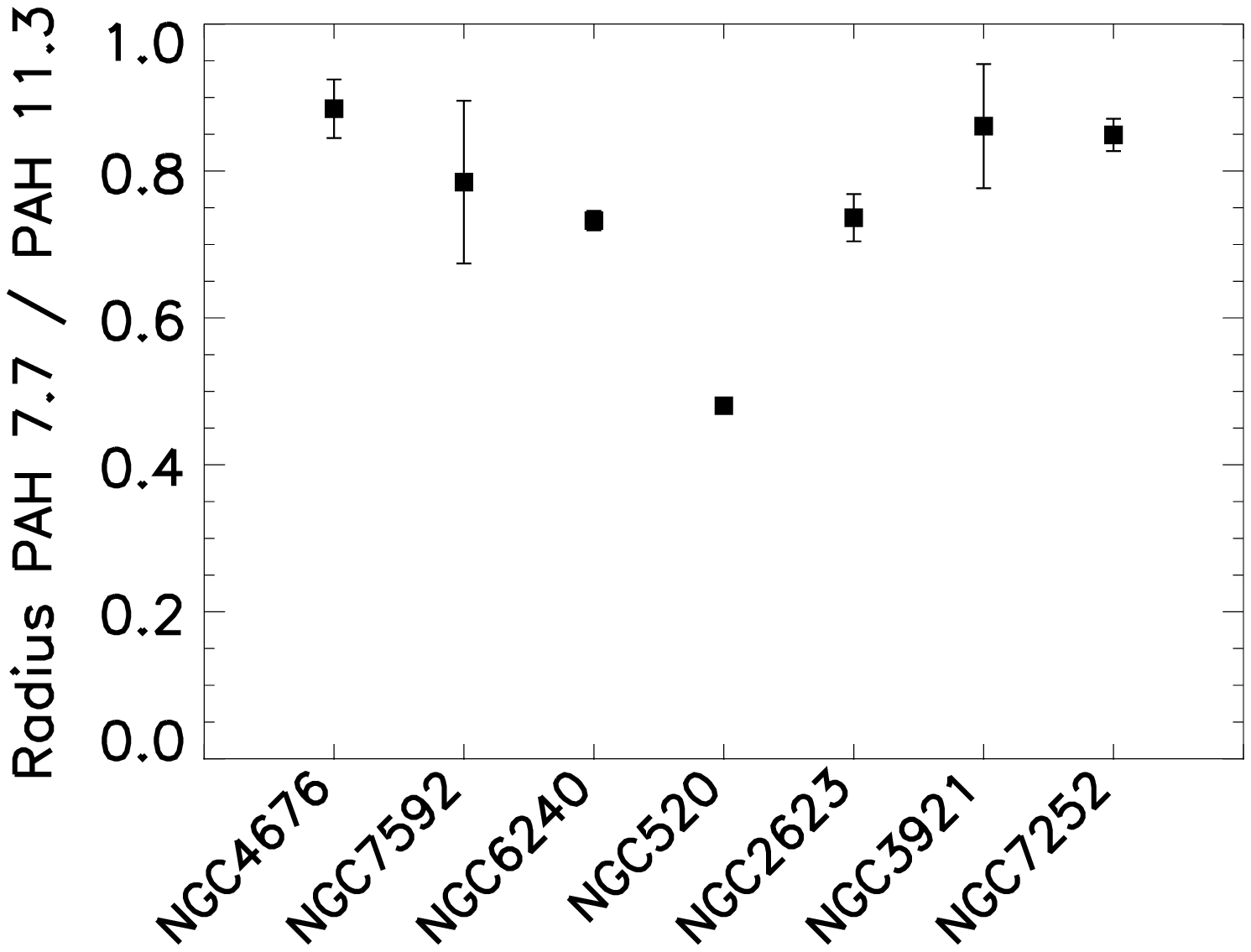}
\end{center}
\caption{\footnotesize{The relative radii of the major merger components as a function of the Toomre Sequence (from the early merger stage NGC~4676 to the late stage NGC~7252). \textbf{Left:} The ratio of the strongest H$_2$ emission line (S(3)) to the mid-IR continuum radius (at 5.5--15~$\mu$m) as a function of merger stage. \textbf{Right:} The ratio of the PAH 7.7~$\mu$m to 11.3~$\mu$m line radius as a function of merger stage.} }
\label{radius_toomre}
\end{figure*}

To investigate possible evolutionary effects on the gas and dust properties of major mergers we have studied several mid-IR features as a function of their merger stage, representing the early (NGC~4676, NGC~7592, NGC~6621 -- distinguishable parent disks and large tidal tails), mid (NGC~2623, NGC~6240, NGC~520 -- highly overlapping disks and/or double nuclei) and late (NGC~3921, NGC~7252 -- single nucleus, elliptical-like light distribution) stages. For instance, a change in dust grain size, ionization stage, neutral to ionized gas fraction, or gas temperatures could be expected. To test such a possible scenario we study the following mid-IR diagnostics as a function of merger stage as shown in Fig.~\ref{prop_toomre}: [Ne III]/[Ne II], the PAH interband strength ratios 11.3/17~$\mu$m and 7.7/11~$\mu$m, the warm gas to PAH ratio H$_2$(S0--S2)/PAH as described in \S~\ref{subsec:PDR}, the dust continuum ratios 30/15~$\mu$m and 15/5.5~$\mu$m as derived in \S~\ref{subsubsec:cont}, and the two warm H$_2$ gas components. The results are presented for the entire merger systems as well as for the individual merger components. Furthermore we explored the radial profile and spatial size (derived from  S\'{e}rsic profile fit) of several mid-IR features as a function of merger stage, such as the radial size ratio of H$_2$/IR continuum, PAH 7.7/11.3$~\mu$m, and H$_2$/PAH. Our main conclusion is that the variety of mid-IR properties in different regions as shown in Fig.~\ref{prop_toomre} is significantly larger than any trend along the merger sequence using the global properties. This does not necessarily imply that an overall trend along the merger sequence does not exist, for which we find indications in the mid-IR continuum slope (Fig.~\ref{prop_toomre}) and the relative radial size of the PAH interband emissions and the H$_2$ emission as shown in Fig.~\ref{radius_toomre}. However, a much larger sample would be required to confirm statistically such a trend in the mid-IR properties \citep[see][for nuclear mid-IR properties as function of merger stage]{Sti11}.\par

Despite the small number of galaxies in our sample we find two galaxies that are outliers in one or more properties: NGC~6240 does not only show a significantly larger H$_2$/PAH flux ratio (see \S~\ref{subsec:PDR}), but also a larger PAH 11.3/17$\mu$m flux ratio as well as a larger ratio of the radii H$_2$/IR cont, H$_2$/PAH and PAH 11.3/7.7$\mu$m than most of the other galaxies in our sample. Previous observations of the center of this galaxy have revealed two nuclei with a separation of 1.8\arcsec \citep{Fri83}  and that the H$_2$ emission peaks between the two nuclei and is thermally excited via slow shocks triggered by the collision of the two nuclei \citep{Wer93,Sug97,Tec00,Arm06}. This may explain the intense H$_2$ emission in the very center of NGC~6240 \citep[unresolved with Spitzer IRS, see][]{Arm06}, but is likely not responsible for the correlation seen in the outer disk between H$_2$ and PAH.  Another galaxy, NGC~2623, is significantly larger in the flux continuum ratio (30/15$\mu$m$=16.3$)  than the rest of the sample (mean 30/15~$\mu$m$\sim6$), which may be due to the star-formation intensity peak that occurs during a relatively short timescale ($<0.5$~Gyrs) at the final coalescence of the nuclei as shown in recent major merger simulations \citep{Cox08, Hop09}. Furthermore, this stage also seems to show a relative change in the radii of the mid-IR features, namely, a maximum in the H$_2$ emission line radius to mid-IR continuum radius (at 5.5--15~$\mu$m) as well as in the PAH 11.3~$\mu$m to 7.7~$\mu$m line ratio radius.

\subsection{Multiwavelength View of the Interstellar Material in Early Merger Stages}
\label{sec:indgalaxies}
In this section we discuss in more detail three ''classical'' major mergers, NGC~4676, NGC~7592, and NGC~6621, and compare their mid-IR properties with the distribution of the stellar (NIR and FUV), cold atomic (HI), and cold molecular \citep[CO, see][]{Yun01} gas components. These three mergers are in particular interesting since they are caught in the early merger stages and have spatially resolved components which allows us to compare in detail the physical properties between the two nuclei in each merger system.

\subsubsection{NGC~4676}
\label{section_4676}
\begin{figure*}
\begin{center}
\includegraphics[scale=0.5]{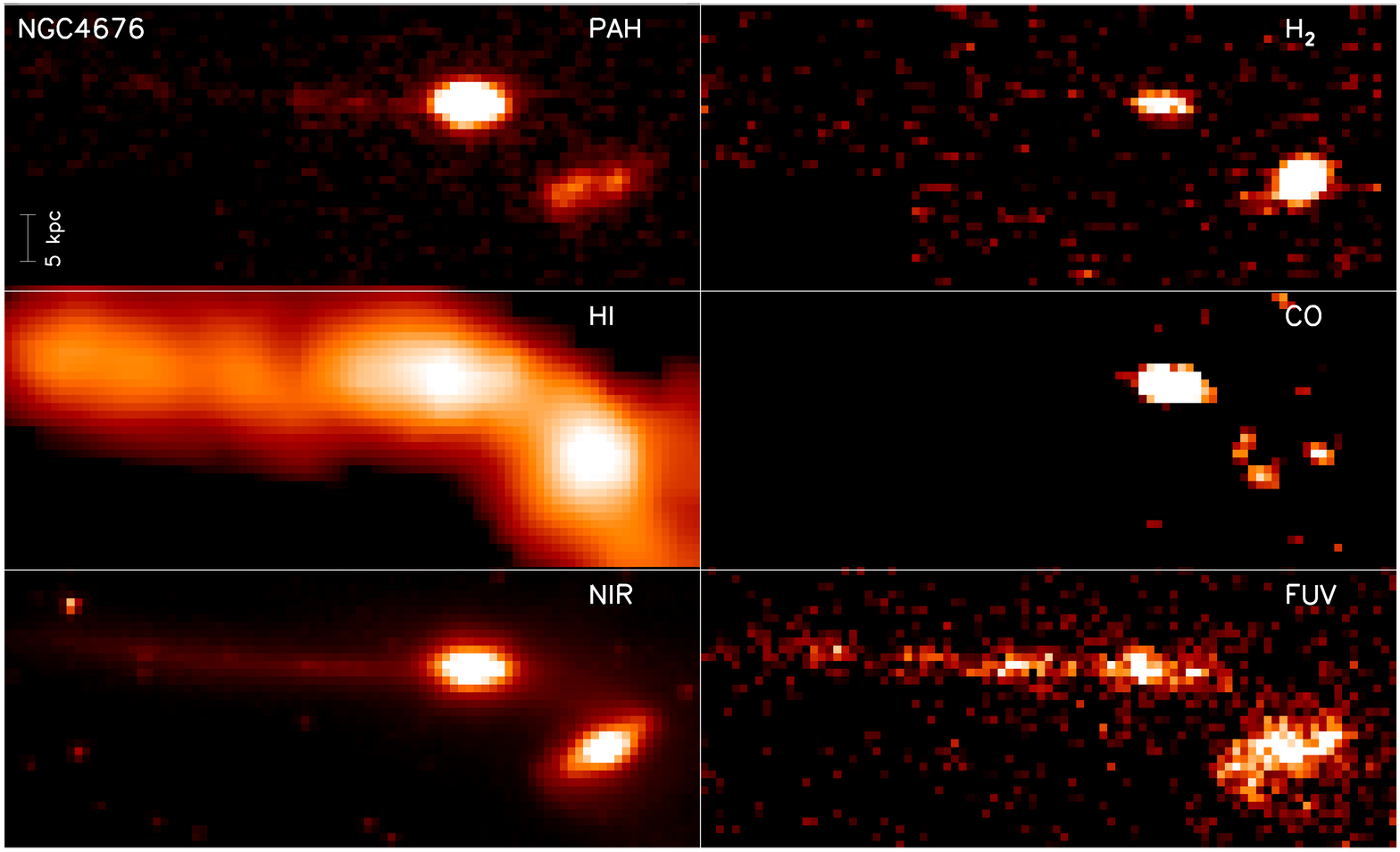}
\end{center}
\caption{\footnotesize{Multiwavelength view of NGC~4676, from left to right and top to bottom: The combined main PAH features, the warm molecular gas (H$_2$ emission), the cold atomic gas (HI), the cold molecular gas (CO), the NIR (Spitzer IRAC 3.6~$\mu$m), and the FUV (GALEX). The North and East directions in each frame of this figure are the same as in the corresponding frame of Fig.~\ref{maps_mid-IR}.}}
\label{multi_NGC4676}
\end{figure*}

Nicknamed the ''Mice'', each of the two galaxies in NGC~4676 possesses a long tail as observed in the optical light. The two galaxies of this merger system have a projected separation distance of 36.8\arcsec ($\sim$17~kpc). In particular, the northern galaxy exhibits a long tidal tail with a projected extent of 48~kpc (measured in the NIR), visible in the FUV, near-infrared, mid-IR and HI emission (see Fig.~\ref{multi_NGC4676}). Material is being exchanged between the two disks as indicated by the presence of a bright optical bridge between them. \cite{Sto74} and \cite{Hib95} found that the tails have widespread star formation which accounts for 16\% of the total H$\alpha$ emission and exhibit a high atomic gas content with on average $M_{HI}/L_R \sim0.4~M_{\sun}L_{\sun}^{-1}$, with the stellar tails exceeding the gaseous tails in extent.\par

Although both galaxies have roughly the same near-IR luminosities (as measured in the H-band), we find significant differences in the distribution of the mid-IR features: As shown in Fig.~\ref{multi_NGC4676}, most of the PAH and CO emission (each $>$90\%) originate from the northern galaxy, while the main H$_2$ line emission ($>$80\%) is found in the southern galaxy. This finding may suggest a significantly different stellar population or distribution of molecular gas and young stars in the merger components. The northern component has a significantly smaller [Ne III]/[Ne II] ratio (0.13) and larger PAH 7.7/11.3~$\mu$m ratio (4.5) than the southern galaxy ($\sim$0.4 and 3.5, respectively), suggesting that the average stellar population is younger and the grains may be more inonized in the northern component. Interestingly, we find that the H$_2$ temperature components ($\sim$300~K for the warm and $\sim$1000~K for the hot component) are roughly the same in both galaxies. Thus, the strong H$_2$ emission in the southern galaxy may be simply due to the presence of a larger amount of warm molecular gas (H$_2$ mass two times larger than in the northern galaxy).\par

Furthermore, we find PAH emission in the northern tidal tail, but no significant H$_2$ emission. Fig.~\ref{PAH_scatter_tail} shows that the interband strength of the PAH 7.7/11~$\mu$m and 6.2/7.7~$\mu$m ratios of the tidal tail regions are all characterized by the ionized phase \citep[using the model by][]{Dra01}. This is perhaps not surprising given the large number of young stars in the tidal tail as seen in the strong FUV emission (Fig.~\ref{multi_NGC4676}) and H$\alpha$ emission \citep{Sto74, Hib95, Hib96}), demonstrating that the grain properties in interacting galaxies as traced by the mid-IR feature ratios can be affected over large scales in interacting systems.

\subsubsection{NGC~7592}

\begin{figure*}
\begin{center}
\includegraphics[scale=0.6]{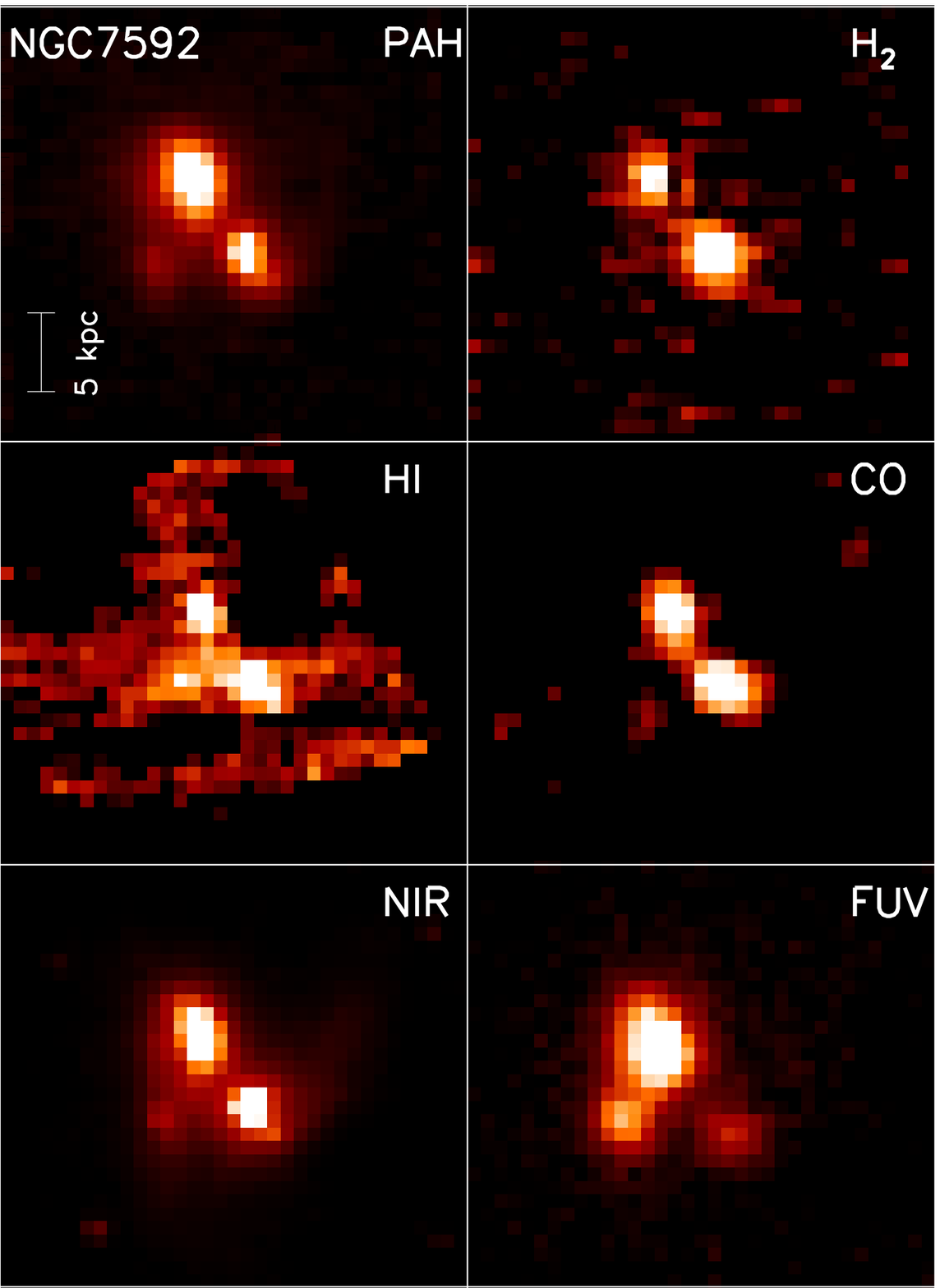}
\end{center}
\caption{\footnotesize{Multiwavelength view of NGC~7592, from left to right and top to bottom: The combined PAH features, the warm molecular gas (H$_2$ emission), the cold atomic gas (HI), the cold molecular gas (CO), the NIR (Spitzer IRAC 3.6~$\mu$m), and the FUV (GALEX). The North and East directions in each frame of this figure are the same as in the corresponding frame of Fig.~\ref{maps_mid-IR}.}}
\label{multi_NGC7592}
\end{figure*}

The two merging galaxies in this LIRG system \citep[L$_{IR}$/L$_{\sun}=10^{11.33}$,][]{San03} have a projected separation of 14\arcsec ($\sim$7~kpc), a NIR luminosity ratio of $\sim$1:1, and show tidal tails in the NIR light. Fig.~\ref{multi_NGC7592} highlights the details of this system at multiple wavelengths. Like NGC~4676, the PAH and H$_2$ emissions have a different distribution, with the PAH emission primarily originating from the eastern starbursting component and the H$_2$ from the western component, which has a Seyfert 2 nucleus \citep{Lai03}.\par 

We find an additional concentration in the overlap region between the two nuclei (with an offset of $\sim$12$\arcsec$ towards south) in the NIR light, FUV light, HI emission and in several mid-IR features (main PAH features, [NeII] and [NeIII]). No significant H$_2$ emission is found at this position. The PAH interband strength and [NeIII]/[NeII] ratios indicate that most of the emission from this intermediate region originates from ionized gas (see Fig.~\ref{PAH_scatter}), which is likely due to star-formation as we find a strong emission peak in the FUV and H$\alpha$ \citep{Lai03} at this position as well.

\subsubsection{NGC~6621}

\begin{figure*}
\begin{center}
\includegraphics[scale=0.6]{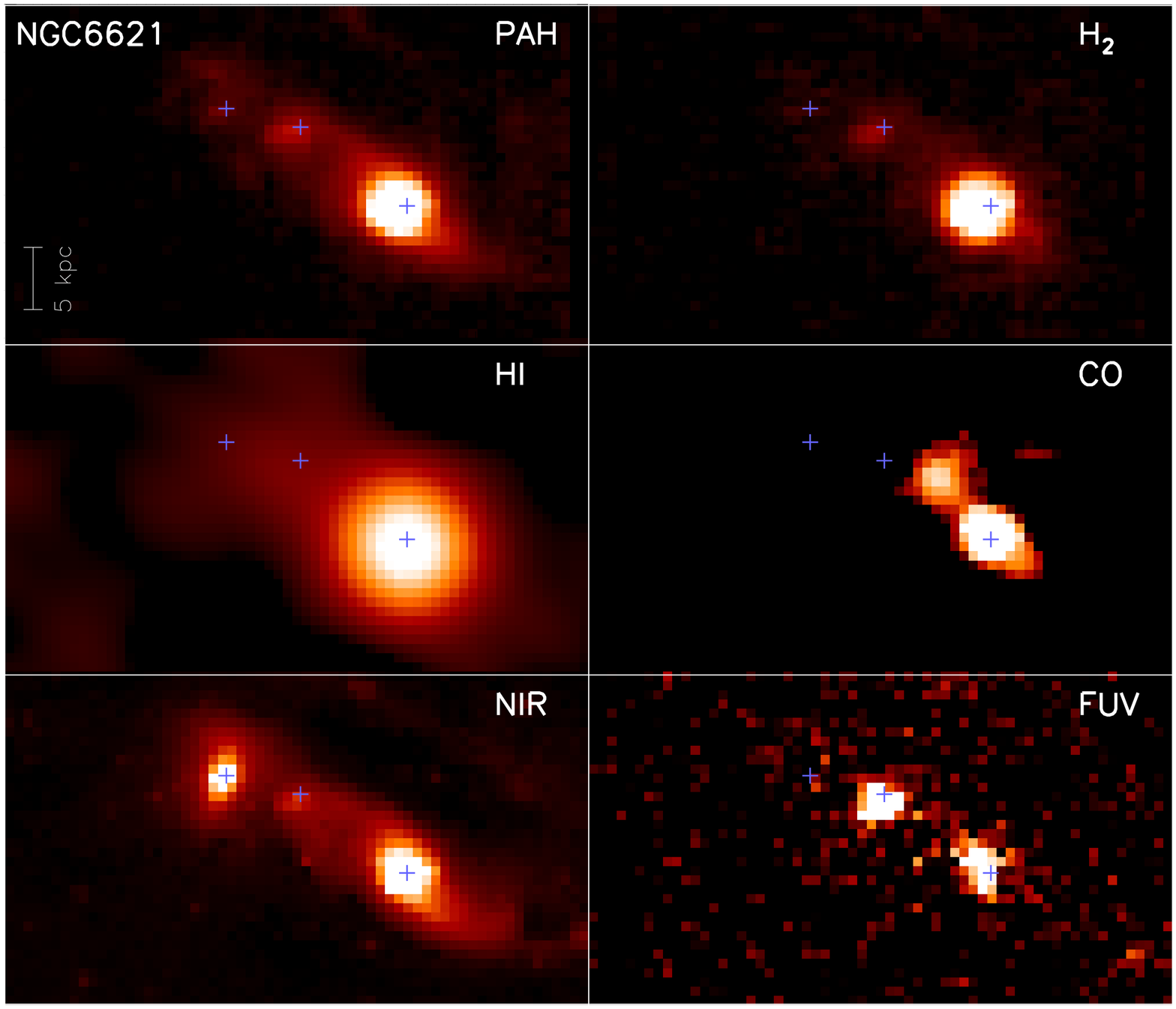}
\end{center}
\caption{\footnotesize{Multiwavelength view of NGC6621, from left to right and top to bottom: The combined PAH features, the warm molecular gas (H$_2$ emission), the cold atomic gas (HI), the cold molecular gas (CO), the NIR (Spitzer IRAC 3.6~$\mu$m), and the FUV (GALEX). The centers of the two galaxies (as seen in the NIR) and the star-formation peak in between are indicated with a blue plus marker. The North and East directions in each frame of this figure are the same as in the corresponding frame of Fig.~\ref{maps_mid-IR}.}}
\label{multi_NGC6621}
\end{figure*}

The two merging components of NGC~6621 have a projected separation of 41\arcsec ($\sim$17~kpc). Unlike the two previous discussed major mergers, NGC~4676 and NGC~7592, which have a NIR light ratio of 1:1, the two components of this system have a ratio of $\sim$1:2. However, this ratio changes dramatically in the light of different wavelengths. The mid-IR light of the smaller component, as given by our IRS continuum measurement, is at least two orders smaller than the larger component (nearly all of the mid-IR emission in NGC~6621 is coming from only one of the two nuclei). As shown in Fig.~\ref{multi_NGC6621}, we find no significant FUV, H$_2$, CO, and HI emission at the position of the smaller galaxy, and only very faint emission from PAH features. Instead, we find significant emission originating from a peak in the overlap region between the two merging galaxies. The interband PAH strength ratios of this peak (see Fig.~\ref{PAH_scatter}) indicate that the PAH component may originate partially from neutral gas layers. However, we also find significant H$_2$ and FUV emission in this overlap region, suggesting a burst of star-formation here.\par 

The distribution of the cold molecular gas, as observed in the CO emission line \citep{Ion05}, is not co-spatial to the stellar (NIR), warm gas (H$_2$) and PAH distribution: besides the CO emission peak of the main galaxy, we find an additional CO concentration ($\sim$30\% of the total CO emission) between the main galaxy and the overlap region as seen in the FUV and NIR light, and mid-IR features. This significant offset suggests a large reservoir of cold molecular gas with no signs of star formation in the mid-IR features.

\section{Conclusions}
\label{sec:sum}
We analyzed the spatial distribution and properties of the main mid-IR features observed with Spitzer/IRS for eight major merger systems of the Toomre Sequence. Maps of the key diagnostic features
of the physical conditions in these interacting starburst galaxies are presented --- such as PAH ratios, ionized emission line regions, the warm molecular gas (H$_2$) distribution and temperatures. Moreover, we have compared the mid-IR features to the cold atomic and molecular gas distribution to obtain a comprehensive view about the variation and evolution of the interstellar material phase space in major mergers. The main results are summarized in the following:
\begin{enumerate}
\item The range and distribution of the global PAH interband strength and [Ne III]/[Ne II] ratios of major mergers are very similar to normal star-forming galaxies. However, regarding the distribution of these properties over all spatial scales within the galaxies, we find significantly different PAH populations (defined by their ionization state and grain size) among the merger systems in our sample.
\item The cold molecular gas component as observed in the CO emission line does not usually correlate with the warm gas in early merger systems, leading to large spatial variations in the ratio of cold-to-warm molecular gas. In particular, the cold molecular gas distribution in NGC~6621 shows a significant offset not only from the warm gas component, but also from the stellar (NIR) and dust distribution (PAH, mid-IR continuum). 
\item Two merger systems (NGC~7592 and NGC~6621) show evidence for star formation in overlap regions between the nuclei, namely strong PAH, [Ne II], [Ne III] and warm gas (H$_2$) emission, but no apparent corresponding CO (J=1--0) emission.
\item The majority of galaxies in our merging sequence sample have similar global H$_2$-to-PAH flux ratios as normal star-forming galaxies with H$_2$ temperatures ranging from $\sim$90K to 1100K. Three starburst galaxies in our sample (NGC~4676 South, NGC~6240, and NGC~3921) have a factor of ten larger ratios, clearly separated from the rest of our sample and normal star-forming galaxies, and even larger than what it is in most AGN galaxies in the SINGS sample \citep{Rou07}. However, regarding the different apertures (locations) within a merger, we find an intrinsic relation between H$_2$ and PAH flux as given by a power law with a very similar exponent ($0.61\pm0.05$) for all galaxies but with a constant that can vary up to a factor of ten between different galaxies. 
Such a tight correlation between PAH and H$_2$ over all spatial locations suggests a common dominant excitation mechanism for H$_2$ emission in major mergers, which is most likely absorption of UV photons in photo-dissociation regions rather than shock excitation, since PAHs are known to arise predominantly from these regions.
\item The minimum of the 11.3/7.7~$\mu$m PAH interband strength ratio is typically located in the nuclei of galaxies, suggesting a larger ionization state of the ISM towards the center, as  expected from central starbursts or low-luminosity AGNs (Seyfert, LINER). The [Ne III/[Ne II] ratio increases with distance from the nucleus, seemingly conflicting the expectation of a decrease of the hardening of the radiation field with distance from the center. A possible explanation for a suppressed [Ne III]/[Ne II] ratio in the center is either a larger contribution of the old stellar population in the nuclei, or a fraction of the most massive stars are hidden in ultra-compact HII regions due to high dust densities in the nuclear regions. Thus, the [Ne III/[Ne II] ratio is likely more sensitive to the average stellar age and/or gas density rather than a hardening of the field, suggesting that the 11.3/7.7$\mu$m PAH ratio is a better diagnostic of the ionization state in all merger components, as seen in the center and outskirts of galaxies, in overlapping regions between merger components, and in tidal tails. 
\item Variations of the internal (local) physical conditions within a merger are much larger than the systematic (global) variations along the Toomre Sequence. Note that this does not necessarily exclude the possibility of an evolutionary trend, for which we find indications in the mid-IR continuum slope and the relative radial size of the PAH interband emissions and the H$_2$ emission.
\end{enumerate}\par

Support for this work was provided by NASA through a grant issued by the Spitzer Science Center and the Jet Propulsion Laboratory. This research has made use of the NASA/IPAC Extragalactic Database, which is operated by the Jet Propulsion Laboratory, California Institute of Technology, under contract with the National Aeronautics and Space Administration. We thank the referee for a careful review and valued suggestions, and Pierre Guillard for a useful discussion.


\begin{figure*}
\begin{center}
\includegraphics[scale=0.4]{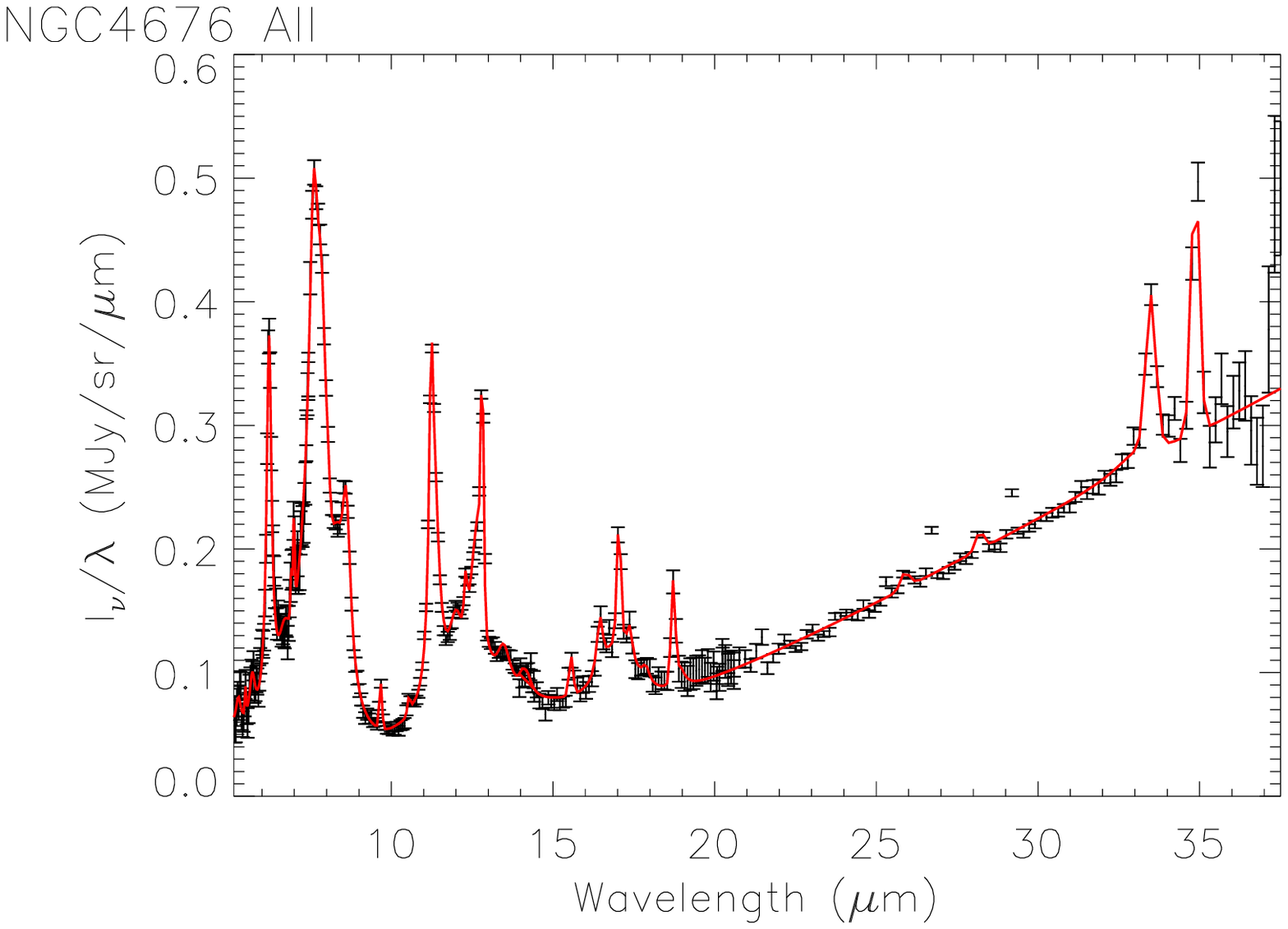}
\includegraphics[scale=0.4]{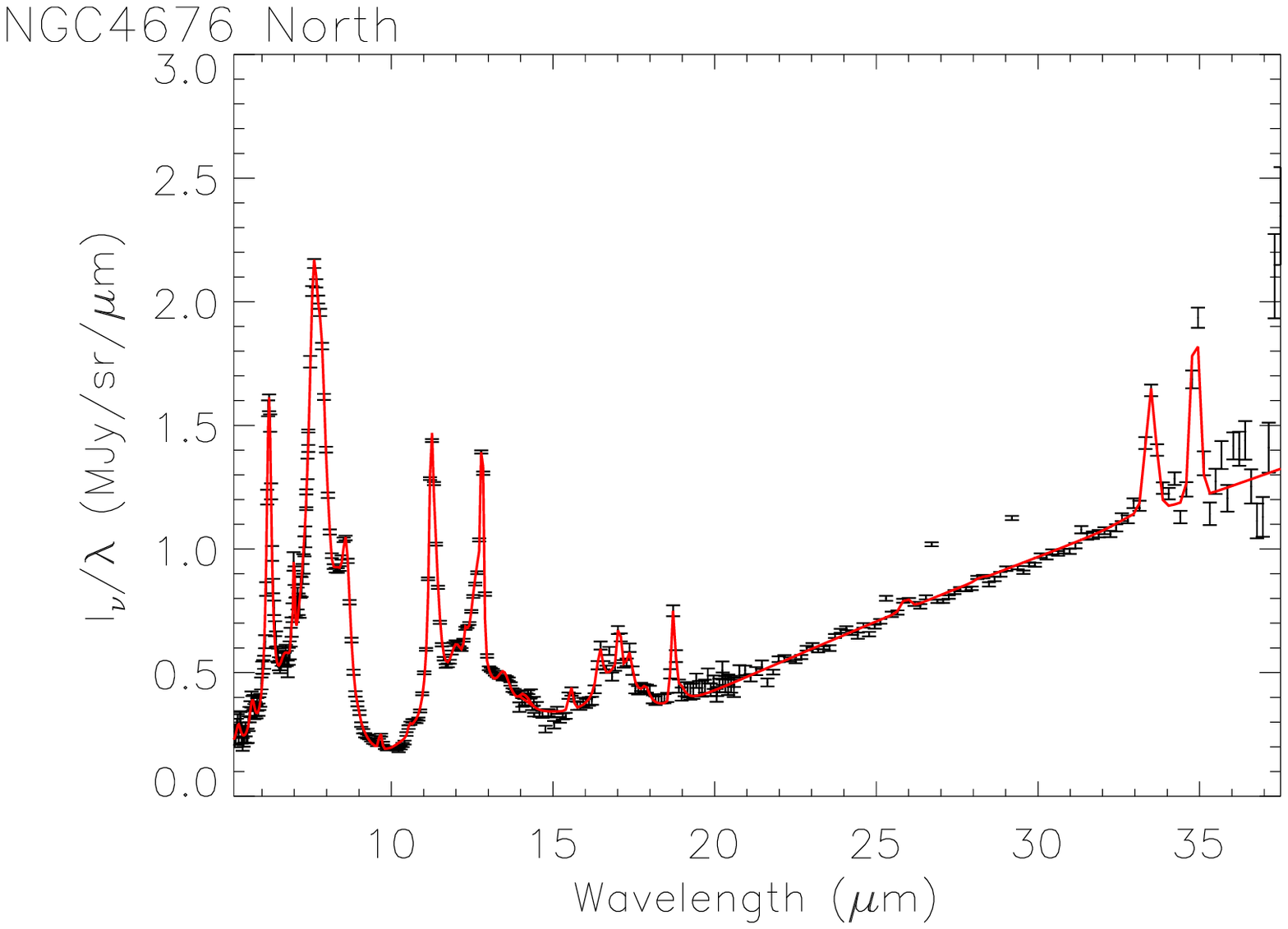}
\includegraphics[scale=0.4]{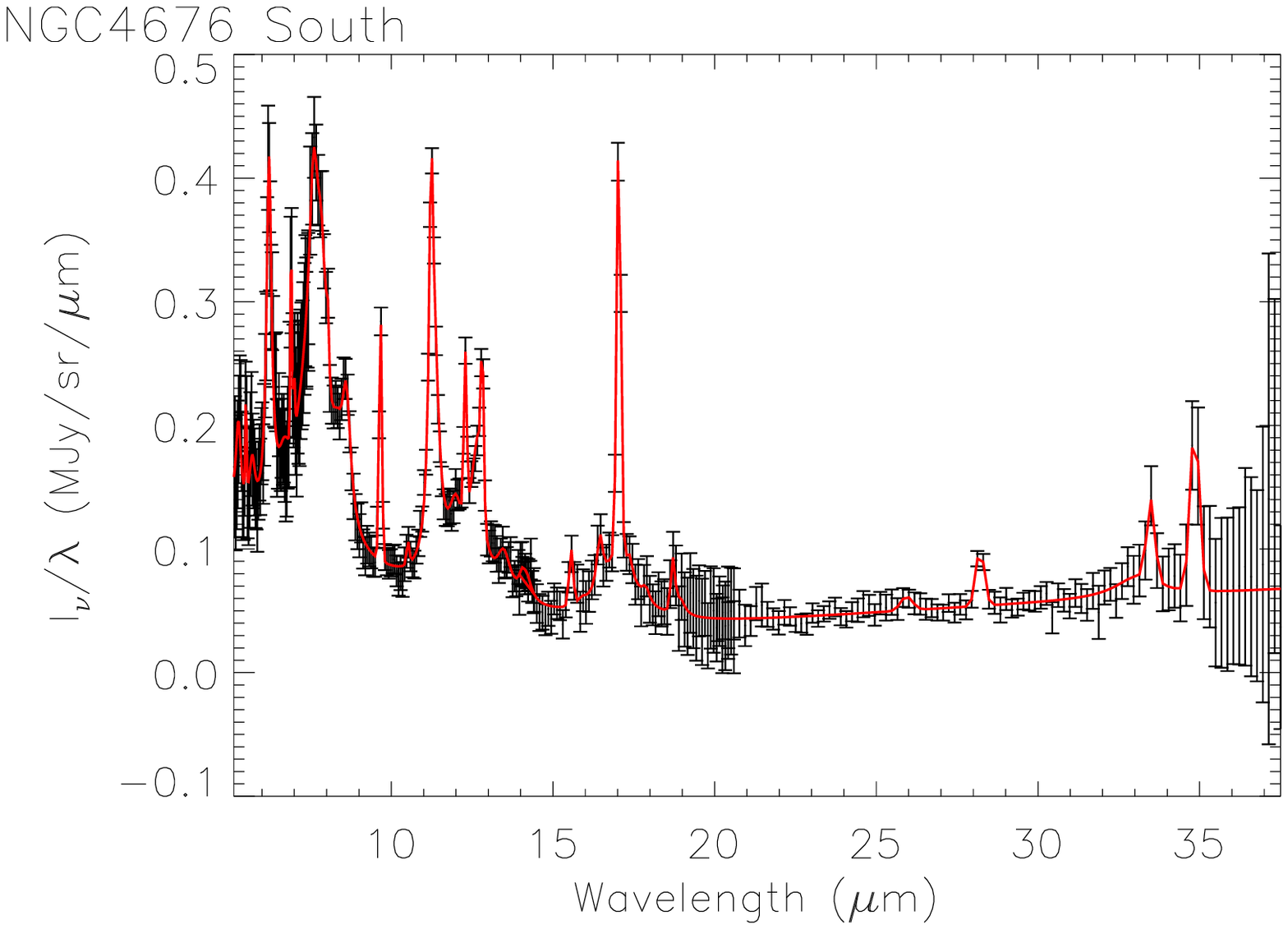}
\includegraphics[scale=0.4]{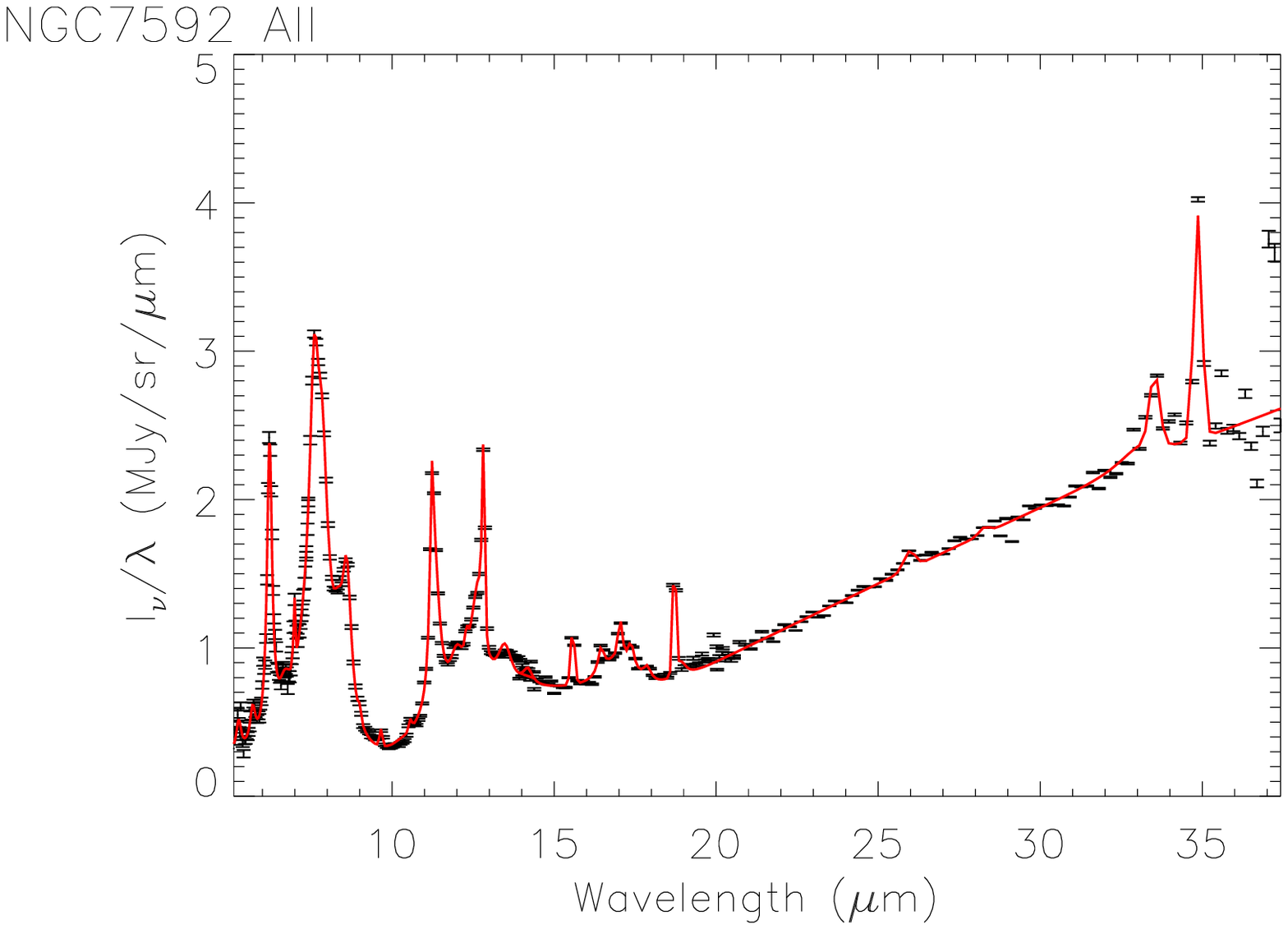}
\includegraphics[scale=0.4]{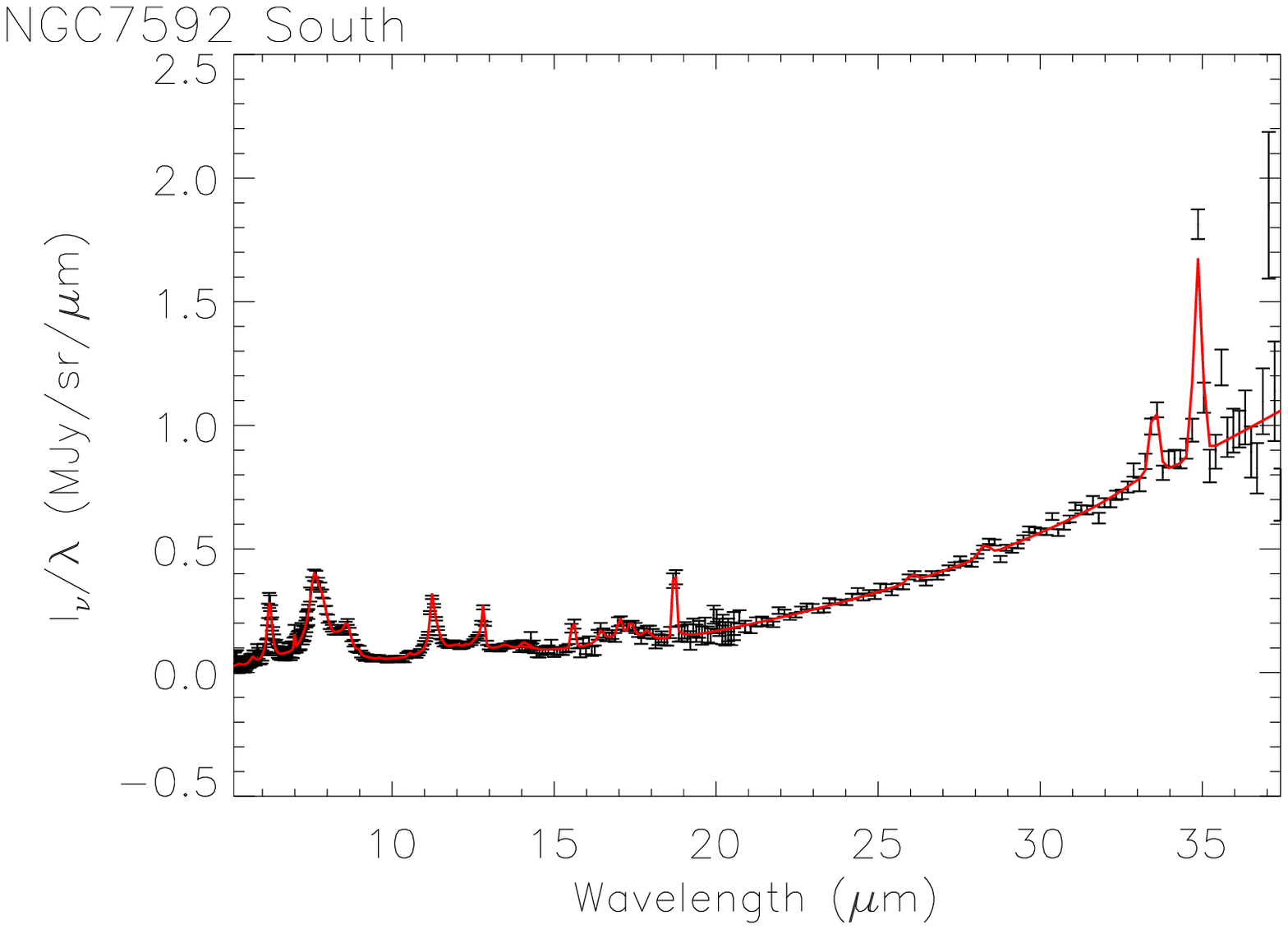}
\includegraphics[scale=0.4]{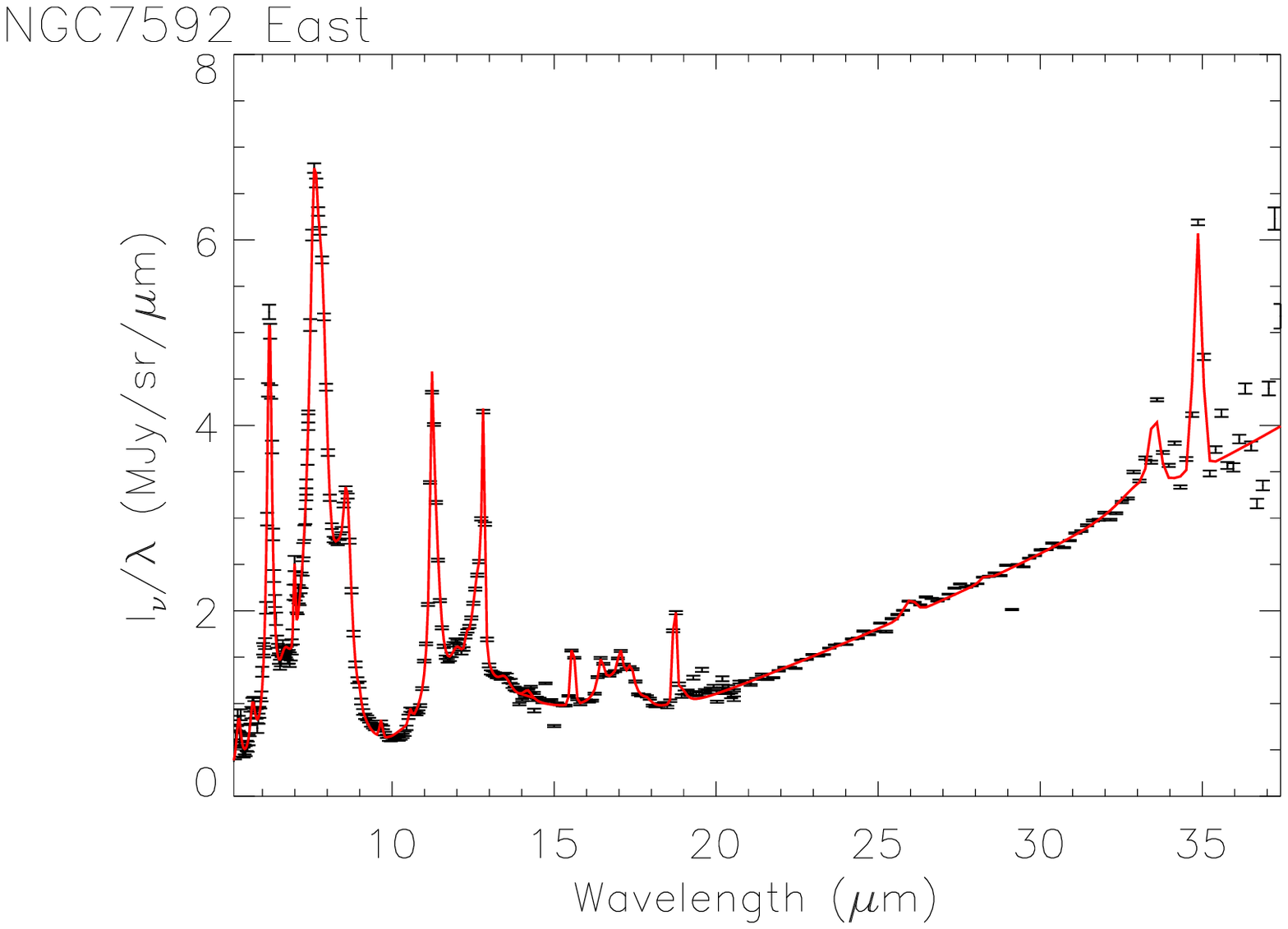}
\includegraphics[scale=0.4]{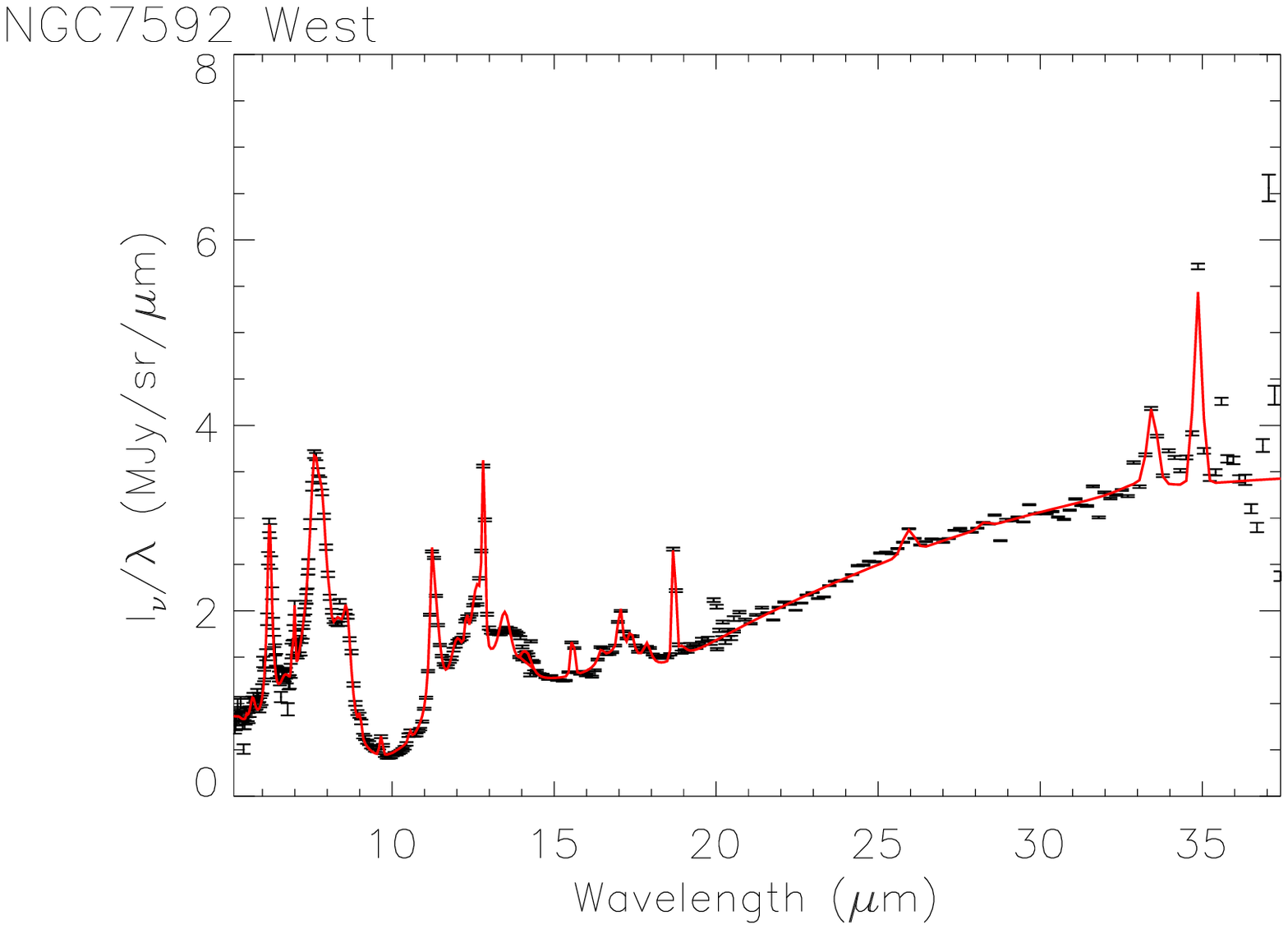}
\includegraphics[scale=0.4]{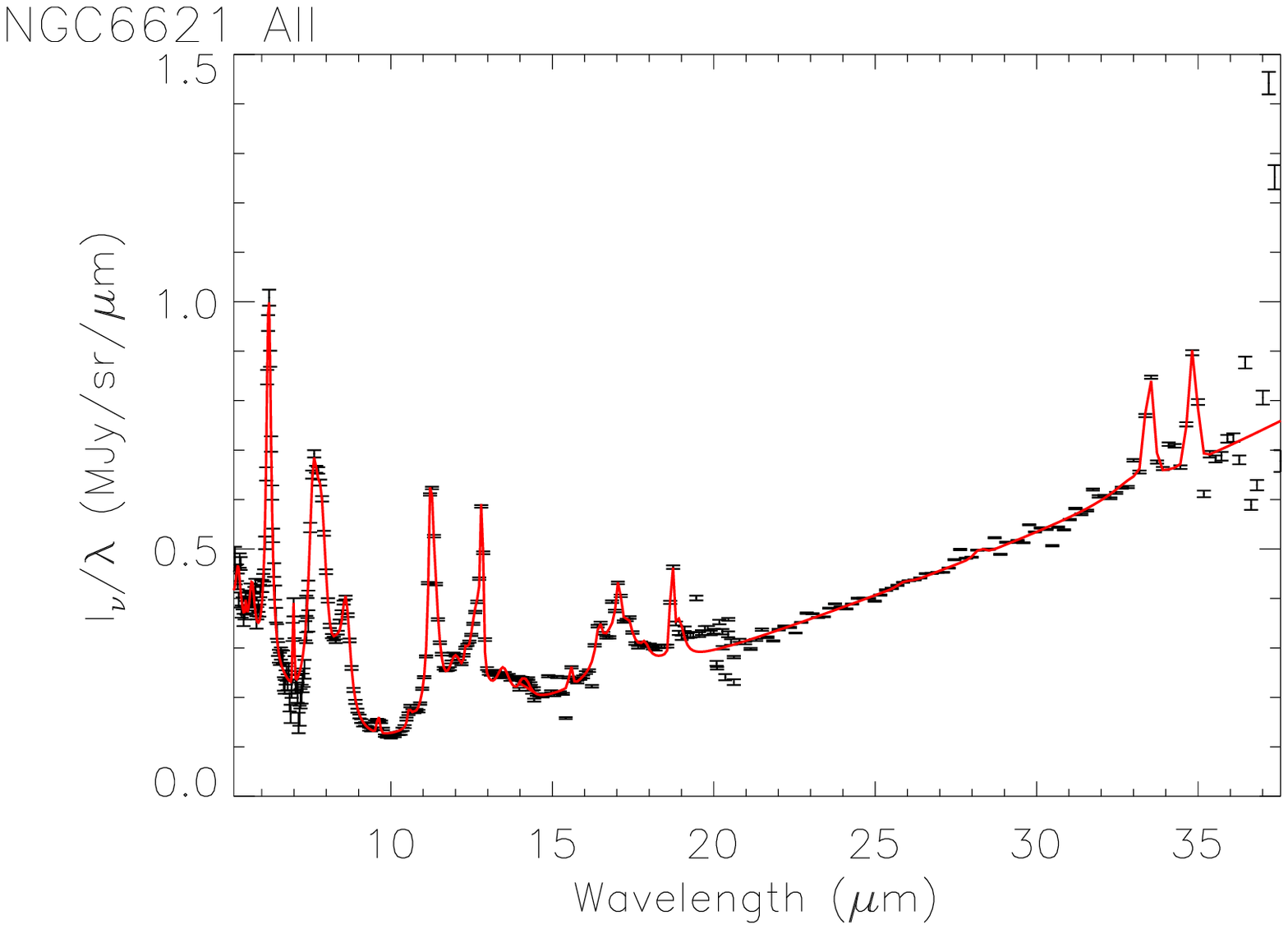}
\end{center}
\caption{\footnotesize{IRS spectra (black errorbars) fitted with PAHFIT (red line). The size and central coordinates of the individual regions are given in Tab.~\ref{tab_PAH}. }}
\label{spectra}
\end{figure*}

\begin{figure*}
\begin{center}
\includegraphics[scale=0.4]{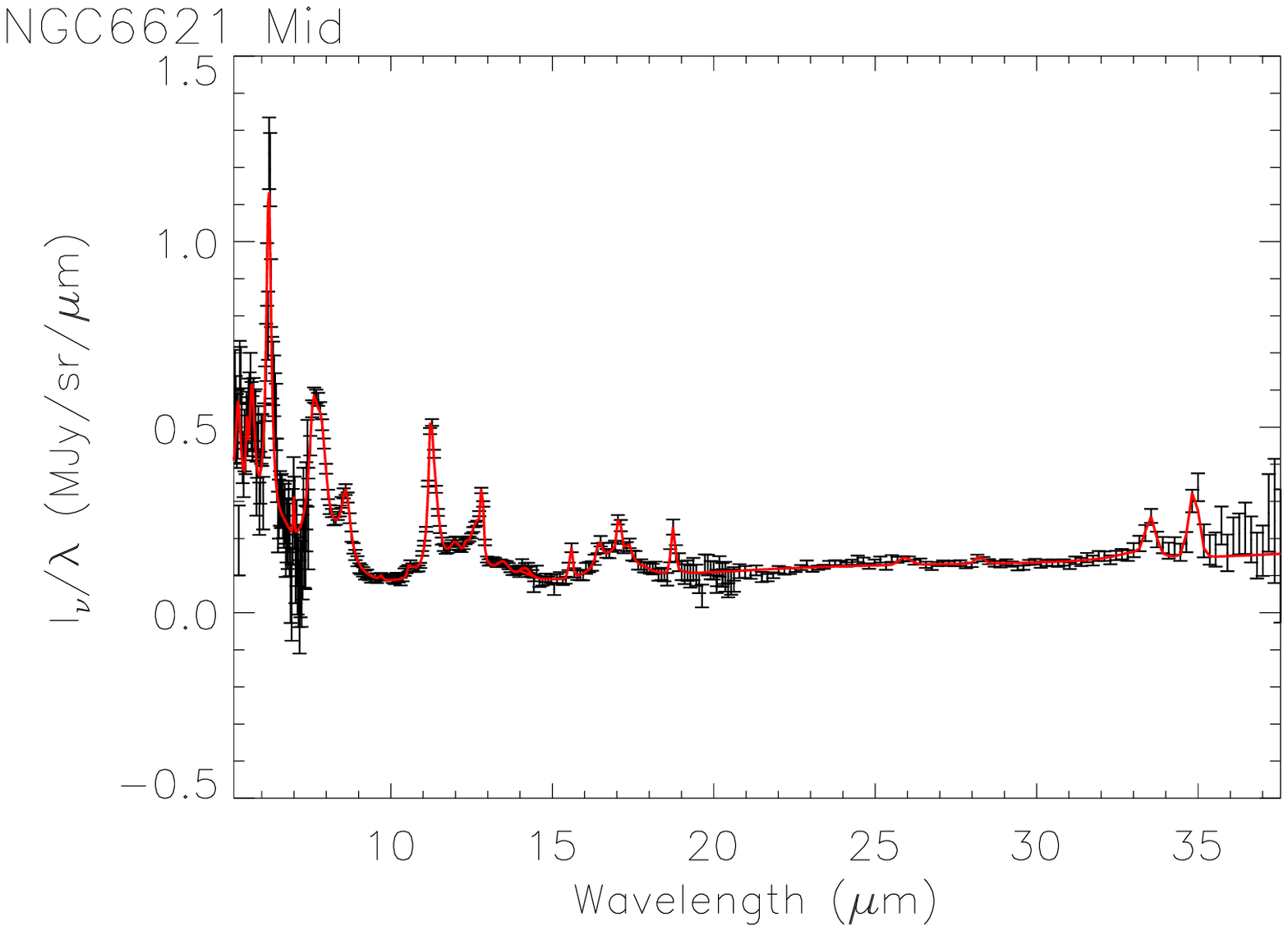}
\includegraphics[scale=0.4]{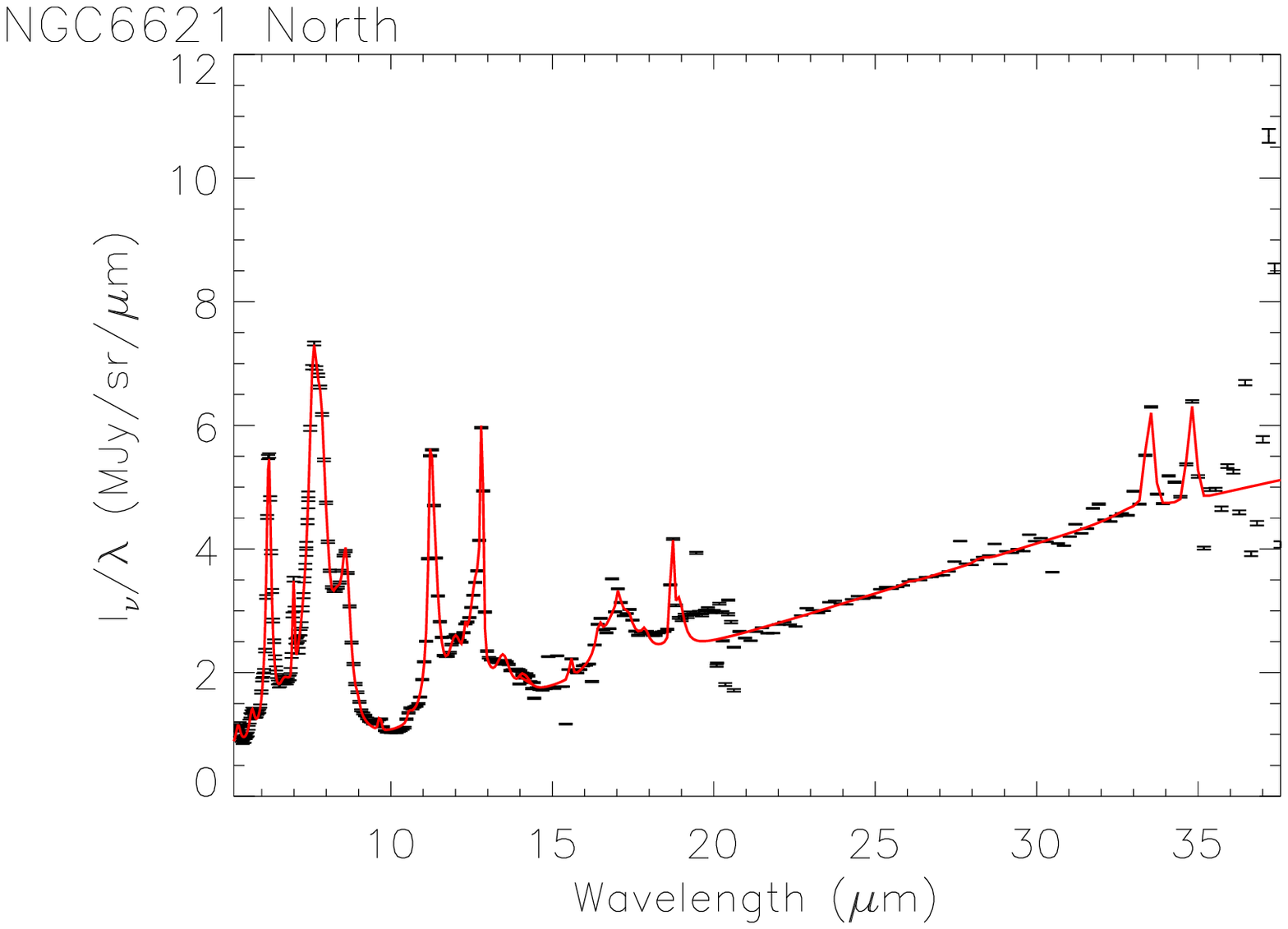}
\includegraphics[scale=0.4]{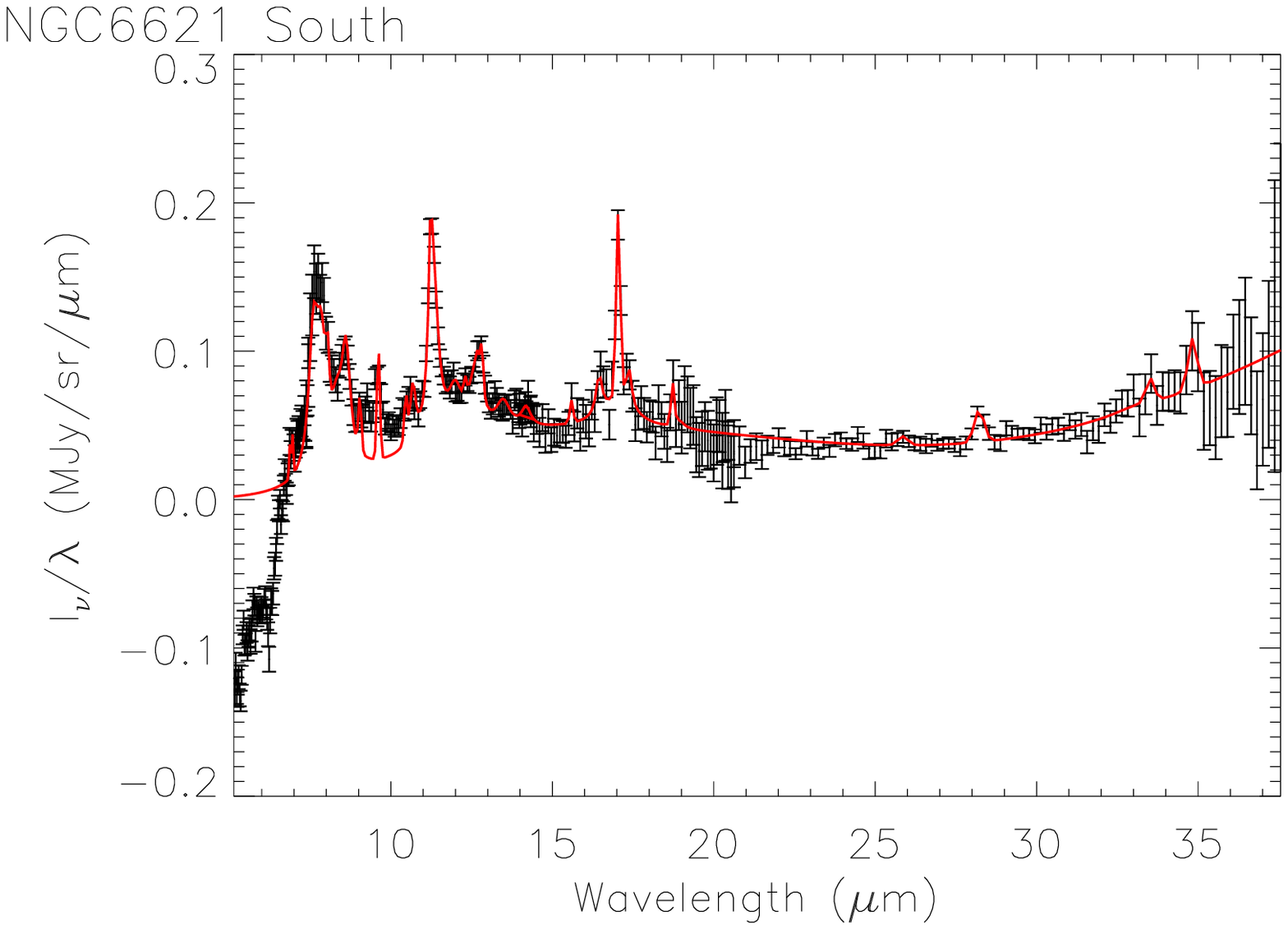}
\includegraphics[scale=0.4]{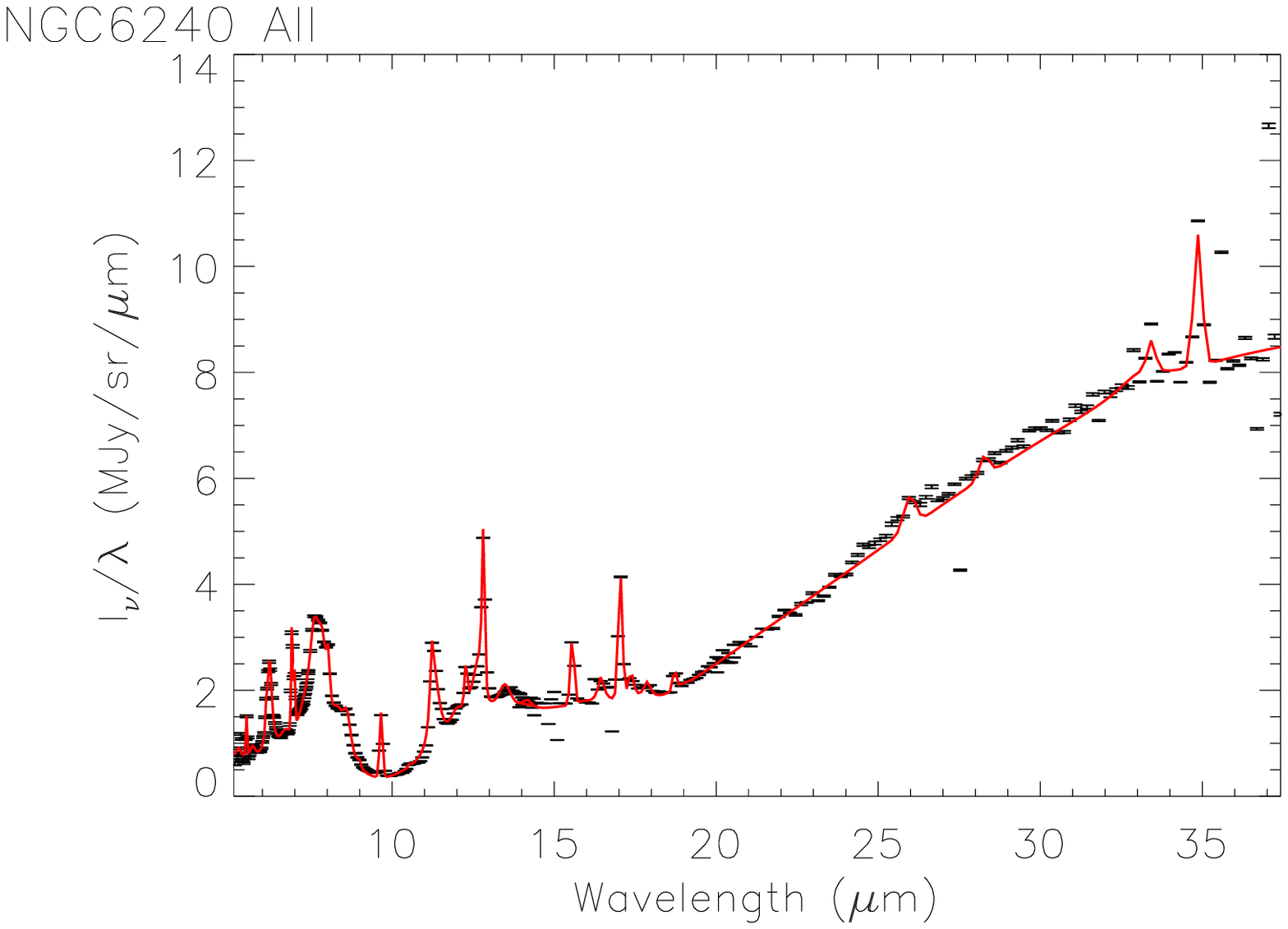}
\includegraphics[scale=0.4]{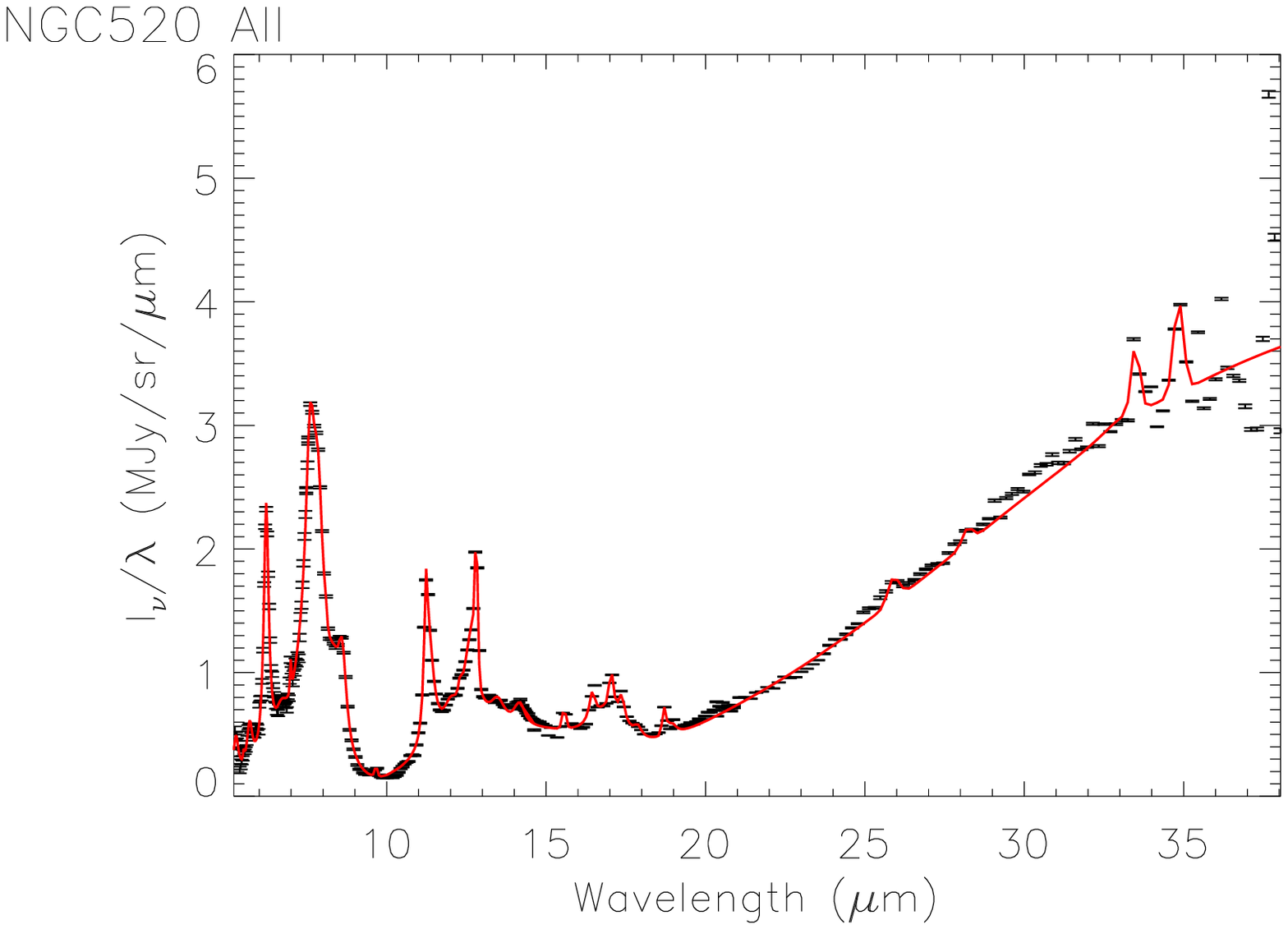}
\includegraphics[scale=0.4]{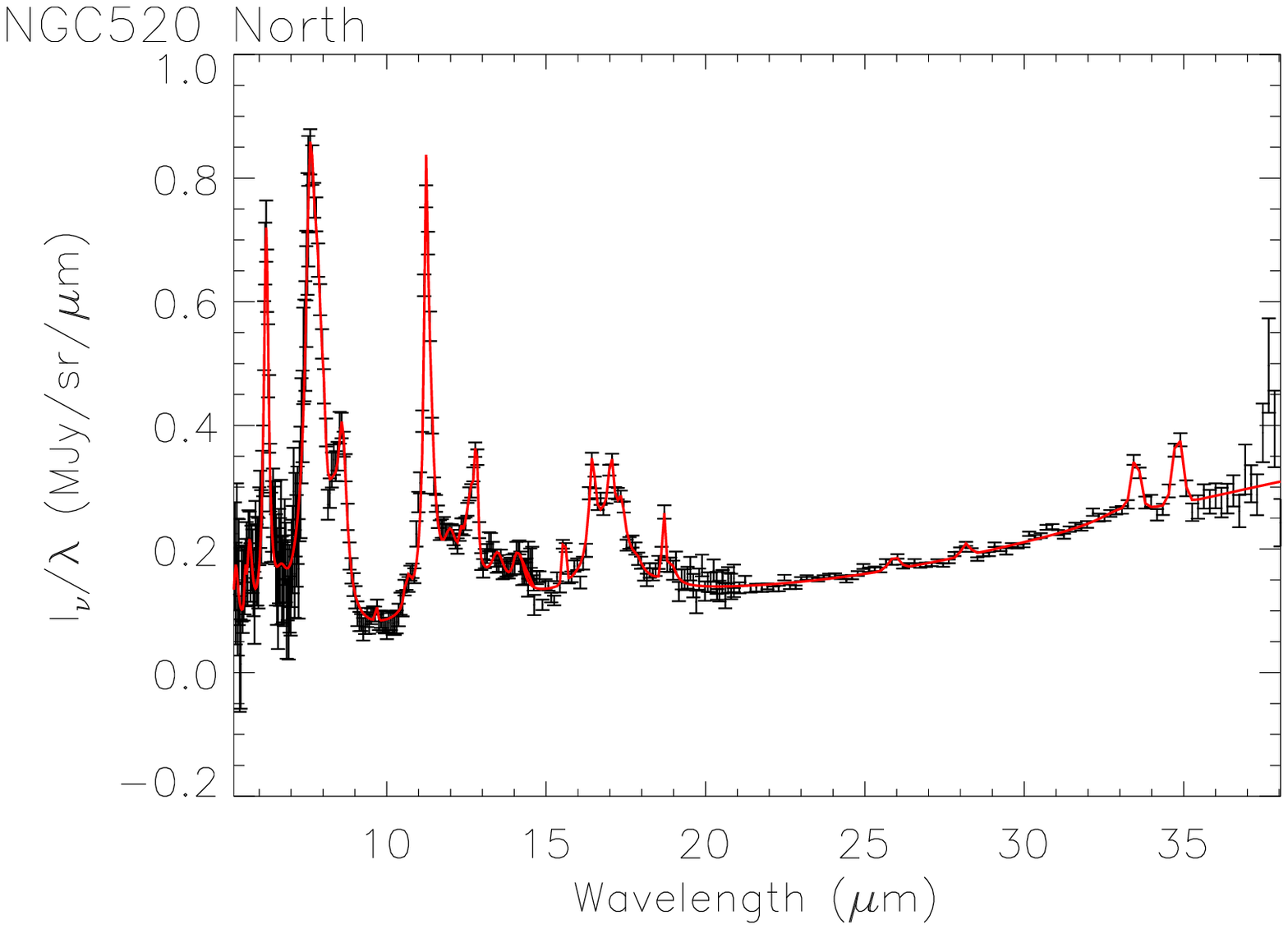}
\includegraphics[scale=0.4]{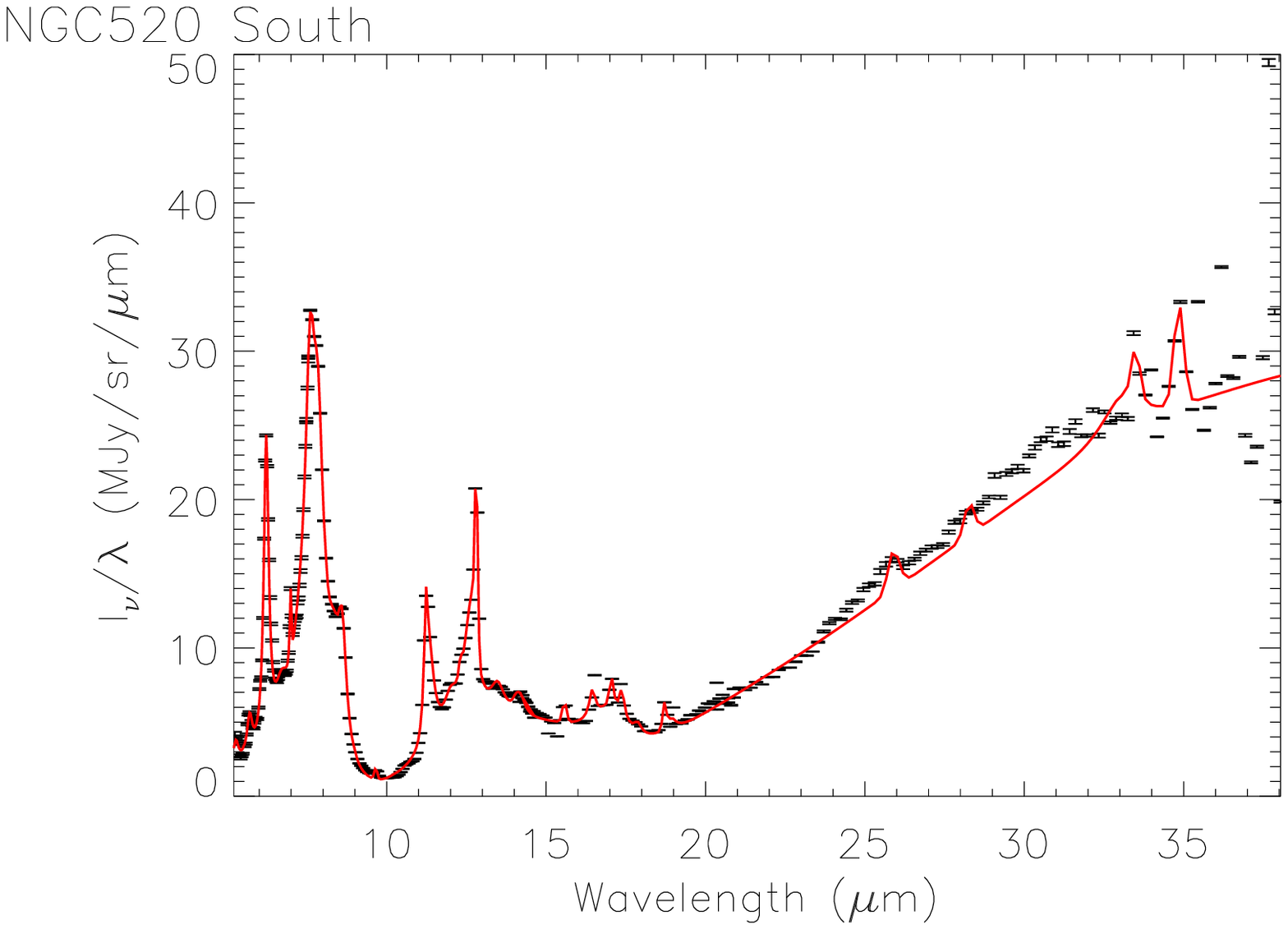}
\includegraphics[scale=0.4]{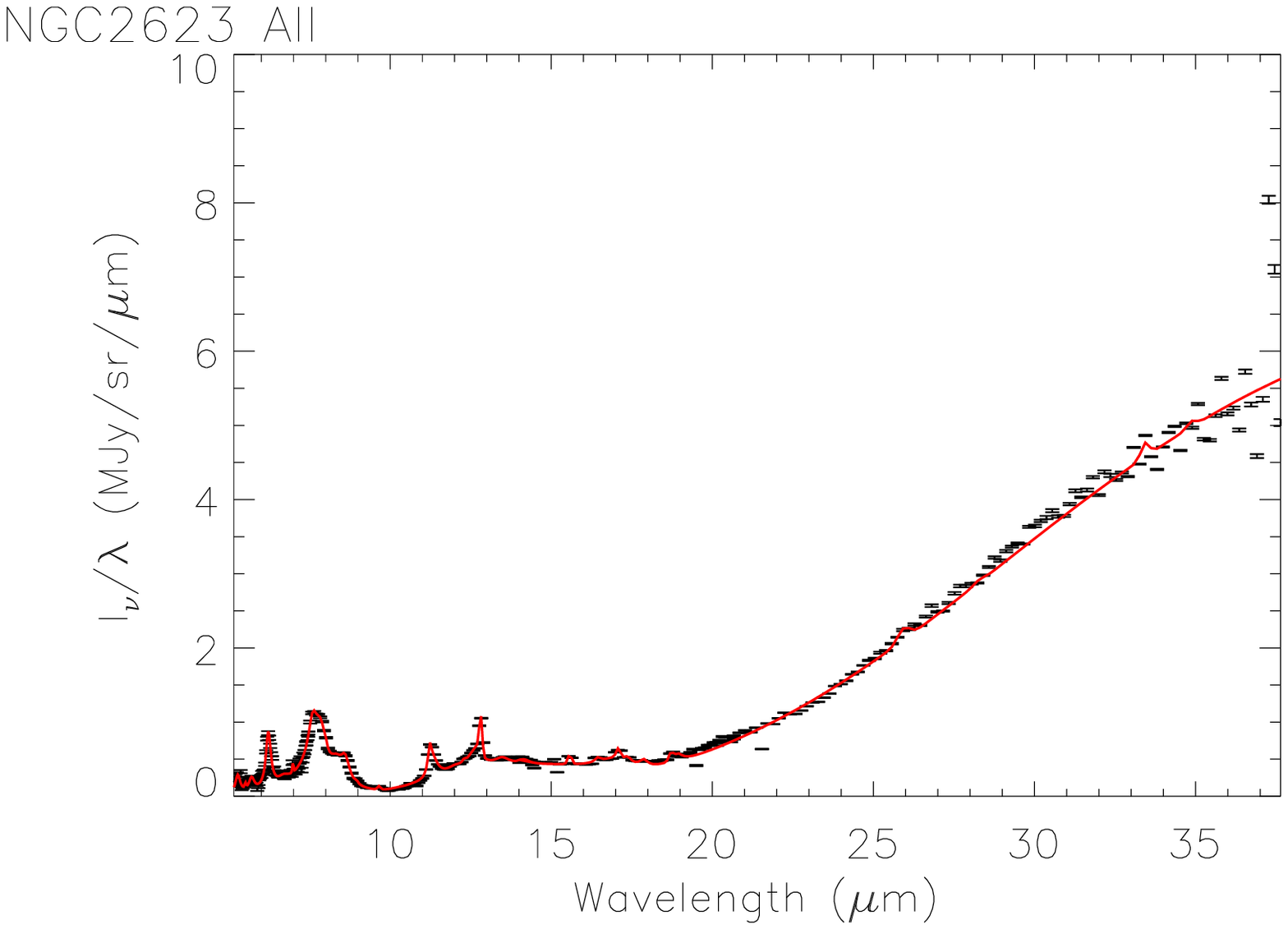}
\end{center}
\figurenum{\ref{spectra}}
\caption{(Continued).}
\end{figure*}

\begin{figure*}
\begin{center}
\includegraphics[scale=0.4]{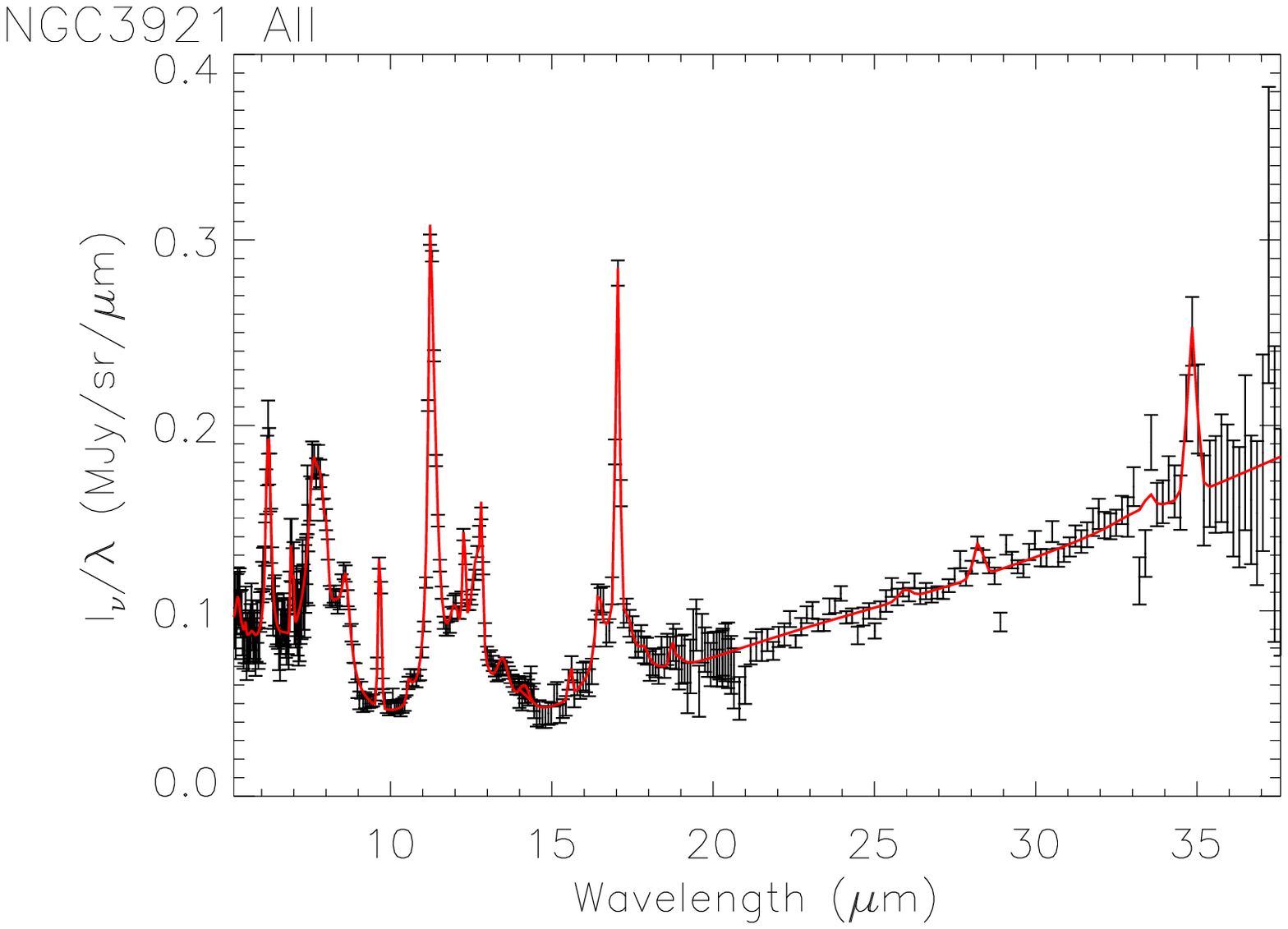}
\includegraphics[scale=0.4]{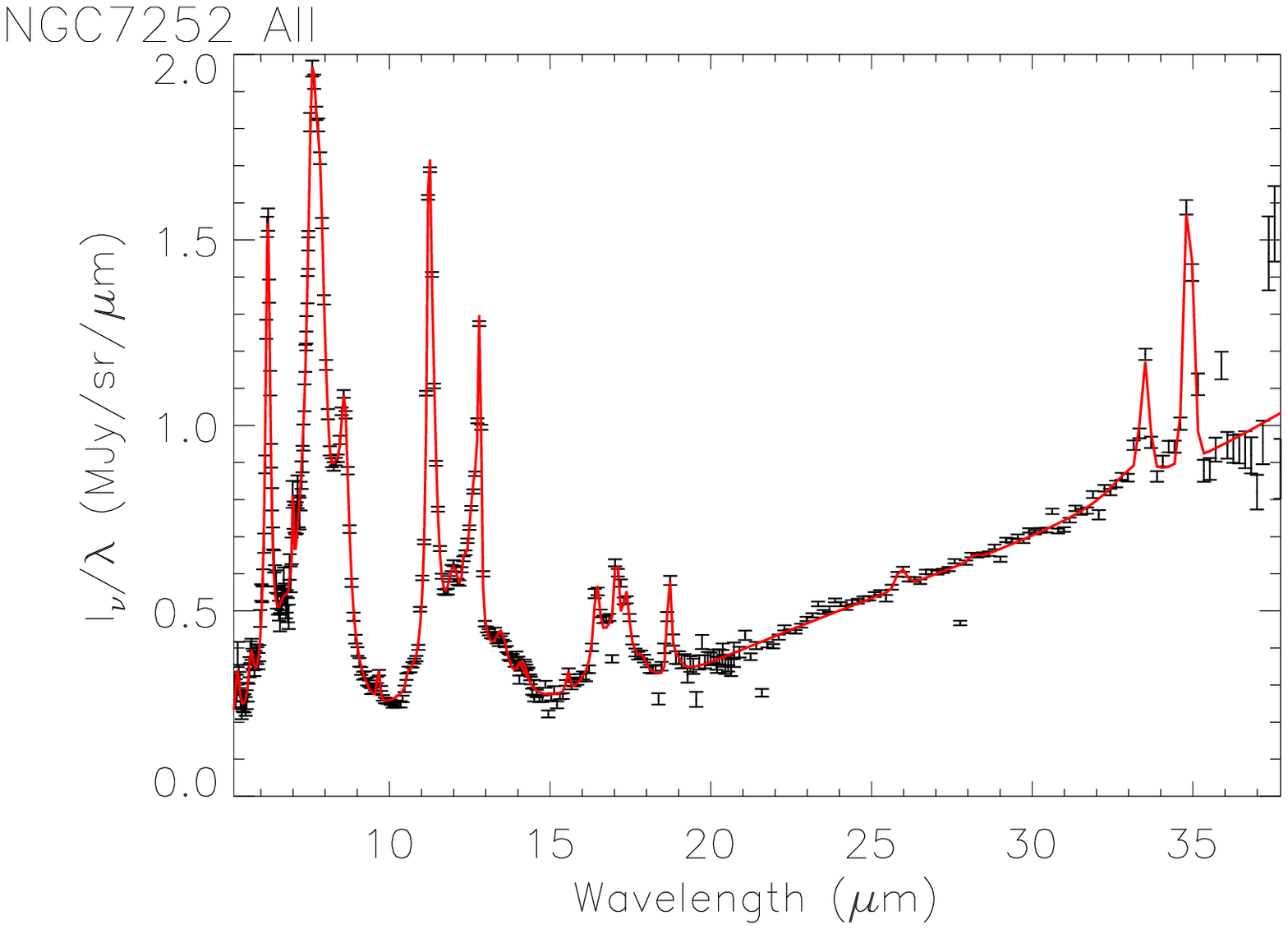}
\end{center}
\figurenum{\ref{spectra}}
\caption{(Continued).}
\end{figure*}

\begin{figure*}
\includegraphics[scale=0.46]{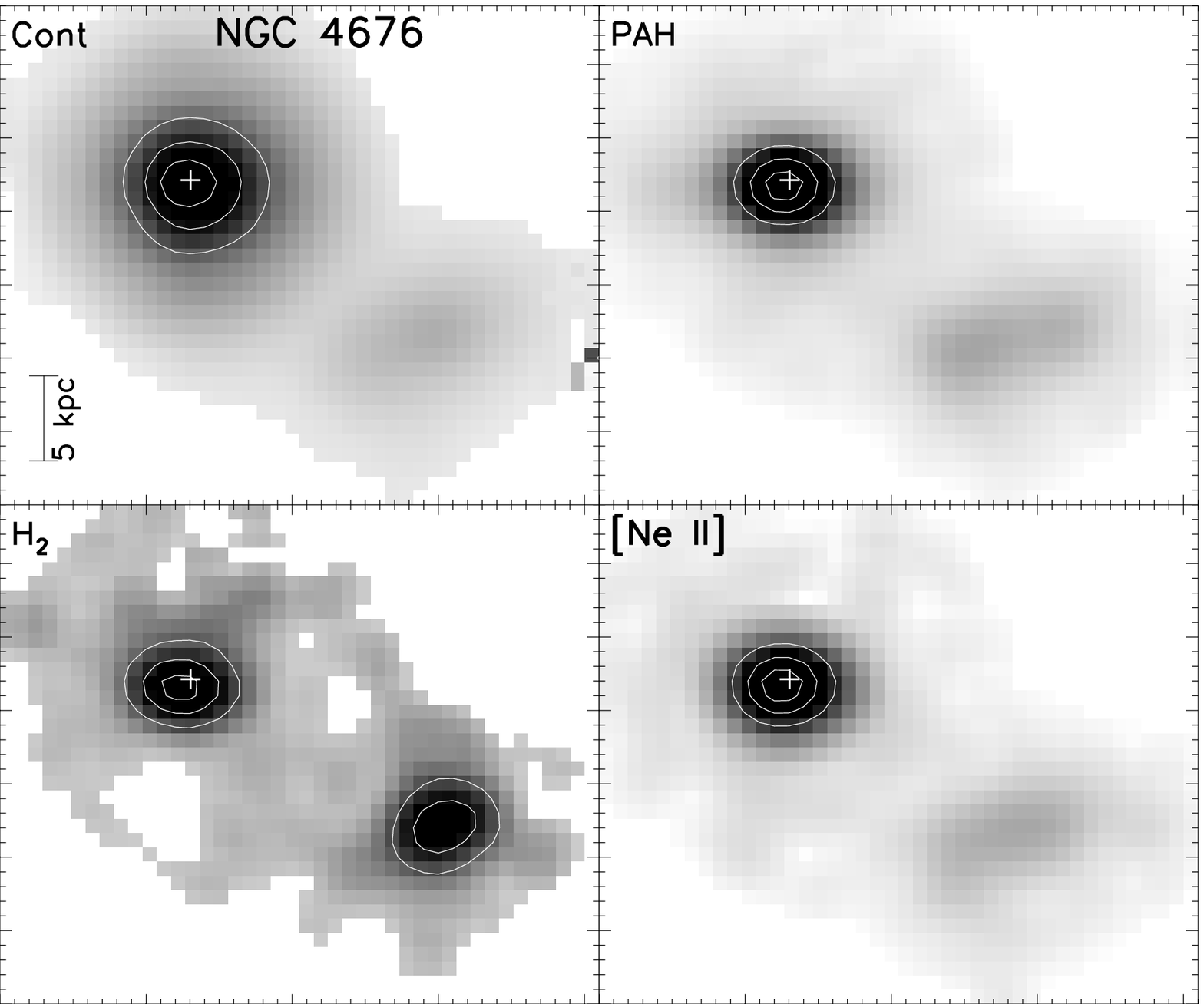}
\includegraphics[scale=0.46]{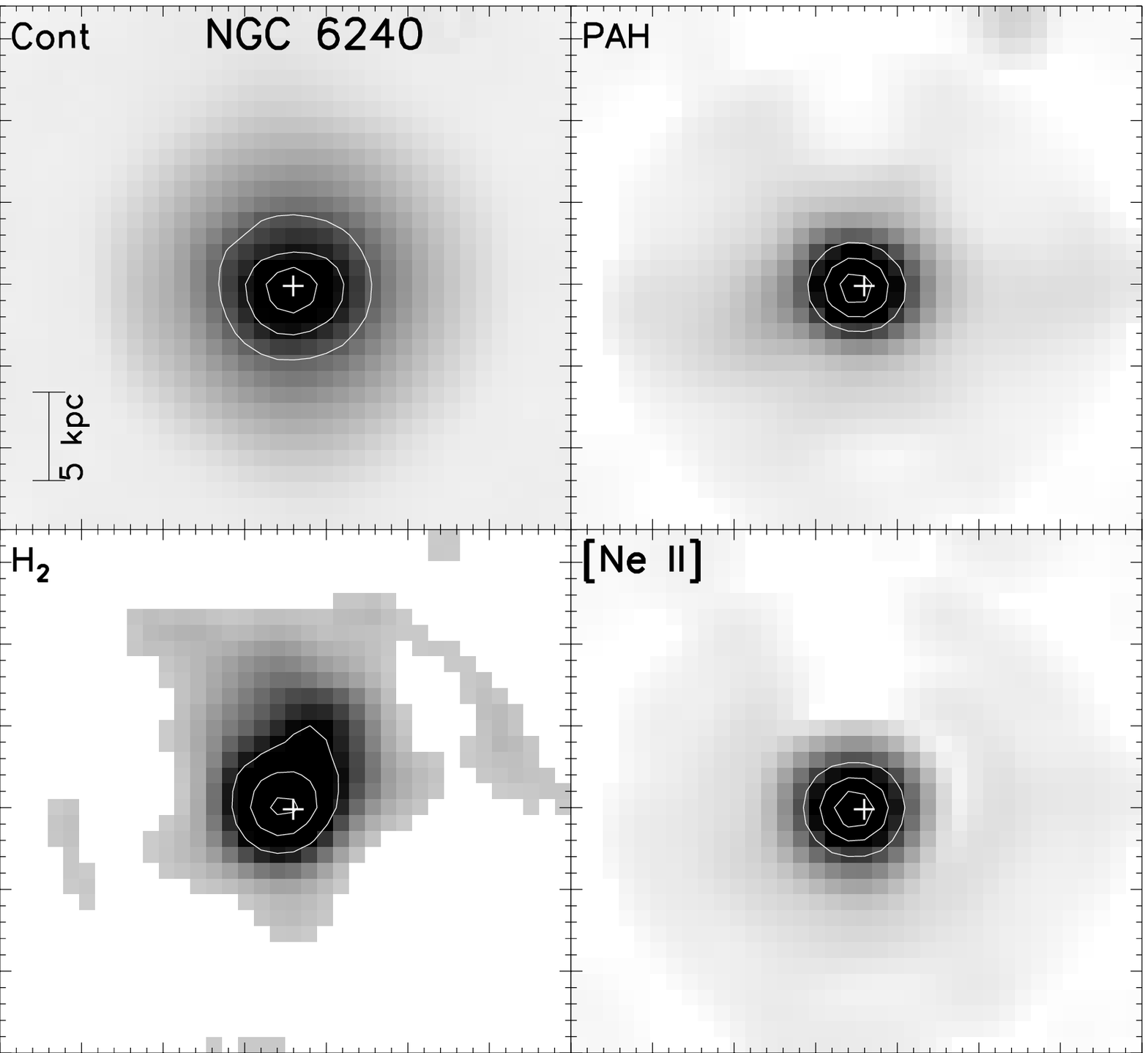}\\

\includegraphics[scale=0.46]{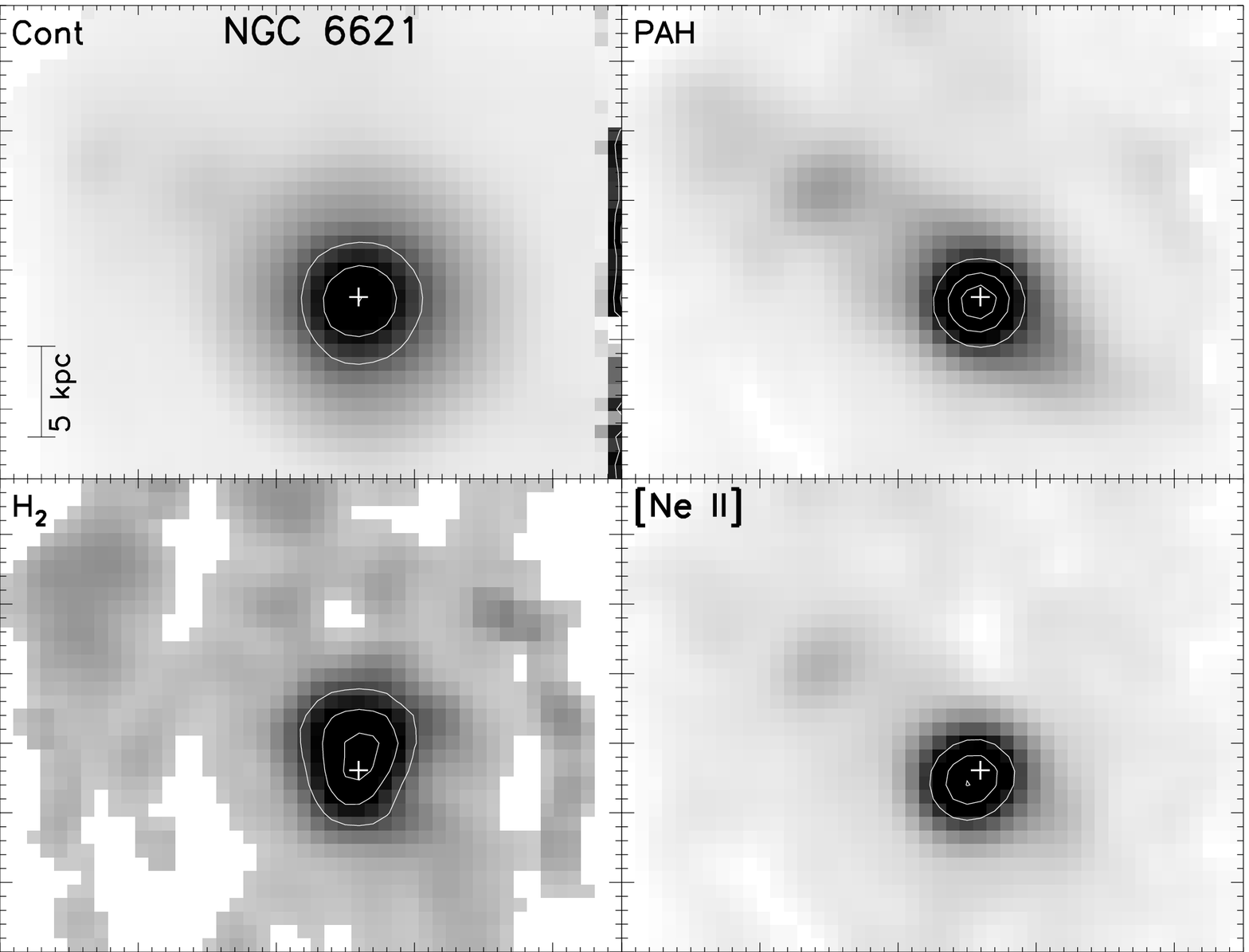}
\includegraphics[scale=0.46]{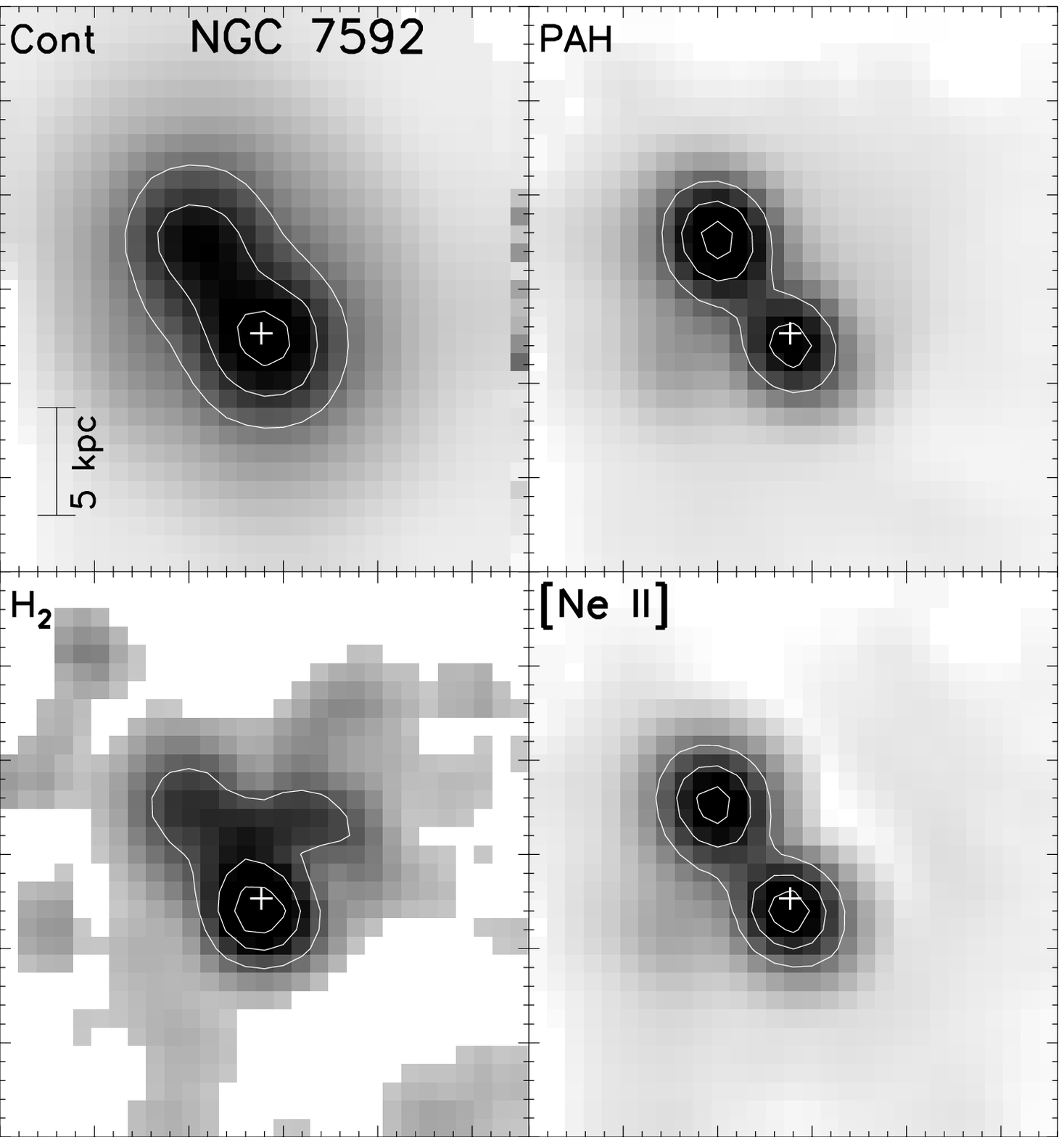}
\caption{\footnotesize{The main mid-IR feature maps with contours, from left to right and top to bottom: the integrated mid-IR continuum (5.3--15~$\mu$m) as measured with PAHFIT, the combined PAH features, the combined H$_2$ emission features, and the [Ne II] emission (see text). The center is marked with a plus symbol signifying the peak in the integrated mid-IR continuum and the tickmarks correspond to 1 pixel-intervals (1.85\arcsec). The North and East directions in each frame of this figure are the same as in the corresponding frame of Fig.~\ref{maps_mid-IR}.}}
\label{maps_feature_4627}
\end{figure*}

\begin{figure*}
\includegraphics[scale=0.46]{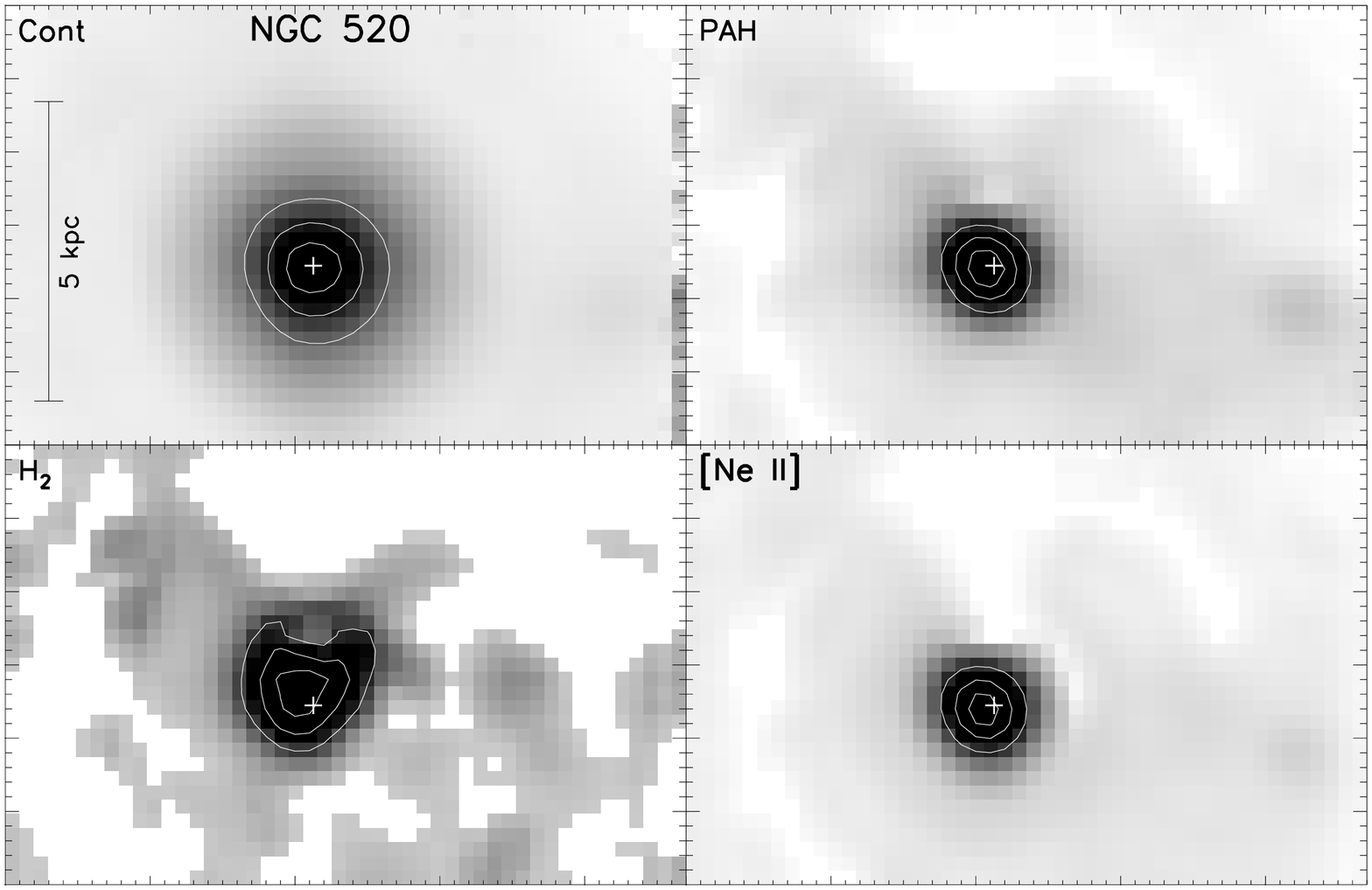}\\
\includegraphics[scale=0.46]{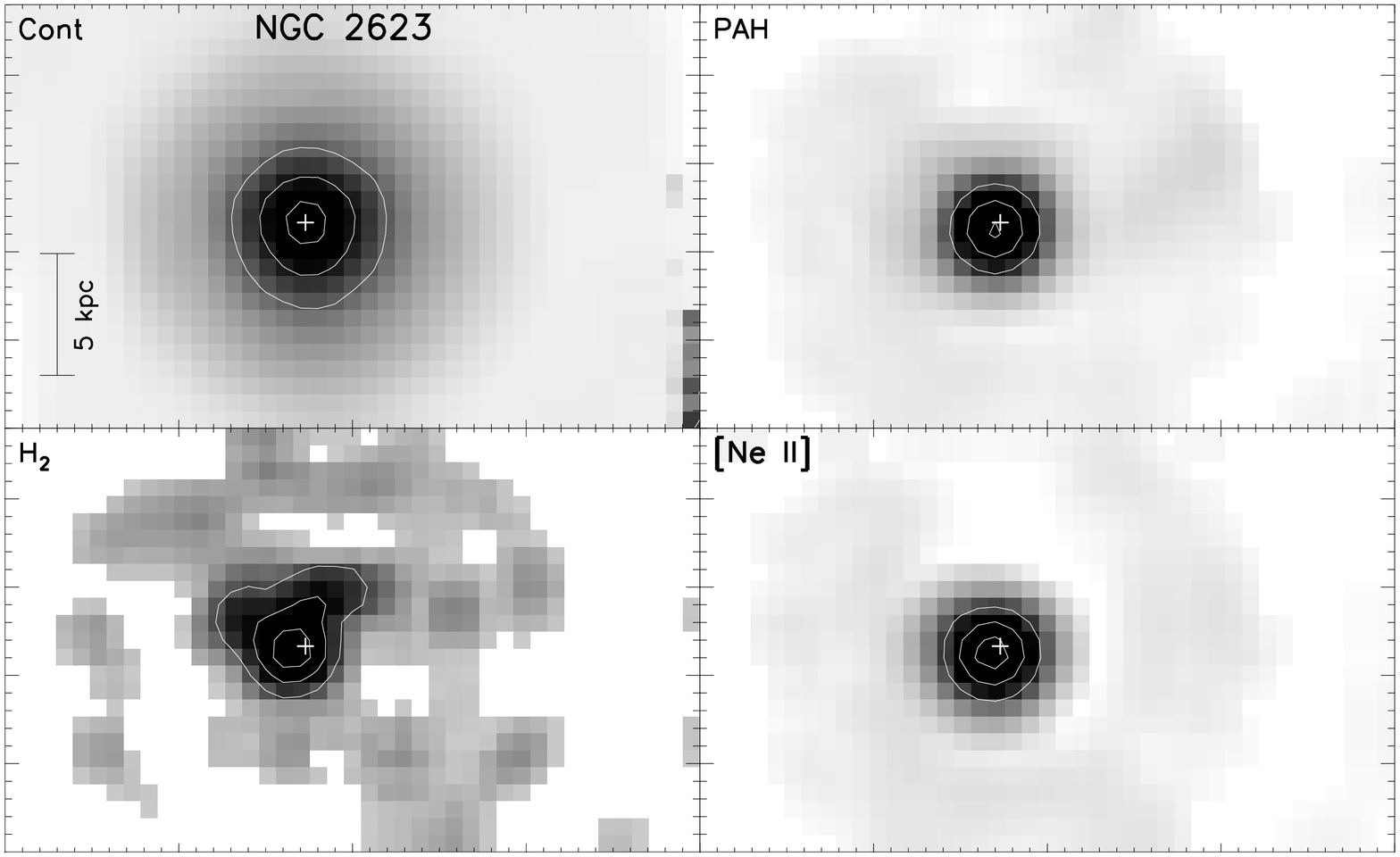}\\
\includegraphics[scale=0.46]{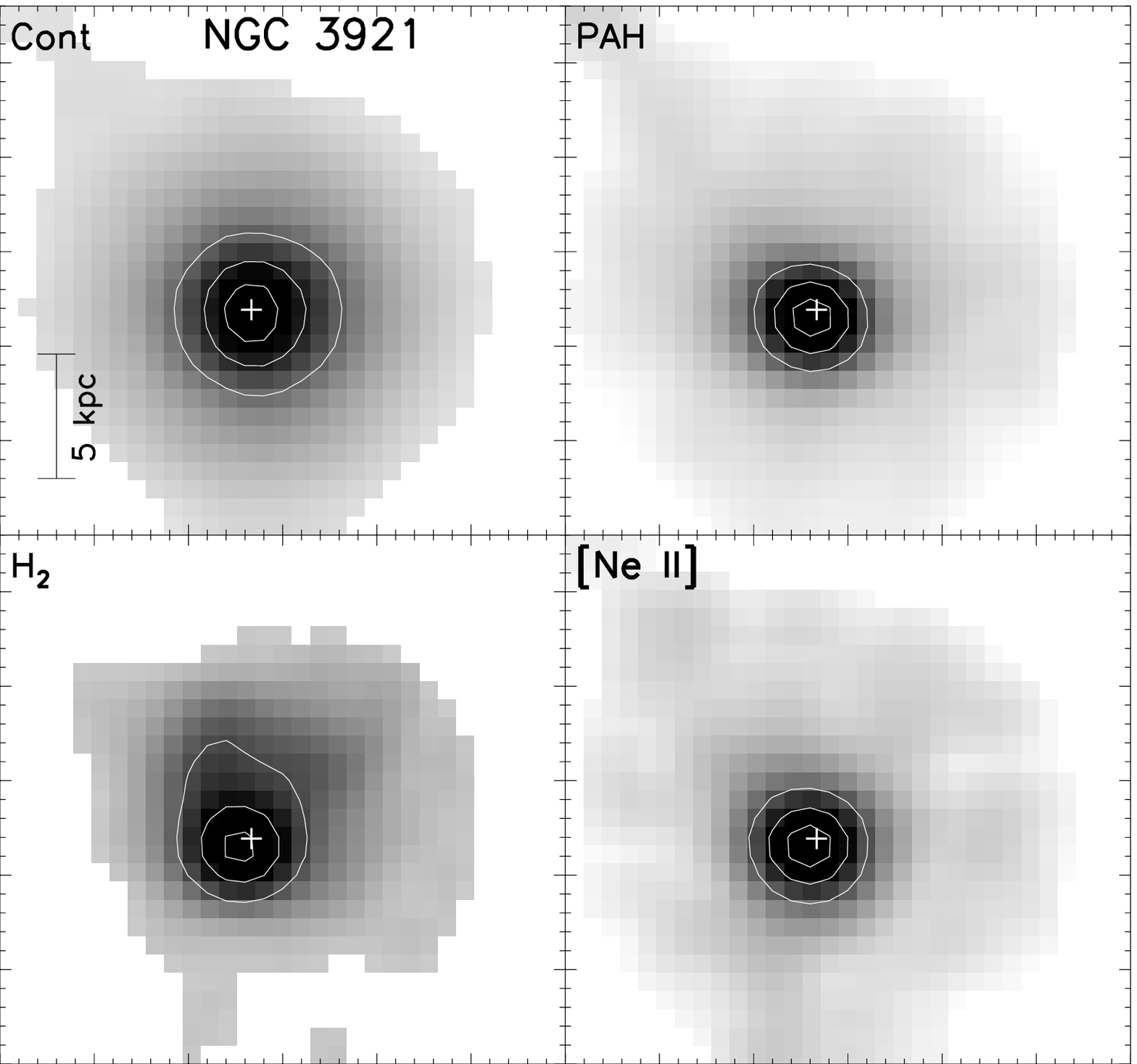}
\includegraphics[scale=0.46]{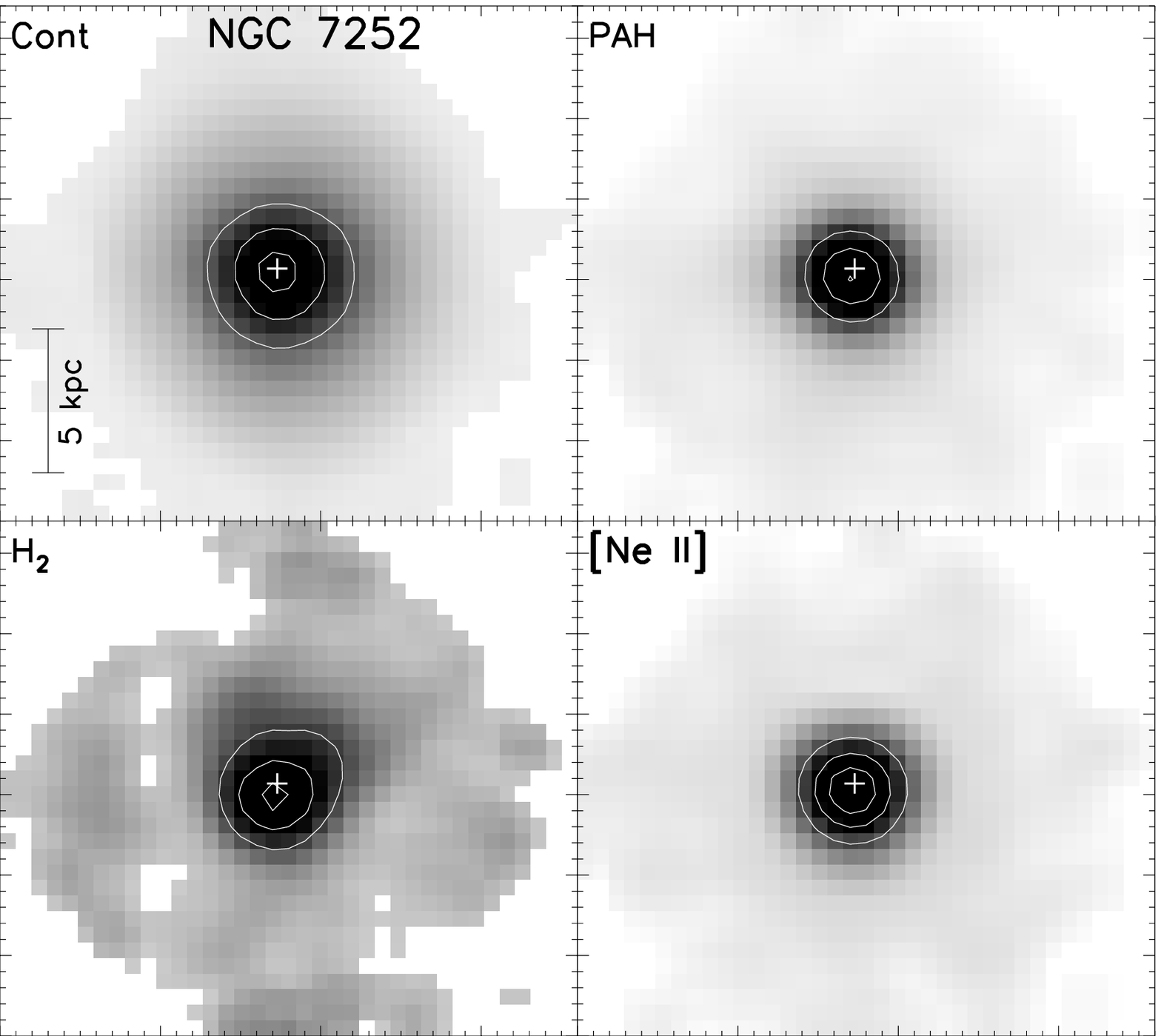}
\figurenum{\ref{maps_feature_4627}}
\caption{\emph{Continued}.}
\end{figure*}

\begin{figure*}
\begin{center}
\includegraphics[scale=0.39]{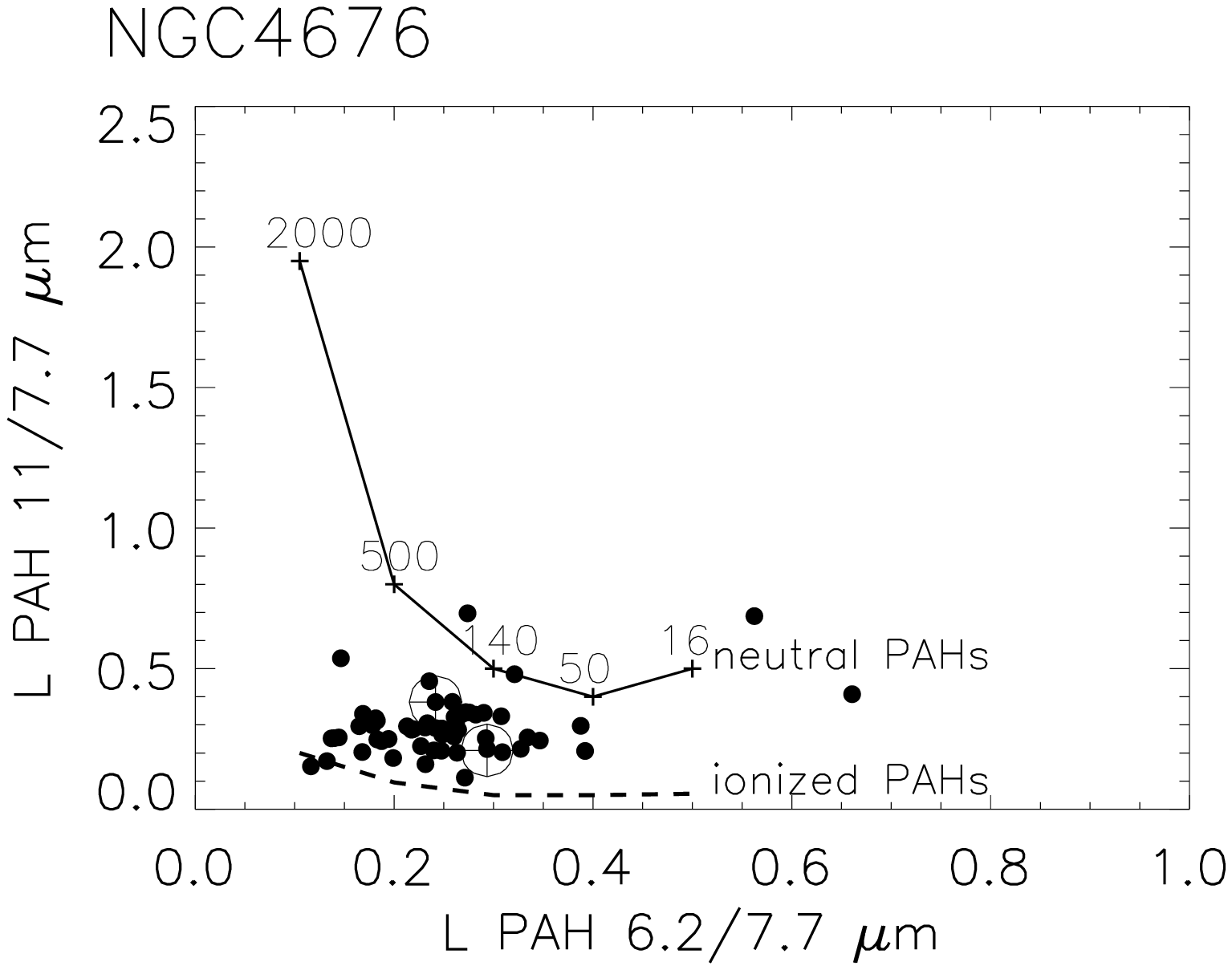}
\includegraphics[scale=0.39]{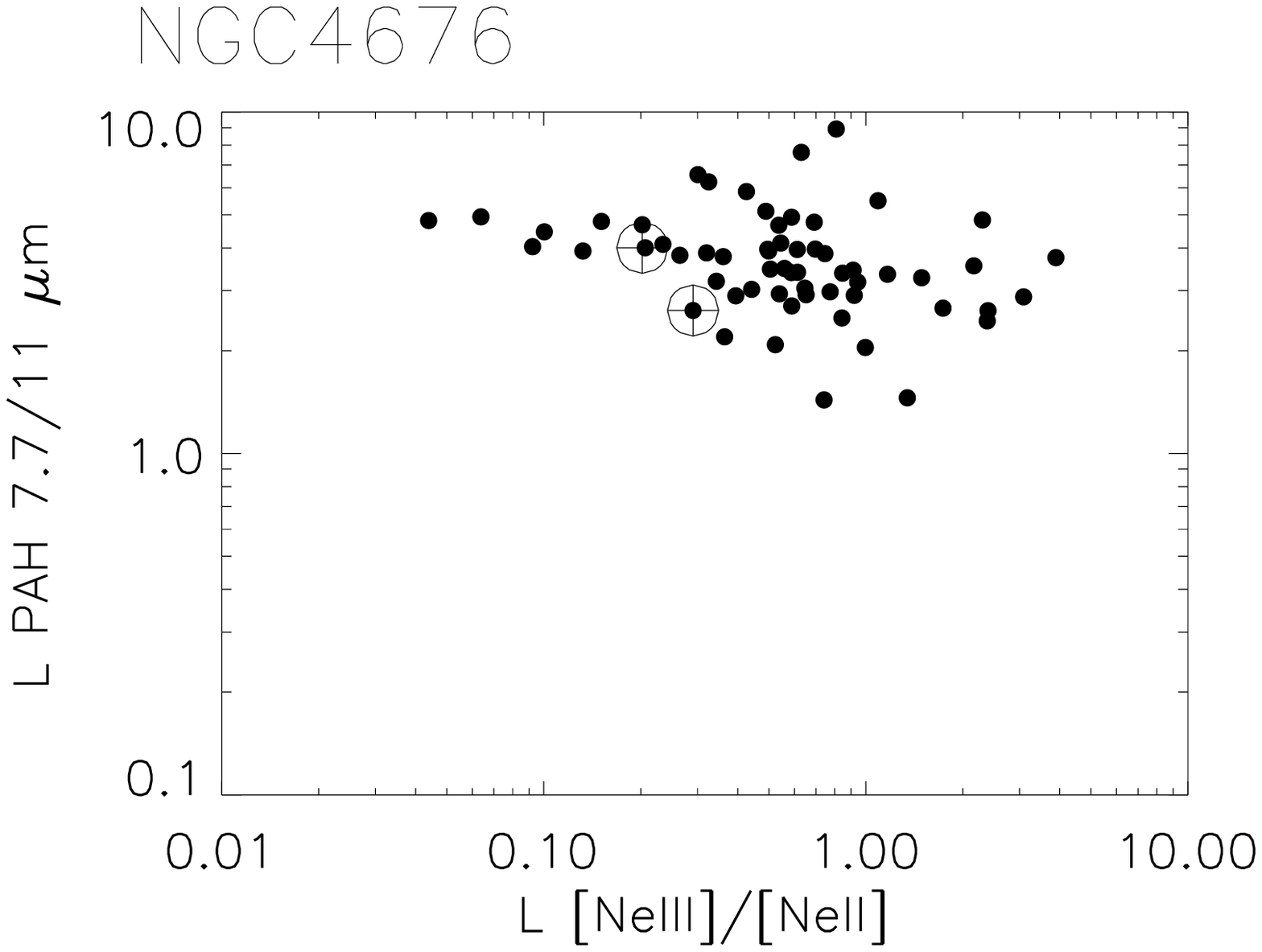}
\includegraphics[scale=0.39]{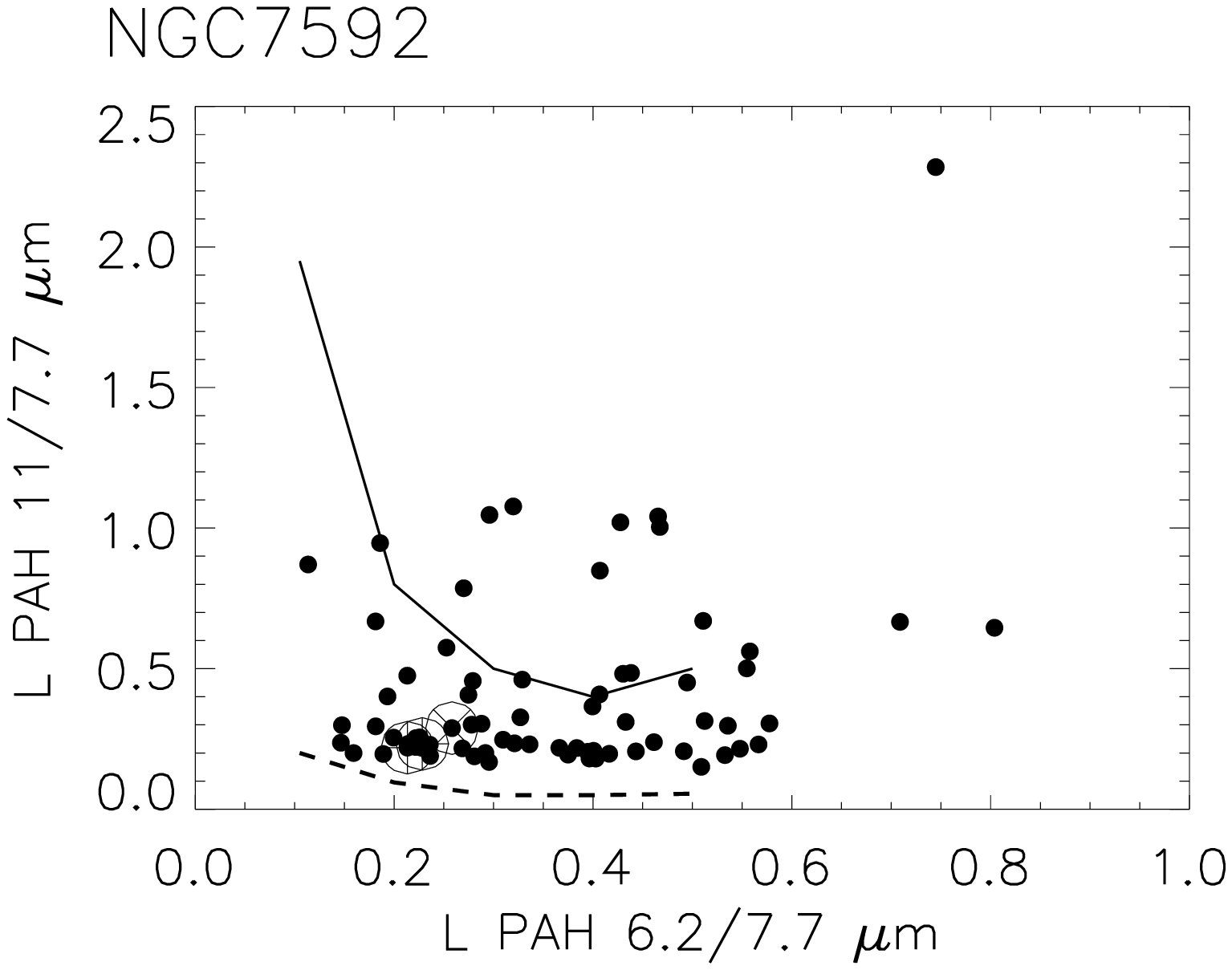}
\includegraphics[scale=0.39]{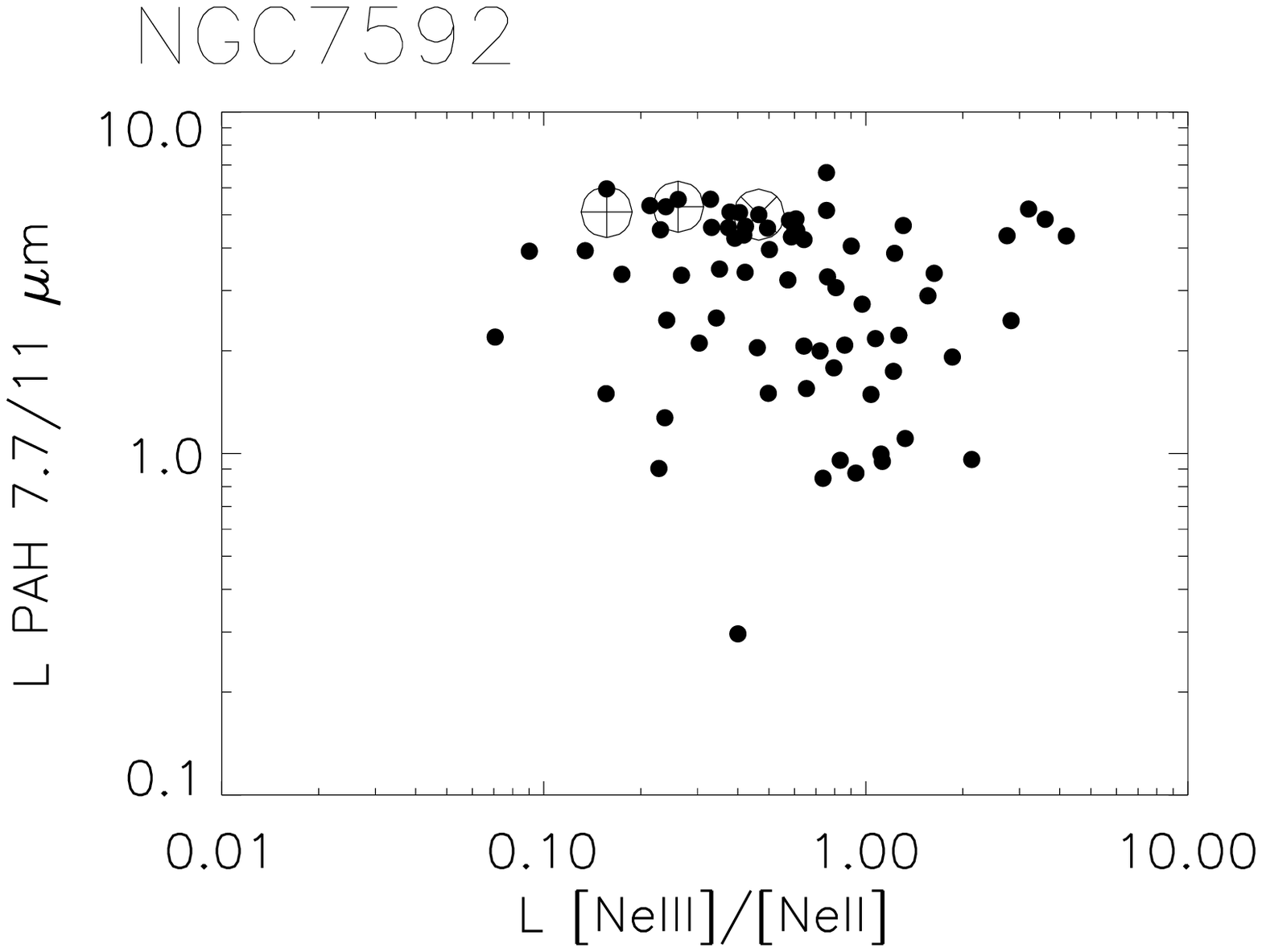}
\includegraphics[scale=0.39]{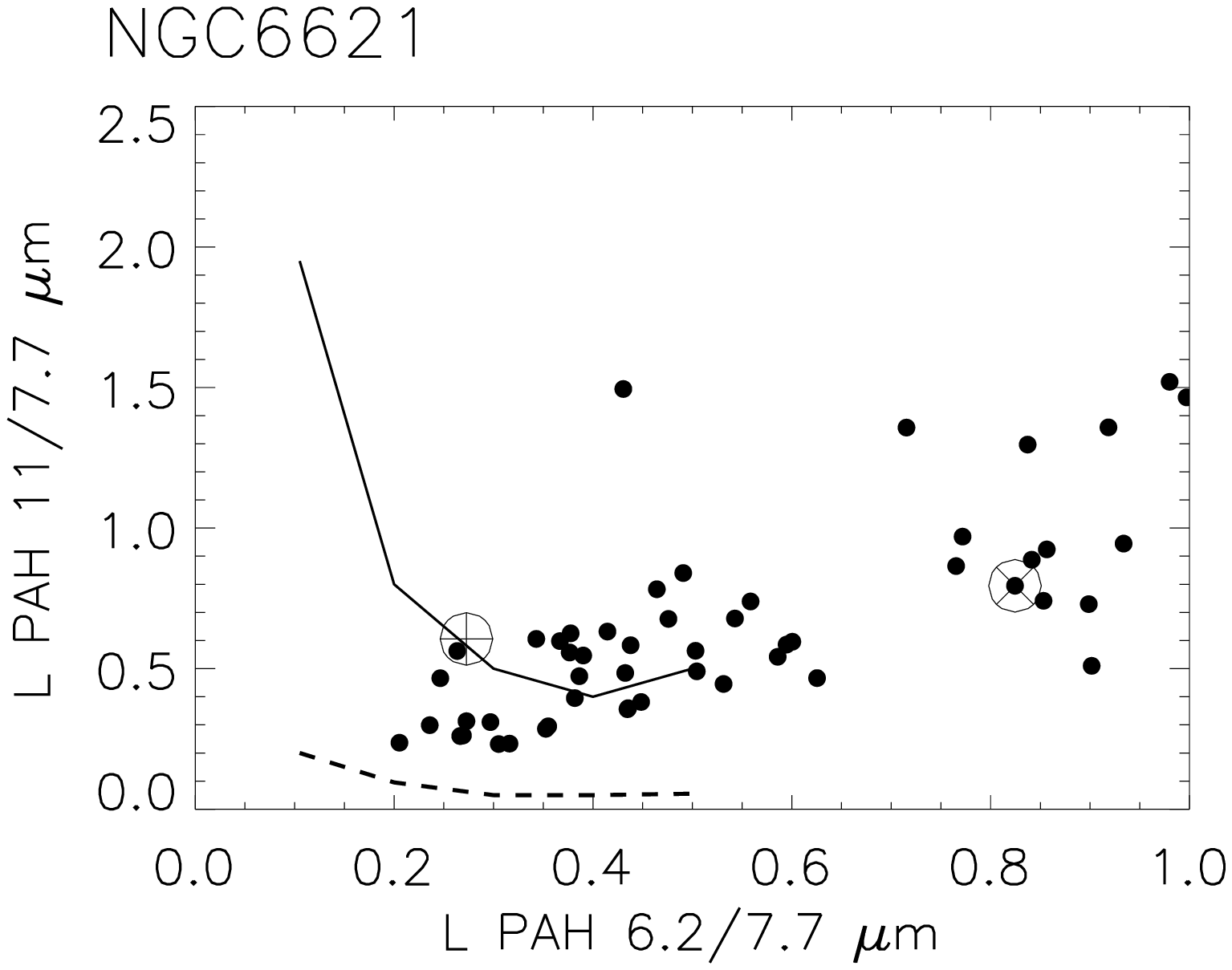}
\includegraphics[scale=0.39]{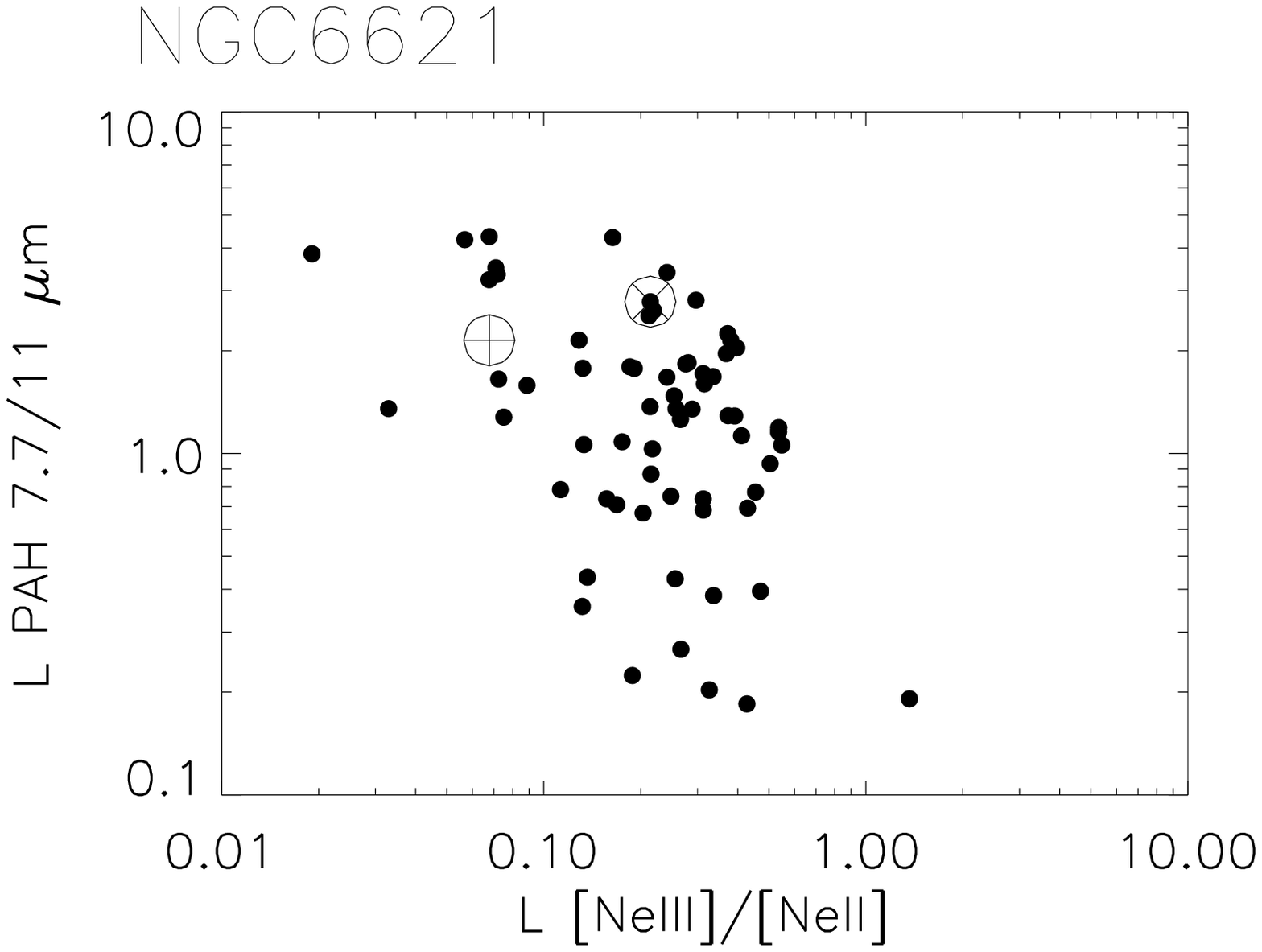}
\includegraphics[scale=0.39]{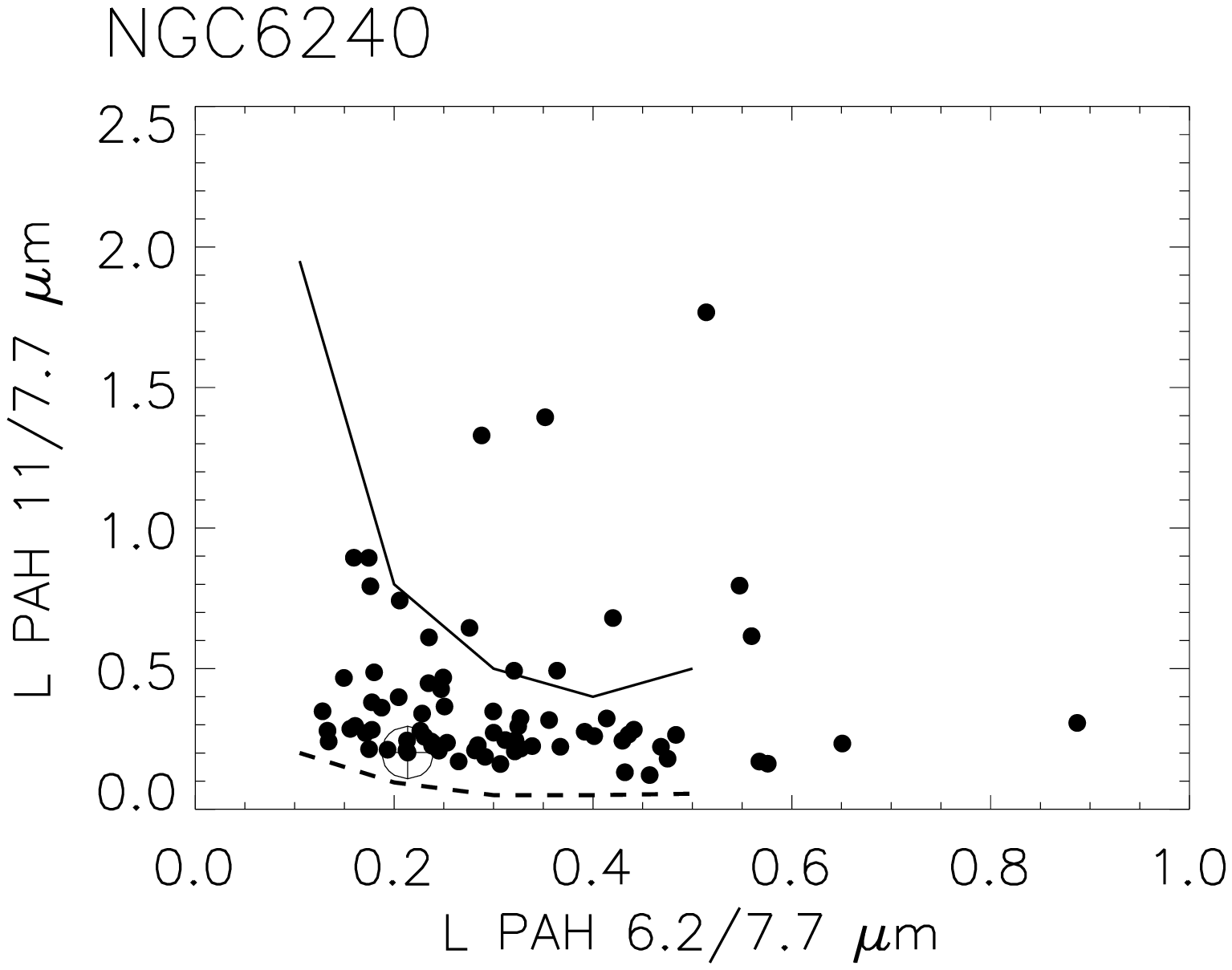}
\includegraphics[scale=0.39]{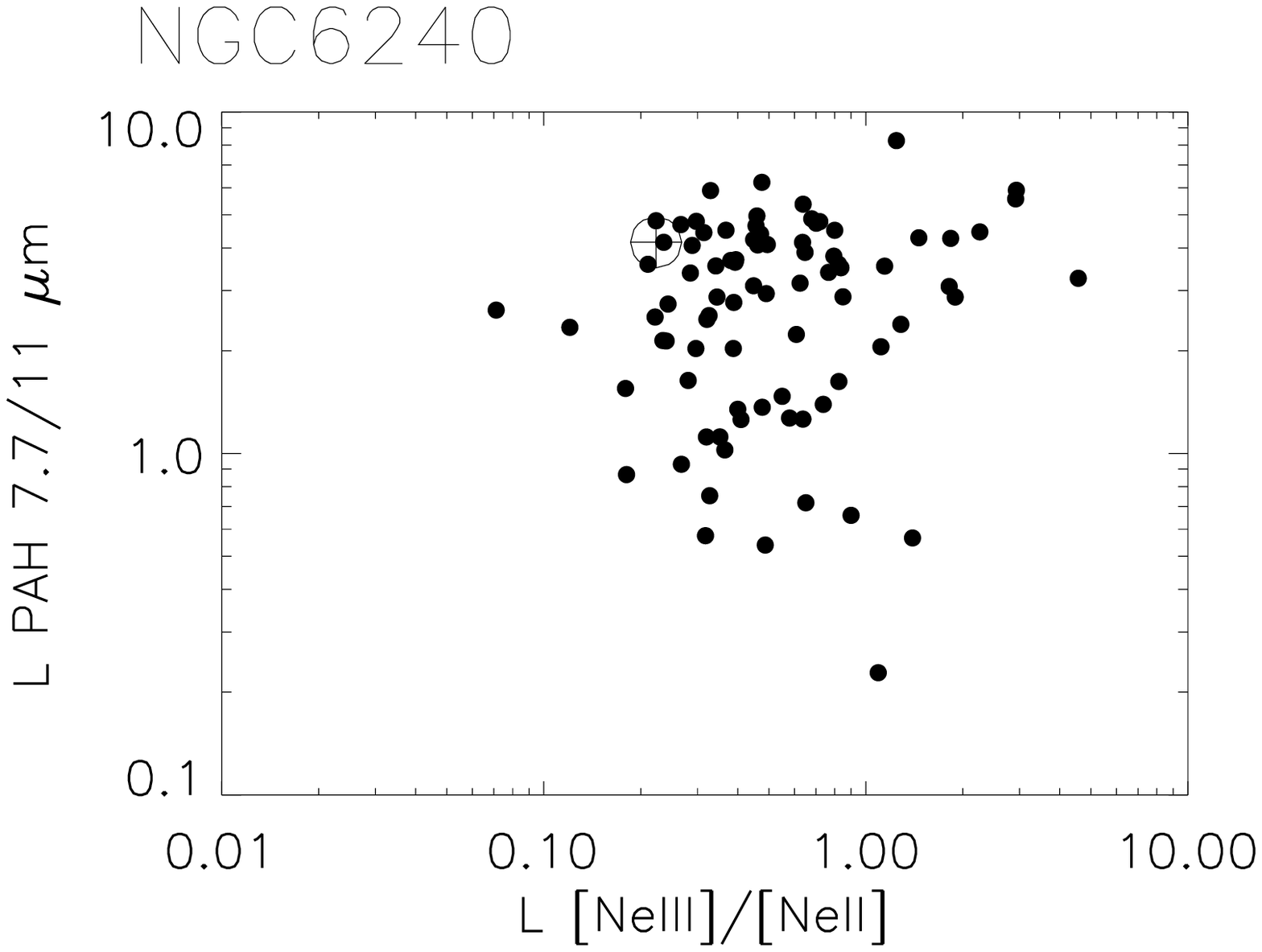}
\end{center}
\caption{\footnotesize{\textbf{Left:} Variation of interband strength ratios of the PAH features (x-axis: PAH 6.2/7.7~$\mu$m, y-axis:PAH 11.3/7~.7$\mu$m) over $3\times3$ pixels ($5.6\times5.6\arcsec$, corresponding to the aperture) within the merger system (see Fig.~\ref{maps_PAH6/7} and  Fig.~\ref{maps_PAH7/11}). The model by \cite{Dra01} for neutral and ionized PAHs (from a few tens to thousands of carbon atoms as marked in the top left panel) is overlaid as solid and dashed line, respectively. The nuclei of the galaxies (central aperture) are indicated as a circle with plus marker. \textbf{Right:} Variation of interband strength ratios of the PAH features at 7.7~$\mu$m relative to the 11.3~$\mu$m band with the line ratio [Ne III]/[Ne II].}}
\label{PAH_scatter}
\end{figure*}
\begin{figure*}
\begin{center}
\includegraphics[scale=0.39]{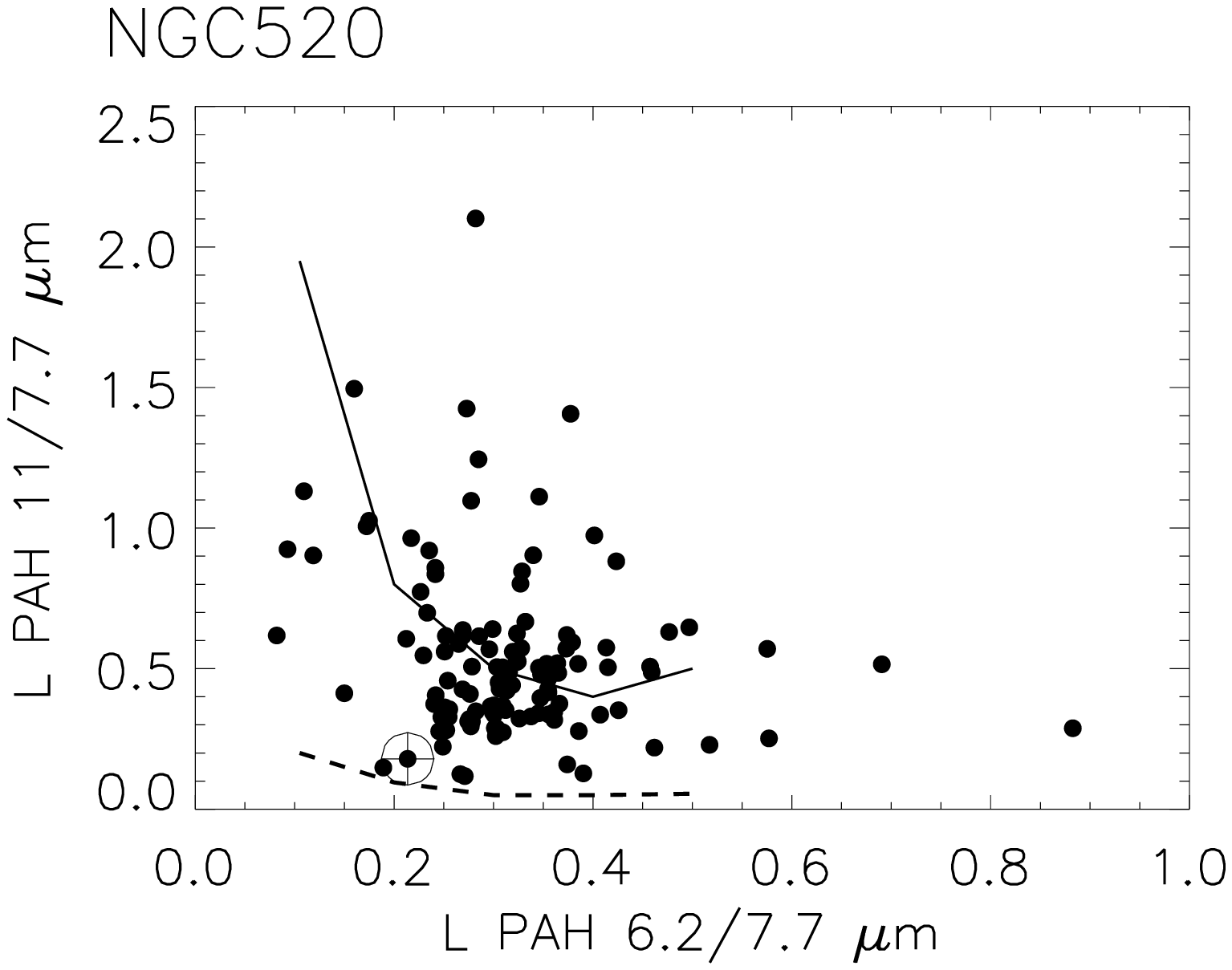}
\includegraphics[scale=0.39]{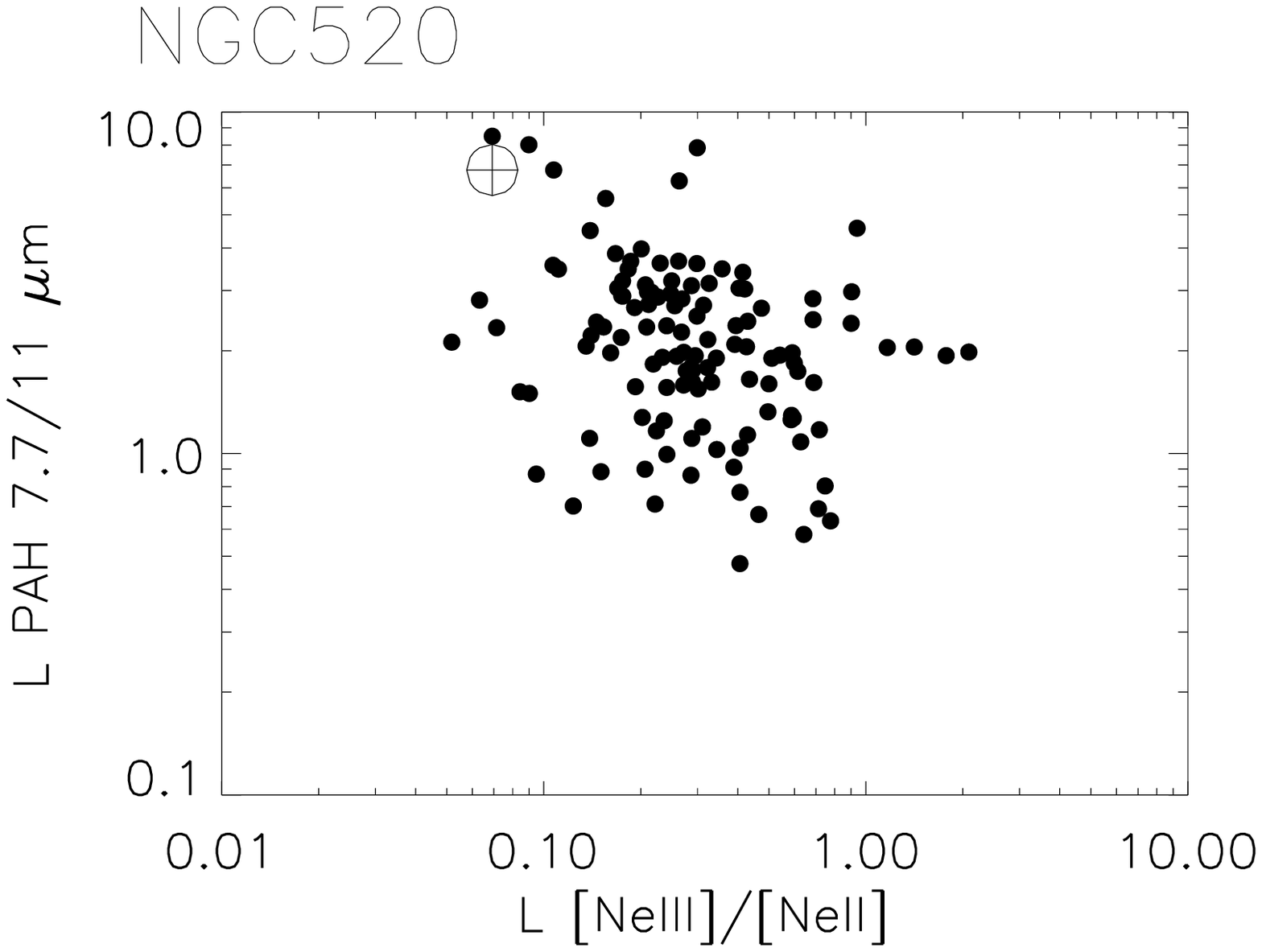}
\includegraphics[scale=0.39]{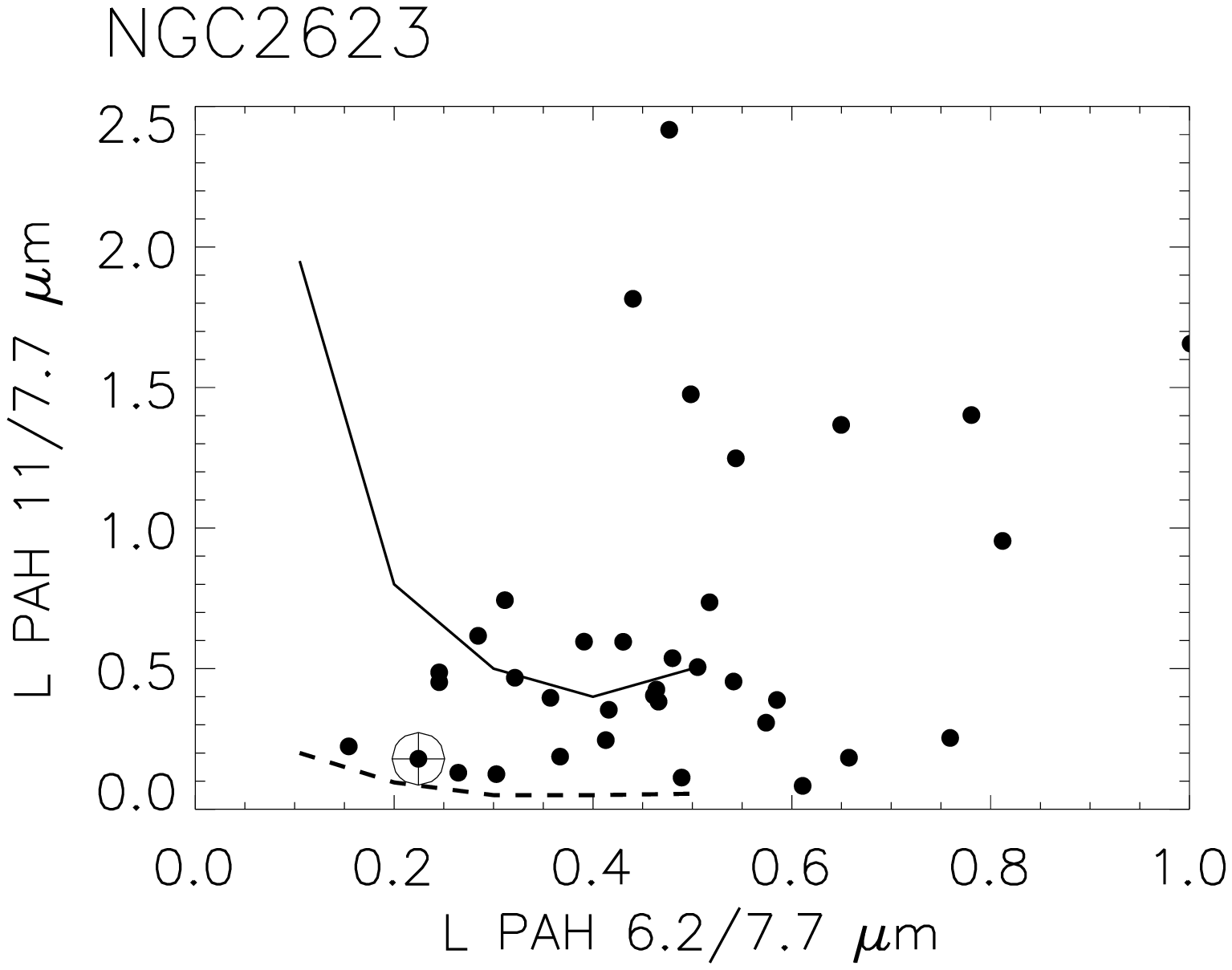}
\includegraphics[scale=0.39]{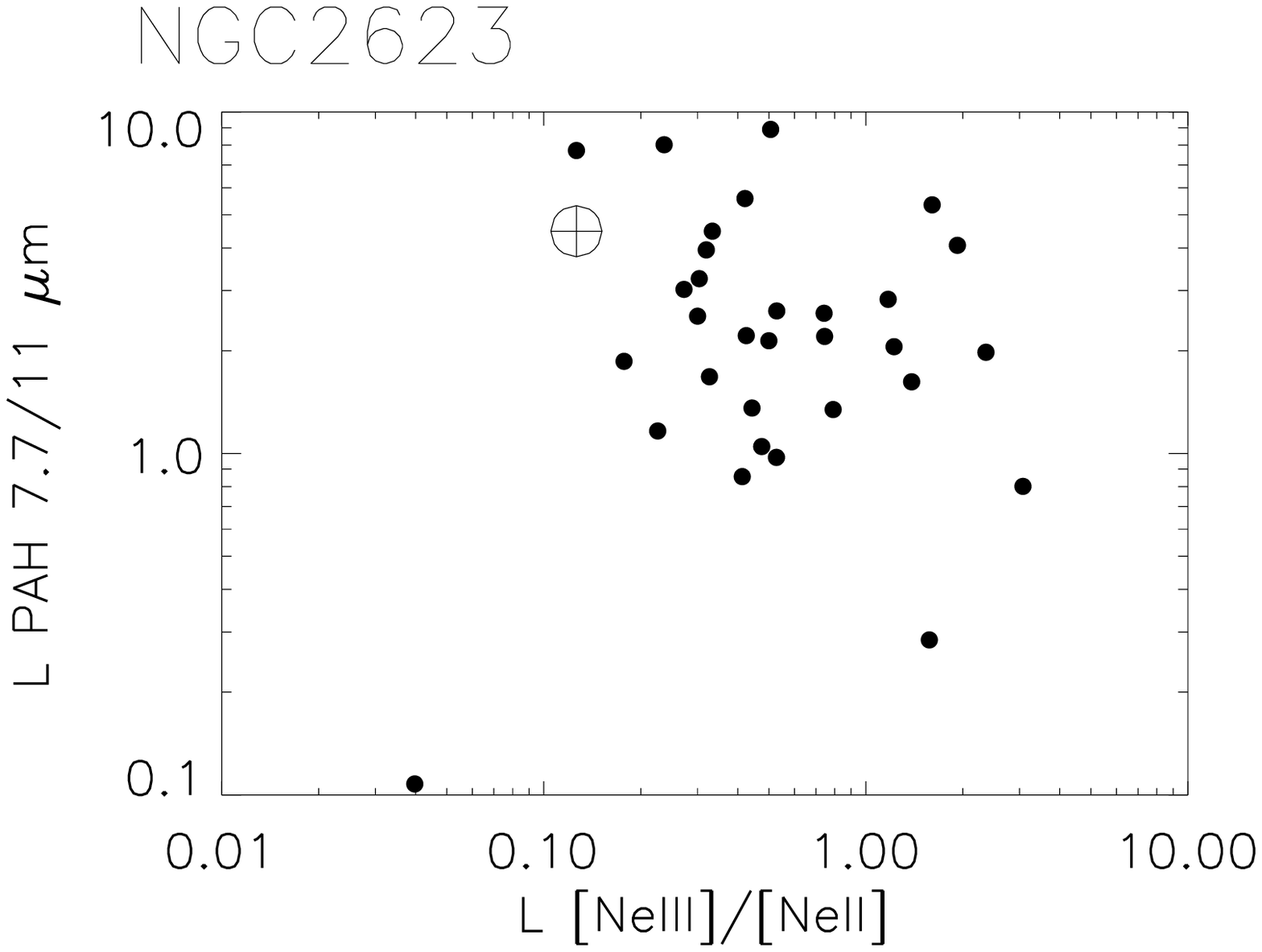}
\includegraphics[scale=0.39]{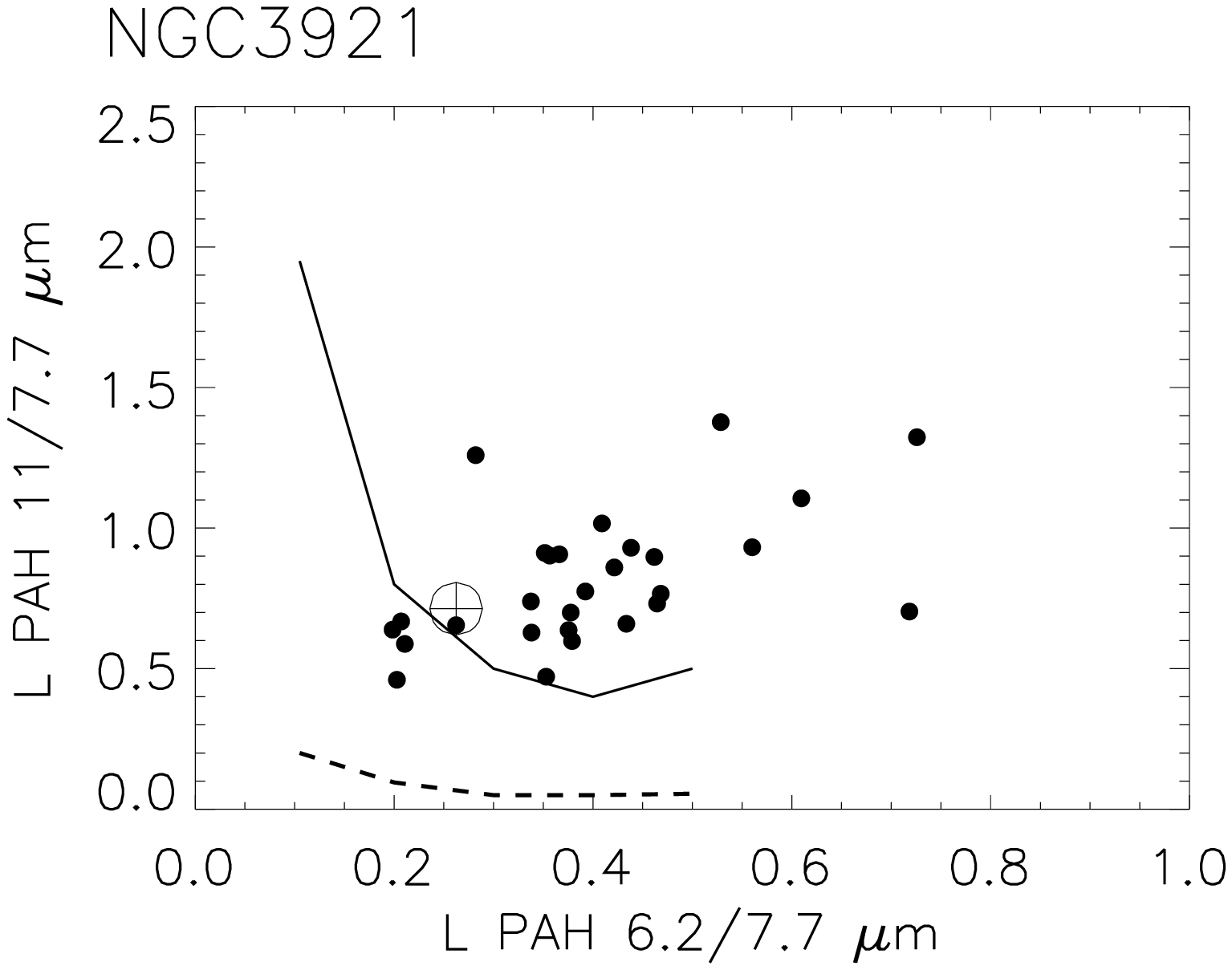}
\includegraphics[scale=0.39]{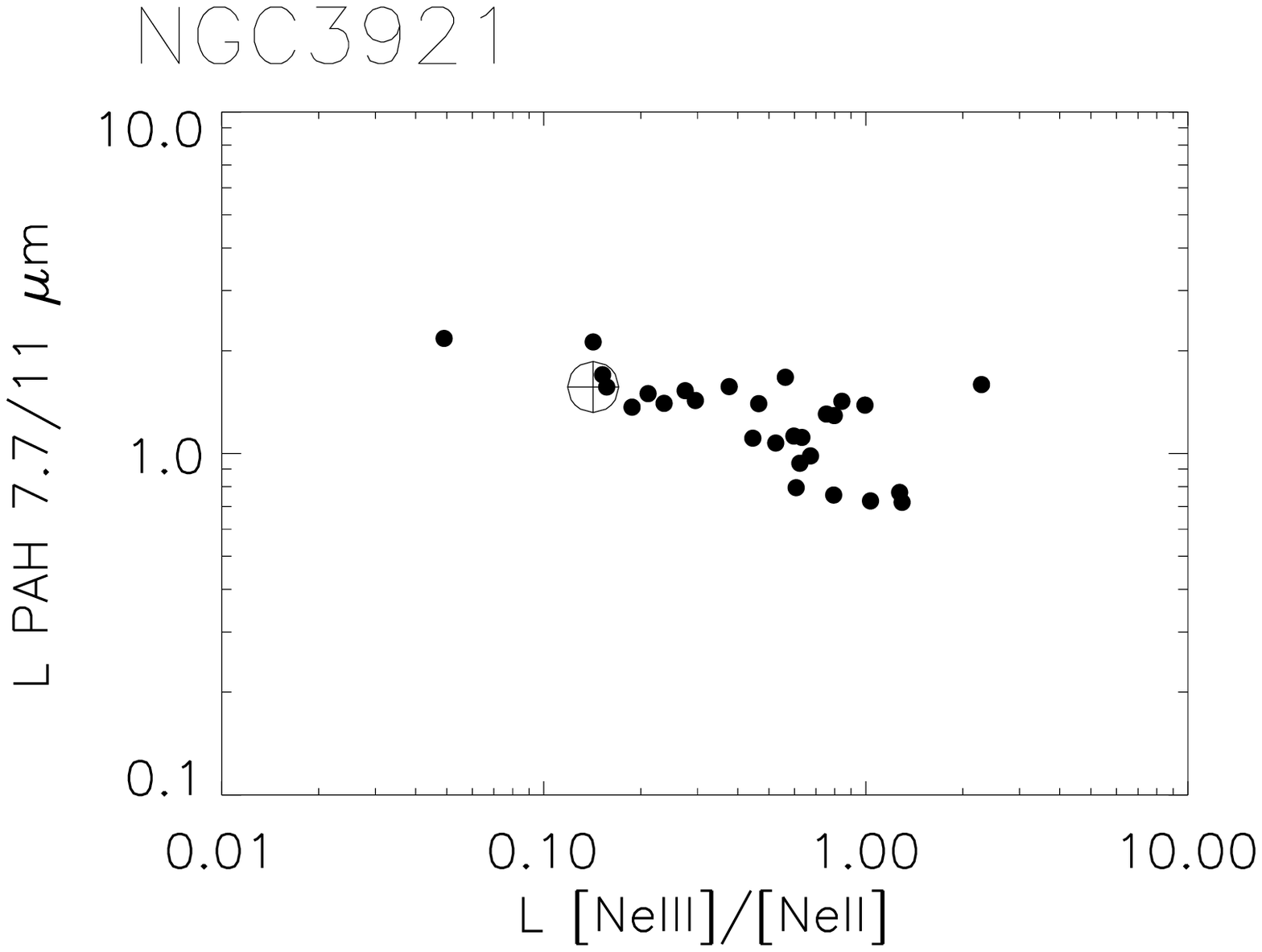}
\includegraphics[scale=0.39]{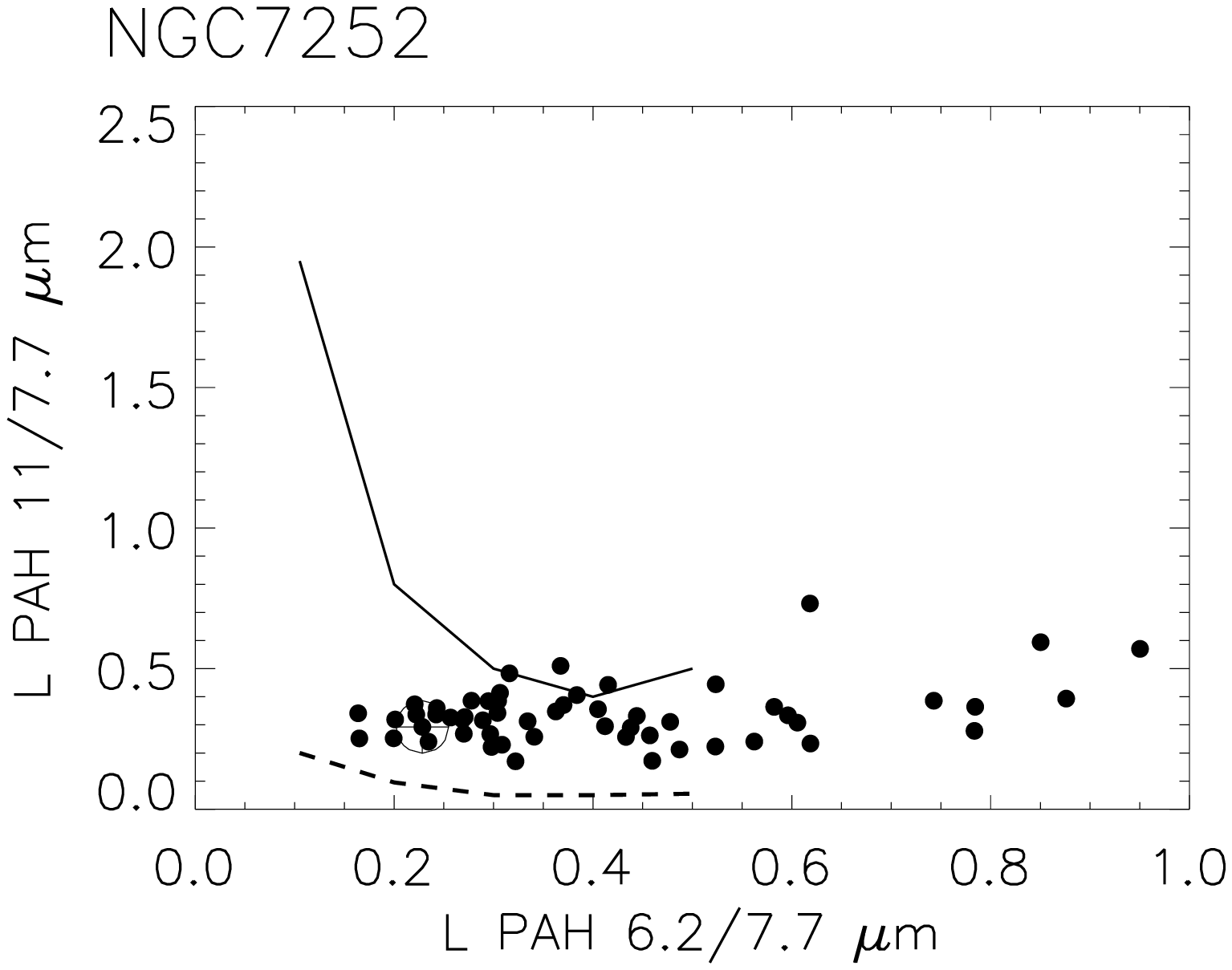}
\includegraphics[scale=0.39]{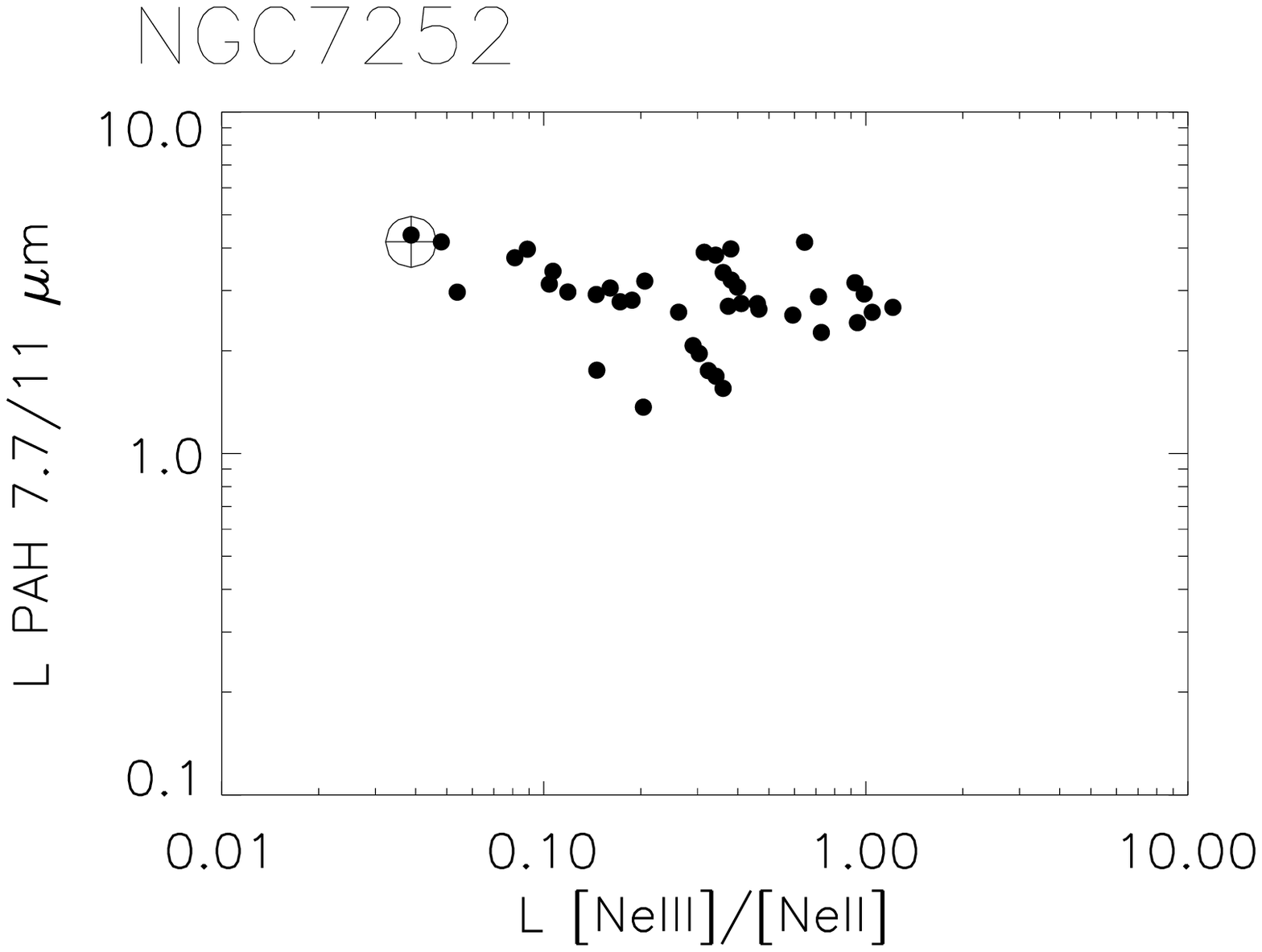}
\end{center}
\figurenum{\ref{PAH_scatter}}
\caption{Continued}
\end{figure*}

\clearpage

 \begin{figure*}
\begin{center}
\includegraphics[scale=0.55]{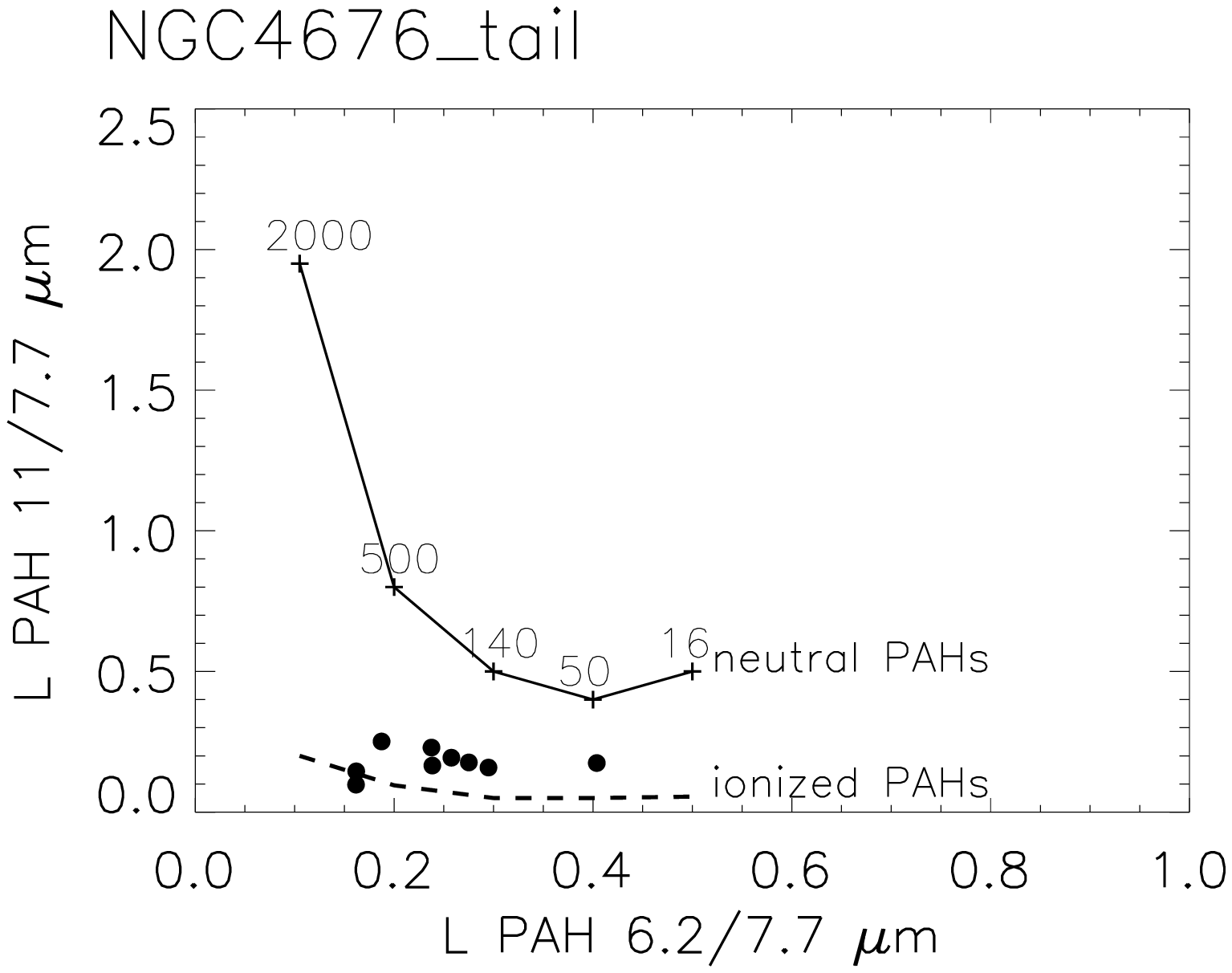}
\end{center}
\caption{\footnotesize{Variation of interband strength ratios of the PAH features in the tidal tail of NGC~4676 indicates emission from ionized PAHs. Same scales, markers and labels as in Fig.~\ref{PAH_scatter}.}}
\label{PAH_scatter_tail}
\end{figure*}

\newpage

\begin{deluxetable}{lccllll}
\tabletypesize{\footnotesize}
\tablecaption{Sample Overview}
\tablehead{\colhead{Name}  & \colhead{RA} & \colhead{DEC} & \colhead{D} &  \colhead{log($L_{IR}/L_{\odot}$)} &  \colhead{T.S.} & \colhead{AGN type} \\
       & (J2000) & (J2000) & [Mpc] & log$_{10}$ [L$_\sun$]  &  & }
\startdata
NGC~520  & 01:24:35.1 & +03:47:33 & 27.2 & 11.91$^1$ & 5 & HII \\
NGC~2623 & 08:38:24.1 & +25:45:17 & 80.6 & 11.54$^1$ & 6 & LINER Sy2 \\
NGC~3921 & 11:51:06.9 & +55:04:43 & 84.5 & $<$10.35$^2$ & 7 &  HII \\
NGC~4676 & 12:46:10.7 & +30:43:38 & 96.0 & 10.88$^3$ & 1 & - \\
NGC~6240 & 16:52:58.9 & +02:24:03 & 103.0 & 11.85$^1$ & 4 & LINER Sy2 \\
NGC~6621 & 18:12:55.3 & +68:21:48 & 85.2 & 11.23$^1$ & 3 & HII \\
NGC~7252 & 22:20:44.7 & -24:40:42 & 62.2 & 10.75$^3$ & 8 & - \\
NGC~7592 & 23:18:22.2 & -04:25:01 & 97.2 & 11.33$^1$ & 2 & - \\
\enddata
\tablecomments{\footnotesize{Summary of the properties of our sample of merger systems. Column (1): Source Name from NASA IPAC Extragalactic Database (NED), Column (2): right ascension from NED (J2000), Column (3): declination from NED (J2000), Column (4): The luminosity distance in Mpc (adopting $H_0=70$~km~s$^{-1}$~Mpc), as provided by NED, Column (5): The total infrared luminosity in log$_{10}$ Solar units (1: \cite{San03}, 2: \cite{Chi02}, 3: \cite{Bra06}), Column (6): The merger stage of the  Toome Sequence (from 1--8), Column (7): The AGN type derived from the optical and UV spectra, as recorded in NED.}}
\label{tab_sample}
\end{deluxetable}


\begin{deluxetable}{lccrrrrrrr}
\tablewidth{8.5in}
\tabletypesize{\footnotesize}
\setlength{\tabcolsep}{0.1in}
\tablecaption{Mid-IR Emission Line and PAH Features}
\tablehead{\colhead{Name}  & \colhead{RA} & \colhead{DEC} &\colhead{Region Size}& \colhead{PAH 6.2$\mu$m} & \colhead{PAH 7.7$\mu$m} & \colhead{PAH 11.3$\mu$m} & \colhead{PAH 17$\mu$m} & \colhead{[Ne II] 12.81$\mu$m} & \colhead{[Ne III] 15.55$\mu$m}\\
    & (J2000) & (J2000) & 10$^{-9}$sterad &  &  & & & & }
\startdata
NGC~520 System &  01:24:35.10  &  +03:47:33.0  & 75.5 & 389.5 $\pm$ 2.4 & 1427.7 $\pm$ 7.1 & 304.9 $\pm$ 1.5 & 207.3 $\pm$ 1.3 & 34.5 $\pm$ 0.2 & 6.1 $\pm$ 0.1\\
NGC~520 North  &  01:24:33.36  &  +03:48:02.5 & 8.8 & 90.8 $\pm$ 4.8 & 264.5 $\pm$ 14.2 & 77.2 $\pm$ 1.4 & 59.3 $\pm$ 2.0 & 4.1 $\pm$ 0.3 & 2.2 $\pm$ 0.2\\
NGC~520 South  &  01:24:34.90  &  +03:47:29.5  & 6.0 & 4171.2 $\pm$ 10.0 & 16167.2 $\pm$ 28.0 & 2677.9 $\pm$ 5.8 & 1670.1 $\pm$ 6.8 & 407.8 $\pm$ 0.7 & 58.2 $\pm$ 0.6\\
NGC~2623  &  08:38:24.11  &  +25:45:16.4 & 33.5 & 145.3 $\pm$ 2.5 & 493.8 $\pm$ 6.4 & 99.9 $\pm$ 1.1 & 95.7 $\pm$ 1.7 & 17.6 $\pm$ 0.2 & 4.8 $\pm$ 0.2\\
NGC~3921  &  11:51:06.86  &  +55:04:43.6 & 46.4 & 16.9 $\pm$ 1.4 & 48.4 $\pm$ 4.2 & 31.7 $\pm$ 1.5 & 18.1 $\pm$ 1.8 & 1.7 $\pm$ 0.1 & 0.5 $\pm$ 0.2\\
NGC~4676 System  &  12:46:10.70  &  +30:43:38.0  & 69.8 & 45.9 $\pm$ 1.4 & 187.5 $\pm$ 5.0 & 41.5 $\pm$ 1.5 & 23.6 $\pm$ 2.1 & 5.2 $\pm$ 0.1 & 1.0 $\pm$ 0.2\\
NGC~4676 North  &  12:46:10.09  &  +30:43:55.1 & 11.2 & 216.0 $\pm$ 3.5 & 830.9 $\pm$ 7.7 & 185.3 $\pm$ 3.0 & 104.1 $\pm$ 4.9 & 23.7 $\pm$ 0.2 & 3.2 $\pm$ 0.5\\
NGC~4676 South  &  12:46:11.28  &  +30:43:21.5 & 9.7 & 40.2 $\pm$ 3.4 & 136.6 $\pm$ 10.1 & 38.4 $\pm$ 1.0 & 20.7 $\pm$ 2.8 & 3.5 $\pm$ 0.3 & 1.4 $\pm$ 0.4\\
NGC~6240  &  16:52:58.92  &  +02:24:03.2 & 32.3 & 314.0 $\pm$ 2.5 & 1490.8 $\pm$ 6.2 & 369.1 $\pm$ 1.0 & 147.2 $\pm$ 1.4 & 88.9 $\pm$ 0.2 & 44.2 $\pm$ 0.2\\
NGC~6621 System  &  18:12:55.30 &  +68:21:48.0 & 340.0 & 108.3 $\pm$ 1.1 & 164.4 $\pm$ 0.9 & 58.6 $\pm$ 0.2 & 46.8 $\pm$ 0.4 & 8.1 $\pm$ 0.1 & 1.0 $\pm$ 0.1\\
NGC~6621 North  &  18:12:55.49  &  +68:21:47.8 & 34.4 & 630.6 $\pm$ 2.8 & 2380.7 $\pm$ 8.5 & 535.6 $\pm$ 1.0 & 368.8 $\pm$ 1.6 & 92.0 $\pm$ 0.3 & 7.7 $\pm$ 0.2\\
NGC~6621 South  &  18:12:59.78  &  +68:21:14.8 & 51.5 &  & 38.0 $\pm$ 1.8 & 19.9 $\pm$ 0.5 & 9.0 $\pm$ 1.4 & 0.8 $\pm$ 0.1 & 0.4 $\pm$ 0.2\\
NGC~6621 Middle  &  18:12:58.58  &  +68:21:28.9 & 10.5 & 135.1 $\pm$ 7.0 & 142.9 $\pm$ 3.4 & 55.9 $\pm$ 1.8 & 34.3 $\pm$ 2.3 & 4.5 $\pm$ 0.2 & 2.2 $\pm$ 0.3\\
NGC~7252  &  22:20:44.77  &  -24:40:41.7 & 34.2 & 188.6 $\pm$ 2.6 & 660.6 $\pm$ 8.8 & 191.8 $\pm$ 3.1 & 92.5 $\pm$ 2.9 & 18.4 $\pm$ 0.2 & 1.4 $\pm$ 0.2\\
NGC~7592 System  &  23:18:22.20  &  -04:25:01.0 & 27.6 & 321.5 $\pm$ 3.4 & 1146.4 $\pm$ 13.0 & 250.9 $\pm$ 1.9 & 127.3 $\pm$ 2.0 & 35.8 $\pm$ 0.2 & 12.6 $\pm$ 0.2\\
NGC~7592 West  &  23:18:21.84  &  -04:24:57.1 & 5.3 & 350.8 $\pm$ 6.5 & 1422.0 $\pm$ 14.9 & 297.4 $\pm$ 2.3 & 171.1 $\pm$ 3.4 & 53.3 $\pm$ 0.4 & 13.6 $\pm$ 0.5\\
NGC~7592 East  &  23:18:22.64  &  -04:24:58.1 & 5.3 & 726.6 $\pm$ 6.8 & 2446.7 $\pm$ 18.6 & 508.8 $\pm$ 4.6 & 224.3 $\pm$ 4.1 & 67.2 $\pm$ 0.6 & 23.9 $\pm$ 0.5\\
NGC~7592 South  &  23:18:22.16  &  -04:25:08.6 & 1.4 & 36.5 $\pm$ 2.5 & 132.3 $\pm$ 6.2 & 29.5 $\pm$ 1.3 & 23.5 $\pm$ 4.0 & 4.2 $\pm$ 0.2 & 3.4 $\pm$ 0.9\\
\enddata
\tablecomments{Overview of the integrated PAH and Mid-IR feature strengths for all merger components. Units are in 10$^{-9}$~W~m$^{-2}$~sr$^{-1}$.}
\label{tab_PAH}
\end{deluxetable}

\begin{deluxetable}{lrrrrrrrr}
\tablewidth{6.5in}
\tabletypesize{\scriptsize}
\setlength{\tabcolsep}{0.06in}
\tablecaption{H$_2$ Fluxes}
\tablehead{\colhead{Name}  & \colhead{H$_2$ S(0)} & \colhead{H$_2$ S(1)} & \colhead{H$_2$ S(2)} & \colhead{H$_2$ S(3)} & \colhead{H$_2$ S(4)} & \colhead{H$_2$ S(5)} & \colhead{H$_2$ S(6)} & \colhead{H$_2$ S(7)}\\
& 28.22$\mu$m &  17.03$\mu$m & 12.28$\mu$m & 9.66$\mu$m & 8.03$\mu$m & 6.91$\mu$m &  6.11$\mu$m &  5.51$\mu$m}
\startdata
NGC~520 System & 5.4 $\pm$ 0.3 & 11.2 $\pm$ 0.2 & 3.4 $\pm$ 0.2 & 10.9 $\pm$ 0.7 & 4.9 $\pm$ 0.5 &     &     & 2.8 $\pm$ 0.7 \\
NGC~520 North  & 0.7 $\pm$ 0.1 & 1.9 $\pm$ 0.3 & 0.7 $\pm$ 0.4 & 0.7 $\pm$ 0.4 & 2.6 $\pm$ 1.2 &     &     & 1.8 $\pm$ 1.7  \\
NGC~520 South  & 96.5 $\pm$ 4.1 & 86.4 $\pm$ 0.8 & 36.8 $\pm$ 0.6 & 139.4 $\pm$ 3.1 & 35.3 $\pm$ 2.0 & 21.5 $\pm$ 4.1 &     & 3.3 $\pm$ 2.9 \\
NGC~2623  & 1.5 $\pm$ 0.3 & 5.5 $\pm$ 0.2 & 1.1 $\pm$ 0.2 & 6.5 $\pm$ 0.8 & 1.9 $\pm$ 0.6 &     &     & 3.2 $\pm$ 0.9 \\
NGC~3921  & 0.7 $\pm$ 0.1 & 5.4 $\pm$ 0.3 & 1.6 $\pm$ 0.1 & 3.5 $\pm$ 0.5 & 0.7 $\pm$ 0.2 & 1.4 $\pm$ 0.4 & 0.5 $\pm$ 0.4 & 0.2 $\pm$ 0.4 \\
NGC~4676 System  & 0.5 $\pm$ 0.1 & 2.9 $\pm$ 0.3 & 1.1 $\pm$ 0.1 & 1.7 $\pm$ 0.2 & 0.8 $\pm$ 0.3 & 1.2 $\pm$ 0.4 &     & 0.6 $\pm$ 0.4 \\
NGC~4676 North  & 0.3 $\pm$ 0.2 & 5.1 $\pm$ 0.6 & 2.5 $\pm$ 0.2 & 4.1 $\pm$ 0.4 & 2.6 $\pm$ 0.5 & 2.9 $\pm$ 1.0 &     &   \\     
NGC~4676 South  & 1.6 $\pm$ 0.2 & 9.6 $\pm$ 0.5 & 3.6 $\pm$ 0.3 & 6.7 $\pm$ 0.4 & 1.6 $\pm$ 0.8 & 4.3 $\pm$ 1.2 &     & 2.2 $\pm$ 1.0\\
NGC~6240  & 16.9 $\pm$ 0.8 & 87.7 $\pm$ 0.3 & 25.1 $\pm$ 0.2 & 103.3 $\pm$ 0.6 & 36.0 $\pm$ 0.6 & 63.8 $\pm$ 0.8 & 12.3 $\pm$ 0.7 & 28.4 $\pm$ 0.8 \\
NGC~6621 System  & 0.5 $\pm$ 0.0 & 2.0 $\pm$ 0.1 & 0.9 $\pm$ 0.1 & 1.2 $\pm$ 0.1 &     &     &     & 0.9 $\pm$ 0.4\\
NGC~6621 North  & 2.0 $\pm$ 0.1 & 5.8 $\pm$ 0.3 & 8.8 $\pm$ 0.3 & 7.6 $\pm$ 0.4 & 6.7 $\pm$ 0.9 & 17.2 $\pm$ 1.4 &     & \\           
NGC~6621 South  & 0.9 $\pm$ 0.1 & 3.5 $\pm$ 0.3 & 0.3 $\pm$ 0.2 & 2.7 $\pm$ 0.2 & 1.3 $\pm$ 0.5 & 0.6 $\pm$ 0.2 &     & \\          
NGC~6621 Mid  & 0.7 $\pm$ 0.2 & 2.3 $\pm$ 0.4 & 0.6 $\pm$ 0.2 & 0.4 $\pm$ 0.3 &     &     &     & 5.0 $\pm$ 2.4 \\
NGC~7252  & 0.4 $\pm$ 0.2 & 4.8 $\pm$ 0.3 & 2.0 $\pm$ 0.2 & 3.7 $\pm$ 0.4 & 1.0 $\pm$ 0.6 & 0.9 $\pm$ 0.8 &     & \\         
NGC~7592 System  & 1.8 $\pm$ 0.1 & 7.4 $\pm$ 0.2 & 3.8 $\pm$ 0.2 & 7.4 $\pm$ 0.5 & 1.8 $\pm$ 0.8 & 2.6 $\pm$ 1.1 &     &  \\        
NGC~7592 West  & 2.4 $\pm$ 0.2 & 14.1 $\pm$ 0.4 & 9.1 $\pm$ 0.4 & 13.0 $\pm$ 1.0 & 6.8 $\pm$ 1.3 & 9.6 $\pm$ 2.0 &     & 1.3 $\pm$ 1.6 \\
NGC~7592 East  & 1.6 $\pm$ 0.2 & 7.3 $\pm$ 0.4 & 4.3 $\pm$ 0.5 & 11.5 $\pm$ 1.1 &     & 6.5 $\pm$ 2.7 &     &    \\         
NGC~7592 South  & 1.8 $\pm$ 0.3 & 2.1 $\pm$ 0.7 & 0.2 $\pm$ 0.2 & 0.2 $\pm$ 0.4 & 0.4 $\pm$ 0.7 &   0.7 &     & \\
\enddata
\tablecomments{Overview of the integrated H$_2$ line strengths of the H$_2$ components for all merger components. The center and size of the regions over which is integrated are the same as in Tab.~\ref{tab_PAH}. All units are in 10$^{-9}$~W~m$^{-2}$~sr$^{-1}$.}
\label{tab_H2}
\end{deluxetable}

\begin{deluxetable}{lrrrr}
\tabletypesize{\scriptsize}
\tablewidth{6.5in}
\setlength{\tabcolsep}{0.02in}
\tablecaption{H$_2$ Temperatures and Masses}
\tablehead{\colhead{Name}  &  \colhead{T$_{warm}$} & \colhead{T$_{hot}$} & \colhead{H$_2$ Mass warm} & \colhead{H$_2$ Mass hot}\\
 &  [K] & [K] & [10$^6$ M$_{\sun}$] & [10$^6$ M$_{\sun}$] }
\startdata
NGC~520 System&221.4$\pm$ 10.5 & 877.1$\pm$ 42.3 & 16.0$\pm$ 0.2 & 1.2$\pm$ 0.3\\
NGC~520 North  &246.1$\pm$ 93.7 & & 0.2$\pm$ 0.1 &\\
NGC~520 South & 87.7$\pm$ 0.2 & 550.1$\pm$ 7.0 & & 10.3$\pm$ 8.9\\
NGC~2623 & 162.6$\pm$ 38.8 & 969.7$\pm$ 67.8 & 118.3$\pm$ 5.1 & 2.6$\pm$ 0.7\\
NGC~3921 & 241.7$\pm$ 17.4 & 664.5$\pm$ 82.5 & 33.4$\pm$ 1.9 & 5.1$\pm$ 10.0\\
NGC~4676 System& 286.5$\pm$ 18.7 & 1041.8$\pm$ 188.0 & 21.0$\pm$ 1.9 & 1.0$\pm$ 0.6\\
NGC~4676 North& 309.4$\pm$ 27.5 & 1091.8$\pm$ 366.2 & 4.9$\pm$ 0.6 & \\          
NGC~4676 South & 291.7$\pm$ 17.0 & 990.2$\pm$ 136.7 & 9.2$\pm$ 0.5 & 0.7$\pm$ 0.3\\
NGC~6240 & 184.9$\pm$ 1.9 & 746.1$\pm$ 3.2 & 1608.1$\pm$ 6.2 & 279.5$\pm$ 8.0\\
NGC~6621 System& 316.8$\pm$ 16.8 & & 43.8$\pm$ 1.8 &\\
NGC~6621 North & 380.8$\pm$ 6.0 &      & 8.6$\pm$ 0.4 &\\            
NGC~6621 Mid & 252.8$\pm$ 27.6 &  & 2.9$\pm$ 0.5 &\\
NGC~7252 & 296.1$\pm$ 37.0 & 663.7$\pm$ 240.8 & 6.5$\pm$ 0.4 &\\          
NGC~7592 System& 333.8$\pm$ 26.2 & 847.3$\pm$ 318.4 & 14.9$\pm$ 0.5 & \\           
NGC~7592 West& 348.1$\pm$ 14.0 & 1078.9$\pm$ 213.3 & 5.0$\pm$ 0.1 & 0.1$\pm$ 0.2\\
NGC~7592 East &339.4$\pm$ 66.3 & 838.2$\pm$ 305.8 & 2.7$\pm$ 0.2 &\\        
\enddata
\tablecomments{Temperatures and masses of the H$_2$ emission derived by fitting the excitation diagram with a two-component model.}
\label{tab_temp}
\end{deluxetable}

\clearpage

\end{document}